\documentclass[pdflatex,sn-mathphys-num]{sn-jnl}

\usepackage{graphicx}%
\usepackage{multirow}%
\usepackage{amsmath,amssymb,amsfonts}%
\usepackage{amsthm}%
\usepackage{mathrsfs}%
\usepackage[title]{appendix}%
\usepackage{xcolor}%
\usepackage{textcomp}%
\usepackage{manyfoot}%
\usepackage{booktabs}%
\usepackage{algorithm}%
\usepackage{algorithmicx}%
\usepackage{algpseudocode}%
\usepackage{listings}%
\usepackage{blindtext}%
\usepackage{hyperref}%
\usepackage{subcaption}%
\usepackage{caption}
\usepackage{adjustbox}
\usepackage{subcaption}
\usepackage{natbib}
\usepackage{bm}
\usepackage{booktabs}
\usepackage{tabularx}
\usepackage{float}
\usepackage{rotating}
\usepackage{listings}
\usepackage{xcolor}
\theoremstyle{thmstyleone}%
\theoremstyle{thmstyletwo}%

\theoremstyle{thmstylethree}%

\begin{document}

\title[Is attention truly all we need? An empirical study of asset pricing in pre-trained RNN sparse and global attention models]{Is attention truly all we need? An empirical study of asset pricing in pre-trained RNN sparse and global attention models}


\author*[1]{\fnm{Shanyan} \sur{Lai}}\email{shanyan.lai@york.ac.uk, annieyanyan125@gmail.com}



\affil*[1]{\orgdiv{Department of Economics and Related Studies}, \orgname{Univiersity of York}, \orgaddress{\street{Heslington}, \city{York}, \postcode{YO10 5DD}, \country{UK}}}




\abstract{This study investigates the pre-trained RNN attention models with the mainstream attention mechanisms such as additive attention, Luong's three attentions, global self-attention and sliding window sparse attention  for the empirical asset pricing research on top 420 large-cap US stocks. This is the first paper on the large-scale state-of-the-art (SOTA) attention mechanisms applied in the asset pricing context. They overcome the limitations of the traditional machine learning-based asset pricing, such as mis-capturing the temporal dependency and short memory. Moreover, the enforced causal masks in the attention mechanisms address the future data leaking issue ignored by the more advanced attention-based models, such as the classic Transformer. The proposed attention models also consider the temporal sparsity characteristic of asset pricing data and mitigate potential overfitting issues by deploying the simplified model structures. This provides some insights for future empirical economic research. All models are examined in three periods, which cover pre-COVID-19, COVID-19 and one year post-COVID-19, for testing the stability of these models under extreme market conditions. The study finds that in value-weighted portfolio back testing, the global self-attention model and the sliding window sparse attention model exhibit excellent capabilities in deriving the absolute returns and hedging downside risks, while they achieve an annualized Sortino ratio of 2.0 and 1.80 respectively in the period with COVID-19 in the static transaction cost scenario. When considering the turnover rate and large-cap adjusted transaction cost rate, the proposed models still achieve the highest Sortino ration in the COVID-19–Inclusive Period, which is 1.65 and 1.51 respectively. Moreover, the sliding window sparse attention model performs more stably than the global self-attention model from the perspective of absolute portfolio returns with respect to the size of stocks' market capitalization.}




\maketitle

\vspace{0.5em}
\noindent
\textbf{Current Version:} June 2, 2026\\
Following insightful comments from Dr Mark Hallam and Dr Chrysovalantis Vaslakis, this version introduces several improvements. First, I correct an error in the out-of-sample $R^2$ calculation caused by the `r2\_score()' function in scikit-learn, which incorrectly aligns predicted returns when computing the denominator. This led to a systematic underestimation of the average OOS $R^2$. I address this by manually computing the mean squared error (MSE) and OOS $R^2$. The denominator now correctly uses the historical mean based on both training and validation samples, consistent with standard asset pricing literature (e.g., Campbell and Thompson, 2008). The updated code is provided in the Appendix. Second, I incorporate a heteroskedasticity and autocorrelation consistent (HAC) estimator into the Diebold-Mariano (DM) test to account for serial correlation in the difference between the two models’ forecast errors. Third, I conduct factor importance analysis and a dynamic transaction-cost robustness examination on the portfolio-wise backtesting. Finally, I extend the abbreviation of the models to improve the readability. `B-Additive', `L-Concat', `L-DotProd' and `L-General' for previous `Batt', `LC', `LD' and `LG', respectively. I am grateful to Dr Mark Hallam and Dr Chrysovalantis Vaslakis for their valuable feedback. All remaining errors are my own.

\vspace{1em}

\section{Introduction}\label{sec:intro_ch2}
\citet{lai2025multilayer} proposed the dynamic MLP models for reducing the dimension of `factor zoo' \citep{Harvey2016Returns} of the asset pricing factor model, which is inspired by the work of \citet{Gu2020EmpiricalLearning} (GKX2020) and \citet{Coqueret2020MachineVersion}. It employs a dynamic structural algorithm from \citet{Coqueret2020MachineVersion} to generate a hidden-layer structure for the proposed MLP models, replacing the fixed pyramidal MLP structure of GKX2020's MLP models. It also substitutes the transformed firm characteristic factors of GKX2020's work to the firm characteristic-sorted portfolio \citep{AndrewY.Zimmermann2020OpenPricing} factors as the input factors to increase the interpretability of the asset pricing factor model in machine learning (ML) methods. The out-of-sample (OOS) fitness of the proposed MLP model with 2 hidden layers significantly outperforms traditional statistical models and GKX2020's benchmarks in both the Pre–COVID-19 Period (1911) and COVID-19–Inclusive Period (2112). Moreover, it provides a significant advantage in mitigating downside risks caused by extreme market fluctuations (e.g. COVID-19). \\

However, as discussed in \citet{zhou2024learning}, although early ML models present excellent capabilities in modelling non-linear structure in financial and economic time series data, fewer authors notice that early ML models, such as MLP, random forest(RF), Support Vector Machine (SVM) and MLP autoencoder, are not originally designed for time series data. They only assume that these models naturally capture the temporal dependencies in the data, or that the data follow independent and identically distributed (i.i.d.) assumptions, or attempt to enforce a temporal sequence via extending or rolling window methods. However, extending or rolling window methods face a dilemma for such tasks: the larger window size has a lower capability to capture the temporal dependency since the temporal dependency inside the window cannot be recognized; the smaller window size has a higher capability to detect the temporal dependency, but increases the possibility of model overfitting due to its small training data size. Therefore, exploring alternative ML structures originally developed the temporal mechanism, such as the recurrent neural networks, RNN\citep{Rumelhart1986LearningErrors}, LSTM\citep{hochreiter1997long}, GRU\citep{Cho2014LearningTranslation} as examples, which relax the aforementioned assumption, could be meaningful for ML-based asset pricing research. Moreover, GKX2020's work extends the dimension of the factors by using the Kronecker product of firm characteristics and macroeconomic indicators in the cross-sectional study and increases the data size for their ML models. This does not assist the `factor zoo' issue but brings their research into serious doubt of pure data-mining exercises. Thus, simplifying the factor configuration and improving model financial and economic interpretability in ML asset pricing research could be another meaningful exercise, even if the study in \citet{lai2025multilayer} has already highly reduced the dimension of factors. Additionally, apart from capturing the sequential information of time series data, how far the model can capture the historical information away from the current time step of a time series, which is known as the `memory' of the model, is an alternative crucial aspect for financial and economic time series modelling. Thus, searching for mechanisms that can extend the memory of recurrent neural network models, which could further improve the model performance without sacrificing the model simplicity for computational efficiency and model interpretability, can be meaningful as well. Furthermore, GKX2020, as the seminal work for the ML-based asset pricing, focuses on the predictability of the ML models instead of evaluating pricing mechanisms, for example, how to explain the pricing error $\alpha$, how to improve the OOS model fitness ($R^2$), and how to ensure the consistency of pricing performance over time. \\

Building on the findings of \citet{lai2025multilayer}, this study introduces a pre-trained recurrent neural network (RNN) architecture enhanced with self-attention mechanisms for addressing the aforementioned problems. Specifically, this study proposed two pre-trained RNN attention models: the RNN global self-attention model (self\_att) and the RNN sliding window sparse attention model (sparse\_att). In the proposed structures, the MLP autoencoder is deployed as the pre-trained method for the factors' dimensional reduction task and factor information extraction, which further compresses the input factors' dimension into 70\% of the original factors' dimension and reconstructs the missing values of the original factors. This process is also known as feature extraction or feature engineering in ML methods. The RNN structure \citep{elman1990finding} was developed to overcome the limitation of the MLP \citep{Rumelhart1986LearningErrors} models, which are the MLP models that process time series data with the sequential assumption. Unlike feedforward network models such as MLP, RNNs develop a hidden state that is updated recurrently over time, which enables the network to process input sequences. Concretely, the hidden state at each time step is calculated based on both the information of the current time step and the previous hidden state. This allows RNNs to learn patterns across time. Nevertheless, the vanilla RNN model does not overcome the issues of gradient vanishing and explosion existing in feedforward network models, especially when it is estimated by the stochastic gradient descent method (SGD). Although such issues can be mitigated through multiple approaches, such as adding L1 regularization to the loss function or using the Adam optimizer with the SGD method, the gradient vanishing issue remains more challenging. It causes the short `memory' of the model, which prevents the RNN model from effectively capturing long-term dependencies in the sequence.\\

Fortunately, LSTM \citep{hochreiter1997long} and GRU \citep{Cho2014LearningTranslation} as RNN variation models extend the memory of the classic RNN to a degree in a general perspective. However, since the temporal sparsity characteristic of the asset pricing data, which presents as a smaller data size with longer time dependency, the LSTM model and the GRU model's high parameterization nature may cause overfitting. Thus, the most prevailing attention mechanism, the dot product based self-attention mechanism \citep{Vaswani2017AttentionNeed} and its sparse version, the sliding window attention mechanism \citep{beltagy2020longformer}, which are designed for the most prevelent large language models (LLM), Transformers, could be a great substitute of LSTM and GRU for balancing the trade-off between the model parameterization complexity and longer-term temporal dependency capturing. The attention mechanism is a specific structure added to neural network models to reinforce pattern recognition. It is inspired by the human brain's attention mechanism. The biological design of the human brain could preferentially process impulsive information (e.g. visual items with stronger colour and louder sound). The ML attention mechanism is the mechanism that introduces heavier weights on the high-impact information and lighter weights on the low-impact information with respect to temporal correlation. The advanced attention mechanisms are confirmed for their capability to significantly improve the predictive power with relatively low computational costs (smaller parameter size) in domains such as natural language processing (NLP) and machine translation \citep{Vaswani2017AttentionNeed,Luong2015EffectiveTranslation}. Given the similar data lengths across early NLP tasks and asset pricing, attention mechanisms may be a suitable approach to improving empirical asset pricing performance. Furthermore, as the attention mechanism was originally designed for sequential data, it does not naturally align with time series data, since attention weights are computed based on correlations between any two sequential points, not on previous points alone. This causes future data leakage when attention mechanisms are applied without the imposed causal masks. However, little of the attention-based asset pricing literature has addressed this issue, for example, the papers of \citet{zhou2024learning} and \citet{Ma2023AttentionApproach}.\\ 

This study also further investigates whether the classic attention mechanisms generally improve the asset pricing and factor investing performance by constructing the pre-trained RNN additive attention model (B-Additive), the pre-trained RNN Luong's general attention model (L-General), the pre-trained RNN Luong's dot product attention model (L-DotProd) and the pre-trained RNN Luong's concatenation attention model (L-Concat) as benchmark models. These attention mechanisms were originally developed for natural language processing (NLP) tasks, and some of them have been examined in asset pricing tasks, such as \citet{zhou2024learning,kelly2025artificial}. The proposed and benchmark models are examined and backtested in three market regimes to demonstrate their ability to handle extreme market fluctuations, covering the Pre–COVID-19 Period (1911), the COVID-19–Inclusive Period (2112), and the Period Including COVID-19 and One-Year After (2212). To the best of my knowledge, this is the first large-scale study focusing on attention mechanisms in the asset pricing context. The MLP-pre-trained RNN global self-attention model and the RNN sliding-window sparse attention model, as innovative time series tailored architectures, are introduced here and examined for the first time in asset pricing research.\\

Therefore, the contributions of this study can be considered from five angles:
\begin{itemize}
  \item Employed MLP autoencoder as the pre-trained method for handling the missing values in firm characteristic-sorted portfolio factors and the dimensional reduction method for improving computational efficiency and mitigating the potential model overfitting issue. \\
  
  \item Examined the RNN structures on firm characteristic-sorted portfolio pre-trained factors as a solution for capturing the temporal dependency of the data.\\
  
  \item Proposed two RNN attention variation models for asset pricing and factor investing contexts, the pre-trained RNN global self-attention model and the pre-trained RNN sliding window sparse attention model. \\
  
  \item Investigated the necessity of RNN classic attention models, such as additive attention and Luong's three attention models, in the context of asset pricing and factor investing.\\
  
  \item Tested models' capabilities of handling the extreme market conditions during the Pre–COVID-19 Period (1911), the COVID-19–Inclusive Period (2112) and the Period Including COVID-19 and One-Year After (2212).\\
  
  \item Enforced the causal mask on the attention mechanism for time series to prevent the future information leaking problem, which is ignored by recent attention-based research on financial and economic time series data.\\
  
\end{itemize}

The study is organized into 6 sections. Section 2 introduces the related work of this research, while Section 3 describes the data employed in this study. Section 4 depicts the models deployed for this research. Section 5 shows the methods for evaluating the model performance and the back-testing performance, as well as an analysis of the empirical results. The conclusion and further discussion are presented in Section 6. \\

\section{Related work}\label{sec:related_work_ch2}
As mentioned in \citet{lai2025multilayer}, researchers focus on the linear and non-linear statistical functions. Later, the burgeoning of AI algorithms provides some promising solutions for the dimensional disaster challenge of asset pricing research. The seminal work of this domain is \citet{Gu2020EmpiricalLearning}'s `empirical asset pricing in machine learning'. It examines the majority of traditional ML models, such as MLP, SVM, RF, and gradient boosting, on 94 firm characteristics and 8 macroeconomic indicators, cross-sectionally from the U.S. stock market, and shows that MLP with 3 hidden layers outperforms all alternative models on high-dimensional factors. This work is followed by \citet{Wang2021CryptocurrenciesLearning}, which tests \citet{Gu2020EmpiricalLearning}'s models on cryptocurrency pricing. It agrees that the MLP with 3 hidden layers is the best model, even in the cryptocurrency market. It generalized the MLP models' application in asset pricing. Later, \citet{Gu2021AutoencoderModels} proposed a structure based on two MLP autoencoders for asset pricing tasks. One for generating $\beta$s for factors and the other for factors' dimensional reduction. By dot-producting the outputs of the two MLP autoencoders, it achieves better performance than \citet{Gu2020EmpiricalLearning}'s best-performing model. Equation~\eqref{eq:additive_ch2} shows the additive form, and Equation~\eqref{eq:general_ch2} shows the general form of the asset pricing factor model. Both of \citet{Gu2021AutoencoderModels,Gu2020EmpiricalLearning} agree that in ML-based empirical asset pricing research, the concentration is on exploring the function form of $g(f_{i,t}; \theta)$. That is the foundation of later research in this direction.\\
\begin{equation}
r_{i,t+1} = \mathbb{E}_t \left[ r_{i,t+1} \right] + \varepsilon_{i,t+1}
\label{eq:additive_ch2}
\end{equation}

\begin{equation}
r_{i,t+1} = g\left(f_{i,t}; \theta\right) + \varepsilon_{i,t+1}
\label{eq:general_ch2}
\end{equation}\\
\citet{Avramov2021MachinePredictability} conduct a comparative study on the MLP model with three hidden layers of \citet{Gu2020EmpiricalLearning}, the generative adversarial neural network model with RNN cores of \citet{Chen2024DeepPricing}, the instrumental PCA model of \citet{Kelly2019CharacteristicsReturn} and the conditional autoencoder extension of \citet{Gu2021AutoencoderModels} in the U.S. stock market from a portfolio strategy-wise perspective. This is discussed extensively in \citet{lai2025multilayer}. \citet{Bagnara2022AssetReview} conduct a comprehensive and critical survey of recent developments at asset pricing in machine learning (ML), with a particular focus on addressing the long-standing `factor zoo' problem. The review mainly investigates the classic ML methods and classifies the literature into five methodological categories: regularization, dimension reduction, regression trees and random forests, neural networks, and comparative analyses, which are applied within two main frameworks: empirical factor modelling with return prediction and stochastic discount factor (SDF) estimation. Its comparative studies find that boosted trees and neural networks outperform linear benchmarks in predictive accuracy. It identifies persistent challenges, including limited economic interpretability of ML models, the risk of data snooping, and weak integration with causal inference, and calls for hybrid approaches that combine economic theory, robust time series validation, and evaluation based on economically meaningful metrics such as portfolio-wise backtesting performance. Moreover, it aligns with the conclusion of \citet{Avramov2021MachinePredictability} that the economic value of ML-based strategies often lies less in boosting unconditional average returns than in the ability to hedge downside risk and preserve capital during downturns or crisis periods.\\ 

Although traditional ML-based asset pricing literature provides substantial evidence that early-stage ML models have advantages in improving prediction accuracy and mitigating market downside fluctuation risks, they face the problems of the compliance of ML models and the asset pricing context. For example, MLP, SVM and RF were not originally designed for sequential or time series data, but early literature imposes the assumption of sequence recognition when they are applied to sequential data. This may affect the prediction accuracy or increase the probability of model misspecification and overfitting, even if extending or rolling window methods are deployed. Thankfully, later literature on the second generation of ML models, such as RNNs and their variants, addressed this problem due to their sequential nature. Unlike traditional empirical asset pricing models, \citet{Chen2024DeepPricing} propose a novel Stochastic Discount Factor (SDF) estimation approach based on a Generative Adversarial Network (GAN) \citep{goodfellow2014generative}. The SDF is a generalised framework for asset pricing factor models under the non-arbitrage assumption. The linear and non-linear factor models, ML models such as MLP, or more advanced, Transformer models, all can be transformed and fitted in the SDF framework. Specifically, their model employs an MLP autoencoder structure for the cross-sectional dimension reduction of firm characteristics, and a Long Short-Term Memory (LSTM) network to extract economic states from large-scale macroeconomic indicators. Within this framework, the generator formulates a candidate SDF, while the discriminator evaluates it. This adversarial process enforces the no-arbitrage condition by driving the pricing error ($\alpha$) towards zero. Namely, it configures the no-arbitrage as the training goal of the entire model. It shows the superiority of traditional ML models and proves that the LSTM macroeconomic states promote the model performance significantly. Their work addresses the suitability for time series data through the LSTM model and statistically defends the no-arbitrage hypothesis. However, if the market exists arbitrage opportunities, which is possibly caused by price limits or liquidity fractions, even the most powerful GAN can only fit an approximate SDF rather than the true pricing kernel. Moreover, although the authors attempt to interpret factor importance, it is difficult to provide clear economic meanings comparable to traditional factors. Additionally, the model may face issues such as mode collapse and non-convergence, which cause training instability, while statistically minimizing the $\alpha$ may raise the suspicion of data snooping. In this sense, the proposed models on an RNN foundation and the empirical factor settings are simpler and more interpretable, thereby respecting real market conditions. \\

However, the application of RNNs and attention mechanisms to asset pricing is still at a very early and cutting-edge stage, and the extant literature remains relatively sparse. The most relevant literature of this study is \citet{zhou2024learning}, which is developed from the work of \citet{Gu2021AutoencoderModels}. It keeps the MLP structure for calculating the factor loadings, but alters the MLP autoencoder to the RNN additive attention structures for factor dimensional reduction. By dot producting the output factor loadings from the MLP structure and the extracted factors from the RNN additive attention structures, the estimated stock excess returns can be computed. It resolves the temporal dependency capturing issue by using RNN, LSTM, and GRU exchange models, while adopting the additive attention mechanism to extend memory and capture long-term temporal dependencies in data. Its LSTM additive attention model with the extracted six factors outperforms all benchmark models, including \citet{Gu2021AutoencoderModels} and \citet{Kelly2019CharacteristicsReturn}. 
Nonetheless, it does not explain how they determine the dimensionality of the extracted factors, nor how to interpret these alternative factors, which leaves their models shrouded in 'black-box' shadows. Furthermore, the additive attention mechanism extends the memory for capturing the longer-term temporal dependency of the data, which potentially improves the prediction accuracy, but the limited data size may cause overfitting due to the complication of the model with high parameter numbers. Moreover, since the attention mechanism computes attention weights across the entire sequence of RNN temporal hidden states, it risks incorporating future data into current estimates. To prevent this forward-looking information leakage, it is crucial to embed a causal mask that restricts the attention focus solely to past and present states. Otherwise, it introduces the look-ahead bias, which may cause spurious performance, loss of economic interpretability, non-tradable strategy and data snooping issues. However, it is a common issue in recent attention-based asset pricing research, including the work of \citet{kelly2025artificial,Ma2023AttentionApproach,Zhang2022AssetLearning} and \citet{Cong2021DeepPricing}. The proposed attention models notice this problem and enforce a causal mask individually to satisfy the single-direction characteristic of the time series data. \\

Interestingly, recent works concentrate on the application of prevailing large language models (LLMs), such as Transformers, which contain multiple attention mechanisms, instead of simplified models that focus on attention mechanisms themselves. For example, \citet{Cong2021DeepPricing} implements a comparative study of ML models with the temporal dependency capture nature, such as RNN, single or bi-directional LSTM, RNN with additive and classic Transformer, for empirical asset pricing on large-scale U.S. stocks. In their study, LSTM achieves the best OOS fitness, but GRU derives the highest annulized Sharpe ratio from the long-short portfolio. \citet{Zhang2022AssetLearning} conducts a similar comparative study to \citet{Cong2021DeepPricing}. It compares MLP and MLP residual, Convolutional Neural Network (CNN) and CNN residual, RNNs, RNN additive attention, classic Transformer model and traditional ML models. It finds that the RNN additive attention model slightly outperforms the classic Transformer model, achieving the highest OOS $R^2$. \citet{Ma2023AttentionApproach} proposes a CNN multi-head self-attention-based encoder-only Transformer model for Chinese stock pricing and volatility forecasting from January 2000 to December 2019. It uses 72 firm characteristics and 8 macroeconomic indicators as the pricing factors and volatility forecasting results for calculating asset allocation weights. The Transformer model with three encoder blocks outperforms all alternative variants and traditional ML benchmarks. It proves that the Transformer model with the multi-head self-attention mechanism performs well in return forecasting and even better in volatility forecasting. Although these works inspire later researchers with the ML function form for asset pricing, the largest problem of these works is neglecting the single-direction characteristic of the time series data, since none of them imposes the causal mask to prevent the future information leaking, even if they are mitigated by sequential methods, such as the rolling window and the extending window. Additionally, most of these works employ the data ending in the year of 2016 \citep{Gu2021AutoencoderModels,Avramov2021MachinePredictability}, which excludes the extreme market conditions caused by the COVID-19 pandemic. Namely, they did not examine the model performance during the market turbulence to show the stability of the models they applied. Instead, this study employs three periods that cover the stages of the Pre–COVID–19 Period (1911), the COVID-19–Inclusive Period (2112), and the Period Including COVID-19 and One-Year After (2212), and investigates how typical RNN-based attention models respond to the effects of the COVID-19 pandemic. Furthermore, most of these works use firm characteristics and macroeconomic indicators directly as input factors, which raises concerns about data snooping or data mining. In this study, the firm characteristics-sorted portfolios are applied as the original input factors, which cross-sectionally consider the information that covers the entire market and improve the economic explainability of the model.\\

Some of the asset pricing literature based on attention mechanisms appears to circumvent the future data leaking issue, as they apply it cross-sectionally rather than temporally. For example, \citet{kelly2025artificial} cross-sectionally applies an encoder-only Transformer model with Luong's general (bilinear form) attention mechanism for generating the best SDF at each time $t$, which derives the highest OOS Sharpe ratio. They examine their model on large-scale U.S. stocks and 132 normalized firm characteristics as factors, while deploying a short rolling training window of 60 observations. Their work covers the entire period of the COVID-19 pandemic. And their model outperforms the MLPs and linear benchmark models. Nevertheless, they simply substitute the missing values in the firm characteristics with zero, but ignore the input factors' data quality. For LLMs, low data quality increases the probability of model hallucination, which introduces untruthful similarity between different firm characteristics and leads to incorrect predictions. Moreover, they generate the SDF at each time $t$, and combine it into a time series-like sequence, but it neglects the temporal dependency to some extent. Still, the limited data size with a highly parameterized model structure increases the risk of model overfitting and lowers the model's economic interpretability. The MLP autoencoder pre-training method of this study is employed for handling the missing values in the input factors and reduces the input dimension to moderate model overfitting issues, which may also assist their asset pricing research. The simplified model structure in this study, with only an RNN and one attention layer, and the extracted factors may conquer the general problems that exist in advanced ML-based asset pricing research. Similarly, \citet{Chatigny2021AssetLearning} proposes a hybrid deep learning framework that combines stacked residual blocks (adapted from the N-BEATS model) for nonlinear feature extraction with a linear self-attention layer. This combined architecture cross-sectionally evaluates the importance of firm characteristics and macroeconomic variables. Specifically, the N-BEATS architecture relies on a deep stack of fully connected layers with forward and backward residual links. In the context of asset pricing, this design has the advantage since it allows the model to capture deep non-linear relationships in noisy financial data while actively preventing gradient vanishing and overfitting. The model directly estimates the SDF by minimizing squared Euler equation errors, and reports remarkably strong out-of-sample performance with annualized Sharpe ratios exceeding 2.8 at the market level, and above 1.0 even after excluding the smallest stocks or restricting the stocks to the Russell 3000 constituents. Their linear self-attention improves economic interpretability by highlighting relevant macroeconomic and firm-level features during different market conditions (e.g., the 2008 crisis). This is an innovative work, but it still faces issues as it practically neglects the transaction costs for short-selling positions, market impact, or leverage constraints, which reduces the implementability in real-world trading. Additionally, although the missing firm characteristics are imputed with cross-sectional medians, which is a typical method commonly applied in the industry, it could induce systematic biases if the missing values are manipulated. It is also worth noting that their attention weights capture the statistical relevance of factors at each time point rather than structural causality and temporal dependencies. Thus, their high Sharpe ratio should be carefully confirmed for robustness, non-overfitting and non-data-snooping suspicion.\\

\section{Data description}\label{sec:Data_description_ch2}
The responses, or labels in ML terminology, used in this research contain 420 large-cap individual stock excess returns from January 1957 to December 2021. These stocks come from the National Association of Securities Dealers Automated Quotations (NASDAQ) and the New York Stock Exchange (NYSE), which are derived from CRSP of Wharton Research Data Services (WRDS). These 420 stocks originally are filtered from the top 15\% of the largest market capitalization (MC), which covers 85\% of total MC for the two markets. They satisfy the conditions of having no missing data in the testing period and less than 50\% of values missing during the training period, hence they can meet the requirement of the ‘going concern’ \citep{Fama1970EfficientMarkets} and data representability. Selecting the large-cap stocks is for avoiding the liquidity risks of market trading, which happens as investors cannot close the positions to stop loss or confirm the trading profit since no counterparty takes their positions. And as a benefit of the assumption of ‘too-big-to-fail’ \citep{Mishkin2006HowBailouts} and ‘going concern’, this setting is more realistic for practitioners since it protects the investors from systemic risks such as companies that are delisted unexpectedly due to insolvency (e.g. Lehman Brothers) or other unexpected issues. The overall coverage of market capitalization from these stocks is 21.38\% in the U.S. stock markets of the NASDAQ and the NYSE. Additionally, excessive missing values can significantly compromise model reliability. They reduce the effective sample size, enlarge estimation variance, and introduce bias if the missingness is not completely at random \citep{little2019statistical}. In this case, missing observations may disrupt the temporal dependency structure,  affect predictive performance and cause the unstable out-of-sample evaluation. Thus, I remove the factors with excessively missing values. The risk-free rate for constructing the excess returns and evaluating the Sharpe ratio (SR) and Sortino ratio (SO) is derived from the Kenneth R. French Data Library. \\

The predictors, also known as features for the ML terminology, are organized by 182 explored firm characteristic-sorted portfolio factors provided by \citet{AndrewY.Zimmermann2020OpenPricing} in their Open Source Asset Pricing (OSAP) database, with at least 60\% of available values in the training period. The 60\% data availability threshold during the training period ensures the historical continuity of the time series from 1957 to 2021. Although the full OSAP dataset provided 202 factors that authors confirmed as high-quality factors, numerous of  them, especially those labeled as `indirect' or `insignificant' ones, suffer from massive data gaps in the early decades. The selected 182 factors provides a high-density information set that covers nearly all `Clear Predictors' across six core clusters: Momentum, Value, Investment, Profitability, Intangibles, and Trading Frictions.\\ 

There are three reasons for employing the firm characteristic-sorted portfolios as factors instead of firm characteristics directly as input factors, which is adopted by literature such as \citet{Gu2020EmpiricalLearning,Ma2023AttentionApproach}. Firstly, compared with the individual firm characteristics as factors, the sorted portfolio factors consider the effects between assets, stocks in this case, and cross-sectionally evaluate the asset excess returns, which improves the economic interpretability of the factor model. Secondly, compared with the factor method of GKX2020 \citep{Gu2020EmpiricalLearning}, the factor method in this study highly reduce the dimension of the input, which potentially causes model overfitting. Thirdly, if firm characteristics are used directly as factors, the data exhibit large disparities across variables, and even after standardization, such heterogeneity remains problematic for model estimation. The details of these 182 factors are presented in the appendix of \citet{lai2025multilayer}. The entire data set is split into training (70\% of the entire data length), validation (15\% of the entire data length) and testing periods (15\% of the entire data length for Period 2112) as shown in Table~\ref{tab:data_des_ch3_data}. It is consistent with the general machine learning data split rule of thumb that the training data length should normally exceed 70\% of the entire data length \citep{Goodfellow2016DeepLearning}. And the validation window and testing period in this case are adjusted to satisfy the research target of the COVID-19 impact investigation. The training window is on an rolling base with a step size of 12 months, while the validation window is fixed with 119 observations. The training window, together with a fixed validation window, moves forward 12 months at a time until it exhausts the entire data length. Thus, the out-of-sample, which contains the full testing period, has 10 years with one parameters renew per year. This configuration balances the computational efficiency and model fitness. It also improves the model's training stability. Moreover, this approach ensures all models operate on the identical information set and avoids the excessive computational workload of monthly retraining. From an economic perspective, since the input predictors are organized by firm characteristics that a substantial number of them update on a quarterly or annual basis, annual re-estimation aligns the model's learning frequency with the natural turnover of these fundamental signals. This prevents the networks from overfitting to short-term monthly trading noise that often lacks long-term predictive power. Between parameter updates, the models generate monthly out-of-sample predictions by applying the most recent estimated parameters to the updated monthly factors. As for the dynamic examination, the study compares the models’ testing periods depending on whether including the COVID-19 period, and separates the testing period into three market regimes: Pre–COVID-19 Period (1911), COVID-19–Inclusive Period (2112) and Period Including COVID-19 and One-Year After (2212). The Pre–COVID-19 Period contains a mild uptrend, the COVID-19–Inclusive Period contains a sharp uptrend with deep drawdowns, and the Period Including COVID-19 and One-Year After contains the high volatile sideway movement, which exhibits in Figure ~\ref{fig:price_index_ch2}.\\

Despite the proposed models deploying the MLP autoencoder as the pre-training method for dimensional reduction and information extraction, excessive missing values can hinder model performance by reducing the effective information available for representation learning and increasing the risk of biased imputations, especially for models with attention mechanisms. The attention mechanisms may concentrate on unreliable or non-informative time steps, weakening local or global feature extractions. The biased imputations pose a risk, particularly in deep learning-based forecasting. If it is not properly constrained, it can introduce hallucinated inputs, values generated from prior correlations rather than observable causality, thereby distorting both training and evaluation. In addition, the risk-free rate for constructing the excess stock returns and evaluating the Sharpe ratio and Sortino ratio is also derived from the Kenneth R. French Data. \\

The entire monthly data covers from January 1957 to December 2022. The separation of the training period, validation period and testing period is listed in Table~\ref{tab:data_des_ch3_data}. Different from the data method of \citet{lai2025multilayer}, this study adopts the rolling window method instead of the prevailing extending window method in previous literature, such as \citet{Gu2020EmpiricalLearning}. The validation window size is approximately 20\% of the training size. Thus, the `Initial in-sample total' in Table~\ref{tab:data_des_ch3_data} includes the training window of 553 observations and the validation window of 119 observations. The step size is 12, which means the parameters are re-estimated every 12 months, and the training window, together with the validation window, moves forward every 12 months. This configuration balances the computational efficiency and model fitness. It also improves the model's training stability. 
\begin{table}[htbp!]
\centering
\begin{tabular}{cccc}
\hline
\textbf{Name} & \textbf{Start date} & \textbf{End date} & \textbf{Observation No.} \\
\hline
Initial in-sample total & 1/1957 & 12/2012 & 672 \\
Testing (OOS) for ‘1911’ & 1/2013 & 11/2019 & 83 \\
Testing (OOS) for ‘2112’ & 1/2013 & 12/2021 & 108 \\
Testing (OOS) for ‘2212’ & 1/2013 & 12/2022 & 120 \\
\hline
\end{tabular}
\caption[Data splits for in-sample and out-of-sample data.]{Data splits for in-sample and out-of-sample data. ‘1911’ means the testing period ends before the pandemic happened, 2112’ means the testing period contains the pandemic period and ‘2212’ means the testing period contains not only the pandemic period but also one year after the pandemic. The `Initial in-sample total' includes the training observation of 553 and the validation observation of 119.}
\label{tab:data_des_ch3_data}
\end{table}

\section{Models}\label{sec:Models_ch2}
\begin{figure}[htbp!]
\centering
\includegraphics[width=0.9\columnwidth]{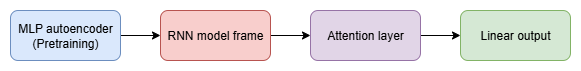}
\caption[The workflow diagram of the RNN attention models.]{The workflow diagram of the RNN attention models.}
\label{fig:entire_workflow_diagram_ch2}
\end{figure}

\begin{figure}[htbp!]
\centering
\includegraphics[width=0.9\columnwidth]{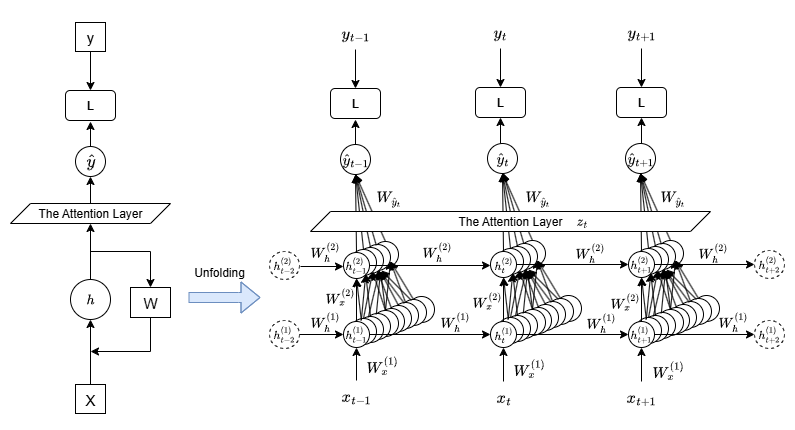}
\caption[The workflow diagram of the RNN attention models.]{The workflow diagram of the RNN attention models. In this figure, $\bm{W}_h^{(1)}$, $\bm{W}_h^{(2)}$, $\bm{W}_x^{(1)}$ and $\bm{W}_x^{(2)}$ are learnable parameter matrices for each layer, $X$ and $y$ are pre-trained input and labels respectively. $h$ indicates the hidden layers of the RNN structure, $\hat{y}$ is the output derived from the attention layer after linear transformation, and $L$ represents the loss function. The `Attention layer' is the attention layer that contains an attention mechanism (e.g. additive attention mechanism or one of Luong's attention mechanisms). The left side of the figure is the main workflow of the RNN attention models, and the right side is the 3-Dimension timeline-based unfolded workflow diagram. The circles indicate the nodes of each hidden layer. Each node is fully connected to all nodes in the next layer.}
\label{fig:workflow_diagram_ch2}
\end{figure}

The pre-trained recurrent neural network (RNN) attention models employ the MLP autoencoder as the pertaining method for the dimensional reduction of the original input factors, which is the firm characteristic-sorted portfolios provided by \citet{AndrewY.Zimmermann2020OpenPricing}, and handling the missing values. The pre-trained factors (extracted features or factors) then feed into the main models shown in Figure~\ref{fig:workflow_diagram_ch2}. The main models are organized by an RNN fundamental structure with an attention layer, which are introduced in Section~\ref{subsec:RNN_ch2} to Section~\ref{subsec:satt_RNN_ln_ch2}. Figure~\ref{fig:entire_workflow_diagram_ch2} shows the entire workflow of the pre-trained RNN attention models. This study investigates the RNN models with the additive attention mechanism \citep{Bahdanau2015NeuralTranslate}, Luong attention mechanisms \citep{Luong2015EffectiveTranslation}, global self-attention mechanism \citep{Vaswani2017AttentionNeed} and the sliding window sparse attention mechanism \citep{beltagy2020longformer}, and compares them with the vanilla RNN model and its variations: long short-term memory (LSTM) neural networks and gated recurrent unit (GRU) neural networks. Among these RNN attention models, the RNN sliding window sparse attention model and the RNN self-attention model are the proposed RNN attention models. The RNN sliding window sparse attention model extracts the sliding window sparse attention mechanism from the Longformer model \citep{beltagy2020longformer}, which is introduced in Section~\ref{subsec:sliding_window_ch2}, while the RNN self-attention model extracts the self-attention mechanism from the classic Transformer model \citep{Vaswani2017AttentionNeed}, which is introduced in Section~\ref{subsec:self_att_RNN_ch2}. Since the sliding window sparse attention belongs to the category of self-attention, in this study, the classic self-attention mechanism from \citep{Vaswani2017AttentionNeed} is denoted as `global self-attention mechanism' for preventing any confusion. Although the pre-trained RNN additive attention model, Luong's attention models are still innovative in the existing literature for asset pricing research, they are configured as the benchmarks, as these attention mechanisms were developed in the early stage. The additive attention, Luong's general attention, and Luong's concatenate attention mechanisms have larger parameter sizes than the dot-product-based attention mechanisms, increasing the risk of overfitting and computational cost. However, Luong's dot-product attention is a non-parameterized attention, which improves the computational efficiency but sacrifices the performance and training stability. Vanilla RNN, LSTM, and GRU models are benchmarks for evaluating whether RNN-based models benefit from attention mechanisms. \\

Figure~\ref{fig:workflow_diagram_ch2} is the workflow diagram of the main models, which assemble a two-hidden-layer vanilla RNN frame with an attention layer. An attention layer contains one attention mechanism, for example, the additive attention mechanism or one of Luong's attention mechanisms. The left part of Figure~\ref{fig:workflow_diagram_ch2} shows the main workflow of the RNN attention models. The input X, which is the pricing factors in this case, goes through the RNN hidden layers and the sliding window sparse attention layer, then derives the output $\hat{y}$. The loss function $L$ measures the difference between the output derived from the RNN attention structure and the label y, which is the stock excess return series in this case. $W$ is the learnable parameter matrix recurrently calculated according to the timeline. The right part exhibits the unfolded 3-dimensional structure of the RNN attention models, where the RNN  frame has two pyramidical hidden layers: $h^{(1)}$ and $h^{(2)}$. The first hidden layer and the second layer configure 64 and 32 neurons, respectively.  From the second dimension scope, the RNN structure collects the current and the one-step previous information and passes them to the next time step as introduced in Section~\ref{subsec:RNN_ch2}. From the third dimension angle of the figure, each time step has a pyramidical MLP network, which achieves the goal of reducing the feature's dimension. Additionally, the attention layer estimates and distributes the attention weights to each time step to depict the relevance of any two time steps in a pre-defined sight range. The attention output $\bm{z}_t$ is used for computing the estimated $\hat{y}$ with a linear connection.\\

\subsection{MLP autoencoder for pertaining}\label{subsec:mlp_auto_ch2}
\begin{figure}[htbp!]
\centering
\includegraphics[width=0.9\columnwidth]{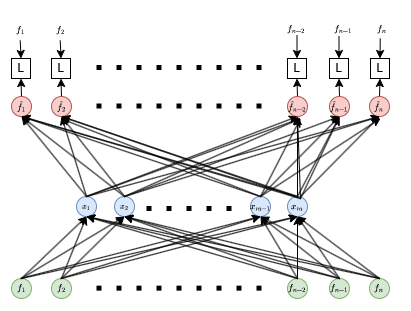}
\caption[MLP autoencoder as the pre-training module.]{MLP autoencoder as the pre-training module. In this figure, $f_{1}$, $f_{2}$ to $f_{n}$, $n\in\mathbb{Z}^+$ are denoted as original factors. $L$ is the loss function, MSE in this case. $\hat{f}_{1}$, $\hat{f}_{2}$ to $\hat{f}_{n}$, $n\in\mathbb{Z}^+$ are the reconstructed factors. The bottom layer with green nodes is the input layer, while the top layer with red nodes is the reconstructed layer. The layer in between with blue nodes named latent space of the autoencoder, which generates the new factors with a lower dimension, where $x_{1}$ to $x_{m}$, $m\in\mathbb{Z}^+$
 denoted as new factors.}
\label{fig:mlp_auto_ch2}
\end{figure}

The MLP autoencoder is commonly used for feature engineering in the machine learning and financial time series forecasting context, especially for dimensional reduction and data denoising purposes \citep{akhtar2017multilayer,sagiraju2021application,bao2017deep,bieganowski2024supervised,heaton2017deep}. This study deploys the MLP autoencoder as a pre-training method for input factors' dimensional reduction and missing value filling. Figure~\ref{fig:mlp_auto_ch2} shows the MLP autoencoder module deployed for this study. The MLP autoencoder encompasses an input layer, an output layer (reconstructed layer in this case) and a latent layer (latent space or bottleneck layer) in between. The process from the input layer to the latent layer is named `encoding', and from the latent layer to the reconstructed (output) layer is named `decoding' in an autoencoder framework \citep{rumelhart1986learning}. In the asset pricing research of this chapter, the original input factors are denoted as $f$, $f\in\mathbb{R}^n$ and the reconstructed factors are recognized as $\hat{f}$, $\hat{f}\in\mathbb{R}^n$. Through an MSE loss function $L$, the reconstructed factors can be trained to closely approach the actual input factors. The layer in between with blue nodes is the latent layer, where it subtracts the new factor with a lower dimension. The parameters which lead to the minimum MSE of the loss function are used to generate new factors. The new factors are denoted as $x$, $x\in\mathbb{R}^m, m<n$. Therefore, the encoding process can be mathematically presented as:
\begin{equation}
\bm{X} = g\left( \bm{W}_{\text{en}} \bm{F} + \bm{b}_{\text{en}} \right)
\label{eq:encoder_ch2}
\end{equation}
where $\bm{F} = [\bm{f}_1, \bm{f}_2, \dots, \bm{f}_n] \in \mathbb{R}^{n \times T}$ is the original input factor matrix, $T$ is the time length, while $\bm{X} = [\bm{x}_1, \bm{x}_2, \dots, \bm{x}_m] \in \mathbb{R}^{m \times T}$ is the latent factor matrix. $W_\text{en}\in\mathbb{R}^{m \times n}$ and $b_\text{en}\in\mathbb{R}^{m \times 1}$ are the weight matrix and bias of the encoder respectively. $g(\cdot)$ is the Rectified Linear Unit (ReLU) activation function $\mathrm{ReLU}(x) = \max(0, x)$. The function of the encoding is to lower the dimension of the original input factors.\\

The function of the decoding is opposite to the function of the encoding. It attempts to reconstruct the original input factors. It utilizes a loss function to measure the accuracy of the reconstructed factors, and it can be presented as:
\begin{equation}
\hat{\bm{F}} = g\left( \bm{W}_{\text{de}} \bm{X} + \bm{b}_{\text{de}} \right)
\label{eq:decoder_ch2}
\end{equation}
where $\bm{W}_{\text{de}} \in \mathbb{R}^{n \times m}$ and $\bm{b}_\text{de}\in \mathbb{R}^{n \times 1}$ are the decoder weight matrix and bias, respectively. $\hat{\bm{F}} = [\hat{\bm{f}}_1, \hat{\bm{f}}_2, \dots, \hat{\bm{f}}_T] \in \mathbb{R}^{n \times T}$ are the estimated original factors. Considering the balance of the information effectiveness and the noise level of the generated latent factors, the neuron number is configured 70\% of the original input. \\

The mean square error (MSE) function is selected as the loss function for estimating the accuracy of factor reconstruction, which is shown in Equation~\eqref{eq:loss_ende_ch2}. Stochastic gradient descent (SGD) algorithm with Adaptive moment estimation (Adam) optimizer is employed for the estimation, and an early-stopping method is applied for improving computational efficiency and preventing model overfitting caused by the training process. They are presented in the appendix.\\
\begin{equation}
\mathcal{L}_{\mathrm{recon}} = \frac{1}{T} \left\| \bm{F} - \hat{\bm{F}} \right\|_F^2
\label{eq:loss_ende_ch2}
\end{equation}

\subsection{Recurrent Neural Networks}\label{subsec:RNN_ch2}
The fundamental structure of the proposed model is the RNN structure ~\citep{Rumelhart1986LearningErrors}, which is shown in the left part of Figure~\ref{fig:workflow_diagram_ch2}. The workflow chart without the sparse attention layer is the naive RNN structure. A fundamental RNN structure includes an input layer, an output layer and a single or multiple hidden layer. In this study, for a better capture of factor patterns, two hidden layers with a pyramid neuron structure are employed, specifically, the first hidden layer with 64 neurons and the second hidden layer with 32 neurons. The mathematical presentation of the multilayer RNN model is presented in Equation~\eqref{rnn_h1_ch2} to Equation~\eqref{rnn_h2_ch2}, where $\bm{h}_t^{(1)}$ and $\bm{h}_t^{(2)}$ mean the output of the first hidden layer and the second hidden layer at time $t$. $\bm{h}_{t-1}^{(1)}$ and $\bm{h}_{t-1}^{(2)}$ are the output of hidden layers at time $t-1$, while $\bm{x}_{t}$ is the abstracted factors from pre-training module. $\bm{W}_h^{(1)}$, $\bm{W}_h^{(2)}$, $\bm{W}_x^{(1)}$ and $\bm{W}_x^{(2)}$ are the estimated parameter matrices of the previous hidden states and current inputs respectively. $\bm{b}_{rnn}^{(1)}$ and $\bm{b}_{rnn}^{(2)}$ are the bias of each hidden layers. \\
\begin{align}
\bm{h}_t^{(1)} = \phi\left( \bm{W}_h^{(1)} \bm{h}_{t-1}^{(1)} + \bm{W}_x^{(1)} \bm{x}_t + \bm{b}_{rnn}^{(1)} \right), \quad \bm{h}_t^{(1)} \in \mathbb{R}^{64}\label{rnn_h1_ch2}\\
\bm{h}_t^{(2)} = \phi\left( \bm{W}_h^{(2)} \bm{h}_{t-1}^{(2)} + \bm{W}_x^{(2)} \bm{h}_t^{(1)} + \bm{b}_{rnn}^{(2)} \right), \quad \bm{h}_t^{(2)} \in \mathbb{R}^{32}\label{rnn_h2_ch2}
\end{align}
And $\phi(\cdot)$ is the hyperbolic tangent (tanh) activation function.\\

\subsection{Attention mechanisms for attention layer}\label{subsec:att_overview_ch2}
The attention mechanism is a powerful neural network component that enables models to dynamically focus on the most relevant parts of a sequence when making predictions. It was originally developed for natural language processing (NLP), such as missing text prediction and machine translation. It employs a function to compute the attention scores to measure the relevance level between any two parts of a textual sequence, and applies the softmax function to normalize the attention scores to attention weights to measure the importance of each word. Thus, the attention mechanism can improve the textual prediction accuracy through the indicator, which shows the importance of a certain word.\\

The earliest attention mechanism is the additive attention mechanism from \citet{Bahdanau2015NeuralTranslate}. Gradually, researchers developed different function forms for the attention mechanism. Luong's three-form attention mechanism \citep{Luong2015EffectiveTranslation} is one of the most significant attention mechanisms. The additive attention is frequently used in recent asset pricing literature, for example, \citet{zhou2024learning} and  \citet{Cong2021DeepPricing}. Later, the Luong's general attention is used in the paper of \citet{kelly2025artificial} as a module for their Transformer model. Inspired by these works, the research in this chapter uses early-developed attention mechanisms as benchmarks for the proposed models. However, the attention mechanisms cannot directly apply to time series data such as factors and stock returns, since the textual data is sometimes bidirectional. For instance, the missing word prediction could depend not only on the previous words but also on the words behind the missing space. However, the time series forecasting only relies on historical data, which implies that bidirectional forecasting may cause the future data leakage issue. Therefore, a compulsory causal mask which can cover future data is crucial for time series attention mechanism applications. For time series attention mechanisms, the attention weights, which measure the relevance level of two time steps, are only computed between the current time step and the historical time steps, but the future time steps are covered by masks. And the attention weights are finally distributed back to the fundamental structure (e.g. RNN or LSTM) output to generate attention output, combining the information of the fundamental structure output and the importance of the information at each time step. The research on applications of attention mechanisms for finance and economics is still in the early stages. As an example, \citet{kelly2025artificial} deploys Luong's general attention mechanism to their Transformer model for the asset pricing research. \\

Attention mechanisms have three essential elements: query, key and value. Query and key are used for computing attention scores via a certain function form, while value is used for computing the attention output. The global self-attention mechanism \citep{Vaswani2017AttentionNeed} and sliding window attention mechanism \citep{beltagy2020longformer} have explicit values since they have a specific function form for the element of `value', but alternative attention mechanisms have hidden value equal to the historical output of the foundation neural network structure.
The additive attention\citep{Bahdanau2015NeuralTranslate}, which employs a feedforward network to compute alignment scores, is introduced in Section~\ref{subsec:additive_att_ch2}. The multiplicative (or general) attention\citep{Luong2015EffectiveTranslation}, which uses linear forms to compute the similarity between queries and keys, is introduced in Section~\ref{subsec:Luong_att_ch2}. Scaled dot-product attention \citep{Luong2015EffectiveTranslation,Vaswani2017AttentionNeed} further improves numerical stability and training efficiency by introducing a scaling factor, and it serves as the foundation of the most prevailing large language model named Transformer. This is introduced in Section~\ref{subsec:self_att_RNN_ch2} and Section~\ref{subsec:sliding_window_ch2}. 

\subsubsection{Additive attention mechanism}\label{subsec:additive_att_ch2}
\citet{Bahdanau2015NeuralTranslate}'s additive attention is the earliest attention mechanism originally designed for natural language processing (NLP) tasks such as machine translation. As a global attention mechanism, it measures the difference between any two time steps and distributes attention weights for every time step. It is suitable for capturing the most relevant information from context, which could increase the prediction accuracy of the missing words within a sentence. However, this causes the future information leakage if the attention mechanism is applied directly to the time series data without a causal mask. Let $\bm{q}_t,\bm{k}_j\in \mathbb{R}^{d_{RNN}}$ be the query and key of at time step of $t$ and $j$, and define:
\begin{equation}
\bm{q}_t = \bm{W}_q \bm{h}^{(2)}_t, \quad \bm{k}_j = \bm{W}_k \bm{h}^{(2)}_j
\end{equation}
where  $\bm{W}_q, \bm{W}_k \in \mathbb{R}^{d_{RNN} \times d_{RNN}}$ are the learnable parameter matrices of $\bm{h}^{(2)}_t,\bm{h}^{(2)}_j \in \mathbb{R}^{d_{RNN}}$. Then the additive attention score for time $t$ and $j$, $e_{t,j}$, can be presented as:
\begin{equation}
e_{t,j} = \bm{v}^\top \tanh(\bm{q}_t + \bm{k}_j)
\end{equation}
where $\bm{v} \in \mathbb{R}^{d_{RNN}}$ is a learnable parameter vector. Considering the input is the time series data, the compulsory causal mask should be applied to force time step $j$ to be earlier than $t$. \\

\subsubsection{Luong's attention mechanism}\label{subsec:Luong_att_ch2}
Luong's attention is developed from the additive attention.  \citet{Luong2015EffectiveTranslation} proposes three attention weights computation methods for boosting the computational efficiency and forecasting accuracy of \citet{Bahdanau2015NeuralTranslate}'s additive attention. It employs the dot product, linear and concatenation for computing the attention weights, respectively, which are defined as `Luong dot-product (L-DotProd) attention', `Luong general (L-General) attention' and `Luong concatenation (L-Concat) attention'. For the convenience of the mathematical presentations, they are denoted as `LD', `LG' and `LC' respectively. \\

For Luong dot-product attention, the query $\bm{q}_t$ and key $\bm{k}_j$ is equal to the RNN output at time $t$ and $j$, which shows in Equation~\eqref{eq:luong_ld_qk_ch2}
\begin{equation}
\bm{q}^{LD}_t=\bm{h}^{(2)}_t, \quad \bm{k}^{LD}_j=\bm{h}^{(2)}_j
\label{eq:luong_ld_qk_ch2}
\end{equation}
The Luong dot-product attention score $e_{t,j}$ is the scaled dot product of query $\bm{q}_t$ and key $\bm{k}_j$,
\begin{align}
e^{\text{LD}}_{t,j} &= \frac{1}{\sqrt{d_{\text{RNN}}}} (\bm{q}^{LD}_t)^\top \bm{k}^{LD}_j \\
                    &= \frac{1}{\sqrt{d_{\text{RNN}}}} (\bm{h}^{(2)}_t)^\top \bm{h}^{(2)}_j
\end{align}

For Luong's general attention mechanism, it distributes a learnable parameter matrix $W\in \mathbb{R}^{{d_{RNN}}\times{d_{RNN}}}$ to the key $\bm{k}_j$ but keep the query $\bm{q}_t$ identical to the one in Luong dot-product attention. It projects key to the query linearly by adding the parameter matrix $W$. Equation~\eqref{eq:luong_lg_qk_ch2} shows the query and key for Luong's general attention mechanism.
\begin{equation}
\bm{q}^{LG}_t=\bm{h}^{(2)}_t, \quad \bm{k}^{LG}_j=\bm{W}\bm{h}^{(2)}_j
\label{eq:luong_lg_qk_ch2}
\end{equation}
Thus, the attention score of Luong's general attention can be:
\begin{equation}
e_{t,j}^{\text{LG}} = \frac{1}{\sqrt{d_{\text{RNN}}}} (\bm{q}^{\text{LG}}_t)^\top \bm{W} \bm{k}^{\text{LG}}_j
\end{equation}

For Luong's concatenation attention mechanism, the query $\bm{q}^{LC}_t$ and key $\bm{k}^{LC}_j$ have their own learnable parameter matrix which denoted as $\bm{W}_q$ and $\bm{W}_k$ respectively, $\bm{W}_q,\bm{W}_k\in \mathbb{R}^{{d_{RNN}\times d_{RNN}}}$. Concretely, the query and key can be presented as:
\begin{equation}
\bm{q}^{LC}_t=\bm{W}_q\bm{h}^{(2)}_t, \quad \bm{k}^{LC}_j=\bm{W}_k\bm{h}^{(2)}_j
\end{equation}
Different from all alternative attention mechanisms, Luong's concatenation attention applies a hyperbolic tangent function on the concatenated query and key for calculating the attention scores. The Luong's concatenation attention score can be computed as:
\begin{equation}
e_{t,j}^{\text{LC}} = \bm{v}^\top \tanh\left( [\bm{q}^{\text{LC}}_t; \bm{k}^{\text{LC}}_j] \right)
\end{equation}
where $\bm{v}\in \mathbb{R}^{2d_{RNN}}$ is the learnable parameter vector for calculating the concatenation attention scores. \\

\subsubsection{Global self-attention RNN model}\label{subsec:self_att_RNN_ch2}
The global self-attention is abstracted from \citet{Vaswani2017AttentionNeed}'s classic Transformer model. In the additive attention mechanism and Luong's attention mechanisms, the format of `query' and `key' varies accordingly, but the third element `value' $v$ is hidden and always equal to the output of the RNN model $h_j$. To prevent any misunderstanding, in the previous sections, the calculation of attention output $z_t$ directly adopts the notation $h^{(2)}_j$ to multiply the attention weights $\alpha_{t,j}$. However, in the self-attention mechanism, the element $v$ has its variable format. Specifically,
\begin{equation}
\bm{q}_t = \bm{W}_q \bm{h}^{(2)}_t, \quad
\bm{k}_j = \bm{W}_k \bm{h}^{(2)}_j, \quad
\bm{v}_j = \bm{W}_v \bm{h}^{(2)}_j
\label{eq:satt_global1_ch2}
\end{equation}
where $\bm{W}_q, \bm{W}_k, \bm{W}_v \in \mathbb{R}^{d_{RNN} \times d_{RNN}}$ are learnable parameter matrix for the query $\bm{q}_t$, key $\bm{k}_j$ and $\bm{v}_j$. And $h^{(2)}_j\in \mathbb{R}^{d_{RNN}}$ is the output at time $j$ of the RNN structure. For each time step $t$ and $j$, the attention score $e_{t,j}$ can be computed through Equation~\eqref{eq:self_att_score_ch2}.
\begin{equation}
e_{t,j} =
\frac{1}{\sqrt{d_{RNN}}} \bm{q}_t^\top \bm{k}_j
\label{eq:self_att_score_ch2}
\end{equation}

\subsubsection{Causal masks, attention weights and attention outputs for global attentions}\label{subsec:mask_w_ch2}
Causal masks are configured for preventing the future data leakage of time series data. After attention scores ($e_{t,j}$) are computed, the causal masks can be mathematically presented as:
\begin{equation}
e_{t,j} =
\begin{cases}
e_{t,j}, & j \leq t \\
-\infty, & j > t
\end{cases}
\end{equation}
Thus, the attention weights $\alpha_{t,j}$ can be computed via softmax function:
\begin{equation}
\alpha_{t,j} =
\frac{\exp(e_{t,j})}{\sum\limits_{i=1}^{t} \exp(e_{t,i})}
\end{equation}
Furthermore, the attention outputs can be computed according to the foundation model outputs, either the hidden status from RNN structure ($h^{(2)}_j$) or the element of value (self-attention) computed from $h^{(2)}_j$ ($v$):
\begin{equation}
\bm{z}_t =
\sum_{j=1}^{t} \alpha_{t,j} \bm{h}^{(2)}_j
\end{equation}
or for self-attentions,
\begin{equation}
\bm{z}_t =
\sum_{j=1}^{t} \alpha_{t,j} \bm{v}
\end{equation}

\subsubsection{Sliding window sparse attention mechanism}\label{subsec:sliding_window_ch2}
\begin{figure}[htbp!]
\centering
\includegraphics[width=0.8\columnwidth]{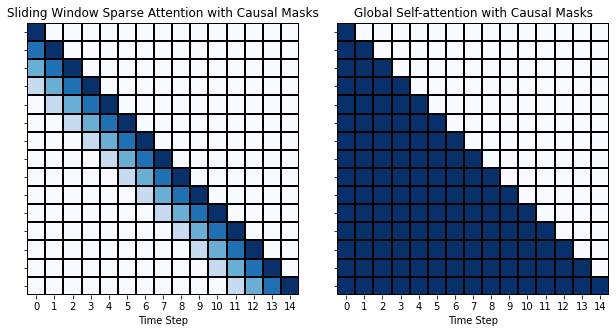}
\caption[The comparison of sliding window sparse attention and global self attention (with causal masks).]{The comparison of sliding window sparse attention and global self attention (with causal masks). This is an example when the attention window is equal to 4. The white grids in both panels represent the time points with causal masks, which means they are not in the calculation of attention weights.}
\label{fig:satt_compare_ch2}
\end{figure}

Sliding window sparse attention belongs to the self-attention family, and it is first proposed by \citet{beltagy2020longformer}. Like other sparse attention mechanisms, it is developed for capturing the longer-term temporal dependency of the sequential data and lowering the computational cost of global attention mechanisms such as Bahanau's attention \citep{Bahdanau2015NeuralTranslate}, Luong's attention  \citep{Luong2015EffectiveTranslation} and Vaswani's self-attention \citep{Vaswani2017AttentionNeed}. The difference between the sliding window attention and Vaswani's self-attention is Vaswani's self-attention calculates the difference between any two time steps, but the sliding window attention of \citet{beltagy2020longformer}'s only calculates the difference between one time step and a range of time steps close to that time step. Figure~\ref{fig:satt_compare_ch2} shows the difference between Vaswani's self-attention and sliding window sparse attention with causal masks for time series data. The left panel of Figure~\ref{fig:satt_compare_ch2} shows an example of a sliding window sparse attention mechanism when the attention window is equal to 4, and the right panel shows the global self-attention mechanism from \citet{Vaswani2017AttentionNeed}. The global self-attention calculates the difference between the series at time step $t$ and all available time steps previous to $t$, but the sliding window sparse attention only measures the difference between the series at time step $t$ and available time steps previous to $t$ up to the restrictions of the attention window, which is $t-3$ in this example.\\

\begin{figure}[htbp!]
\centering
\includegraphics[width=0.5\columnwidth]{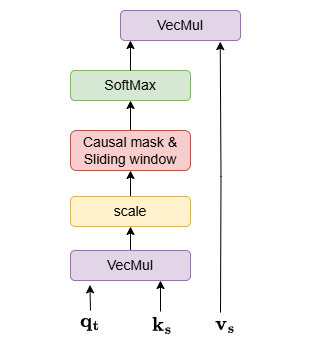}
\caption[Sliding window sparse attention mechanism workflow diagram.]{Sliding window sparse attention mechanism workflow diagram. $q_t,k_s,v_s\in\mathbb{R}^{d_{RNN}}$ are query, key, value at time $t$ and $s$ respectively.'VecMul' is denoted as the vector-wise multiplication.}
\label{fig:satt_ch2}
\end{figure}

The sliding window sparse attention is developed from the global self-attention,which is shown in Figure~\ref{fig:satt_ch2}. Mathematically, to each time step $t$, defining the attention window $\mathcal{S}_t = \{ s \mid \max(0, t - w) \leq s \leq t \}$, $s$ is the time step previous to time $t$. Thus, the query (q), key (k) and value (v) here of sliding window sparse attention in time step $t$ and $s$ are:
\begin{equation}
q_t = W_q h_t,\quad k_s = W_k h_s,\quad v_s = W_v h_s
\end{equation}
where $h_t,h_s\in \mathbb{R}^{d_{RNN}}$ are the output from RNN second hidden layer at time $t$ and $s$ respectively. $W_q,W_k,W_v\in\mathbb{R}^{d_{RNN}\times d_{RNN}}$ are the learnable weight matrices of $q_t,k_s,v_s$. Therefore, the attention score between time $t$ and $s$ can be:
\begin{equation}
\alpha_{t,s} = \frac{ \exp\left( \bm{q}_t^\top \bm{k}_s / \sqrt{d_{RNN}} \right) }{ \sum\limits_{j \in \mathcal{S}_t} \exp\left( \bm{q}_t^\top \bm{k}_j / \sqrt{d_{RNN}} \right)}
\end{equation}
Further, the attention output of time $t$ denoted as $\bm{z_t}$ can be:
\begin{equation}
\bm{z}_t = \sum_{s \in \mathcal{S}_t} \alpha_{t,s} \bm{v}_s
\end{equation}
Due to the limited data size of this study, the proposed model only considers the single-head sliding window sparse attention. With the attention window, the computational cost of this attention becomes $\bm{O(T\times \mathcal{S}_t)}$.\\

The main difference between the classic Transformer model \citep{Vaswani2017AttentionNeed} and the Longformer model \citep{beltagy2020longformer} is that the latter employs the assembled attention mechanism which encompasses sparse attention of sliding window attention, dilated sliding window attention (optional) and global attention. For simplifying the model structure for the temporal sparse data, such as the monthly stock pricing data, the proposed model only deploys the sliding window sparse attention layer from the Longformer model and assembles it on an RNN frame for a better capture of the temporal longer-term dependency that exists in the stock pricing data.\\ 

To summarize, the calculation of attention output follows the procedure of computing the attention scores, applying the softmax function on attention scores for generating the attention weights and utilizing attention weights to multiply the `value' for the attention output.\\ 

\subsection{Linear output of the RNN attention models}\label{subsec:satt_RNN_ln_ch2}
Finally, the attention outputs are utilized to compute the estimated label through the linear function. It is shown in Equation~\eqref{eq:ln_yhat_ch2_yt}, where $\hat{y}_t$ is the linearly estimated label, which is the excess return of an individual stock at time $t$. 
\begin{equation}
\hat{y}_t = {\bm{W}_{\hat{y}_t}} {\bm{z}_{t}} + b_{\hat{y}_t}, \quad \hat{y}_t \in \mathbb{R}\label{eq:ln_yhat_ch2_yt}
\end{equation}

\subsection{Long short-term memory (LSTM) and gated recurrent unit (GRU) as benchmark models}\label{subsec:LSTM_GRU_ch2}
The RNN variation models, LSTM and GRU, serve as benchmark models in this study. Due to the gradient explosion and vanishing issues that exist in the naive RNN model, which was suggested by \citet{bengio1994learning}, \citet{hochreiter1997long} proposes the first LSTM model for capturing a longer dependency of sequential data. They design the input gate, output gate, and cell structure, and embed them in the naive RNN framework to moderate the gradient explosion and vanishing issues. It is improved by \citet{gers2000learning}, who adds a forget gate in the previous LSTM structure. This LSTM is the most commonly applied LSTM in the literature of different domains. Further, people find LSTM is over-parameterized since it has three gates, one cell and one hidden state, albeit it moderates the gradient explosion and vanishing issues to a degree. Thus, \citet{Cho2014LearningTranslation} proposes the GRU which simplifies the structure of LSTM. GRU only configure two gates: update gate and reset gate, which highly reduces the parameters of a recurrent neural network model. \\

\begin{figure}[htbp!]
\centering
\includegraphics[width=0.9\columnwidth]{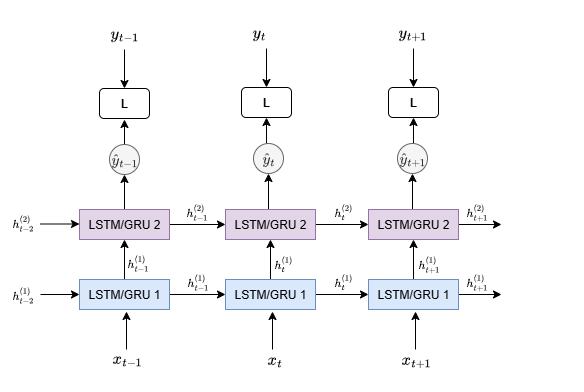}
\caption[Entire workflow chart for two-layer LSTM and GRU model.]{Entire workflow chart for two-layer LSTM and GRU model. $x$ is the abstracted factors from the MLP pretraining process, while $h$ is the output from the LSTM or GRU layer. $\hat{y}$ is the estimated output from the linear layer, and $L$ means the loss function. $y$ is the actual value of the label, the stock excess returns in this case.}
\label{fig:LSTM_GRU_entire_model_workflow_ch2}
\end{figure}

In the early stage, LSTM was developed for speech recognition, sequential labelling, time series forecasting and handwritten character recognition, while GRU was developed for machine translation originally. However, researchers have proven that GRU performs similarly to LSTM in prediction accuracy from different angles \citep{jozefowicz2015empirical,galphade2022comparative,pudikov2020comparison}, especially in the time series forecasting context. Figure~\ref{fig:LSTM_GRU_entire_model_workflow_ch2} exhibits the entire workflow of the two-layer LSTM adopted in this study from a bird's-eye view. The first LSTM or GRU layer ('LSTM/GRU 1') contains 64 neurons; thus, the dimension of the first layer is 64. And the second LSTM or GRU layer ('LSTM/GRU 2') contains 32 neurons, which indicates the dimension of the second LSTM or GRU layer is 32. For time series forecasting tasks, both LSTM and GRU frame connect a linear layer as the output layer.
Briefly, a two-layer LSTM with distinctive neuron numbers can be generalized mathematically presented as:
\begin{align}
\bm{i}^{(l)}_t &= \sigma\big(W^{(l)}_i \bm{h}^{(l-1)}_t + U^{(l)}_i \bm{h}^{(l)}_{t-1} + \bm{b}^{(l)}_i\big) \\
\bm{f}^{(l)}_t &= \sigma\big(W^{(l)}_f \bm{h}^{(l-1)}_t + U^{(l)}_f \bm{h}^{(l)}_{t-1} + \bm{b}^{(l)}_f\big) \\
\bm{o}^{(l)}_t &= \sigma\big(W^{(l)}_o \bm{h}^{(l-1)}_t + U^{(l)}_o \bm{h}^{(l)}_{t-1} + \bm{b}^{(l)}_o\big) \\
\tilde{\bm{c}}^{(l)}_t &= \tanh\big(W^{(l)}_c \bm{h}^{(l-1)}_t + U^{(l)}_c \bm{h}^{(l)}_{t-1} + \bm{b}^{(l)}_c\big) \\
\bm{c}^{(l)}_t &= \bm{f}^{(l)}_t \odot \bm{c}^{(l)}_{t-1} + \bm{i}^{(l)}_t \odot \tilde{\bm{c}}^{(l)}_t \\
\bm{h}^{(l)}_t &= \bm{o}^{(l)}_t \odot \tanh\big(\bm{c}^{(l)}_t\big)
\end{align}
The $l$-th layer, where $l \in \{1, 2, \dots, L\}$. The $\bm{h}^{(l-1)}_t$ denotes the input to the $l$-th layer at time $t$. For the first layer ($l=1$), the input is the original sequence data, meaning $\bm{h}^{(0)}_t = \bm{x}_t$. The $\bm{i}^{(l)}_t$, $\bm{f}^{(l)}_t$, and $\bm{o}^{(l)}_t$ are the input gate, forget gate, and output gate of the $l$-th layer, respectively. $\tilde{\bm{c}}^{(l)}$ and $\bm{c}^{(l)}_t$ are the candidate cell state and the cell state. $\sigma(\cdot)$ is the sigmoid activation function, and $\tanh(\cdot)$ is the hyperbolic tangent function. $W^{(l)}_i,\, W^{(l)}_f,\, W^{(l)}_o,\, W^{(l)}_c \in \mathbb{R}^{d_l \times d_{l-1}}$ and $U^{(l)}_i,\, U^{(l)}_f,\, U^{(l)}_o,\, U^{(l)}_c \in \mathbb{R}^{d_l \times d_l}$ are the learnable parameter matrices for each gate and cell, where $d_l$ is the dimension of the $l$-th layer (the number of neurons). Note that $d_0 = d_{\text{in}}$, indicating the dimension of the original input, which are the extracted factors from the pretraining process. $\bm{b}^{(l)}_i,\, \bm{b}^{(l)}_f,\, \bm{b}^{(l)}_o,\, \bm{b}^{(l)}_c \in \mathbb{R}^{d_l}$ are the bias vectors for input gate, forget gate, output gate and cell status, respectively. And $\odot$ denotes the Hadamard product. Extensively, the paper \citet{gers2000learning} discussed the technical details for the LSTM model.\\

GRU simplifies the LSTM by substituting the input gate, forget gate, output gate and cell states system with the update gate and reset gate system. A multi-layer GRU can be presented as:
\begin{align}
\bm{up}_t^{(l)} &= \sigma\big( W_{up}^{(l)} \bm{h}_t^{(l-1)} + U_{up}^{(l)} \bm{h}_{t-1}^{(l)} + \bm{b}_{up}^{(l)} \big)\\
\bm{re}_t^{(l)} &= \sigma\big( W_{re}^{(l)} \bm{h}_t^{(l-1)} + U_{re}^{(l)} \bm{h}_{t-1}^{(l)} + \bm{b}_{re}^{(l)} \big) \\
\tilde{\bm{h}}_t^{(l)} &= \tanh\big( W_h^{(l)} \bm{h}_t^{(l-1)} + U_h^{(l)} \big( \bm{re}_t^{(l)} \odot \bm{h}_{t-1}^{(l)} \big) + \bm{b}_h^{(l)} \big) \\
\bm{h}_t^{(l)} &= \big(1 - \bm{up}_t^{(l)}\big) \odot \bm{h}_{t-1}^{(l)} + \bm{up}_t^{(l)} \odot \tilde{\bm{h}}_t^{(l)}
\end{align}
The $l$-th layer, where $l \in \{1, 2, \dots, L\}$ and $L=2$ in the benchmark GRU architecture. $\bm{h}^{(l-1)}_t$ denote the input to the $l$-th layer at time $t$. For the initial layer ($l=1$), the input is the original sequence data, meaning $\bm{h}^{(0)}_t = \bm{x}_t$. Where $\bm{up}_t^{(l)}$ and $\bm{re}_t^{(l)}$ represent the update gate and reset gate of the $l$-th layer, respectively. $\tilde{\bm{h}}_t^{(l)}$ denotes the candidate hidden state, and $\bm{h}_t^{(l)}$ is the final hidden state output of the $l$-th layer at time $t$. $\sigma(\cdot)$ is the sigmoid activation function, and $\tanh(\cdot)$ is the hyperbolic tangent function. The weight matrices $W^{(l)}_{up},\, W^{(l)}_{re},\, W^{(l)}_h \in \mathbb{R}^{d_l \times d_{l-1}}$ and $U^{(l)}_{up},\, U^{(l)}_{re},\, U^{(l)}_h \in \mathbb{R}^{d_l \times d_l}$ are the learnable parameters of the gates and hidden states for the $l$-th layer, where $d_l$ is the dimension (number of neurons) of the $l$-th layer. It is important to note that $d_0 = d_{\text{in}}$, indicating the dimension of the input of the extracted factors. The terms $\bm{b}^{(l)}_{up},\, \bm{b}^{(l)}_{re},\, \bm{b}^{(l)}_h \in \mathbb{R}^{d_l}$ correspond to the bias vectors for the $l$-th layer, and $\odot$ denotes the Hadamard product.\\

Similar to LSTM, GRU employs a gated system to control the memory length. It removes the cell states of LSTM but embeds the candidate hidden state for a substitution of candidate cell state of LSTM which highly reduces the parameter numbers. In the first GRU layer, the updated gate and reset gate are computed according to the input factors ($x_t$) computed via the MLP autoencoder pretraining process and previous hidden states with the sigmoid activation function. The candidate hidden state is computed through the input and reset gated previous hidden state with the hyperbolic tangent activation function. The output hidden state is computed via update gated hidden state and candidate hidden state. Subsequently, the hidden state computed from the first layer is passed to the second layer for the process to compute the final hidden state which is for derivation of estimated label $\hat{\bm{y}}_t$ through a linear function shows in Equation~\eqref{eq:ln_output_gru_lstm_ch2}. \\
\begin{equation}
\hat{\bm{y}}_t = {\bm{W}_{\hat{y}_{t}}} \bm{h}^{(2)}_t + \bm{b_{\hat{y}_{t}}}
\label{eq:ln_output_gru_lstm_ch2}
\end{equation}
where $\bm{h}_t^{(2)} \in \mathbb{R}^{d_2}$ is the output hidden state of second LSTM or GRU layer, while $\bm{W}_y \in \mathbb{R}^{d_{\text{out}} \times d_2}$ and $\bm{b}_y \in \mathbb{R}^{d_{\text{out}}}$ are the parameter matrix and bias of linear layer, where $d_{out}$ is the output dimension. $\hat{\bm{y}}_t \in \mathbb{R}^{d_{\text{out}}}$ is the estimated label $\hat{\bm{y}}_t$ at time $t$. In the context of asset pricing, the estimated output $\hat{\bm{y}}_t$ is the individual stock excess returns $\hat{\bm{r}}_t$.\\ 

\subsection{Estimation}\label{subsec:estimation_ch2}
For estimating the main module, RNN with an attention layer, of the proposed model, the MSE function is selected as the loss function, and the L1 regularization term is added to the loss function as a regularization penalty. Concretely, it can be presented as:
\begin{equation}
\mathcal{L} = \frac{1}{T} \sum_{t=1}^{T} \left( y_t - \hat{y}_t \right)^2 + \lambda \left\| \boldsymbol{W} \right\|_1
\end{equation}
where $\mathcal{L}$ is denoted as loss function, $T$ is the length of the entire time window and $y_t$ is the actual value of the label at time $t$. $\boldsymbol{W}$ represents all parameters in the model, and $\lambda$ is the regularization coefficient which requires pre-setting. The best parameters for prediction can be derived from minimizing the loss function via the stochastic gradient descent (SGD) method or the adaptive moment estimation (Adam) optimizer for improving the model training efficiency and mitigating gradient explosion and vanishing problems. The SGD and Adam optimizer are recruited for the estimation task, which is shown in the appendix of this paper. \\ 
\begin{equation}
\boldsymbol{W}^* = \arg\min_{\boldsymbol{W}} \left\{ \frac{1}{T} \sum_{t=1}^{T} \left( y_t - \hat{y}_t \right)^2 + \lambda \left\| \boldsymbol{W} \right\|_1 \right\}
\label{eq:argmin_mse_ch2}
\end{equation}
Equation~\eqref{eq:argmin_mse_ch2} depicts how to derive the best parameters from minimizing the MSE loss function, where $\boldsymbol{W}^*$ is denoted as the best parameters selected for forecasting.\\ 

\subsection{Forecasting}\label{subsec:LSTM_GRU_ch2}
One-step ahead forecasting for the RNN attention models can be derived from Equation~\eqref{eq:ln_yhat_ch2}
\begin{equation}
\hat{y}_{t+1} = {\bm{W}_{\hat{y}_{t}}} {\bm{z}_{t+1}} + b_{\hat{y}_{t}}\label{eq:ln_yhat_ch2}
\end{equation}
And the one-step ahead forecasting for the RNN, LSTM and GRU models is
\begin{equation}
\hat{\bm{y}}_{t+1} = {\bm{W}_{\hat{y}_t}} \bm{h}^{(2)}_{t+1} + b_{\hat{y}_{t}}
\end{equation}
It applies the best weights and biases to the outputs from the main models to derive the estimated stock excess return at time $t+1$. \\

Similar to \citet{lai2025multilayer}, the recurrent neural network models require intensive computational power; they are still infeasible in laptops or desktops, which are only equipped with CPUs. Therefore, they were implemented via Python (v3.8) using PyTorch \citep{paszke2019pytorch} packages, such as `nn.RNN', `nn.LSTM' and `nn.GRU'. The sliding window sparse attention mechanism is from the `longformer' package of the `transformers' \citep{wolf2020transformers} provided by Hugging Face, specifically the function of the `LongformerSelfAttention'. Training was conducted on NVIDIA A40 and H100 GPUs provided by the Viking HPC cluster at the University of York.\\

\section{Empirical results}\label{subsec:Empirical_results_ch2}
The proposed models and benchmark models are compared from two angles: model performance (predictive power) and backtesting performance. Model performance is measured by the mean of 420 stocks' the out-of-sample (OOS) coefficient of determination ($R_{OOS}^2$), mean square error (MSE), residual $\alpha$ and its statistics, which are shown in Table~\ref {tab:model_perform_ch2}. The algorithms of these indicators are presented in Equation~\eqref{eq:r2_ch2} to Equation~\eqref{eq:alpha_ch2}. The research in this chapter still employs the one-step-ahead forecasting method.
\begin{equation}
R^2_{\text{oos}} = 1 - \frac{\sum \left( r_{i,t+1} - \hat{r}_{i,t+1} \right)^2}{\sum \left( r_{i,t+1} - \overline{r}_{\text{in}} \right)^2}
\label{eq:r2_ch2}
\end{equation}
\begin{equation}
MSE = \frac{1}{TN} \sum_{t=1}^{T} \left( r_{i,t} - \hat{r}_{i,t} \right)^2
\label{eq:mse_ch2}
\end{equation}
\begin{equation}
\alpha_{avg} := \frac{1}{N} \left[ E(r_{i}) - E(\hat{r}_{i}) \right]\label{eq:alpha_ch2}
\end{equation}
Where $r_{i,t+1}$ are the real out-of-sample excess returns, $\hat{r}_{i,t+1}$ is the predicted excess returns, which is equal to $\hat{y}_{t+1}$ in Section~\ref{sec:Models_ch2} and $\overline{r}_{\text{in}}$ is the mean of the excess returns in the training and validation (in-sample) period. $i$ indicates the stock $i$, and $TN$ is the time length observation number multiplied by the number of stocks. Here, $\alpha$ denotes the OOS average estimation error in model evaluation, but it has a specific meaning in the asset pricing context, as introduced in the general introduction section.\\

In the traditional asset pricing concept, $\alpha$ is the intercept of a linear factor model. It means the risk-adjusted return of a stock. If the $\alpha=0$, the stock excess return is perfectly explained by the selected or constructed factors; if $\alpha>0$, it indicates the extra capital gain exists after controlling the risks brought by factors, which is also known as the `arbitrage opportunities'; if $\alpha<0$, it implies the stock underperforms the model's prediction. The latter two circumstances indicate the existence of pricing errors, but practitioners chase the positive $\alpha$ as the abnormal returns beyond the factors. In traditional asset pricing literature, those with a linear model frame, factor exploration research pursues $\alpha$ approaching zero to verify that their factors can perfectly explain the asset returns, which satisfies the non-arbitrage hypothesis. That is slightly different in ML-based asset pricing literature, especially in this case. The linear output layer still makes the asset pricing model explainable, but forces the restriction of the non-arbitrage condition by removing the intercept term of $\alpha$. The computation of $\alpha$ for the ML-based asset pricing is also known as residual $\alpha$, which is employed by \citet{Gu2020EmpiricalLearning,Ma2023AttentionApproach} and is shown in Equation~\eqref{eq:alpha_ch2}. That means if the assumption stands for the linear output layer form, the mean of both pricing error $\alpha$ and model bias $\bm{b}$ should be equal to zero; otherwise, it implies that the $\alpha$, the abnormal return or the arbitrage opportunity exists, or there are unexplored factors. Exploring the structure of $\alpha$, how to minimise $\alpha$, and the relation to the bias of the ML frame are interesting directions for future research. In that sense, the residual $\alpha$ includes the traditional $\alpha$ and the risk premium of the linear and non-linear model structures. The mathematical explanation of the residual $\alpha$ can be found in the Appendix.\\

The Diebold-Mariano (DM) \citep{Diebold1995ComRacy} test with the HAC estimator (preventing loss differential series autocorrelation causes spurious statistical significance) is adopted for evaluating the difference between models in this study. Equation~\eqref{eq:dm1_ch2} to Equation~\eqref{eq:dm4_ch2} presents the mechanism of the DM test. DM test here using the Mean Absolute Error (MAE) as the foundation for its robustness compared with the MSE option. MAE is less sensitive to outliers than MSE, which is particularly suitable for financial time series in AI models that contain more complicated and variable structure in their prediction errors (e.g. non-Gaussian, heavy-tailed, etc.). Therefore, MAE-DM here reflects typical difference forecasting performance and avoids being affected by a few large deviations, which causes false detection of the difference between models. Hence, MAE-DM statistics provide a more realistic and reliable assessment of model superiority in this study.
\begin{align}
d_{t+1} &= \frac{1}{h} \sum_{i=1}^{h} \left( \left| e^{(m)}_{i,t+1} \right| - \left| e^{(n)}_{i,t+1} \right| \right) \label{eq:dm1_ch2} \\
\bar{d} &= \frac{1}{T} \sum_{t=1}^{T} d_{t+1} \label{eq:dm2_ch2} \\
DM_{\text{statistics}} &= \frac{\bar{d}}{d_{\text{standard error}}} \label{eq:dm3_ch2} \\
d_{\text{standard error}} &= \sqrt{\widehat{V}_{HAC} / T} \label{eq:dm4_ch2}
\end{align}
where $|e^{(m)}_{i,t+1}|$, $|e^{(n)}_{i,t+1}|$ are the absolute forecasting errors derived from $(r_{i,t+1} - \hat{r}_{i,t+1})$ of model $m$ and $n$ respectively. $i$ indicates the stock $i$. $h$ is the number of stocks, which is 420 in this case. $d_{t+1}$ is the cross-sectional average difference between model $m$ and model $n$’s absolute forecasting errors, $\bar{d}$ is the mean of $d_{t+1}$. The denominator, $d_{\text{standard error}}$, is the HAC-consistent standard error of the loss differential series $\{d_t\}$. $\widehat{V}_{HAC}$ denotes the heteroskedasticity and autocorrelation consistent (HAC) variance of the loss differential series $d_{t+1}$.\\

For a higher economic and financial interpretation of the model, this study also calculates the factor importance for the MLP autoencoder abstracted factors and the correlation between the abstracted factors and the original factors. Mathematically, the permutation variable importance (VI) can be exhibit as:
\begin{align}
L_{\text{baseline}} &= \mathcal{L}(r_{i,t}, f(X)) \label{eq:perm1_ch2} \\
L_{\text{perm}}(X_i) &= \mathcal{L}(r_{i,t}, f(X_{\text{perm}})) \label{eq:perm2_ch2} \\
\text{Importance}(X_i) &= L_{\text{perm}}(X_i) - L_{\text{baseline}} \label{eq:perm3_ch2}
\end{align}
where $L_{baseline}$ is baseline loss or original out-of-sample loss, $X$ is the original feature matrix, and $f(\cdot)$ is the model forms. $L_{perm(X_i)}$ is the loss that an individual feature $X_i$ is permutated. Then, the importance of $X_i$ is equal to the difference between the permutated loss and the baseline loss. \\

This study uses the trend-following sign signal trading strategy as the market timing approach, and the transaction cost is configured as 50 base points via a traditional static transaction cost calculation method \citep{jegadeesh1993returns,lesmond2004illusory,grundy2001understanding}, and employs a 20 bps \citep{frazzini2018trading,novy2016taxonomy,Hou2020ReplicatingAnomalies} concerning turnover rate (shows in Equation ~\eqref{eq:TC_turnover}) as a robustness examination for exploring how transaction cost and turnover rate affect the backtesting performance of the models in this chapter. For measuring the backtesting performance, it follows the traditional asset pricing indicators, which are annualized returns, Sharpe ratio, Sortino ratio and maximum drawdown (MDD). Equation~\eqref{eq:ar_ch2} to Equation~\eqref{eq:mdd_ch2} presents the calculation of these indicators.
\begin{align}
Annualized\ Return &= \left\{ 1 + \left[ \prod_{i=1}^{n} (1 + r_{i,t}) - 1 \right]^{\frac{12}{n}} \right\} - 1 \label{eq:ar_ch2} \\
Sharpe\ Ratio &= \frac{E(r_p) - r_f}{\sigma_p}\label{eq:sr_ch2} \\
Sortino\ Ratio &= \frac{E(r_p) - r_f}{\sigma_d}\label{eq:so_ch2} \\
Maximum\ Drawdown &= \max_{t \in [0,T]} \left( \frac{c_{\max}(t) - c(t)}{c_{\max}(t)} \right)\label{eq:mdd_ch2}
\end{align}
where $n$ is the number of the observations, $E(r_p)$ is the expected portfolio return and $r_f$ represents the risk-free rate. $\sigma_p$ and $\sigma_d$ are the portfolio and portfolio downside standard deviation, respectively. $C_{\max}(t)$ is the highest value during time $t$, and $C(t)$ is the value at time $t$. For a better comparison of models that perform under the extreme market fluctuations caused by the COVID-19 pandemic, all results are displayed in three periods: Pre–COVID-19 Period (1911), COVID-19–Inclusive Period (2112) and Period Including COVID-19 and One-Year After (2212). \\

The static transaction cost (TC) of 50bps \citep{jegadeesh1993returns,lesmond2004illusory,grundy2001understanding} is used for backtesting, and the turnover-based transaction cost of 20bps (large-cap stocks) \citep{frazzini2018trading,novy2016taxonomy,Hou2020ReplicatingAnomalies} is used for the robustness examination to investigate how the turnover rate influences the profitability of these models. Specifically, the dynamic transaction cost can be mathematically presented as:
\begin{equation}
TC_t = \phi \cdot \sum_{i=1}^{N} |w_{i,t} - w_{i,t-}^{+}|
\label{eq:TC_turnover}
\end{equation}
where $\phi = 0.002$ represents the cost rate (20 basis points), $w_{i,t}$ is the target weight of asset $i$ at time $t$, and $w_{i,t-}^{+}$ is the weight immediately before rebalancing. \\

\begin{figure}[htbp!]
\centering
\includegraphics[width=1\columnwidth]{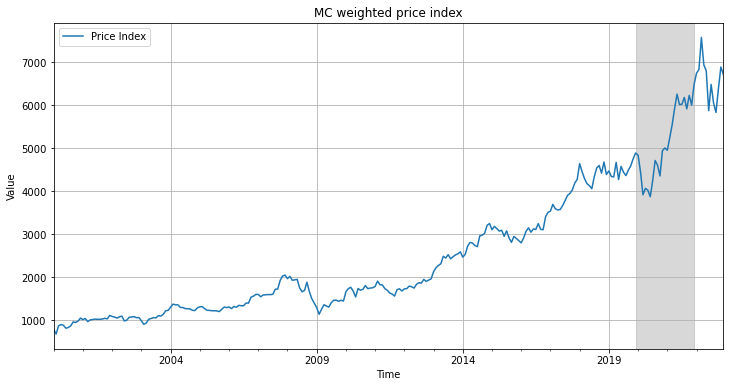}
\caption[Market-cap-weighted (MC) Price Index.]{Market-cap-weighted (MC) Price Index.}
\label{fig:price_index_ch2}
\end{figure}

In addition, Figure~\ref{fig:price_index_ch2} shows the price index of the three periods, which depicts the market fluctuations during the three periods. Concretely, Pre–COVID-19 Period (1911) includes a mild uptrend, COVID-19–Inclusive Period (2112) encompasses Pre–COVID-19 Period (1911) plus a volatile but sharp uptrend, and Period Including COVID-19 and One-Year After (2212) covers COVID-19–Inclusive Period (2112) plus a phase with no significant trend but extreme volatile sideways movement.\\

\subsection{Model performance}
Table~\ref{tab:r2_alpha_ch2} displays the model performance indicators for all periods. It includes the average OOS $R^2$, $MSE$, $\alpha$ and its t-statistics to show whether it is significantly different from zero. From the average OOS $R^2$ of all periods, which indicates the out-of-sample fitness of the models, we can see that all models have positive OOS $R^2$ values. Surprisingly, the vanilla RNN model obtains the highest OOS $R^2$, which is 11.37\% in Period `1911', followed by the RNN global self-attention (self\_att) model (2.8\%) and the RNN sliding window attention (sparse\_att) model (2.45\%). In Period `2112', all models' fitness is slightly lower, apart from the vanilla RNN model and GRU. The RNN model (14.01\%) still outperforms the alternative models significantly, followed by the RNN global self-attention model (2.02\%) and the RNN sliding window attention model (1.94\%). Model GRU's $R^2$ increases considerably from 0.91\% to 1.71\%. For Period `2212', the vanilla RNN model remains the model with the highest OOS $R^2$ value, which is 14.32\% followed by the RNN global attention model (1.64\%) and the RNN sliding window attention model (1.89\%). The OOS $R^2$ of GRU and RNN continuously increases across the three periods. They exhibit persistence in the extreme market turbulence. The lowest OOS $R^2$ is shown by Luong's concatenated attention model (L-Concat) in all periods, worse than the LSTM model. The vanilla RNN exhibits the best predictive power during the extreme market conditions, and even shows strong persistence in the period with steep drawdown (Period 2212). which contains a mild uptrend. Interestingly, the proposed self-attention models (self\_att and sparse\_att) show great fitness in all periods, but self\_att decays quickly during the extreme fluctuation of markets. Instead, the sparse\_att presents significant stability during the extreme conditions, and performs better than the self\_att model when a sharp drawdown (Period 2212) happens. Among the rest of the benchmark attention models, Luong's dot-product (L-DotProd) attention model is superior to the alternative benchmarks, followed by Luong's general attention (L-General) model and additive attention (B-Additive) model, but similar to self\_att, it decays slightly during the extreme market fluctuations. This provides strong evidence that higher-parameterized models have a higher risk of overfitting. It is crucial to control the parameter size in ML-based asset pricing research. Overall, the neural network models with the temporal-capturing mechanism exhibit noticeable persistence during extreme market turbulence. Average OOS MSE agrees with the findings from the average OOS $R^2$.\\

\begin{sidewaysfigure}[htbp!]
\centering
\begin{subfigure}{0.32\textwidth}
  \includegraphics[width=\linewidth, height=0.22\textheight, keepaspectratio]{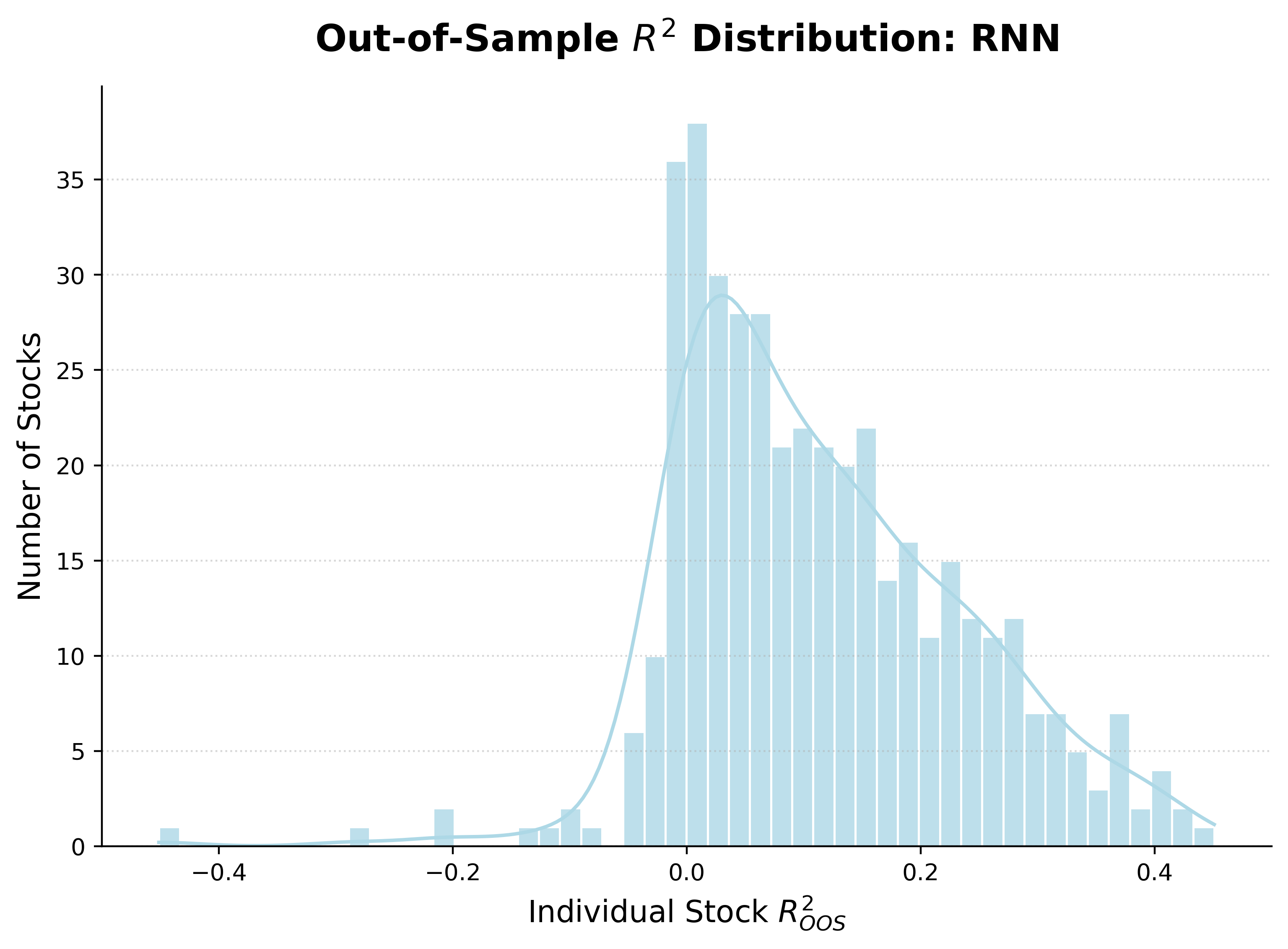}
  \caption{RNN(1911)}
\end{subfigure}
\hfill
\begin{subfigure}{0.32\textwidth}
  \includegraphics[width=\linewidth, height=0.22\textheight, keepaspectratio]{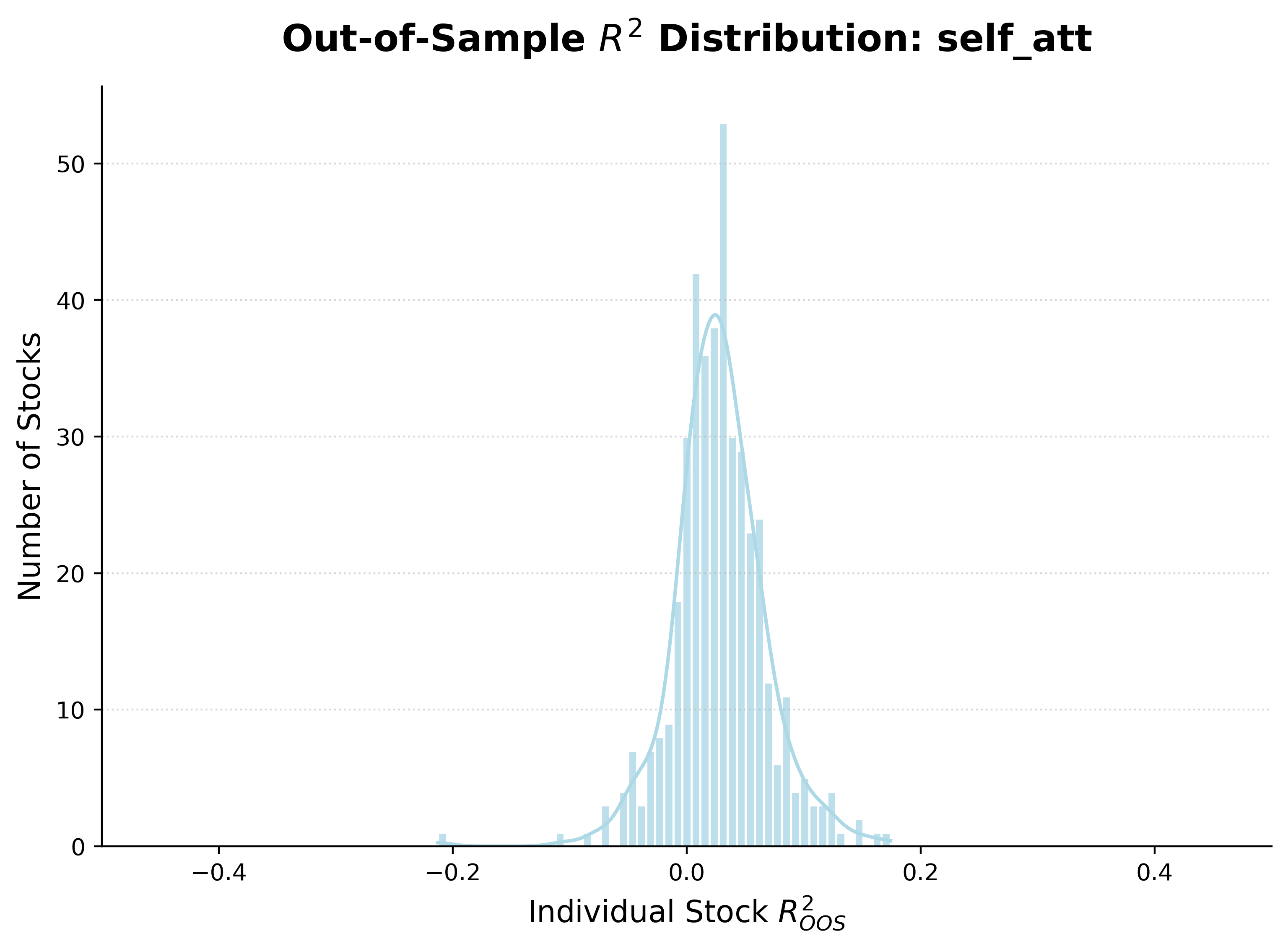}
  \caption{self\_att(1911)}
\end{subfigure}
\hfill
\begin{subfigure}{0.32\textwidth}
  \includegraphics[width=\linewidth, height=0.22\textheight, keepaspectratio]{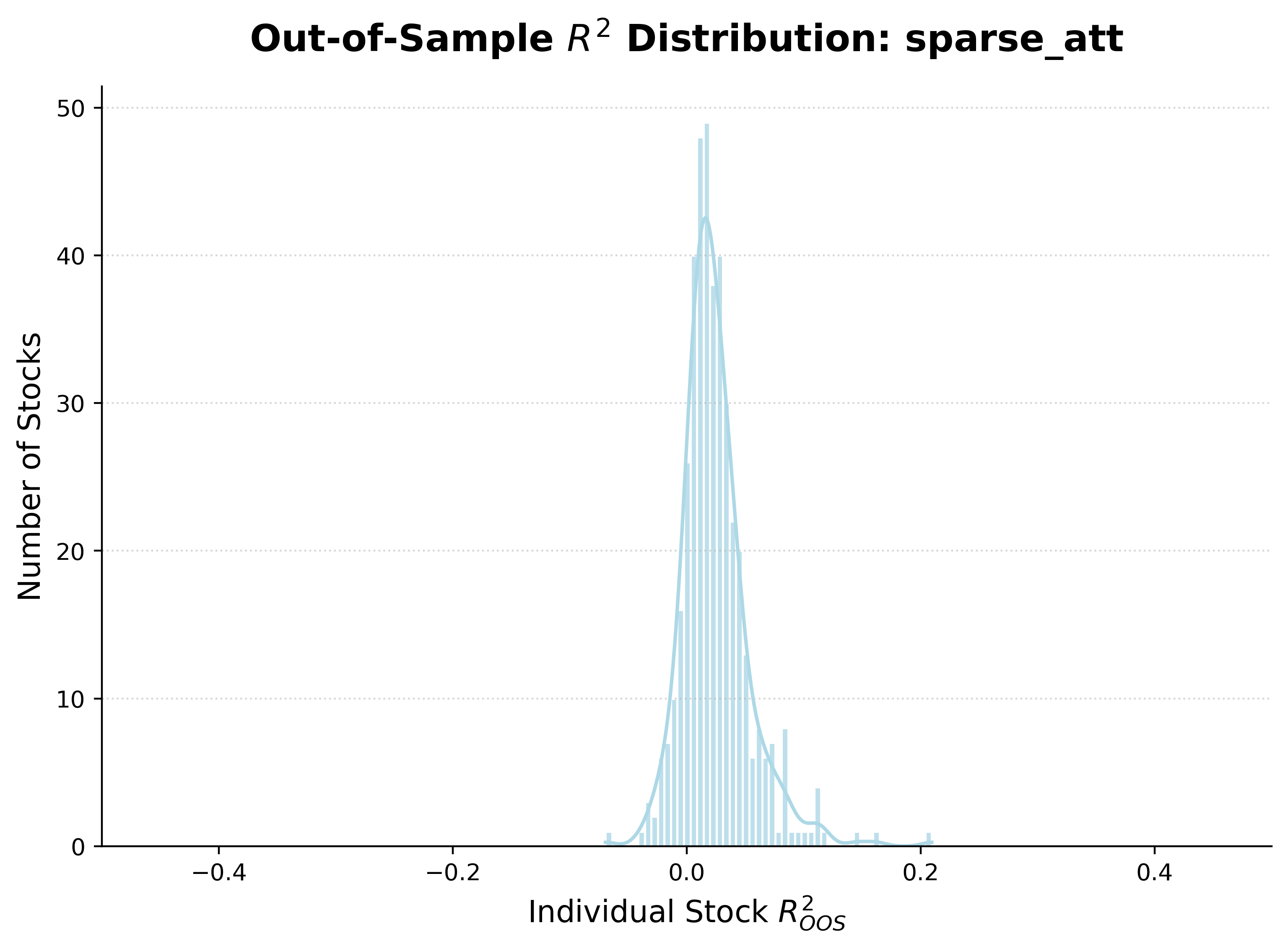}
  \caption{sparse\_att(1911)}
\end{subfigure}

\vspace{0.1em} 

\begin{subfigure}{0.32\textwidth}
  \includegraphics[width=\linewidth, height=0.22\textheight, keepaspectratio]{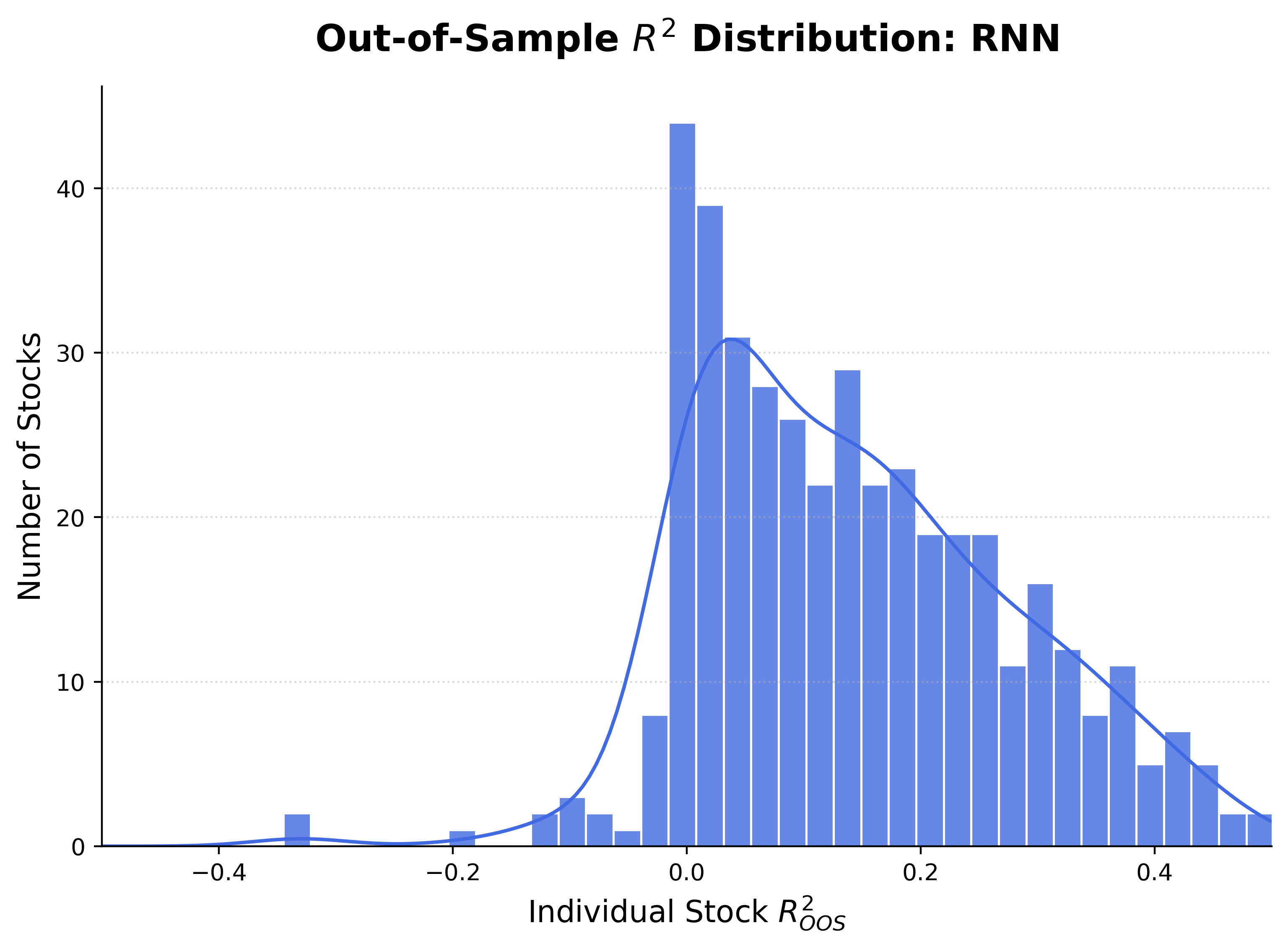}
  \caption{RNN(2112)}
\end{subfigure}
\hfill
\begin{subfigure}{0.32\textwidth}
  \includegraphics[width=\linewidth, height=0.22\textheight, keepaspectratio]{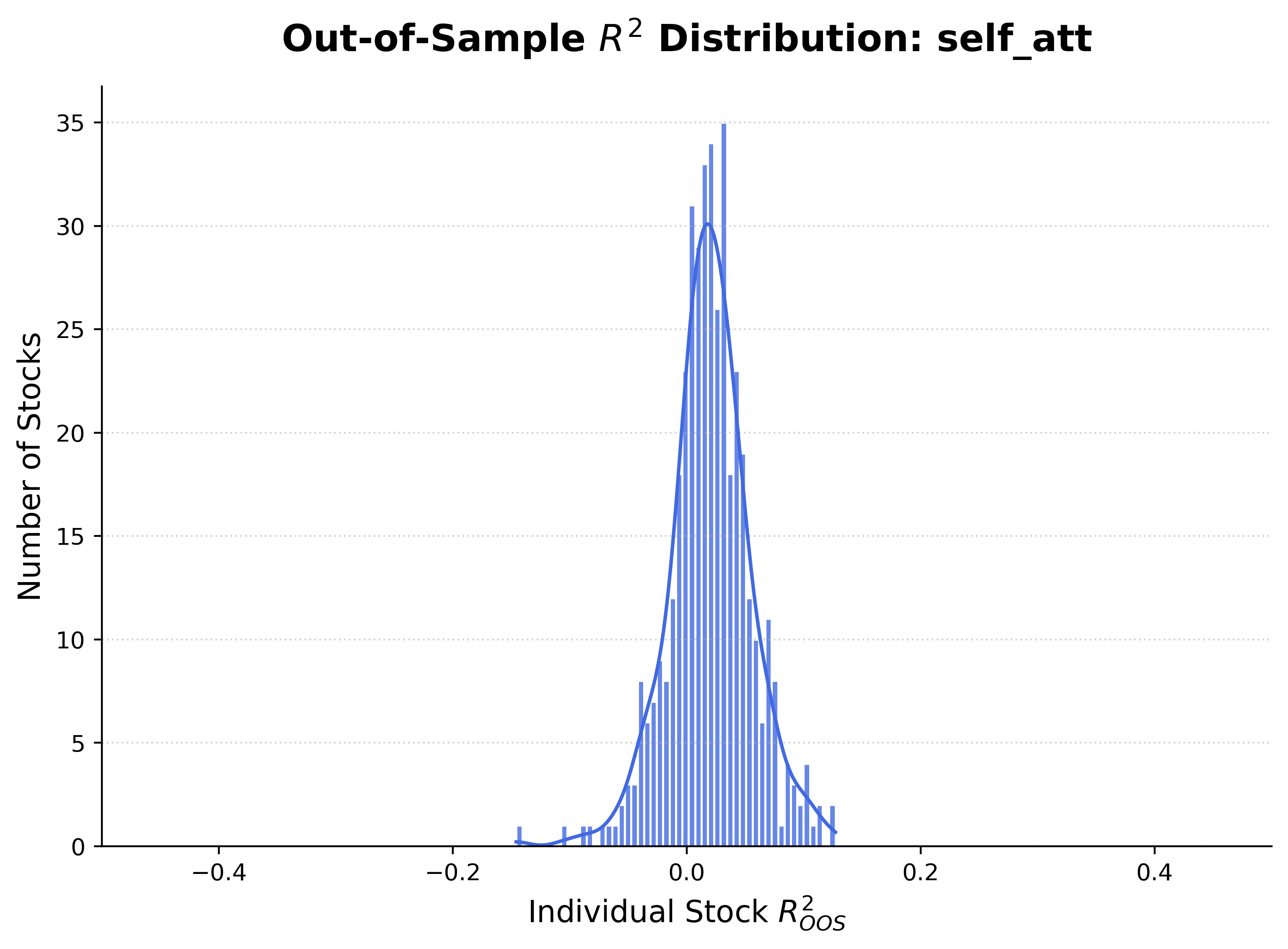}
  \caption{self\_att(2112)}
\end{subfigure}
\hfill
\begin{subfigure}{0.32\textwidth}
  \includegraphics[width=\linewidth, height=0.22\textheight, keepaspectratio]{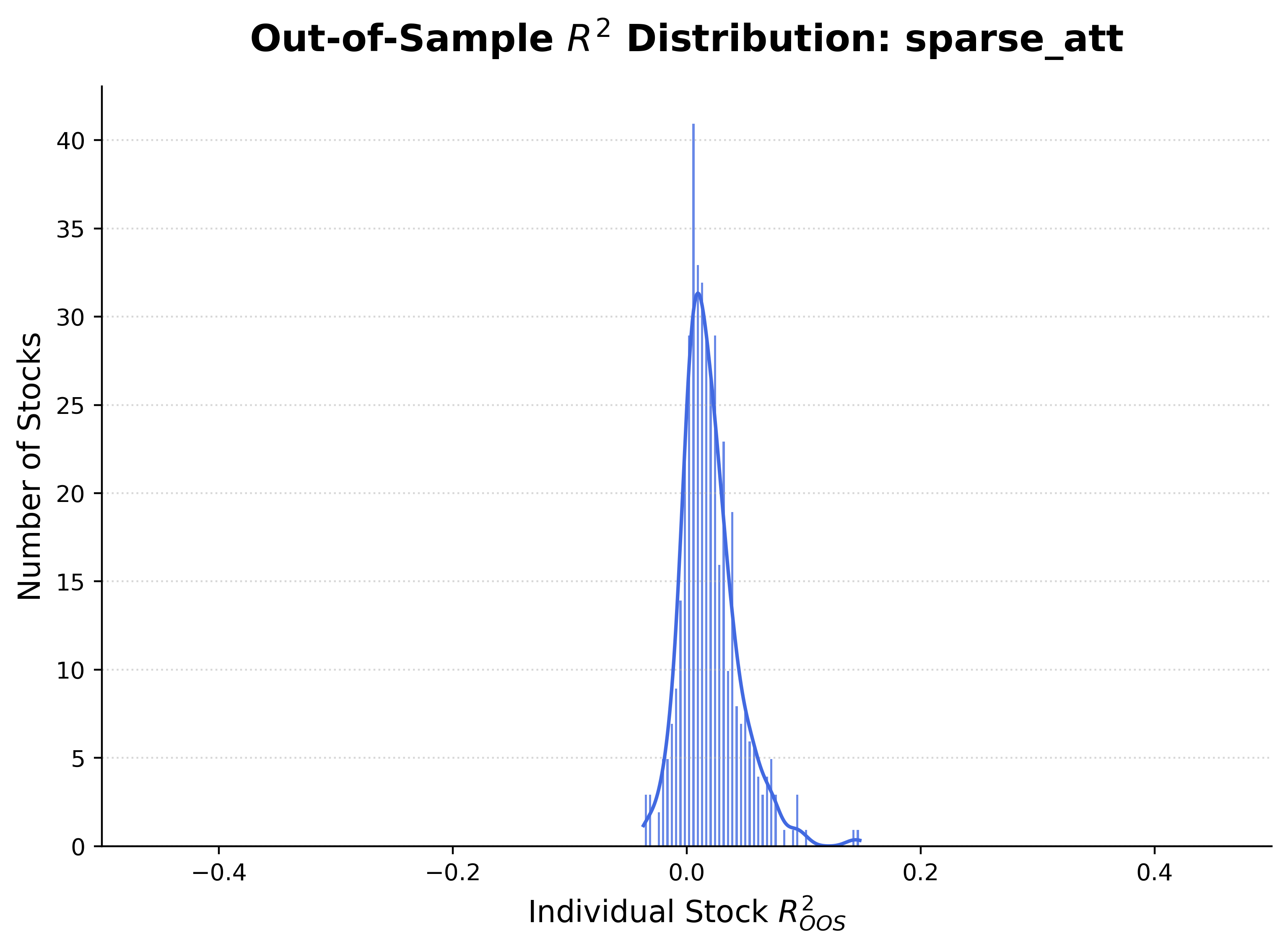}
  \caption{sparse\_att(2112)}
\end{subfigure}

\vspace{0.1em} 

\begin{subfigure}{0.32\textwidth}
  \includegraphics[width=\linewidth, height=0.22\textheight, keepaspectratio]{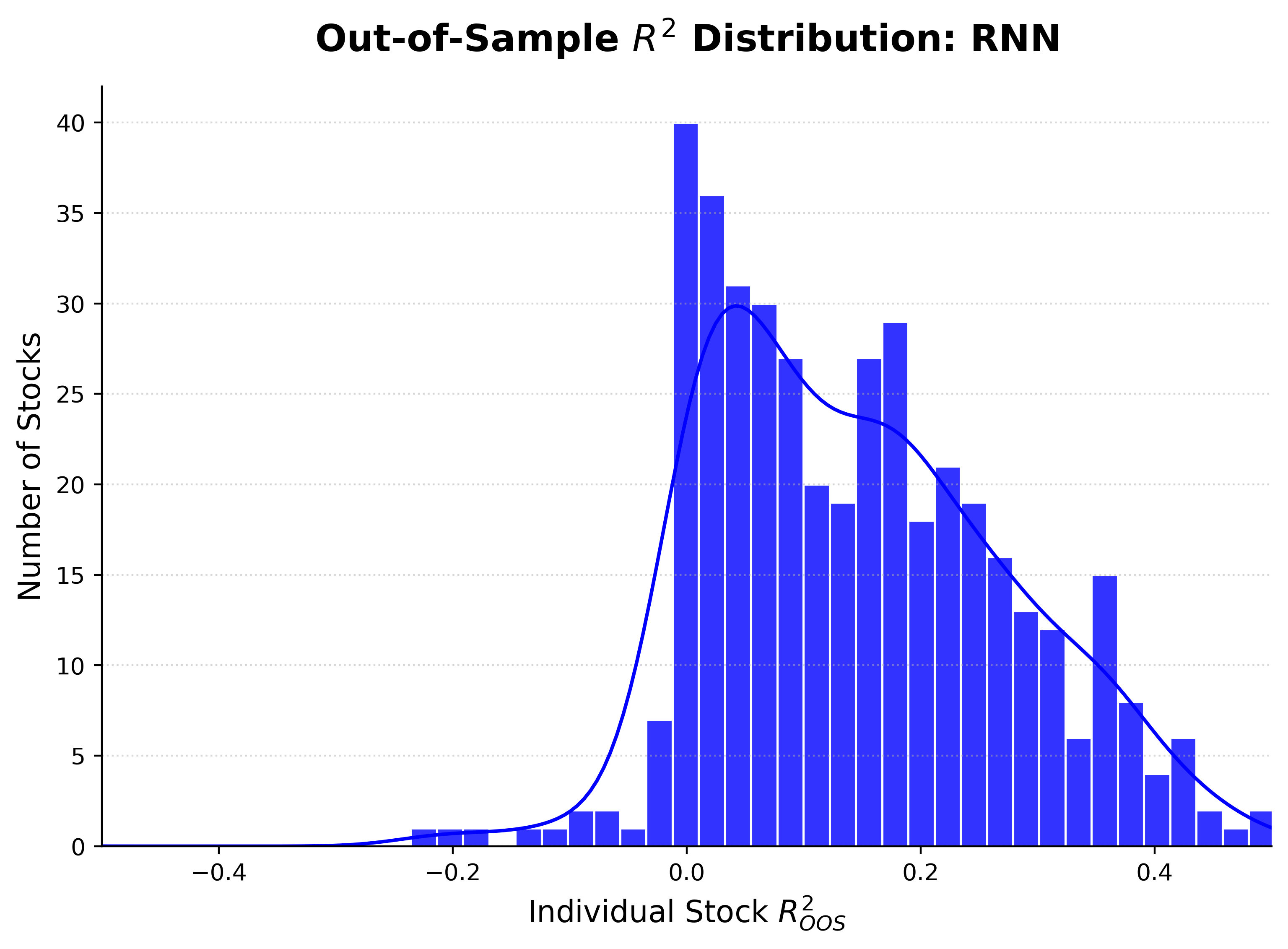}
  \caption{RNN(2212)}
\end{subfigure}
\hfill
\begin{subfigure}{0.32\textwidth}
  \includegraphics[width=\linewidth, height=0.22\textheight, keepaspectratio]{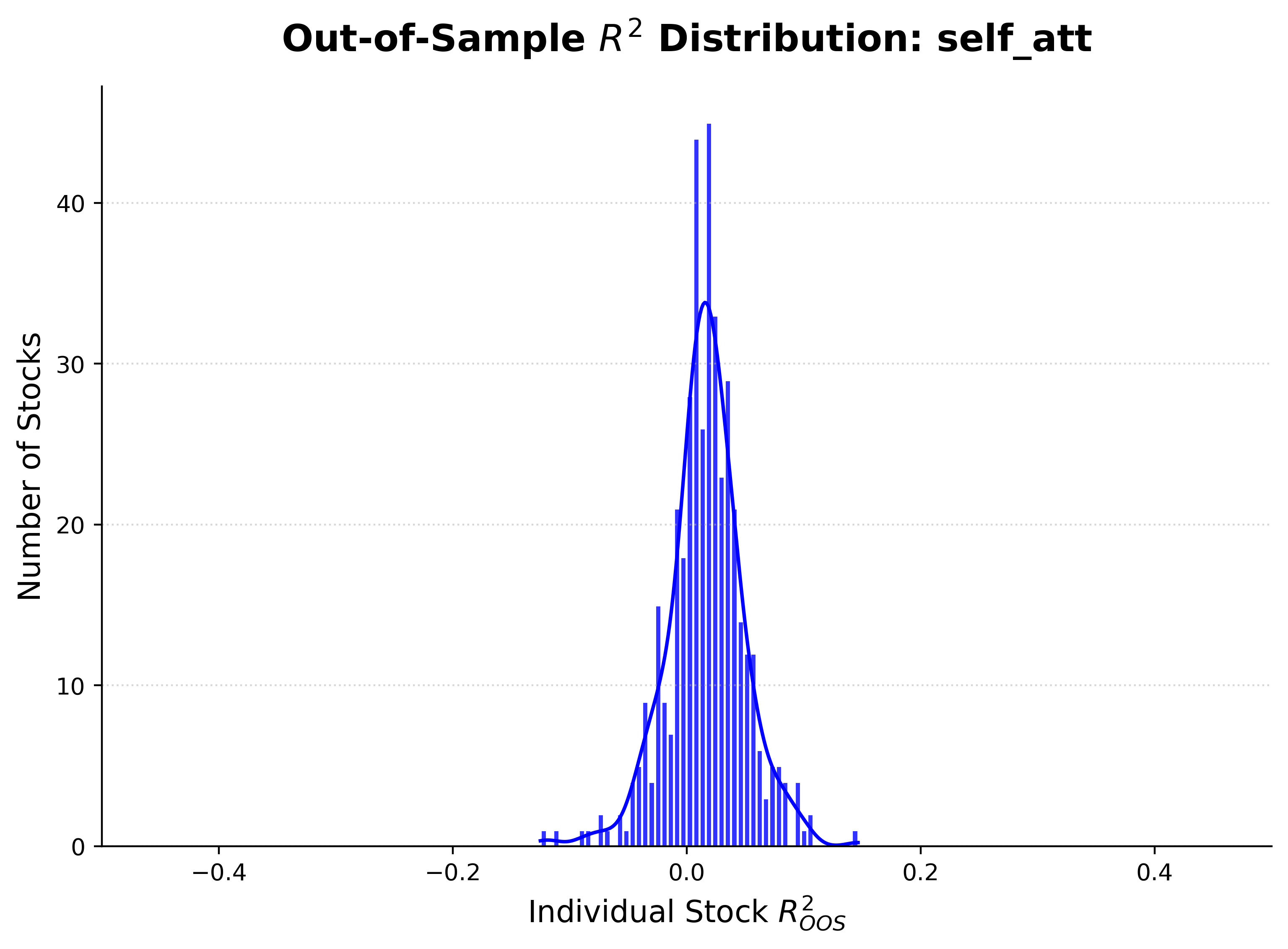}
  \caption{self\_att(2212)}
\end{subfigure}
\hfill
\begin{subfigure}{0.32\textwidth}
  \includegraphics[width=\linewidth, height=0.22\textheight, keepaspectratio]{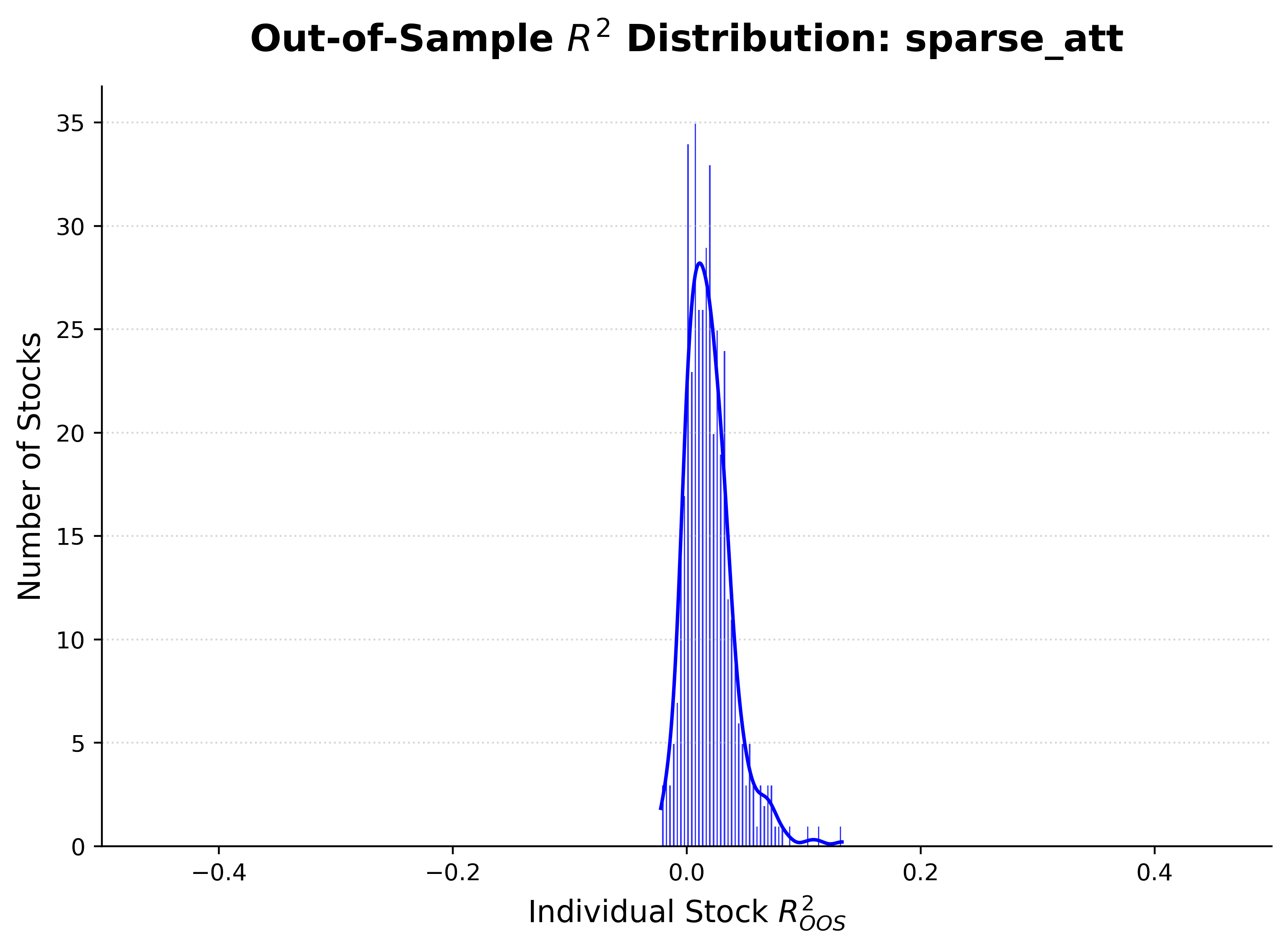}
  \caption{sparse\_att(2212)}
\end{subfigure}
\caption[Out-of-sample $R^2$ distribution of the best fitted models.]{Out-of-sample $R^2$ distribution of the best fitted models}
\label{fig:r2_distribution_ch2}
\end{sidewaysfigure}

\begin{sidewaysfigure}[htbp!]
\centering
\begin{subfigure}{0.32\textwidth}
  \includegraphics[width=\linewidth, height=0.22\textheight, keepaspectratio]{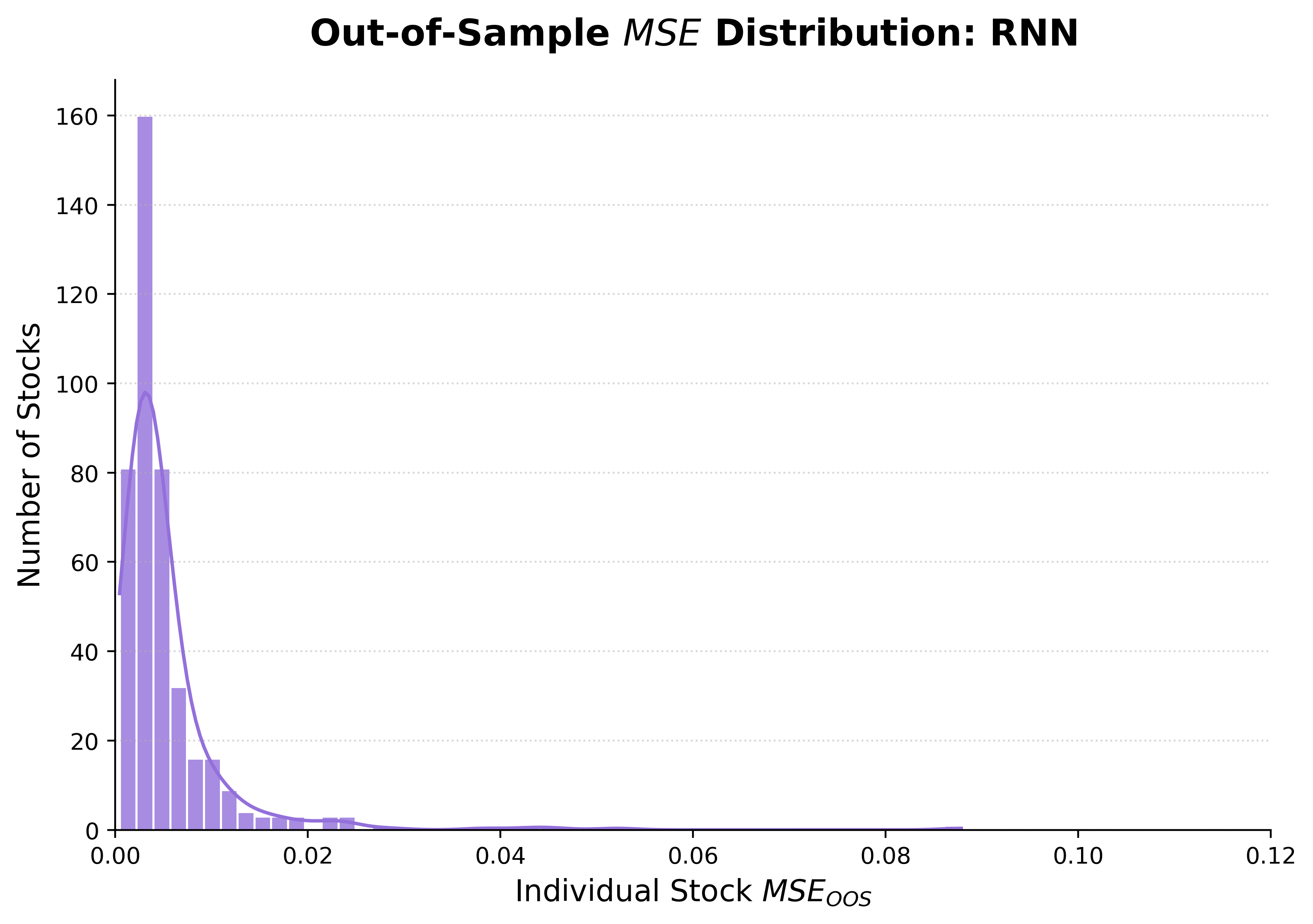}
  \caption{RNN(1911)}
\end{subfigure}
\hfill
\begin{subfigure}{0.32\textwidth}
  \includegraphics[width=\linewidth, height=0.22\textheight, keepaspectratio]{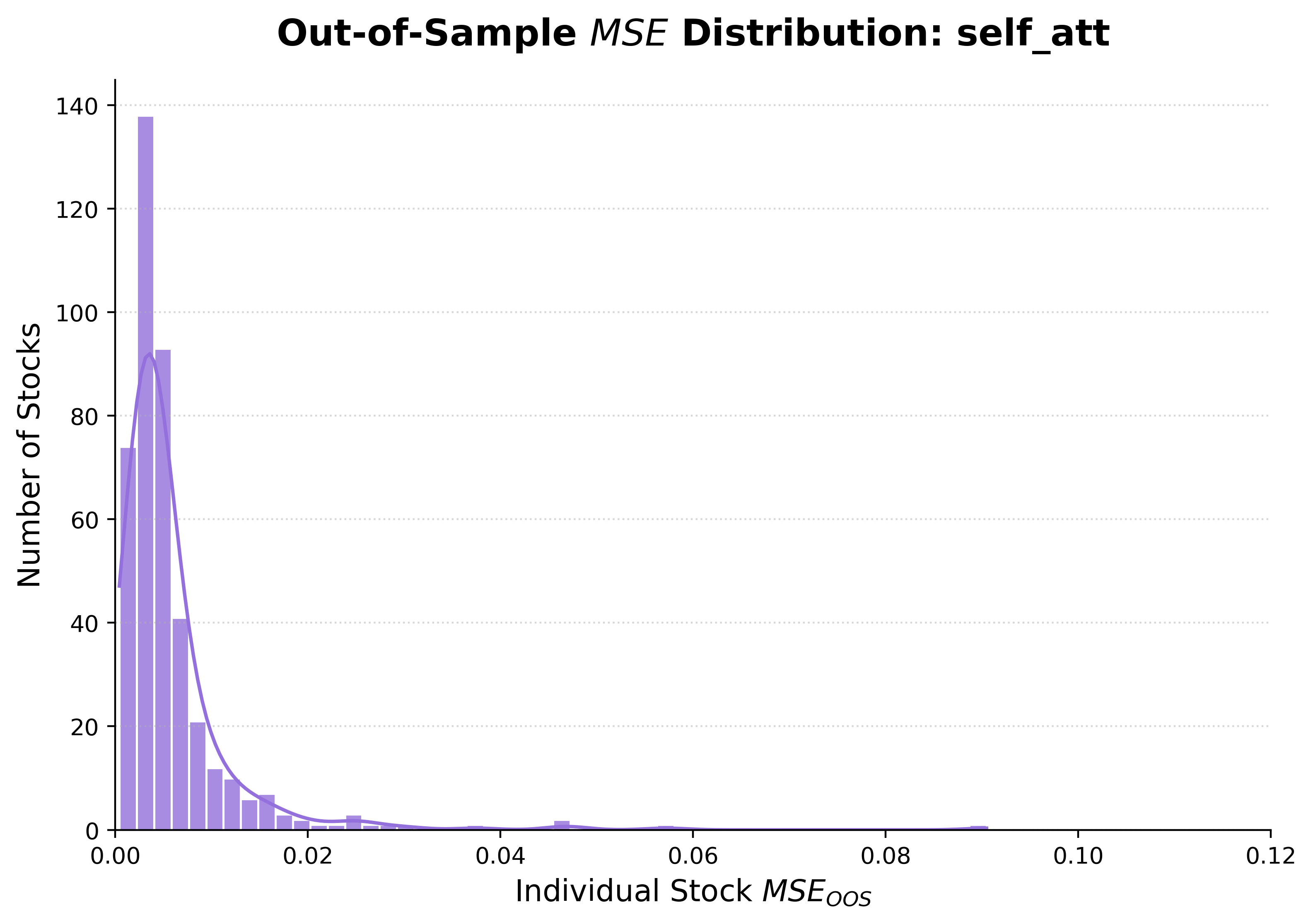}
  \caption{self\_att(1911)}
\end{subfigure}
\hfill
\begin{subfigure}{0.32\textwidth}
  \includegraphics[width=\linewidth, height=0.22\textheight, keepaspectratio]{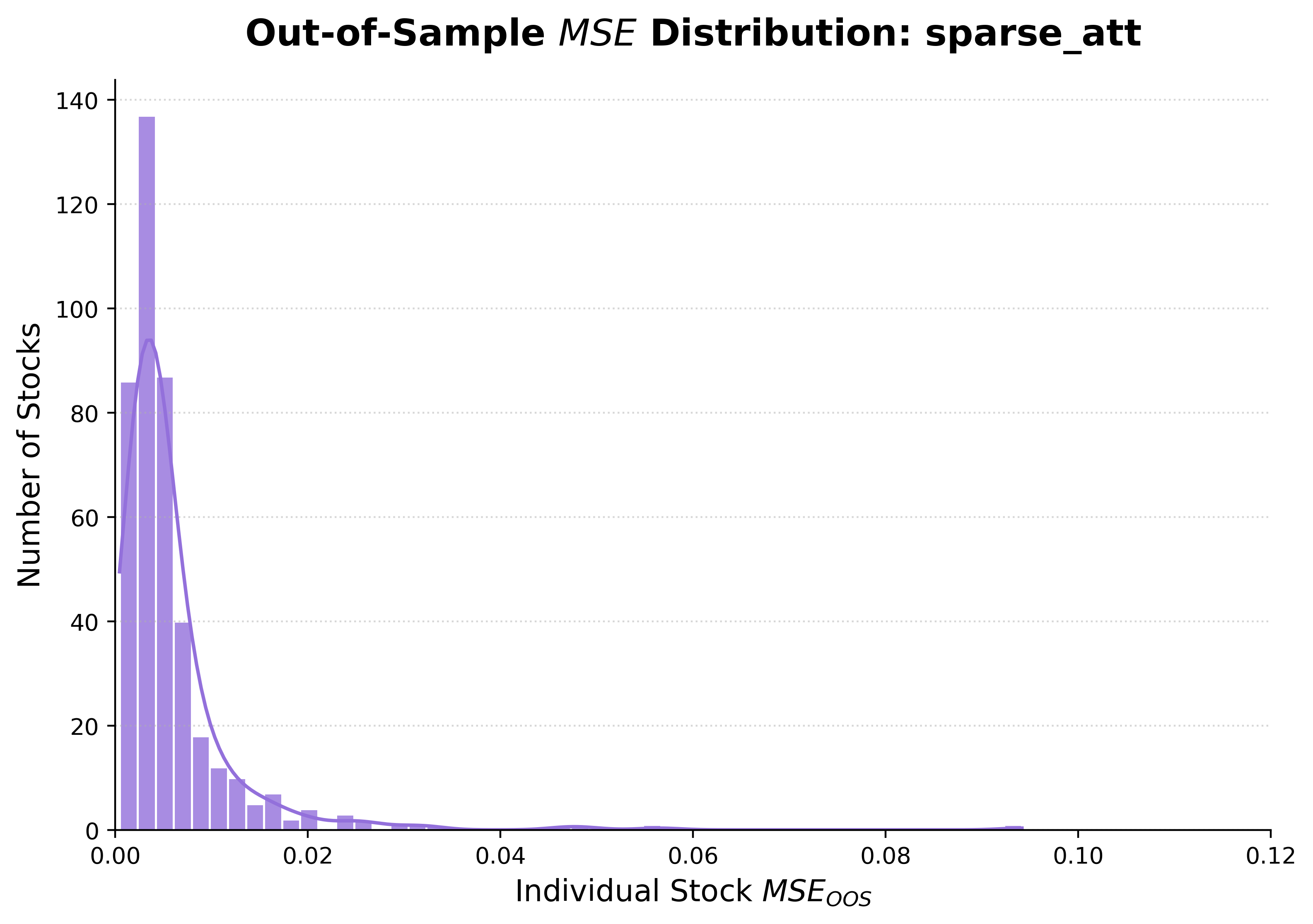}
  \caption{sparse\_att(1911)}
\end{subfigure}

\vspace{0.1em} 

\begin{subfigure}{0.32\textwidth}
  \includegraphics[width=\linewidth, height=0.22\textheight, keepaspectratio]{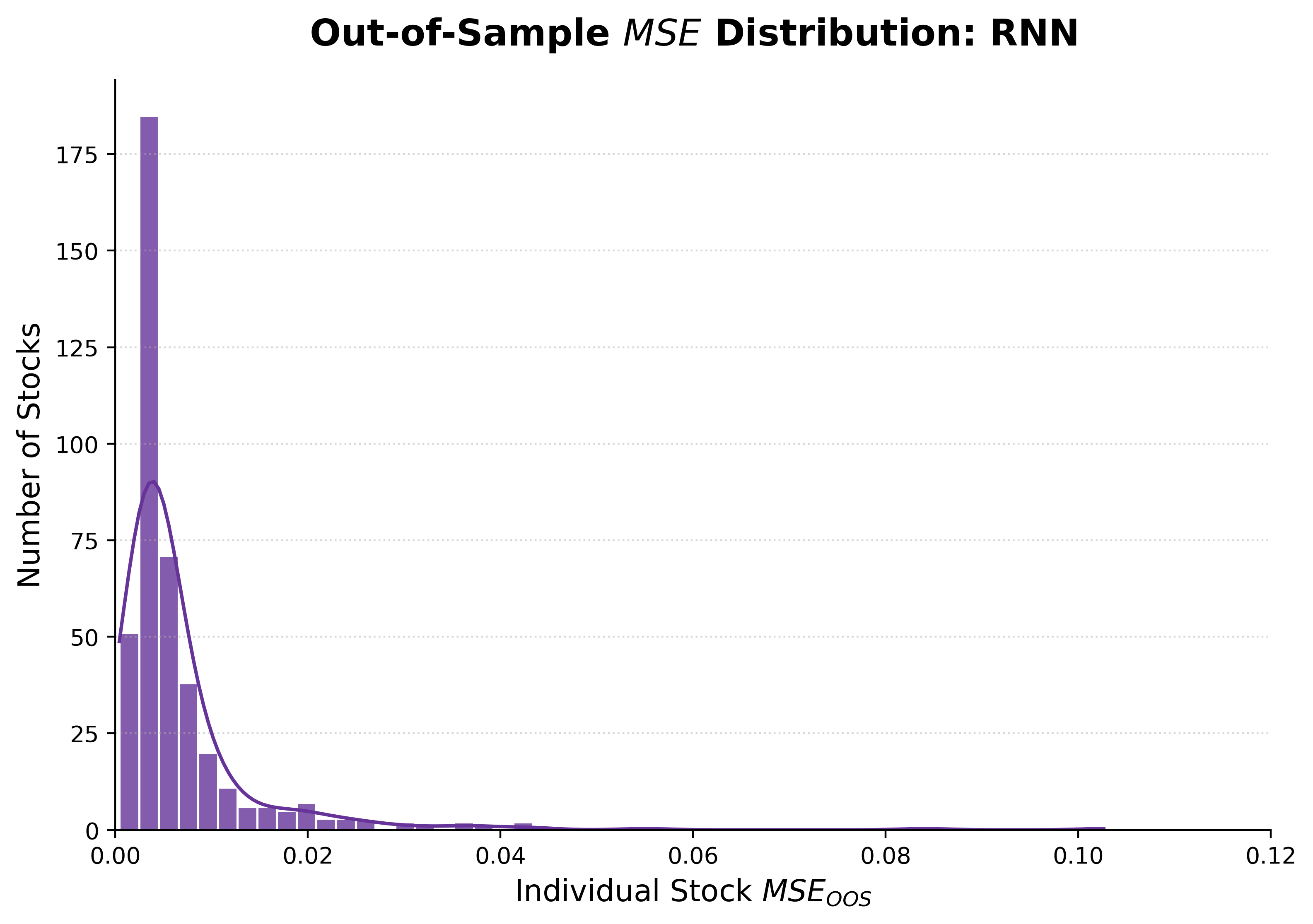}
  \caption{RNN(2112)}
\end{subfigure}
\hfill
\begin{subfigure}{0.32\textwidth}
  \includegraphics[width=\linewidth, height=0.22\textheight, keepaspectratio]{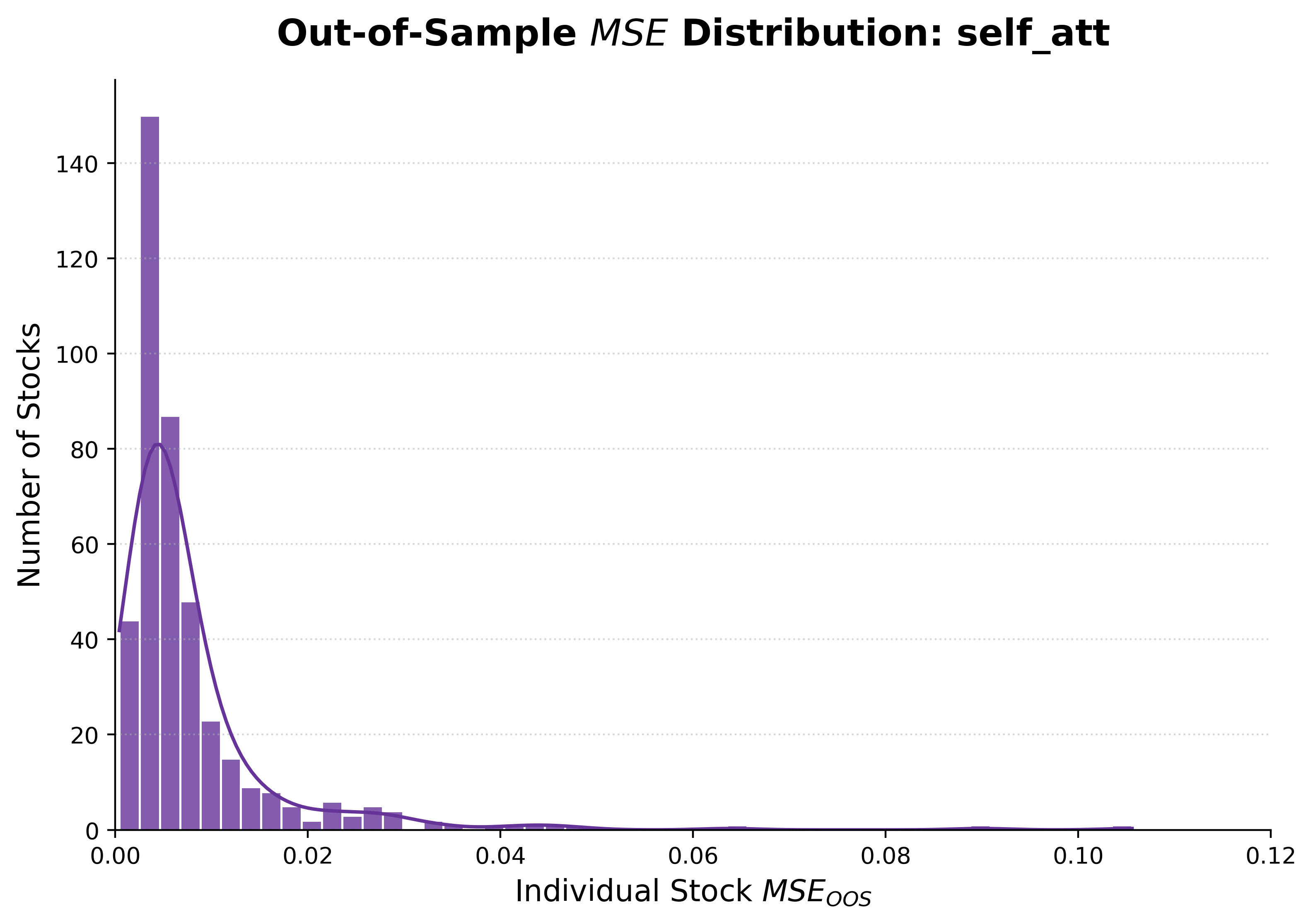}
  \caption{self\_att(2112)}
\end{subfigure}
\hfill
\begin{subfigure}{0.32\textwidth}
  \includegraphics[width=\linewidth, height=0.22\textheight, keepaspectratio]{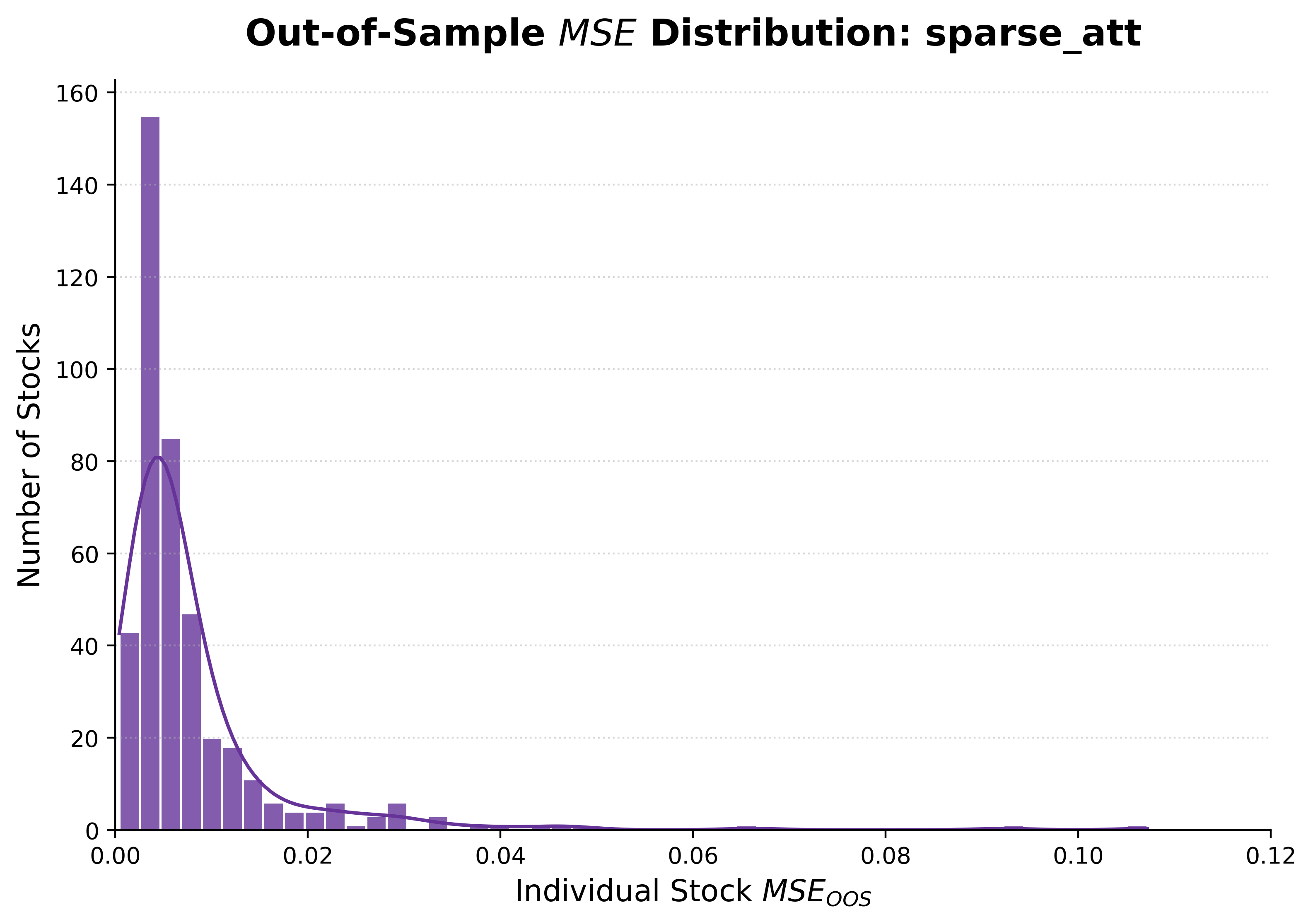}
  \caption{sparse\_att(2112)}
\end{subfigure}

\vspace{0.1em} 

\begin{subfigure}{0.32\textwidth}
  \includegraphics[width=\linewidth, height=0.22\textheight, keepaspectratio]{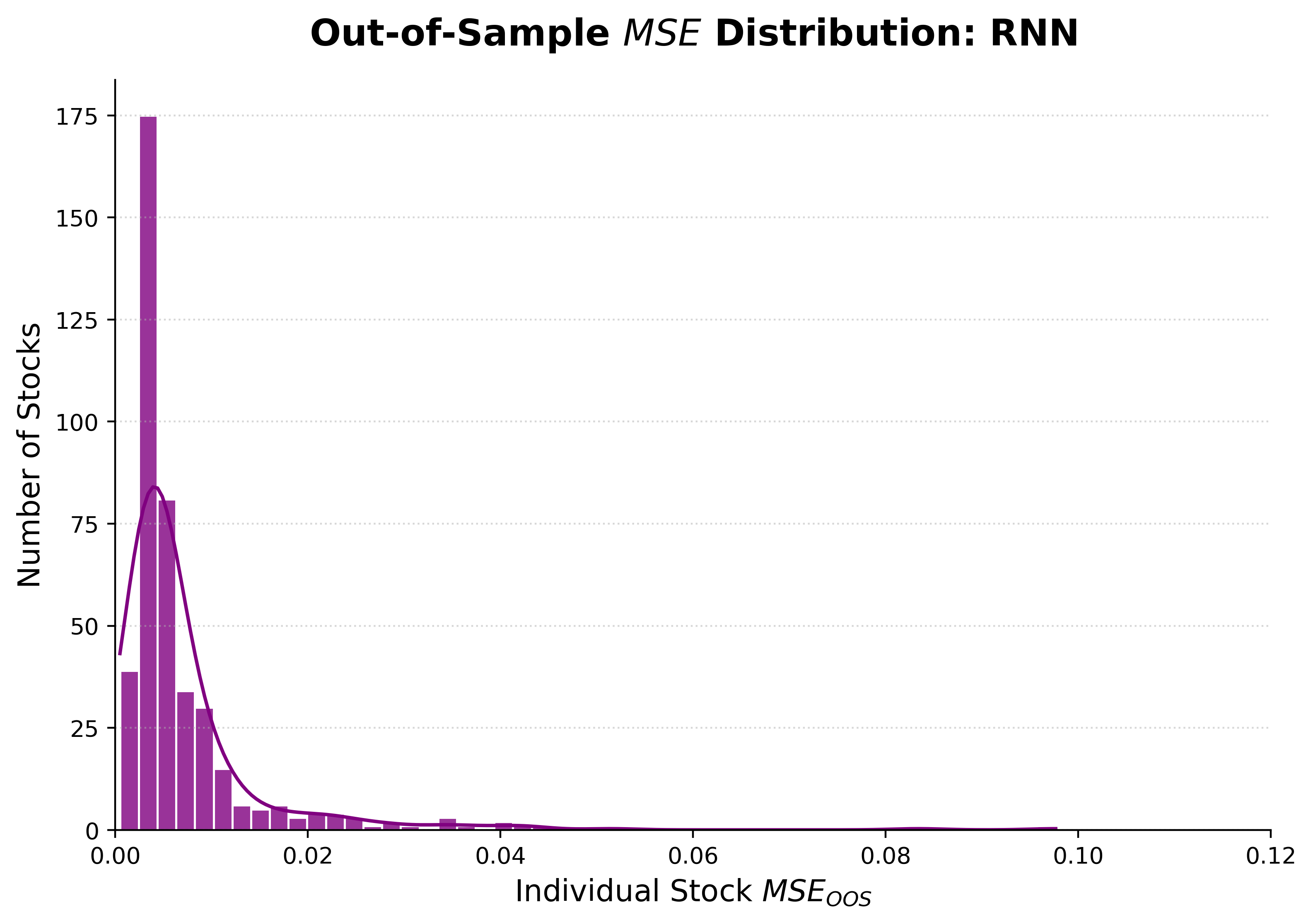}
  \caption{RNN(2212)}
\end{subfigure}
\hfill
\begin{subfigure}{0.32\textwidth}
  \includegraphics[width=\linewidth, height=0.22\textheight, keepaspectratio]{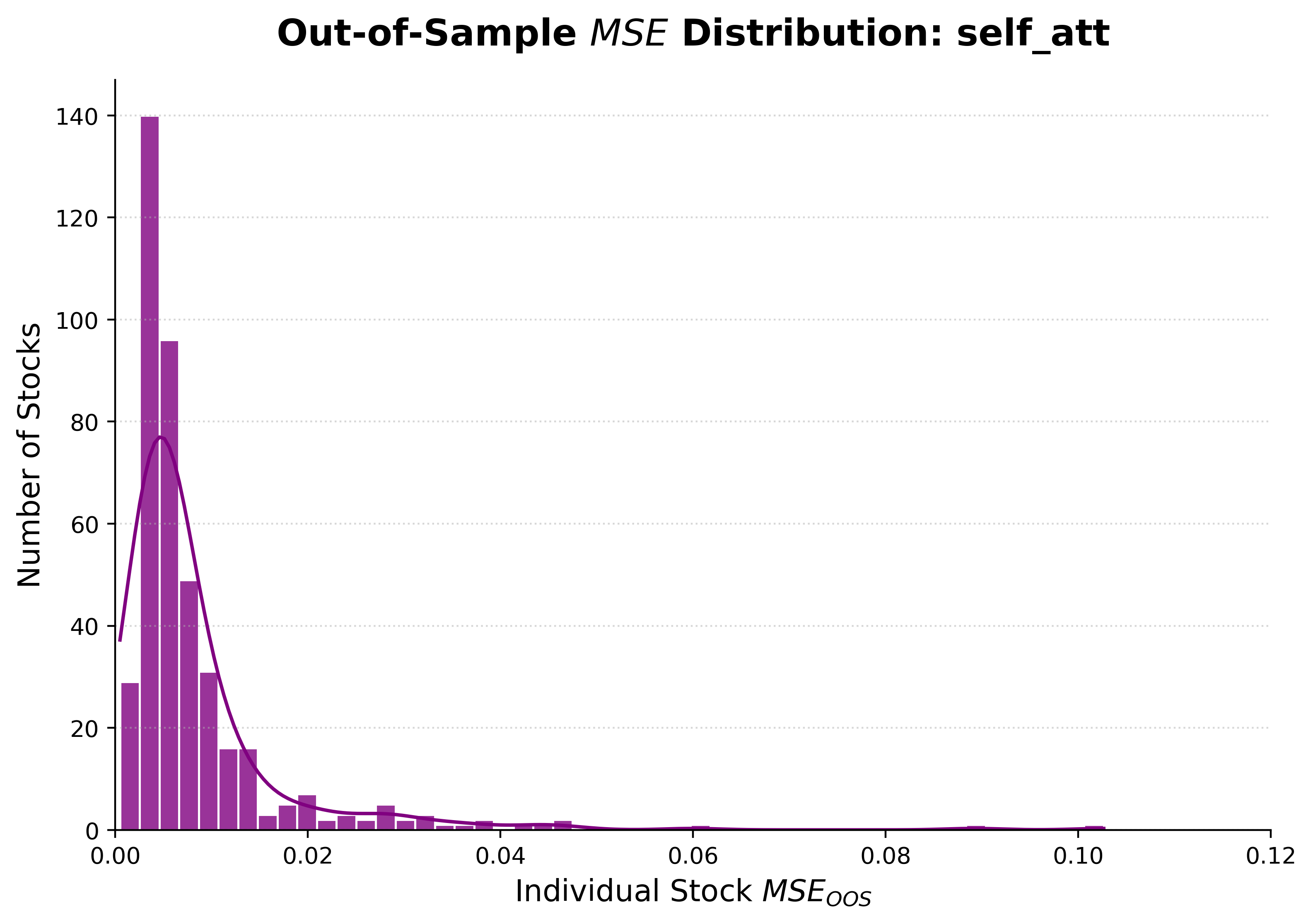}
  \caption{self\_att(2212)}
\end{subfigure}
\hfill
\begin{subfigure}{0.32\textwidth}
  \includegraphics[width=\linewidth, height=0.22\textheight, keepaspectratio]{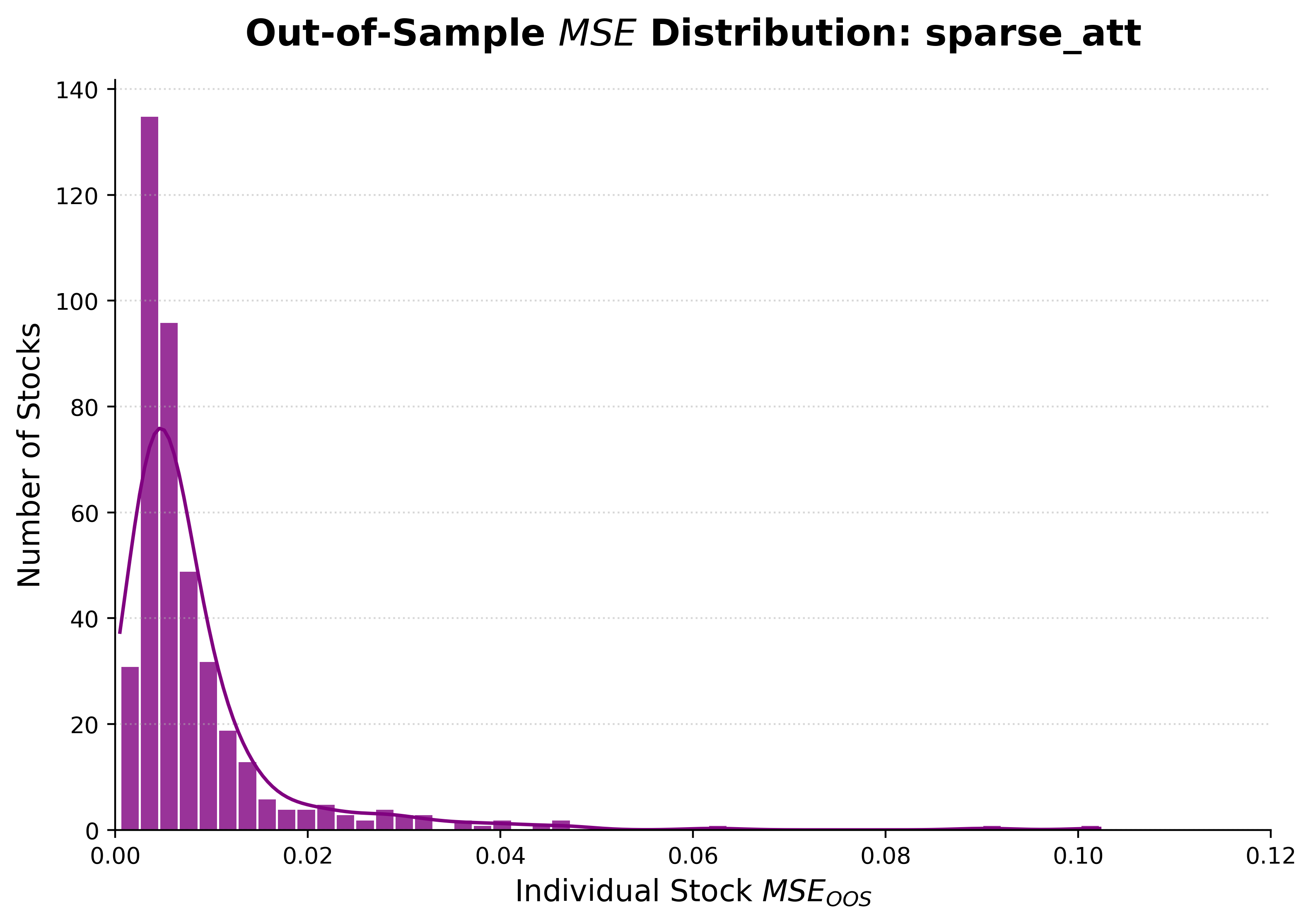}
  \caption{sparse\_att(2212)}
\end{subfigure}
\caption[Out-of-sample MSE distribution of the best fitted models.]{Out-of-sample MSE distribution of the best fitted models}
\label{fig:mse_distribution_ch2}
\end{sidewaysfigure}

\begin{table}[htbp]
  \centering
  \small 
  \begin{tabular}{lccccc}
    \toprule
    Model & Avg\_R2 & Avg\_MSE & avg\_$\alpha$ & Ann\_avg\_$\alpha$ & t\_statistic \\
    \midrule
    \multicolumn{6}{c}{\textbf{Pre-COVID-19 Period (1911)}} \\
    \midrule
    B-Additive & 0.0063 & 0.0060 & 0.0035 & 0.0417 & 9.4624*** \\
    GRU & 0.0091 & 0.0060 & 0.0036 & 0.0429 & 9.5687*** \\
    L-Concat & 0.0016 & 0.0061 & 0.0035 & 0.0423 & 9.5809*** \\
    L-DotProd & 0.0128 & 0.0060 & 0.0035 & 0.0426 & 9.4785*** \\
    L-General & 0.0067 & 0.0060 & 0.0037 & 0.0450 & 10.2364*** \\
    LSTM & 0.0045 & 0.0060 & 0.0038 & 0.0455 & 10.2192*** \\
    RNN & 0.1137 & 0.0053 & 0.0040 & 0.0478 & 10.5572*** \\
    self\_att & 0.0280 & 0.0059 & 0.0028 & 0.0340 & 7.9699*** \\
    sparse\_att & 0.0245 & 0.0059 & 0.0030 & 0.0363 & 8.5817*** \\
    \midrule
    \multicolumn{6}{c}{\textbf{COVID-19--Inclusive Period (2112)}} \\
    \midrule
    B-Additive & 0.0043 & 0.0080 & 0.0053 & 0.0632 & 16.5328*** \\
    GRU & 0.0171 & 0.0079 & 0.0049 & 0.0586 & 14.3459*** \\
    L-Concat & 0.0011 & 0.0080 & 0.0053 & 0.0632 & 15.8036*** \\
    L-DotProd & 0.0102 & 0.0079 & 0.0053 & 0.0641 & 16.5363*** \\
    L-General & 0.0056 & 0.0080 & 0.0054 & 0.0645 & 17.0102*** \\
    LSTM & 0.0023 & 0.0080 & 0.0055 & 0.0660 & 17.2300*** \\
    RNN & 0.1401 & 0.0068 & 0.0039 & 0.0467 & 11.7454*** \\
    self\_att & 0.0202 & 0.0078 & 0.0045 & 0.0540 & 14.2660*** \\
    sparse\_att & 0.0194 & 0.0079 & 0.0048 & 0.0581 & 15.3381*** \\
    \midrule
    \multicolumn{6}{c}{\textbf{Period Including COVID-19 and One-Year After (2212)}} \\
    \midrule
    B-Additive & 0.0058 & 0.0084 & 0.0038 & 0.0453 & 13.0552*** \\
    GRU & 0.0176 & 0.0083 & 0.0036 & 0.0429 & 11.7015*** \\
    L-Concat & 0.0028 & 0.0084 & 0.0038 & 0.0456 & 13.1388*** \\
    L-DotProd & 0.0099 & 0.0083 & 0.0038 & 0.0473 & 13.4858*** \\
    L-General & 0.0069 & 0.0084 & 0.0039 & 0.0469 & 13.8141*** \\
    LSTM & 0.0044 & 0.0084 & 0.0040 & 0.0480 & 14.3907*** \\
    RNN & 0.1432 & 0.0071 & 0.0025 & 0.0303 & 8.5837*** \\
    self\_att & 0.0164 & 0.0083 & 0.0038 & 0.0461 & 13.7656*** \\
    sparse\_att & 0.0189 & 0.0083 & 0.0035 & 0.0416 & 11.8082*** \\
    \bottomrule
  \end{tabular}
\caption[Out-of-sample Predictive Indicators and $\alpha$.]{Out-of-sample Predictive indicators and $\alpha$. `Ann' is the notation of annualization, and `avg' is short for `average'. $*$, $**$ and $***$ depict the statistical significance level of 90\%, 95\% and 99\% respectively.}
\label{tab:r2_alpha_ch2}
\end{table}
Figure~\ref{fig:r2_distribution_ch2} and Figure~\ref{fig:mse_distribution_ch2} are the OOS $R^2$ and MSE distribution plots of models with the best OOS fitness. The OOS $R^2$ distribution plots show that the vanilla RNN model has the majority of stocks with the positive OOS $R^2$ during all periods, where they are distributed between 0 to 0.45. There are some similarities between the distribution of the RNN global self-attention and the RNN sliding window attention model: more than 60\% of the stocks have the positive OOS $R^2$ values, and both of them are right-skewed in all periods. However, the RNN sliding window attention (sparse\_att) model is more concentrated than the RNN global attention model (self\_att), which explains why the sparse\_att model performs more stably than the self\_att model. The MSE distribution plots agree with the findings from the OOS $R^2$ distribution plots.\\

\begin{sidewaysfigure}[htbp!]
\centering
\begin{subfigure}{0.32\textwidth}
  \includegraphics[width=\linewidth, height=0.22\textheight, keepaspectratio]{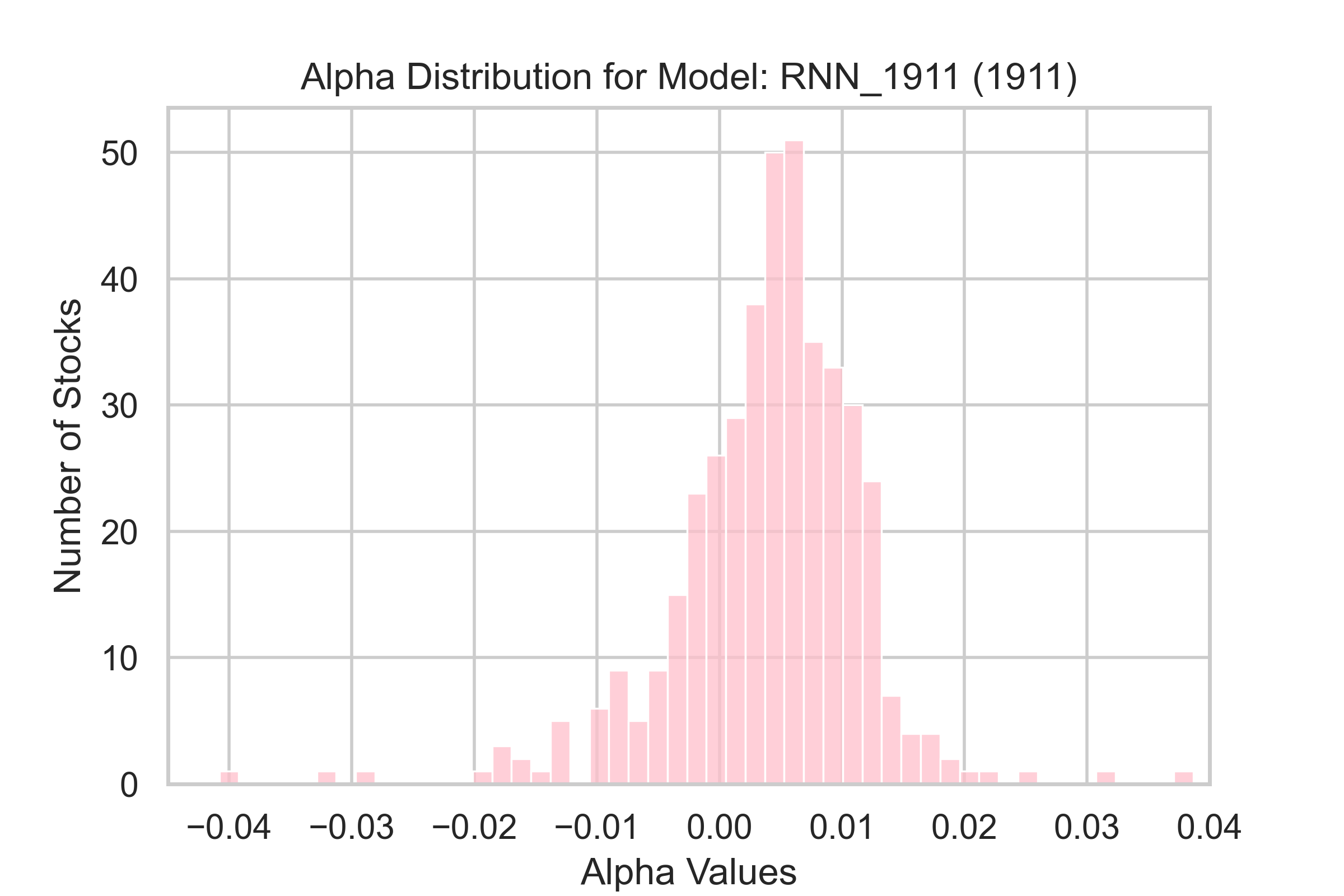}
  \caption{RNN(1911)}
\end{subfigure}
\hfill
\begin{subfigure}{0.32\textwidth}
  \includegraphics[width=\linewidth, height=0.22\textheight, keepaspectratio]{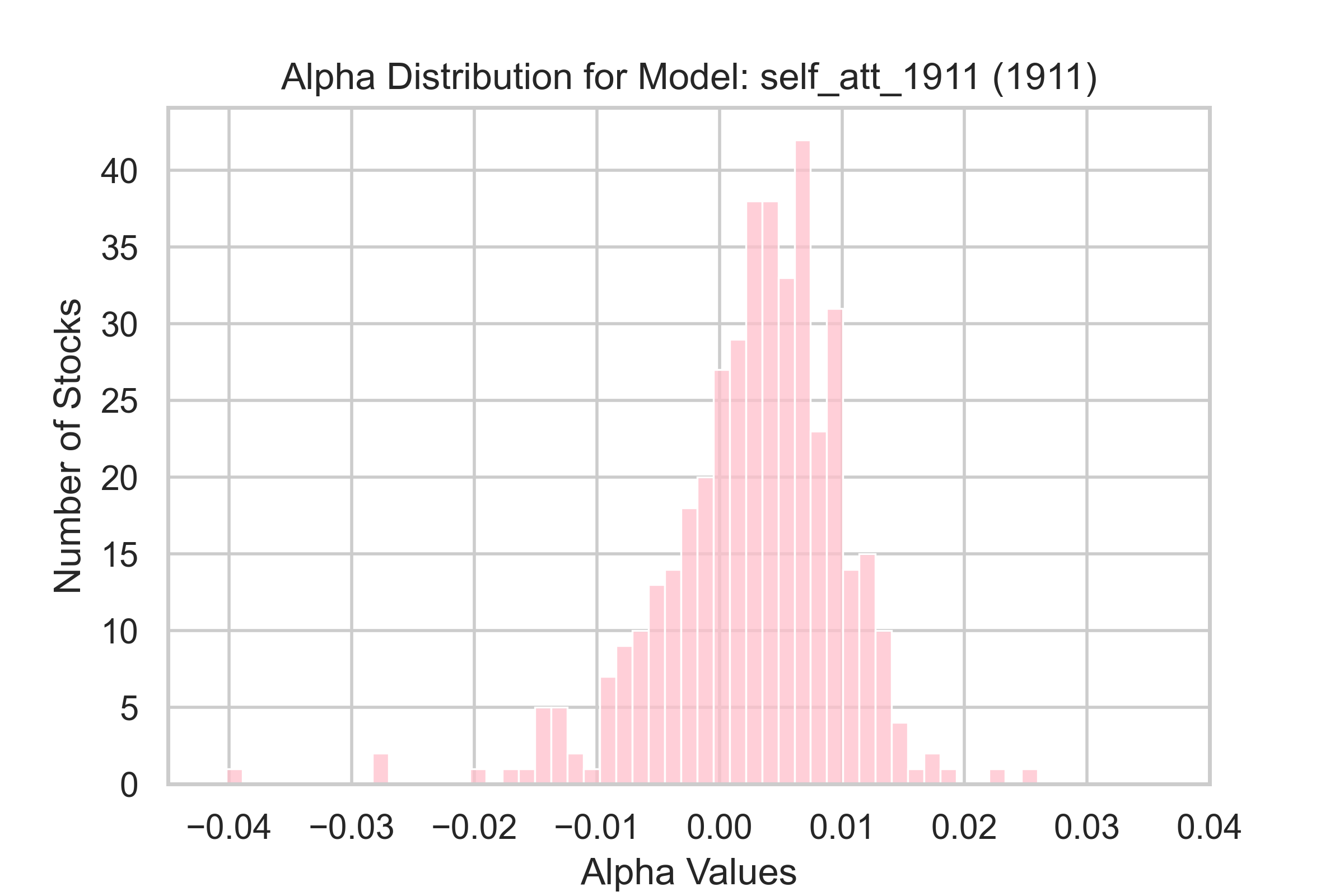}
  \caption{self\_att(1911)}
\end{subfigure}
\hfill
\begin{subfigure}{0.32\textwidth}
  \includegraphics[width=\linewidth, height=0.22\textheight, keepaspectratio]{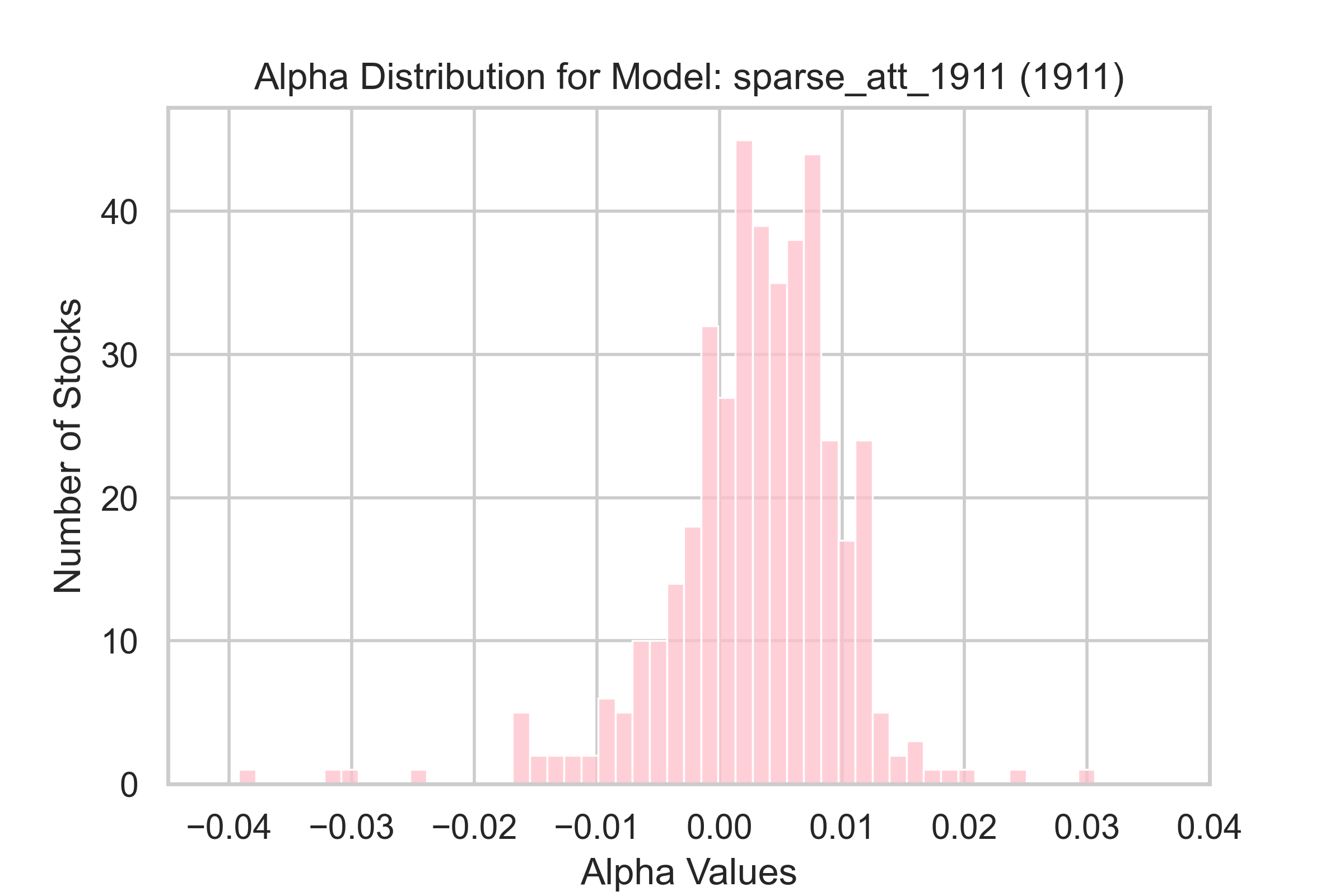}
  \caption{sparse\_att(1911)}
\end{subfigure}

\vspace{0.1em} 

\begin{subfigure}{0.32\textwidth}
  \includegraphics[width=\linewidth, height=0.22\textheight, keepaspectratio]{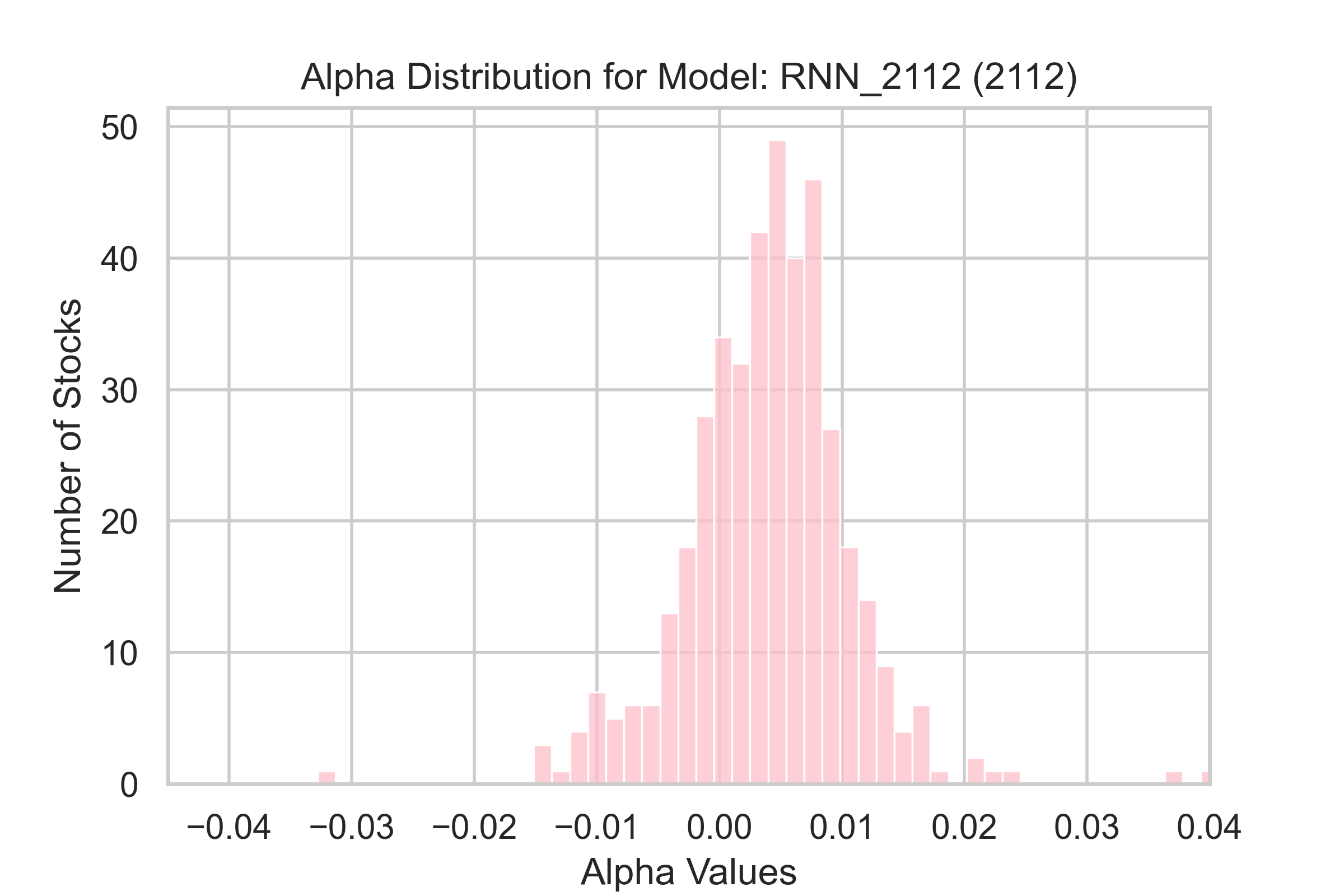}
  \caption{RNN(2112)}
\end{subfigure}
\hfill
\begin{subfigure}{0.32\textwidth}
  \includegraphics[width=\linewidth, height=0.22\textheight, keepaspectratio]{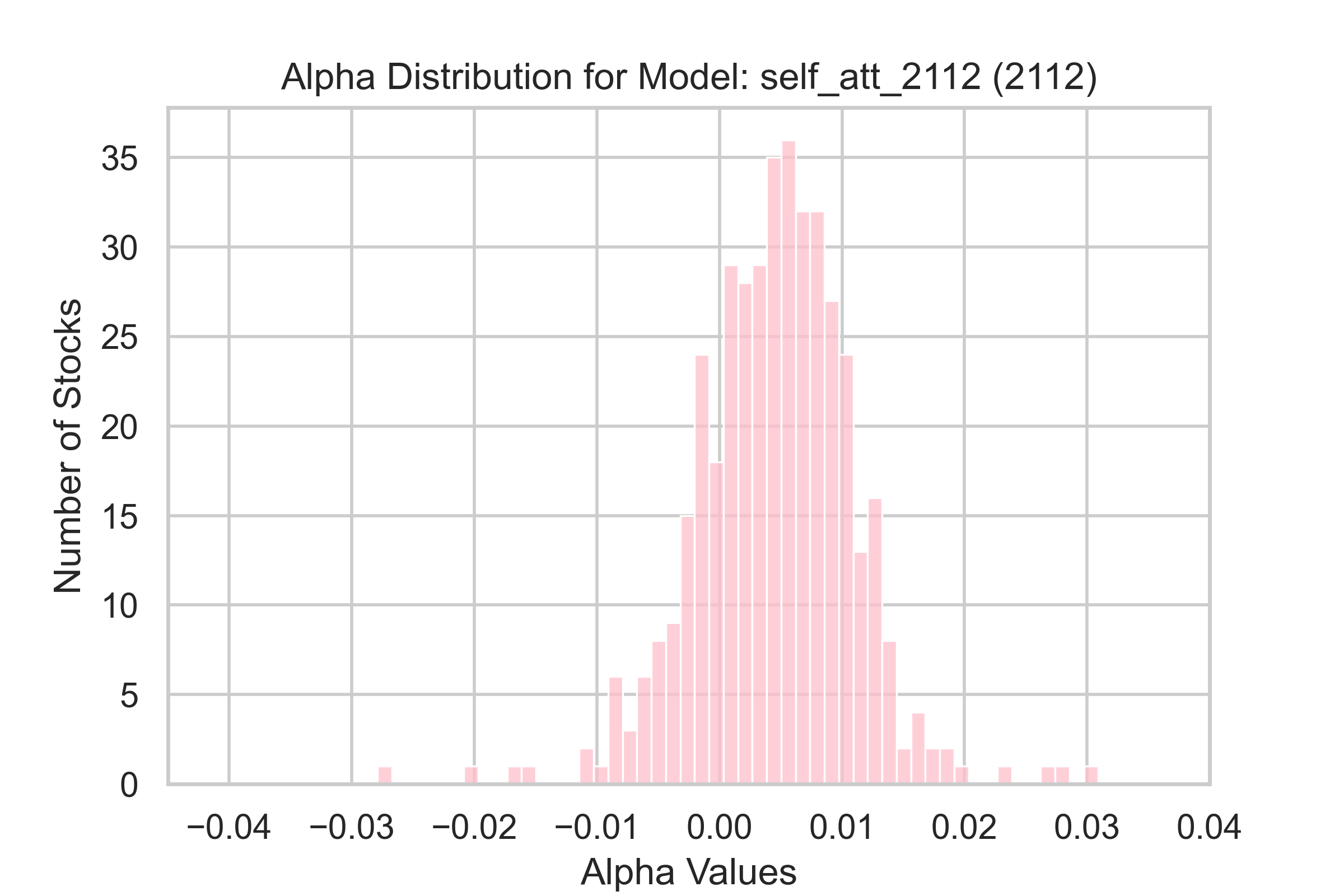}
  \caption{self\_att(2112)}
\end{subfigure}
\hfill
\begin{subfigure}{0.32\textwidth}
  \includegraphics[width=\linewidth, height=0.22\textheight, keepaspectratio]{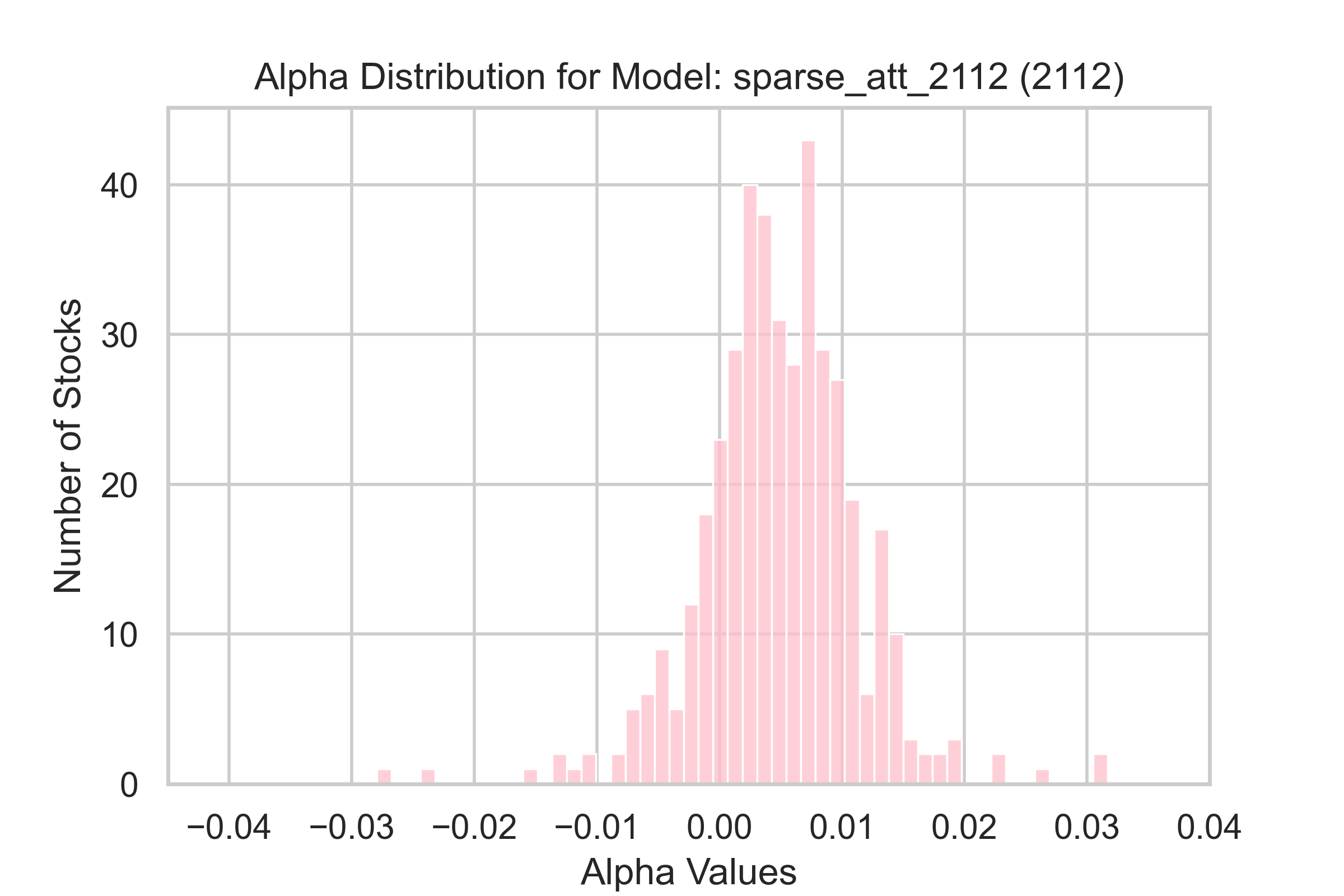}
  \caption{sparse\_att(2112)}
\end{subfigure}

\vspace{0.1em} 

\begin{subfigure}{0.32\textwidth}
  \includegraphics[width=\linewidth, height=0.22\textheight, keepaspectratio]{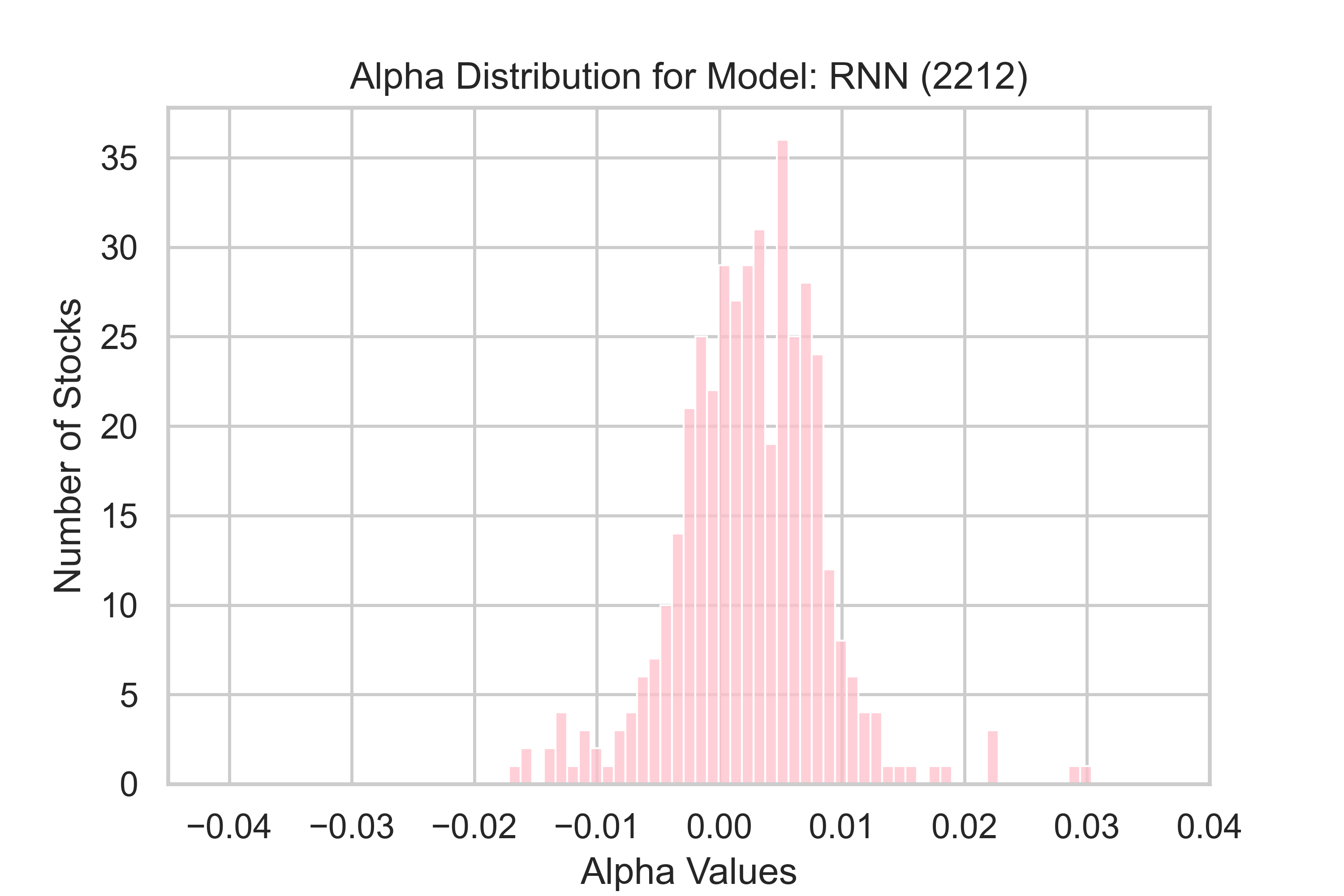}
  \caption{RNN(2212)}
\end{subfigure}
\hfill
\begin{subfigure}{0.32\textwidth}
  \includegraphics[width=\linewidth, height=0.22\textheight, keepaspectratio]{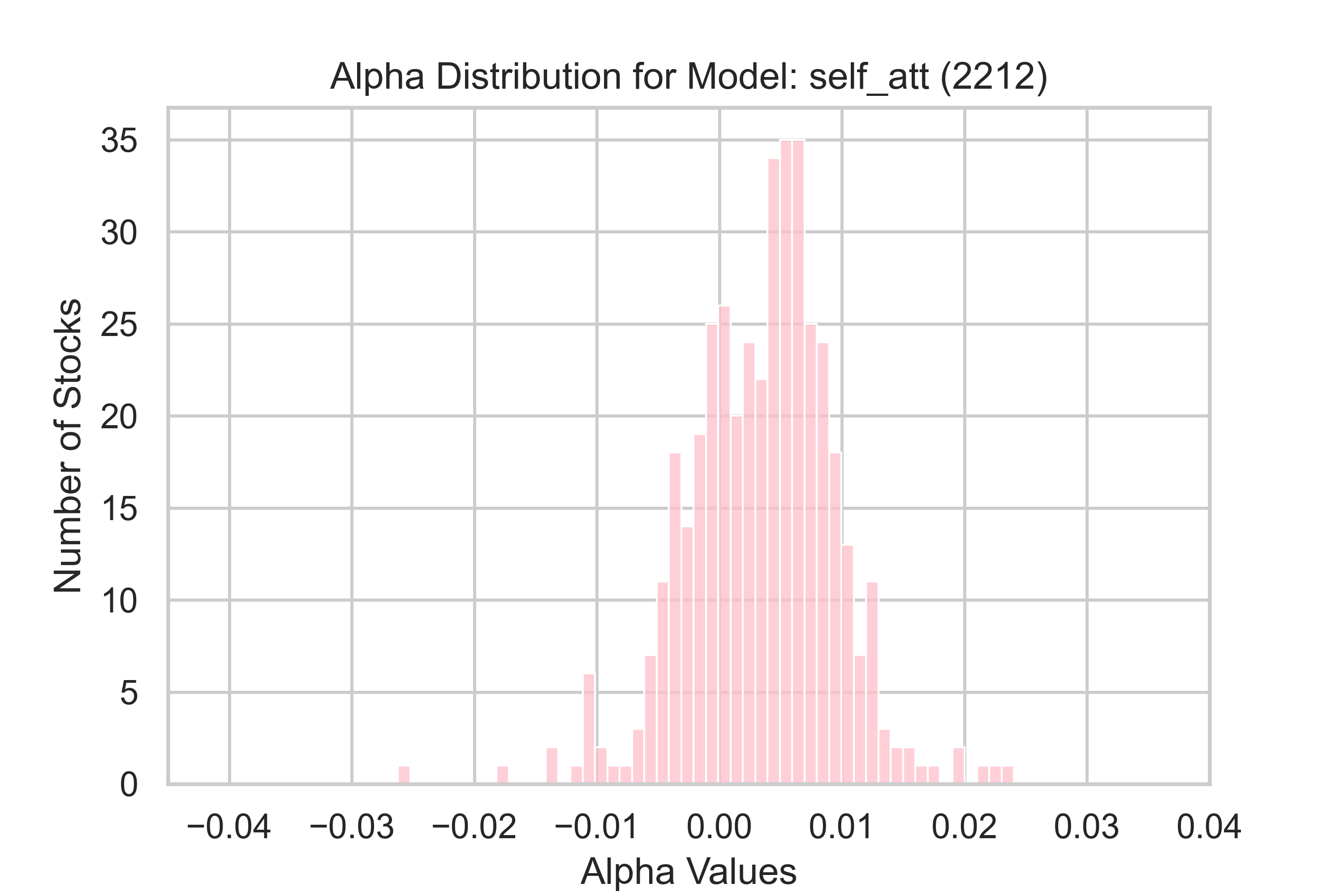}
  \caption{self\_att(2212)}
\end{subfigure}
\hfill
\begin{subfigure}{0.32\textwidth}
  \includegraphics[width=\linewidth, height=0.22\textheight, keepaspectratio]{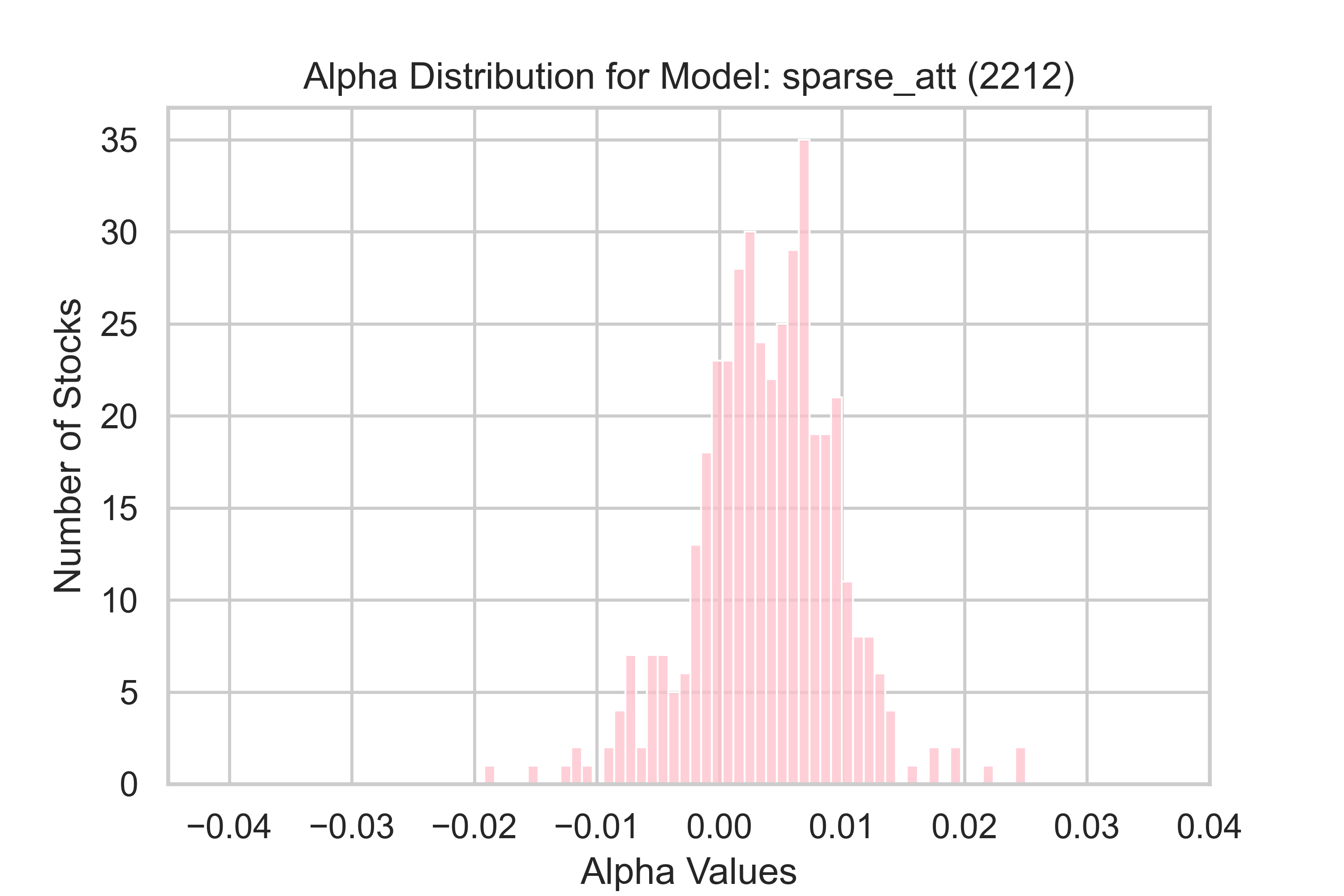}
  \caption{sparse\_att(2212)}
\end{subfigure}
\caption[Out-of-sample $\alpha$ distribution of the best fitted models.]{Out-of-sample $\alpha$ distribution of the best fitted models.}
\label{fig:alpha_distribution_ch2}
\end{sidewaysfigure}

All models have significant positive average $\alpha$s during all periods. This demonstrates the persistence of the extra capital gain beyond the factors. Economically, this indicates that there are excess returns that cannot be fully explained by the factors, which implies the unexplored or omitted factors or the existence of persistent arbitrage opportunities in the market, but it is unlikely to be caused by model misspecification. Three reasons stand for this interpretation. Firstly, the notably high $\alpha$ values observed during the `2112' period were characterized by a sharp market uptrend, which likely contributed to the model's tendency to underpredict returns systematically. In contrast, the relatively moderate $\alpha$ values in `1911' and `2212', which correspond to a mild uptrend and a sideways market, respectively, suggest that excess return potential tends to concentrate in highly intensive trending environments. Secondly, apart from the vanilla RNN model, there is no incredible difference between the average OOS $\alpha$ of models in a single period. Thirdly, at least 60\% of stocks have positive OOS $\alpha$ individually in different models with distinctive periods. An example $\alpha$ distribution plot is shown in Figure~\ref{fig:alpha_distribution_ch2}. Periodically, the vanilla RNN model has the highest annualized OOS residual $\alpha$ (4.78\%) in Period `1911', while the LSTM model has the highest values in Period `2112' (6.6\%) and Period `2212'(4.8\%). Nevertheless, in each period, the difference between models is modest. \\

The DM test evaluates the null hypothesis that two competing forecasts have equal predictive accuracy. Table~\ref{tab:dm_ch2} shows the results of the DM test. A negative value between two models means the model that shows as the row label outperforms the model that shows as the column label. The DM test results agree with the OOS $R^2$ results from Table~\ref{tab:r2_alpha_ch2}. The proposed attention models (self\_att and sparse\_att) significantly outperform alternative models in Period 1911, except for the vanilla RNN model. However, with the HAC estimator moderating the issue of autocorrelation between the differences of pair-wise models' prediction errors, the significance of the differences between the proposed RNN global self-attention model (self\_att) and alternative models is reduced significantly in COVID-19–Inclusive Period (2112) (significantly different from Model B-Additive, L-Concat and RNN) and Period Including COVID-19 and One-Year After (2212) (only RNN). In contrast, the sparse\_att are significantly different from alternatives in all periods except self\_att and GRU (in Period 2112 and Period 2212). The advantage of the proposed models against the GRU becomes less pronounced in Period 2112, when the market surged sharply. In Period 2212, when the market experienced a sharp decline followed by a period of sideways movement, the advantage of the proposed RNN global self-attention model (self\_att) nearly vanished, but the advantage of the proposed RNN sliding window attention model (sparse\_att) remains strong. On the other hand, the difference between the proposed models is insignificant. Among the benchmark models, RNN Luong's cancatenate attention model (L-Concat) achieves the worst-performing model during all periods, followed by LSTM and the additive attention model (B-Additive). The L-DotProd model, as the foundation of the proposed models, performs significantly better than RNN Luong's attention model (L-General) in Period 1911, but shows less advantage in the market turbulence periods. GRU insignificantly underperforms the RNN L-DotProd model in Period 1911, but apart from this, it outperforms all benchmark attention models (B-Additive, L-DotProd, L-Concat, L-General), excluding the proposed models, in all periods. The performance of GRU in all periods is better and more stable than that of LSTM. Overall, the robust DM test results support the findings in Table ~\ref{tab:r2_alpha_ch2}.\\

\begin{table}[htbp]
  \centering
  \small
  \setlength{\tabcolsep}{2.5pt} 
  \begin{tabular*}{\textwidth}{@{\extracolsep{\fill}} l rrrrrrrr}
    \toprule
    & B-Additive & GRU & L-Concat & L-DotProd & L-General & LSTM & RNN & self\_att \\
    \midrule
    \multicolumn{9}{c}{\textbf{Pre--COVID-19 Period (1911)}} \\
    \midrule
    B-Additive       &       &       &       &       &       &       &       &       \\
    GRU        & $-$0.78 &       &       &       &       &       &       &       \\
    L-Concat         & 2.40** & 2.47** &       &       &       &       &       &       \\
    L-DotProd         & $-$2.62*** & $-$1.13 & $-$3.45*** &       &       &       &       &       \\
    L-General         & 0.77  & 1.40  & $-$1.79* & 2.68*** &       &       &       &       \\
    LSTM       & 0.79  & 1.10  & $-$2.14** & 2.83*** & $-$0.11 &       &       &       \\
    RNN        & $-$4.62*** & $-$4.80*** & $-$4.73*** & $-$4.40*** & $-$4.61*** & $-$4.48*** &       &       \\
    self\_att  & $-$3.20*** & $-$2.86*** & $-$3.46*** & $-$2.38** & $-$3.22*** & $-$2.99*** & 4.32*** &       \\
    sparse\_att & $-$3.09*** & $-$2.84*** & $-$3.40*** & $-$1.98** & $-$2.95*** & $-$2.72*** & 4.57*** & 0.60  \\
    \midrule
    \multicolumn{9}{c}{\textbf{COVID-19--Inclusive Period (2112)}} \\
    \midrule
    B-Additive       &       &       &       &       &       &       &       &       \\
    GRU        & $-$1.91* &       &       &       &       &       &       &       \\
    L-Concat         & 1.51  & 2.18** &       &       &       &       &       &       \\
    L-DotProd         & $-$1.97** & 1.14  & $-$2.12** &       &       &       &       &       \\
    L-General         & $-$1.30 & 1.55  & $-$1.92* & 1.30  &       &       &       &       \\
    LSTM       & $-$0.10 & 1.58  & $-$1.79* & 1.34  & 0.78  &       &       &       \\
    RNN        & $-$3.90*** & $-$4.09*** & $-$3.88*** & $-$3.95*** & $-$3.86*** & $-$3.79*** &       &       \\
    self\_att  & $-$1.83* & $-$0.20 & $-$1.98** & $-$1.25 & $-$1.64 & $-$1.51 & 4.04*** &       \\
    sparse\_att & $-$3.96*** & $-$0.45 & $-$3.68*** & $-$2.69*** & $-$3.66*** & $-$2.82*** & 3.74*** & $-$0.29 \\
    \midrule
    \multicolumn{9}{c}{\textbf{Period Including COVID-19 and One-Year After (2212)}} \\
    \midrule
    B-Additive       &       &       &       &       &       &       &       &       \\
    GRU        & $-$1.92* &       &       &       &       &       &       &       \\
    L-Concat         & 1.29  & 2.26** &       &       &       &       &       &       \\
    L-DotProd         & $-$2.19** & 1.06  & $-$2.27** &       &       &       &       &       \\
    L-General         & $-$1.58 & 1.46  & $-$2.07** & 1.00  &       &       &       &       \\
    LSTM       & $-$0.37 & 1.62  & $-$2.19** & 1.32  & 0.80  &       &       &       \\
    RNN        & $-$4.26*** & $-$4.45*** & $-$4.25*** & $-$4.21*** & $-$4.18*** & $-$4.16*** &       &       \\
    self\_att & $-$1.32 & 0.63  & $-$1.51 & $-$0.41 & $-$0.86 & $-$0.99 & 4.42*** &         \\
    sparse\_att & $-$3.61*** & $-$0.18 & $-$3.38*** & $-$2.07** & $-$2.96*** & $-$2.50** & 4.12*** &  1.08 \\
    \bottomrule
  \end{tabular*}
\caption[DM test statistics.]{DM test statistics. Values in parentheses indicate the negative values. A negative value means the model named with the row label outperforms the model with the column label. For example, the first value in the first column of the '1911' panel, which is -0.78, indicates that Model GRU outperforms Model B-Additive. $*$, $**$ and $***$ depict the statistical significance level of 90\%, 95\% and 99\%, respectively.}
\label{tab:dm_ch2}
\end{table}

\begin{figure}[htbp!]
\centering
\includegraphics[width=1\columnwidth]{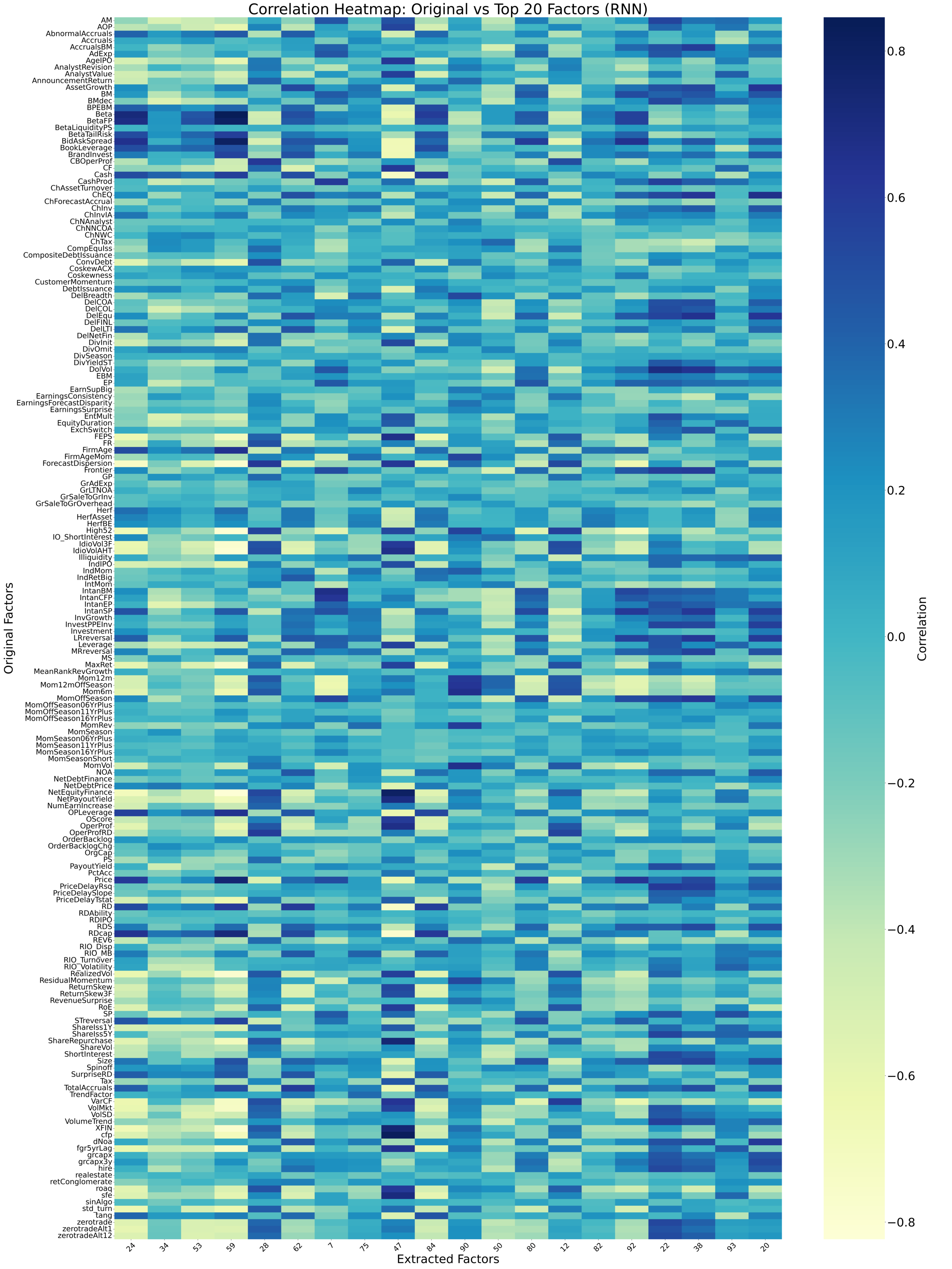}
\caption[Correlation heatmap of top 20 extracted factors and original factors for the vanilla RNN model.]{Correlation heatmap of top 20 extracted factors and original factors for the vanilla RNN model. The Y-axis is the original factors, and the X-axis is the name of the extracted factors ranked as the top 20 most important factors.}
\label{fig:corr_rnn_ch3}
\end{figure}

\begin{figure}[htbp!]
\centering
\includegraphics[width=1\columnwidth]{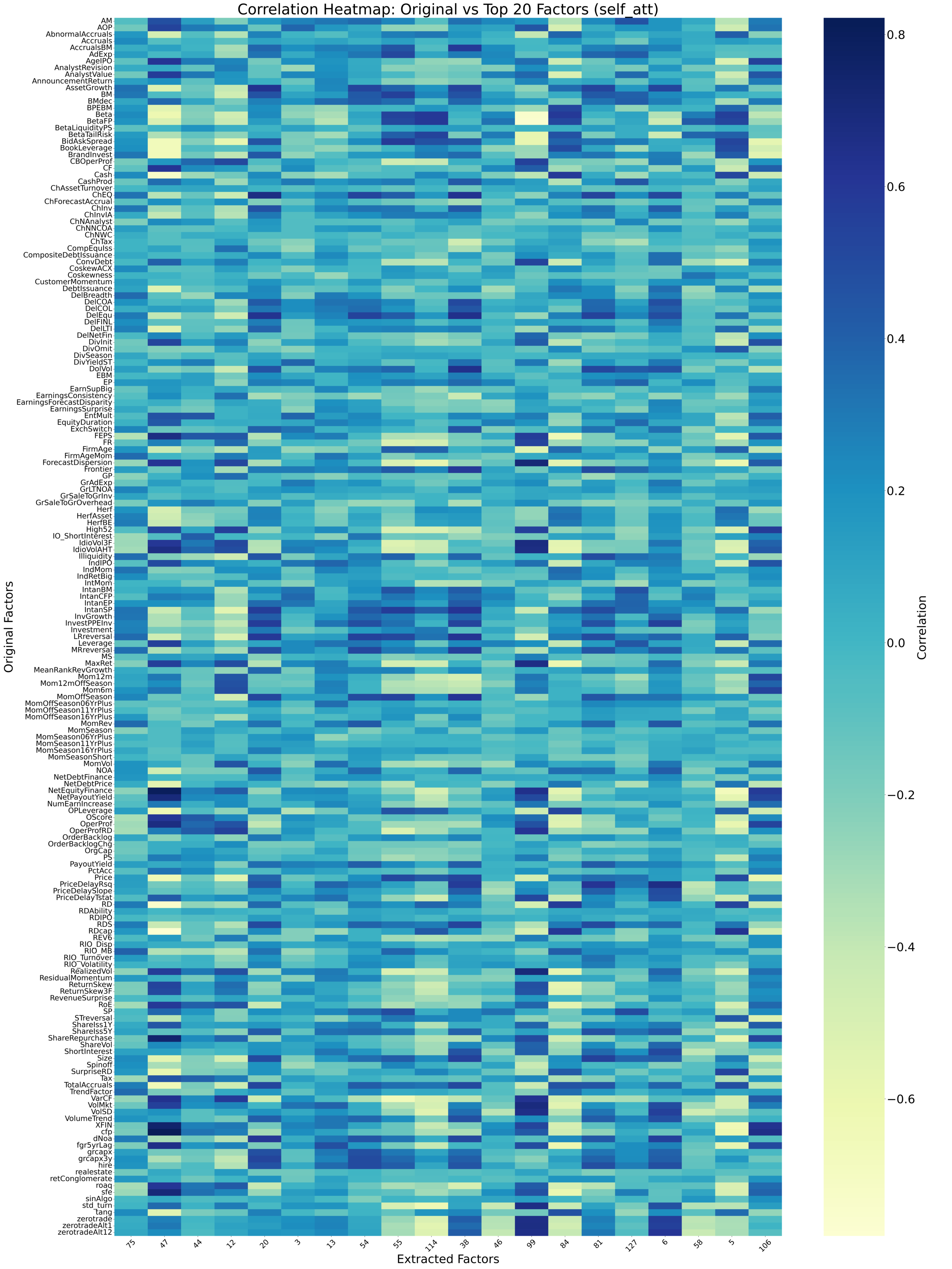}
\caption[Correlation heatmap of top 20 extracted factors and original factors for the global self-attention model.]{Correlation heatmap of top 20 extracted factors and original factors for the global self-attention model. The Y-axis is the original factors, and the X-axis is the name of the extracted factors ranked as the top 20 most important factors.}
\label{fig:corr_self_att_ch3}
\end{figure}

\begin{figure}[htbp!]
\centering
\includegraphics[width=1\columnwidth]{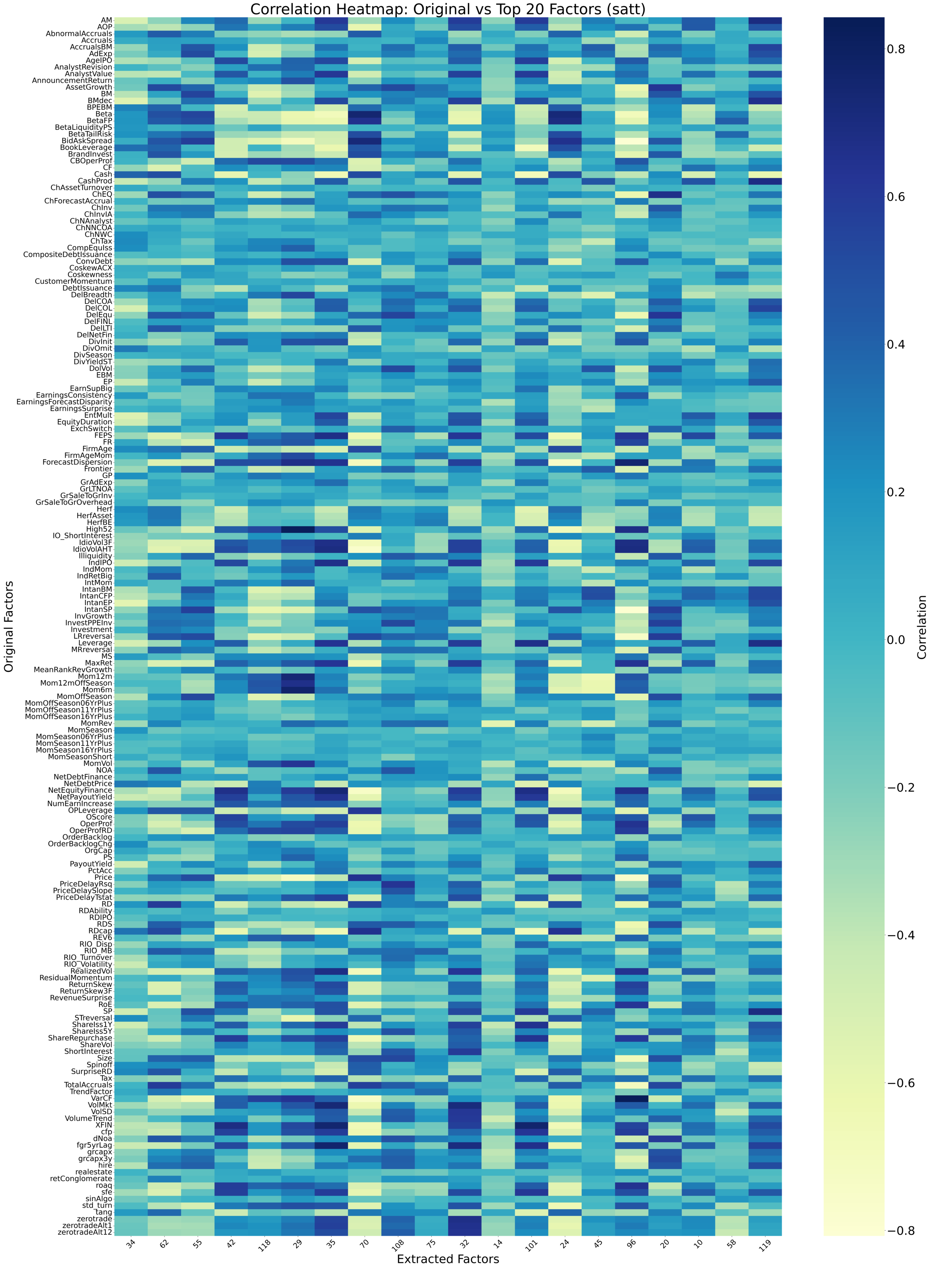}
\caption[Correlation heatmap of top 20 extracted factors and original factors for the sliding window sparse attention model.]{Correlation heatmap of top 20 extracted factors and original factors for the sliding window sparse attention (sparse\_att) model. The Y-axis is the original factors, and the X-axis is the name of the extracted factors ranked as the top 20 most important factors.}
\label{fig:corr_satt_ch3}
\end{figure}

Since the MLP pre-training process extracted the information from the original factors and compressed the 182 factors into 128 extracted factors, which reduces the financial and economic interpretability of the models, the variable (factor) importance of the extracted factors and the correlation between the top 20 extracted factors and original factors are also computed to investigate which factors have the highest impact on the stock pricing. Figure~\ref{fig:corr_rnn_ch3} to Figure~\ref{fig:corr_satt_ch3} illustrates the models with the highest OOS $R^2$ (RNN, self\_att and sparse\_att) of the correlation between the top 20 extracted factors and the original factors. This is a widely applied post-hoc interpretability method in ML for the asset pricing domain, for example, \citet{Gu2020EmpiricalLearning,Chen2024DeepPricing} used the method on latent (extracted) factors and original input factors. The method uses the sum of the latent factors correlation to capture the overall economic meanings of an original factor, and the max of the latent factors correlation to measure the individual latent factor's impact. From the sum of the correlation, the most highly impactful factors are quite concentrated; the top 5 factors are Bid Ask Spread (BidAskSpread), NetEquityFinance (NetEquityFinance), Forecast Dispersion (ForecastDispersion), Price and volatilities (IdioVol). This means under the framework of RNN, self\_att, sparse\_att, the market frictions (BidAskSpread), Equity Financing capability (NetEquityFinance), analyst disagreement (ForecastDispersion), lowest price stocks (Price) and Idiosyncratic volatilities (IdioVol) are the top 5 factors that drive the large-cap stock returns. These are followed by the factors of the highest point for the last 52 weeks (High52), cashflow variance (VarCF), systematic risk (Beta), intangible assets (IntanSP), forecast earnings per share (FEPS), long-term reversal (LRreversal) and operating profit (OperProf). These factors are highly consistent with classic literature, for example, Bid Ask Spread (BidAskSpread) ~\citep{amihud2002illiquidity,brennan1998alternative}, Equity Financing capability (NetEquityFinance) \citep{baker2000equity,fama2005financing}, analyst disagreement (ForecastDispersion) \citep{diether2002differences,johnson2004forecast} and Idiosyncratic volatilities (IdioVol) \citep{ang2009high}. For the maximum correlation value of individual latent factors, the top 10 latent factors of these models agree that Bid Ask Spread (BidAskSpread), Equity Financing capability (NetEquityFinance), analyst disagreement (ForecastDispersion), the last 52 weeks (High52), lowest price stocks (Price), Idiosyncratic volatilities (IdioVol) and intangible assets (IntanSP) are the most important factors, which aligns with literature as well. The results of variable (factors) importance examination verified the conclusion of \citet{AndrewY.Zimmermann2020OpenPricing} that most of the pricing ability is concentrated in a few factors, even in the context of ML neural network models. Overall, the factor importance and correlation examination highly improve the transparency of the ML models in the asset pricing context.\\

\subsection{Back-testing performance}
Backtesting refers to the process of evaluating a trading strategy or predictive model by applying it to OOS returns to assess its performance. The predictive models provide the predictive returns; henceforth, a market timing strategy or trading signal generating system would be deployed to the real returns and predictive returns to generate trading signals for guiding trading behaviours (open or close positions). In this study, the sign signal generation method from the typical trend-following trading strategy is adopted for the market timing assignment. Concretely, if the predictive return at a time point is positive, which matches the real return in that time point, the circumstance is considered a buy-side (long) signal. The simulated buy transaction takes place, and the real returns from the next time point would be counted as a gain or loss until the sell-side signal (negative predictive return matches negative real return) shows up as the indicator of closing position. This is also known as the long-only strategy. The short-only strategy is in contrast to the long-only strategy, which opens a short position when both real return and predictive return have negative values at one time point, while closing this position when they both have positive values. The back-testing in this study only simulates the long-only transaction by considering the practical difficulties in real stock markets and follows the suggestion of \citet{Avramov2021MachinePredictability}. Specifically, real stock market trading has restrictions for the sell-side positions, for example, the availability of the stocks for selling, variable borrowing rates for individual stocks, as well as the liquidity of the stock. Two simple asset allocation methods are applied for portfolio-wise backtesting, which are equal-weighted and value-weighted. An equal-weighted portfolio distributes weights to 420 stocks evenly, while the value-weighted portfolio allocates the weights to stocks based on the size of the stocks' market capitalization. In this sense, in a value-weighted portfolio, stocks with larger market capitalization would be distributed with heavier weights, and the summation of all weights for these 420 stocks is equal to one. Thus, the computation of indicators in the value-weighted portfolio concentrates more on larger market capitalization stocks.\\ 

Table~\ref{tab:ew_perf} presents the back testing results of the equal-weighted portfolio of each model in the three periods after deducting the static transaction cost (50bps). All models in all periods consistently outperform the buy-and-hold (BHE) strategy except for the vanilla RNN model and RNN Luong's concatenate attention model (L-Concat). The proposed models have the highest annulized returns during all periods. When considering the standard deviation risks (Sharpe Ratio) and downside risks (Sortino Ratio) of the returns, the proposed models achieve first place (RNN global self-attention model) and second place (RNN sliding window attention model) in Period 1911 and Period 2212. However, in Period 2112, the vanilla RNN model shows a marginal advantage on the Sharpe ratio (SR) and the Sortino ratio (SO). The Sharpe ratio of the vanilla RNN model lies between the RNN global self-attention model (self\_att) and the RNN sliding window attention model (sparse\_att), while the Sortino ratio is higher than both proposed models. This demonstrates that the vanilla RNN model has a higher risk protection capability, especially downside risks, when it faces extreme market conditions such as a volatile market with a significant trend. GRU and LSTM models perform quite similarly in all periods. For the rest of the RNN attention models, including the RNN additive attention model (B-Additive) and the RNN Luong's attention models (L-Concat, L-DotProd, L-General), their performance varies depending on the market conditions. In Period 1911, Model B-Additive is superior to Model L-Concat, L-DotProd, L-General, as well as GRU and LSTM models, and becomes the most profitable RNN attention model following the proposed models. However, when considering volatility risks, including standard deviation and downside risks, it falls behind the RNN Luong's dot product attention model (L-DotProd). The performance of Model B-Additive in Period 2112 and Period 2212 is not as well as in Period 1911 when comparing with alternative RNN attention benchmark models (L-Concat, L-DotProd, L-General) and RNN variation models (GRU, LSTM) in the same period. It falls behind the L-DotProd model but still outperforms RNN variation models in both periods. For RNN Luong's three attention models (L-Concat, L-DotProd, L-General), L-DotProd performs best in all periods which is followed by L-General. These phenomena imply that all RNN attention models, apart from the L-Concat model, are superior to the RNN variation models, especially when encountering extreme market conditions. Nonetheless, the vanilla RNN model shows a great capability to control downside risks during the market turbulence, but sacrifices its absolute capital gain. \\

\begin{table}[htbp!]
\centering
\captionsetup{justification=raggedright,singlelinecheck=false,position=bottom}
  \small
\begin{tabularx}{\textwidth}{lXXXXXX}
\toprule
\textbf{Strategy} & \textbf{Max Drawdown} & \textbf{Ann Return} & \textbf{Sharpe Ratio (SR)} & \textbf{Sortino Ratio (SO)}  & \textbf{Ann SR} & \textbf{Ann SO} \\
\midrule
\multicolumn{7}{c}{\textbf{Pre–COVID-19 Period (1911)}} \\
\midrule
B-Additive      & (0.3128) & 0.1229 & 0.2834 & 0.3993  & 0.9816 & 1.3832 \\
GRU       & (0.3241) & 0.1162 & 0.2748 & 0.3853  & 0.9520 & 1.3349 \\
L-Concat        & (0.3564) & 0.1168 & 0.2681 & 0.3760  & 0.9288 & 1.3024 \\
L-DotProd        & (0.3077) & 0.1289 & 0.2986 & 0.4259  & 1.0343 & 1.4754 \\
L-General        & (0.3180) & 0.1212 & 0.2795 & 0.3954  & 0.9681 & 1.3698 \\
LSTM      & (0.3562) & 0.1166 & 0.2680 & 0.3798  & 0.9283 & 1.3155 \\
RNN       & (0.2588) & 0.1041 & 0.2733 & 0.3650  & 0.9466 & 1.2643 \\
self\_att & (0.3008) & 0.1357 & 0.3159 & 0.4570  & 1.0944 & 1.5832 \\
sparse\_att & (0.3154) & 0.1298 & 0.3020 & 0.4298 & 1.0462 & 1.4889 \\
BHE	& (0.3701) & 0.1139 &	0.2588 & 0.3631  & 0.8965 &	1.2578\\ 
\midrule
\multicolumn{7}{c}{\textbf{COVID-19–Inclusive Period (2112)}} \\
\midrule
B-Additive      & (0.4953) & 0.1459 & 0.2672 & 0.3323  & 0.9255 & 1.1512 \\
GRU       & (0.4932) & 0.1392 & 0.2650 & 0.3279  & 0.9180 & 1.1357 \\
L-Concat        & (0.5212) & 0.1367 & 0.2530 & 0.3100  & 0.8765 & 1.0739 \\
L-DotProd        & (0.4820) & 0.1471 & 0.2751 & 0.3370  & 0.9530 & 1.1675 \\
L-General        & (0.4886) & 0.1423 & 0.2660 & 0.3317  & 0.9215 & 1.1489 \\
LSTM      & (0.4895) & 0.1395 & 0.2617 & 0.3287  & 0.9067 & 1.1385 \\
RNN       & (0.4123) & 0.1300 & 0.2838 & 0.3862  & 0.9833 & 1.3379 \\
self\_att & (0.4427) & 0.1540 & 0.2913 & 0.3823  & 1.0091 & 1.3242 \\
sparse\_att & (0.4584) & 0.1503 & 0.2811 & 0.3598  & 0.9739 & 1.2463 \\
BHE	& (0.5310) & 0.1349 & 0.2472 & 0.2988  & 0.8563 &	1.0351\\ 
\midrule
\multicolumn{7}{c}{\textbf{Period Including COVID-19 and One-Year After (2212)}} \\
\midrule
B-Additive      & (0.4960) & 0.1241 & 0.2218 & 0.2911  & 0.7684 & 1.0083 \\
GRU       & (0.4968) & 0.1177 & 0.2177 & 0.2835  & 0.7541 & 0.9820 \\
L-Concat        & (0.5208) & 0.1165 & 0.2092 & 0.2707  & 0.7248 & 0.9377 \\
L-DotProd        & (0.4731) & 0.1279 & 0.2323 & 0.3021  & 0.8046 & 1.0465 \\
L-General        & (0.4933) & 0.1215 & 0.2197 & 0.2875  & 0.7609 & 0.9961 \\
LSTM      & (0.4980) & 0.1181 & 0.2145 & 0.2812  & 0.7430 & 0.9742 \\
RNN       & (0.4191) & 0.1012 & 0.2117 & 0.2896  & 0.7334 & 1.0033 \\
self\_att & (0.4409) & 0.1334 & 0.2433 & 0.3336  & 0.8429 & 1.1558 \\
sparse\_att   & (0.4545) & 0.1304 & 0.2354 & 0.3178  & 0.8154 & 1.1010 \\
BHE & (0.5308) & 0.1141 & 0.2148 & 0.2760  & 0.7441 & 	0.9561 \\
\bottomrule
\end{tabularx}
\caption[Equal-weighted portfolio performance with deduction of static transaction cost.]{Equal-weighted portfolio performance with deduction of static transaction cost. `Ann' is the notation for annulization, while `SR' and `SO' are short for Sharpe ratio and Sortino ratio respectively. `BHE' means buy-and-hold strategy in Equal-weighted portfolios. The values in the parentheses indicate negative values. `Ann Return' and `Max Drawdown' can be transformed to a percentage presentation by multiplying by 100.}
\label{tab:ew_perf}
\end{table}

\begin{figure}[htbp!]
    \centering
    
    \begin{subfigure}{\textwidth}
        \centering
        \includegraphics[width=0.9\linewidth]{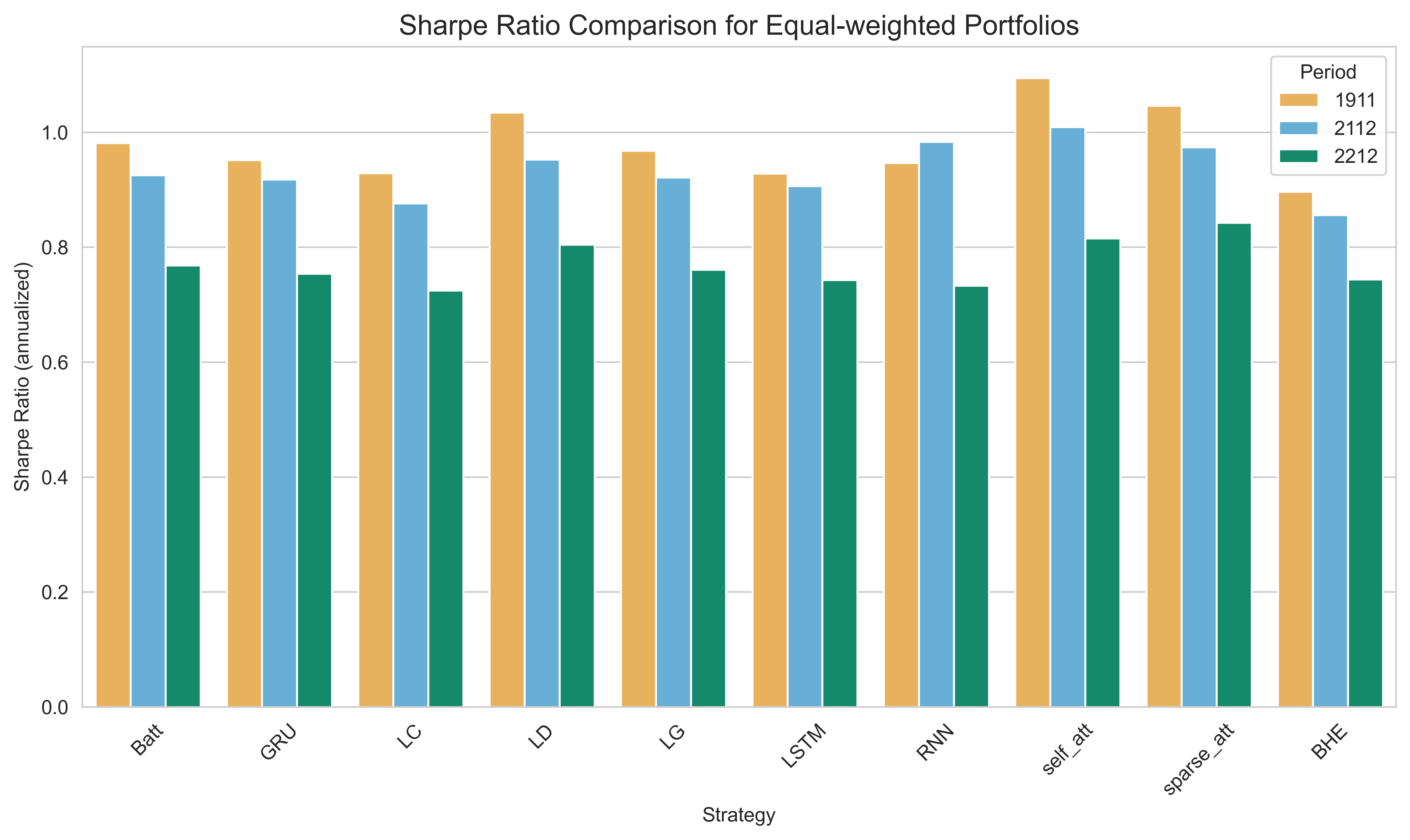}
        \caption{Annualized Sharpe ratio comparison}
        \label{fig:sr_bar_ew_ch2}
    \end{subfigure}
    
    \vspace{0.1cm} 
    
    \begin{subfigure}{\textwidth}
        \centering
        \includegraphics[width=0.9\linewidth]{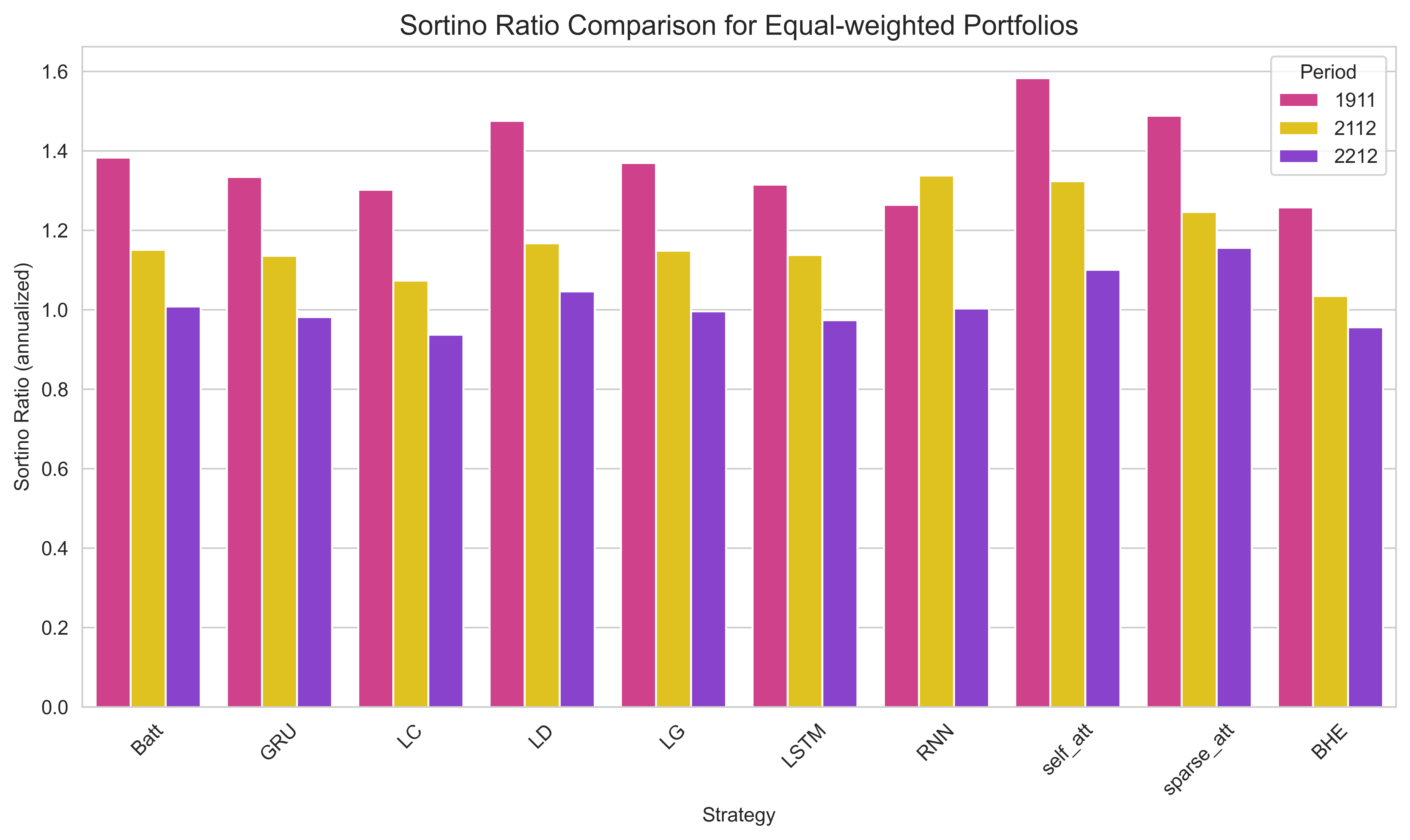}
        \caption{Annualized Sortino ratio comparison}
        \label{fig:so_bar_ew_ch2}
    \end{subfigure}
    \caption[Annualized Sharpe ratio and Sortino ratio for Equal-weighted with static transaction cost of 50bps.]{Annualized Sharpe ratio and Sortino ratio for Equal-weighted with static transaction cost of 50bps. `Batt' indicates the additive attention model (B-Additive), `LC', `LD' and `LG' mean L-Concat, L-DotProd and L-General attention model, respectively.}
    \label{fig:sr_so_ew_ch2}
\end{figure}

Figure~\ref{fig:sr_so_ew_ch2} is the diagram for the comparison of the annulized Sharpe ratio and Sortino ratio in the equal-weighted portfolios. The upper panel shows the Sharpe ratio, and the lower panel shows the Sortino ratio. It is apparent that the Sharpe ratio of the self\_att model achieves the highest value in the pre-pandemic period (1911) and the pandemic period (2112) whereas during the period that covers one year after the pandemic (2212), which contains a rapid drop and a highly fluctuating sideways movement of the price shows in Figure~\ref{fig:price_index_ch2}, the proposed sparse\_att wins the first place. It is also worth noting that the L-DotProd model follows closely to the performance of the proposed models and achieves third place. This implies that the attention mechanisms based on the dot product, including L-DotProd, self\_att and sparse\_att, demonstrate superior performance compared to other attention variants. The diagram of the Sortino ratios enlarges the difference between models. It supports the conclusion drawn from the diagram of the Sharpe ratios in Period 1911 and Period 2212. However, in Period 2112, the vanilla RNN model achieves the highest value with an insignificant advantage compared to the proposed models, especially the self\_att model, which implies that the vanilla RNN model has a great capability to hedge the downside risks during the extreme market fluctuations. \\

\begin{sidewaysfigure}[htbp!]
\centering
\includegraphics[width=1\columnwidth]{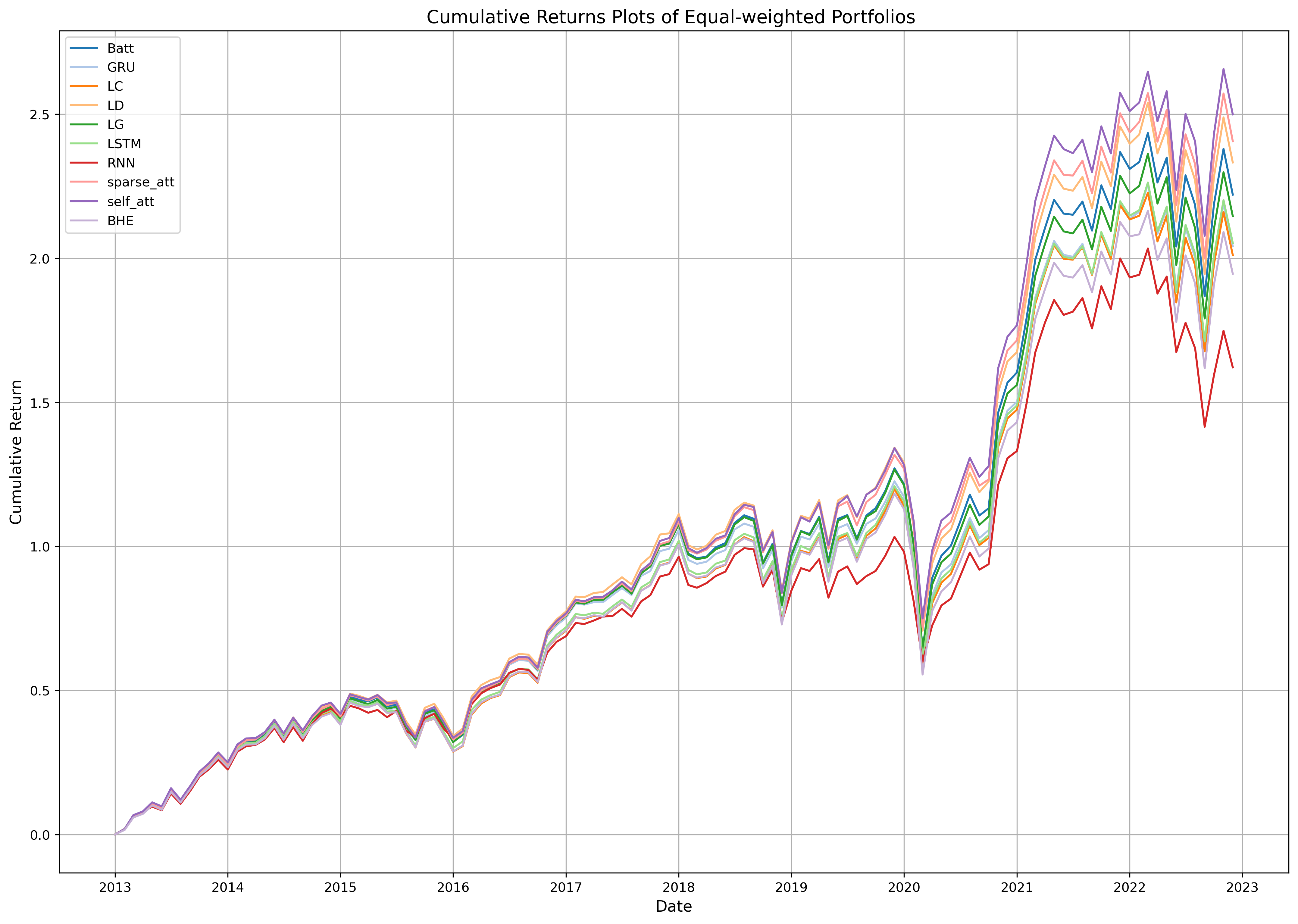}
\caption[Cumulative return plot for Equal-weighted portfolios with deduction of static transaction cost (50bps).]{Cumulative return plot for Equal-weighted portfolios with deduction of static transaction cost (50bps). `Batt' indicates the additive attention model (B-Additive), `LC', `LD' and `LG' mean L-Concat, L-DotProd and L-General attention model, respectively.}
\label{fig:cum_ret_ew_ch2}
\end{sidewaysfigure}

The equal-weighted portfolio cumulative return plot, Figure~\ref{fig:cum_ret_ew_ch2}, agrees with the findings from Table~\ref{tab:ew_perf}. In Period 1911, the cumulative returns of the L-DotProd model and the RNN sliding window attention model (sparse\_att) are similar, and the L-DotProd model outperforms the sparse\_att model in a short period, from late 2015 to the middle of 2018. However, since 2019, the proposed RNN global self-attention model has shown a non-negligible advantage over alternatives, closely followed by the proposed RNN sliding window attention model. All models' cumulative returns outperform the buy-and-hold benchmark (BHE) and the vanilla RNN model since 2017. Apart from the L-Concat model, all attention models show superiority to those models without attention mechanisms during the market shock periods. Model L-DotProd, B-Additive, L-General rank in third, fourth and fifth place, respectively. Although the vanilla RNN model exhibits strong OOS fitness and ability for controlling downside risks, it sacrifices its absolute capital gain during the extreme market turbulence. In this respect, the vanilla RNN model is possibly more suitable for funds that require high-capability hedging against downside risk rather than pursuing high absolute capital gains, or for modelling market volatility. \\ 

The robust examination is conducted to investigate how the turnover rate of strategies based on these models affects profitability, which is shown in Table~\ref{tab:ew_perf_turnover_ch3}. The self-attention (self\_att) model consistently delivers superior risk-adjusted performance across all three periods. In both tables, self\_att achieves the highest annualized Sharpe ratio (Ann.SR) and annualized Sortino ratio (Ann.SO), together with the competitive annualized returns (Ann Return/AR), while maintaining mild maximum drawdowns (MDD). For instance, in the pre-COVID period under static costs, self\_att has an Ann.SR of 1.0944 and Ann.SO of 1.5832, outperforming the buy-and-hold benchmark (BHE) by substantial margins (0.8965 and 1.2578, respectively). This dominance persists even after incorporating dynamic turnover costs, where self\_att’s moderate turnover rate (approximately 0.1156 to 0.1196) limits cost deterioration compared with higher-turnover models. The sliding window sparse attention model (sparse\_att) ranks a close second, highlighting the capability of advanced attention-based architectures in capturing return dynamics stock pricing data.\\

In contrast, traditional recurrent neural network models such as LSTM, GRU, and benchmark attention models (L-Concat, L-General) exhibit moderate performance, generally trailing the attention mechanisms in risk-adjusted returns, particularly in the COVID-19–Inclusive Period (2112) and Period Including COVID-19 and One-Year After (2212), where Ann.SR and Ann.SO declines significantly across the board due to the extreme market turbulence. The Bahdanau attention model (B-Additive) performs respectably in the middle tier but is outperformed by both self\_att, sparse\_att and L-DotProd, highlighting the advantages of dot-product-based attention over additive attention in equal-weighted portfolio configurations. Notably, the RNN model exhibits the lowest MDD in Table~\ref{tab:ew_perf}, suggesting strong downside protection; however, its high turnover (exceeding 0.49 across periods) leads to severe performance deterioration once transaction costs are fully accounted for in Table~\ref{tab:ew_perf_turnover_ch3}, resulting in the lowest Ann.SR and Ann.SO among all strategies. All models substantially outperform the buy-and-hold benchmark apart from the vanilla RNN model in both transaction cost scenarios, confirming the presence of extra profitability beyond the benchmark.\\

Dynamic transaction costs (turnover with 20bps for large-cap stocks), especially when combined with high portfolio turnover, significantly enlarge the differences between backtesting performances across models. The self-attention model (self\_att) and sliding-window sparse attention (sparse\_att) not only demonstrate superior performance in static transaction cost environments but also exhibit the best net cumulative returns and robustness to turnover-based realistic market frictions. These findings highlight that, in equal-weighted portfolio construction, controlling turnover is critical for maintaining the profitability of these models. In this perspective, dot-product-based attention mechanisms have significant advantages in generating more stable and selective trading signals. On the other hand, although the RNN model shows persistence in risk control, its profitability significantly decays due to the high turnover. This highlights a key insight: high forecasting accuracy does not mean high real-world returns.\\

\begin{table}[htbp]
  \centering
  \small 
  \begin{tabular}{lcccccccc}
    \toprule
    \textbf{Model} & \textbf{AR} & \textbf{Ann.SR} & \textbf{SR} & \textbf{Ann.SO} & \textbf{SO} & \textbf{MDD} & \textbf{Turnover} \\
    \midrule
    \multicolumn{8}{c}{\textbf{Pre-COVID-19 Period (1911)}} \\
    \midrule
    BHE        & 0.1141 & 0.8892 & 0.2567 & 1.2303 & 0.3552 & -0.1483 & 0.0120 \\
    B-Additive       & 0.1217 & 0.9564 & 0.2761 & 1.3208 & 0.3813 & -0.1466 & 0.0672 \\
    GRU        & 0.1145 & 0.9233 & 0.2665 & 1.2691 & 0.3664 & -0.1407 & 0.0833 \\
    L-Concat         & 0.1160 & 0.9075 & 0.2620 & 1.2478 & 0.3602 & -0.1481 & 0.0496 \\
    L-DotProd         & 0.1275 & 1.0063 & 0.2905 & 1.4066 & 0.4060 & -0.1366 & 0.0761 \\
    L-General         & 0.1199 & 0.9425 & 0.2721 & 1.3084 & 0.3777 & -0.1478 & 0.0682 \\
    LSTM       & 0.1159 & 0.9077 & 0.2620 & 1.2610 & 0.3640 & -0.1453 & 0.0450 \\
    RNN        & 0.0916 & 0.8266 & 0.2386 & 1.0701 & 0.3089 & -0.1307 & 0.4955 \\
    self\_att  & 0.1332 & 1.0577 & 0.3053 & 1.4998 & 0.4330 & -0.1319 & 0.1156 \\
    sparse\_att& 0.1277 & 1.0134 & 0.2925 & 1.4132 & 0.4079 & -0.1365 & 0.1011 \\
    \midrule
    \multicolumn{8}{c}{\textbf{COVID-19-Inclusive Period (2112)}} \\
    \midrule
    BHE        & 0.1353 & 0.8512 & 0.2457 & 1.0181 & 0.2939 & -0.2890 & 0.0093 \\
    B-Additive       & 0.1448 & 0.9063 & 0.2616 & 1.1085 & 0.3200 & -0.2806 & 0.0628 \\
    GRU        & 0.1369 & 0.8917 & 0.2574 & 1.0847 & 0.3131 & -0.2743 & 0.1058 \\
    L-Concat         & 0.1360 & 0.8604 & 0.2484 & 1.0376 & 0.2995 & -0.2857 & 0.0463 \\
    L-DotProd         & 0.1458 & 0.9320 & 0.2690 & 1.1231 & 0.3242 & -0.2774 & 0.0721 \\
    L-General         & 0.1411 & 0.9019 & 0.2604 & 1.1065 & 0.3194 & -0.2748 & 0.0632 \\
    LSTM       & 0.1388 & 0.8900 & 0.2569 & 1.1000 & 0.3175 & -0.2714 & 0.0465 \\
    RNN        & 0.1174 & 0.8823 & 0.2547 & 1.1683 & 0.3373 & -0.2166 & 0.4922 \\
    self\_att  & 0.1514 & 0.9800 & 0.2829 & 1.2626 & 0.3645 & -0.2578 & 0.1182 \\
    sparse\_att& 0.1483 & 0.9490 & 0.2739 & 1.1930 & 0.3444 & -0.2632 & 0.0964 \\
    \midrule
    \multicolumn{8}{c}{\textbf{Period Including COVID-19 and One-Year After (2212)}} \\
    \midrule
    BHE        & 0.1143 & 0.7017 & 0.2026 & 0.8883 & 0.2564 & -0.2890 & 0.0083 \\
    B-Additive       & 0.1229 & 0.7522 & 0.2171 & 0.9736 & 0.2810 & -0.2798 & 0.0616 \\
    GRU        & 0.1151 & 0.7300 & 0.2107 & 0.9370 & 0.2705 & -0.2760 & 0.1130 \\
    L-Concat         & 0.1158 & 0.7116 & 0.2054 & 0.9083 & 0.2622 & -0.2858 & 0.0451 \\
    L-DotProd         & 0.1266 & 0.7869 & 0.2272 & 1.0094 & 0.2914 & -0.2733 & 0.0688 \\
    L-General         & 0.1203 & 0.7446 & 0.2149 & 0.9611 & 0.2775 & -0.2777 & 0.0625 \\
    LSTM       & 0.1174 & 0.7295 & 0.2106 & 0.9438 & 0.2725 & -0.2744 & 0.0442 \\
    RNN        & 0.0881 & 0.6382 & 0.1842 & 0.8543 & 0.2466 & -0.2210 & 0.5192 \\
    self\_att  & 0.1308 & 0.8170 & 0.2359 & 1.1033 & 0.3185 & -0.2554 & 0.1196 \\
    sparse\_att& 0.1284 & 0.7941 & 0.2292 & 1.0566 & 0.3050 & -0.2608 & 0.0941 \\
    \bottomrule
  \end{tabular}
\caption[Equal-weighted portfolio performance considering the dynamic transaction cost.]{Equal-weighted portfolio performance considering the dynamic transaction cost. 'Ann' is the notation for annulization, while 'SR' and 'SO' are short for Sharpe ratio and Sortino ratio respectively. The notation of 'MDD' means Maximum Drawdown, and 'BHV' means buy-and-hold strategy in value-weighted portfolios. The Annulized Return (AR) and 'MDD' can be transformed to a percentage presentation by multiplying by 100.}
\label{tab:ew_perf_turnover_ch3}
\end{table}

\begin{figure}[htbp!]
    \centering
    
    \begin{subfigure}{\textwidth}
        \centering
        \includegraphics[width=0.9\linewidth]{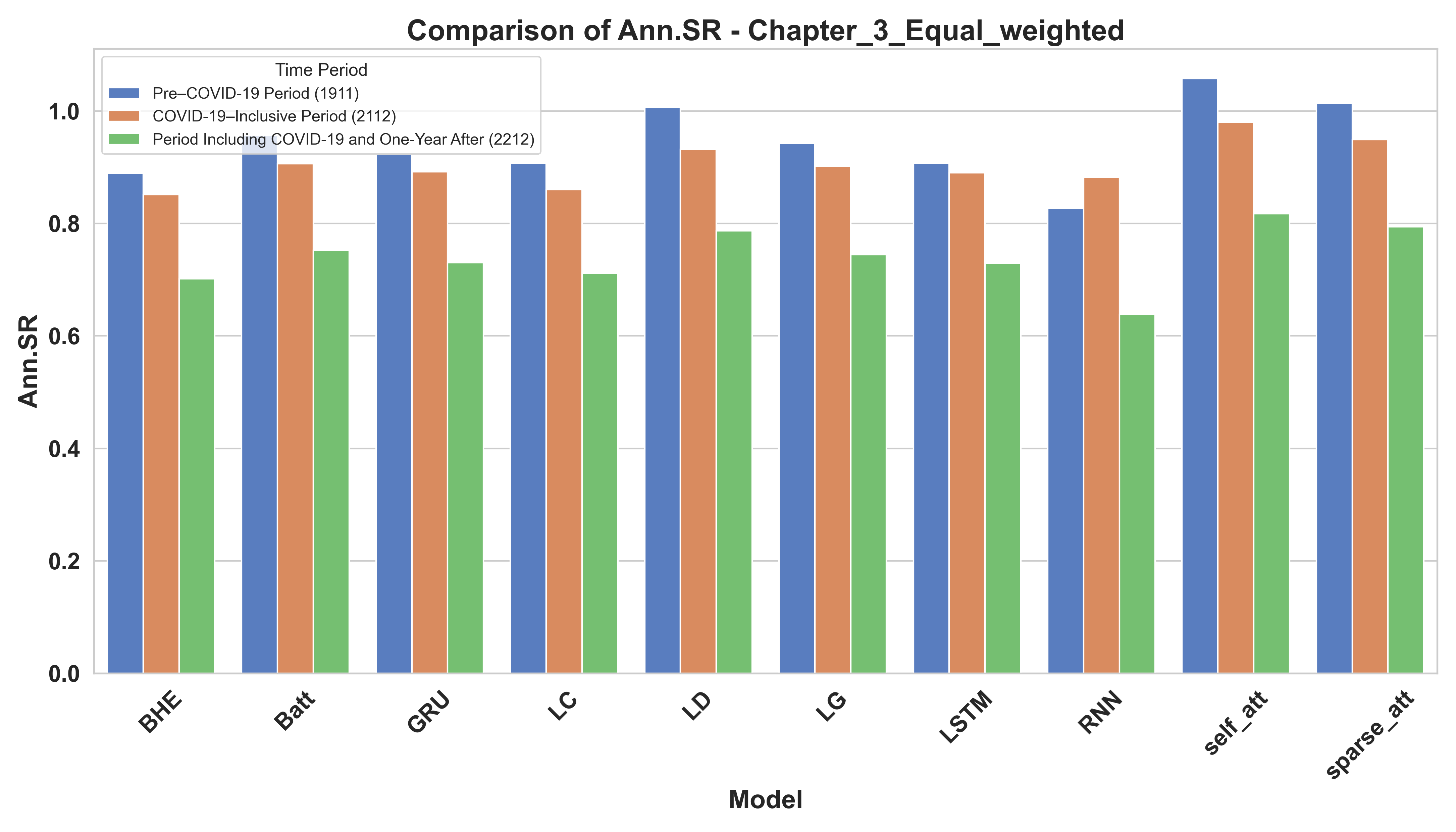}
        \caption{Annualized Sharpe ratio comparison}
        \label{fig:sr_bar_ew_ch2}
    \end{subfigure}
    
    \vspace{0.1cm} 
    
    \begin{subfigure}{\textwidth}
        \centering
        \includegraphics[width=0.9\linewidth]{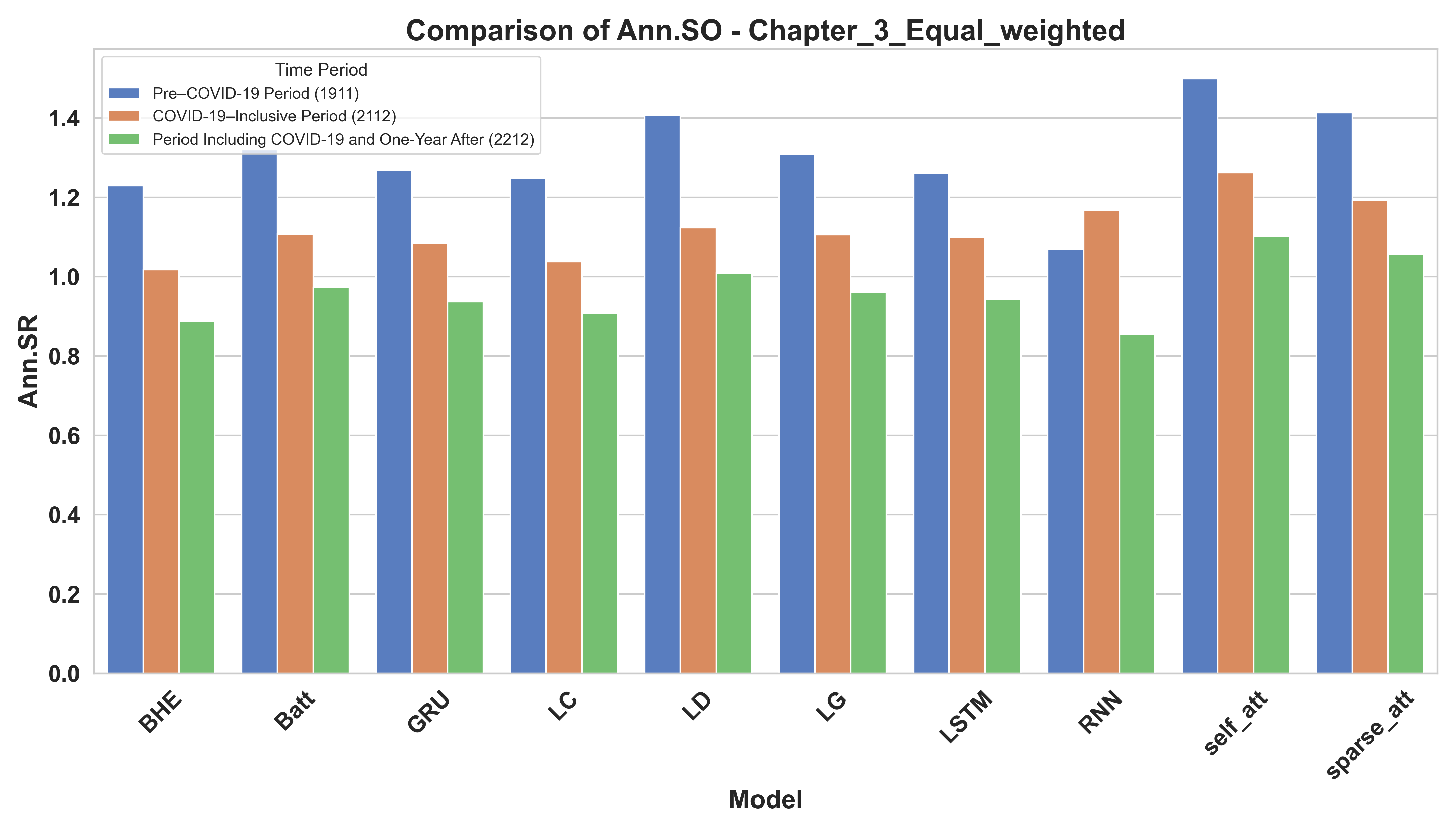}
        \caption{Annualized Sortino ratio comparison}
        \label{fig:so_bar_ew_ch2}
    \end{subfigure}

    \caption[Annualized Sharpe ratio and Sortino ratio for Equal-weighted considering the dynamic transaction cost.]{Annualized Sharpe ratio and Sortino ratio for Equal-weighted the dynamic transaction cost. `Batt' indicates the additive attention model (B-Additive), `LC', `LD' and `LG' mean L-Concat, L-DotProd and L-General attention model, respectively.}
    \label{fig:sr_so_ew_ch3_turnover}
\end{figure}

\begin{sidewaysfigure}[htbp!]
\centering
\includegraphics[width=1\columnwidth]{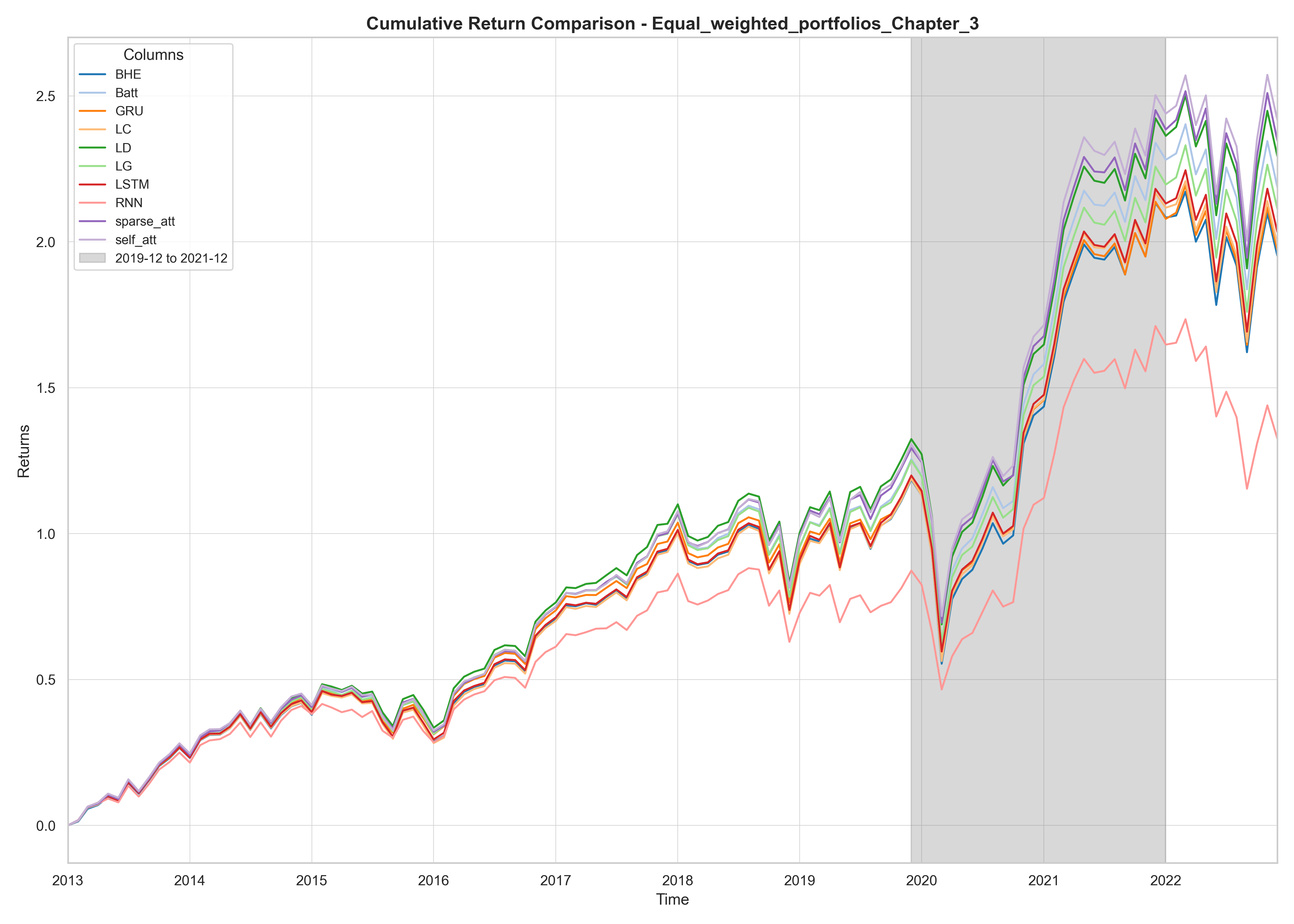}
\caption[Cumulative return plot for Equal-weighted portfolios considering the dynamic transaction cost.]{Cumulative return plot for Equal-weighted portfolios considering the dynamic transaction cost. `Batt' indicates the additive attention model (B-Additive), `LC', `LD' and `LG' mean L-Concat, L-DotProd and L-General attention model, respectively.}
\label{fig:cum_ret_ew_ch3_turnover}
\end{sidewaysfigure}

In the value-weighted portfolios, which are shown in Table~\ref{tab:value_weight_perf}, apart from the proposed models and the vanilla RNN model, the annualized returns of all models in all periods are significantly higher than those in the equal-weighted portfolio. Annulized returns of the vanilla RNN model in all periods are considerably lower than those in the equal-weighted portfolio, but annulized returns of the proposed models vary according to the periods. In Period 1911 and Period 2212, the RNN global self-attention (self\_att) model's annualized return is lower than the one in the same period of the equal-weighted portfolio, and the RNN sliding window attention model's annualized return is higher than the one in the same period of the equal-weighted portfolio. However, in Period 2212, the annualized returns of the proposed models become lower than those in equal-weighted portfolios. Specifically, the difference of the annualized return in the RNN sliding window attention model (sparse\_att) is quite insignificant, but the one in self\_att is significant. Nevertheless, when returns are assessed in light of volatility risks, which are shown as the Sharpe ratio and the Sortino ratio, all models in all periods illustrate significant advantages compared with those in equal-weighted portfolios. This phenomenon implies that all models work better on stocks with larger market capitalization when considering risks, but the annualized returns variation also indicates that the vanilla RNN model and self\_att model are less preferred by larger market-cap stocks. \\

\begin{sidewaysfigure}[htbp!]
\centering
\includegraphics[width=1\columnwidth]{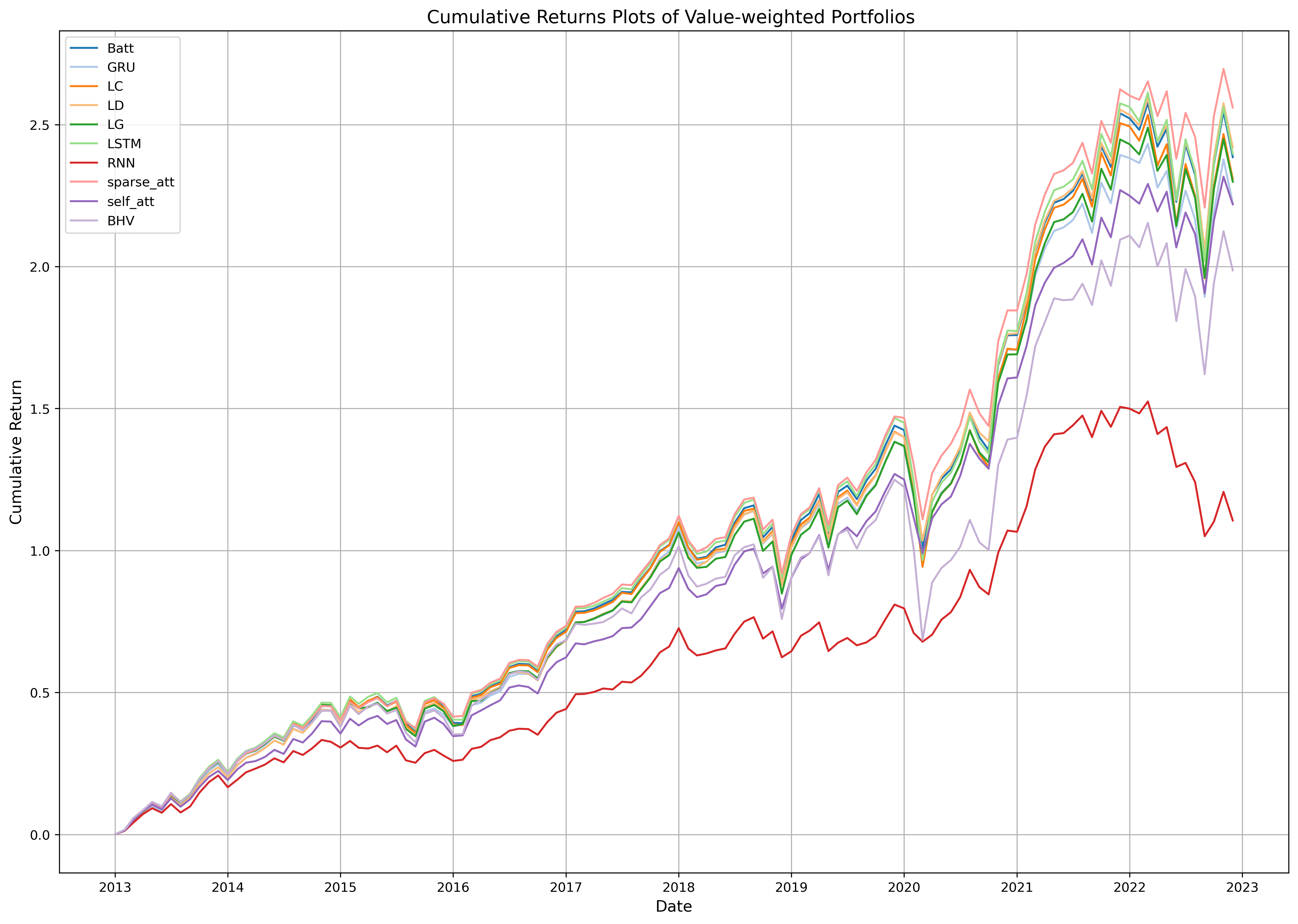}
\caption[Cumulative return plot for value-weighted portfolios considering the static transaction cost.]{Cumulative return plot for value-weighted portfolios considering the static transaction cost. `Batt' indicates the additive attention model (B-Additive), `LC', `LD' and `LG' mean L-Concat, L-DotProd and L-General attention model, respectively.}
\label{fig:cum_ret_vw_ch3}
\end{sidewaysfigure}

The rest of the findings from equal-weighted portfolios still stand for the value-weighted portfolios. All models in all periods outperform the buy-and-hold (BHV) strategy in value-weighted portfolios, excluding the vanilla RNN model in Period 1911, whose Sortino ratio is lower than that of BHV. From a full angle, value-weighted portfolios have a higher capability to moderate market fluctuation risks, and most of the models have higher annualized returns than equal-weighted portfolios. It is noteworthy to notice the rank change of the GRU model in Period 2112 of the value-weighted portfolio according to its Sortino ratio; it outperforms the RNN benchmark attention models (B-Additive, L-Concat, L-DotProd, L-General), demonstrating that the GRU model works more efficiently on handling the downside risk during the market turbulence for larger-cap stocks. The value-weighted portfolio cumulative return plots in Figure~\ref{fig:cum_ret_vw_ch3} support the conclusions drawn from Table~\ref{tab:value_weight_perf}.\\ 
\begin{table}[htbp!]
\centering
\small
\captionsetup{justification=raggedright,singlelinecheck=false,position=bottom}
\begin{tabularx}{\textwidth}{lXXXXXXX}
\toprule
\textbf{Strategy} & \textbf{Max Drawdown} & \textbf{Ann Return} & \textbf{Sharpe Ratio (SR)} & \textbf{Sortino Ratio (SO)} & \textbf{Returns Std} & \textbf{Ann SR} & \textbf{Ann SO} \\
\midrule
\multicolumn{8}{c}{\textbf{Pre–COVID-19 Period (1911)}} \\
\midrule
B-Additive        & (0.2526) & 0.1338 & 0.3507 & 0.4787 & 0.0300 & 1.2149 & 1.6582 \\
GRU         & (0.2494) & 0.1335 & 0.3578 & 0.4922 & 0.0293 & 1.2394 & 1.7049 \\
L-Concat          & (0.2639) & 0.1319 & 0.3400 & 0.4699 & 0.0306 & 1.1777 & 1.6278 \\
L-DotProd          & (0.2196) & 0.1343 & 0.3634 & 0.5135 & 0.0290 & 1.2589 & 1.7787 \\
L-General          & (0.2460) & 0.1302 & 0.3499 & 0.4692 & 0.0292 & 1.2122 & 1.6252 \\
LSTM        & (0.2477) & 0.1363 & 0.3496 & 0.4806 & 0.0307 & 1.2112 & 1.6649 \\
RNN         & (0.2848) & 0.0827 & 0.3021 & 0.3955 & 0.0212 & 1.0466 & 1.3701 \\
self\_att   & (0.2560) & 0.1299 & 0.3800 & 0.5616 & 0.0267 & 1.3164 & 1.9455 \\
sparse\_att & (0.2258) & 0.1339 & 0.3713 & 0.5048 & 0.0282 & 1.2864 & 1.7488 \\
BHV    & (0.2963) & 0.1197 & 0.2956 & 0.4243 & 0.0323 & 1.0240 	& 1.4698 \\
\midrule
\multicolumn{8}{c}{\textbf{COVID-19–Inclusive Period (2112)}} \\
\midrule
B-Additive        & (0.3034) & 0.1515 & 0.3383 & 0.4636 & 0.0355 & 1.1720 & 1.6060 \\
GRU         & (0.2684) & 0.1474 & 0.3488 & 0.5016 & 0.0333 & 1.2083 & 1.7377 \\
L-Concat          & (0.3394) & 0.1487 & 0.3224 & 0.4270 & 0.0367 & 1.1167 & 1.4792 \\
L-DotProd          & (0.2888) & 0.1524 & 0.3511 & 0.4881 & 0.0343 & 1.2161 & 1.6907 \\
L-General          & (0.2895) & 0.1473 & 0.3446 & 0.4656 & 0.0338 & 1.1936 & 1.6129 \\
LSTM        & (0.3264) & 0.1539 & 0.3322 & 0.4414 & 0.0368 & 1.1509 & 1.5292 \\
RNN         & (0.2811) & 0.1071 & 0.3497 & 0.5364 & 0.0238 & 1.2112 & 1.8581 \\
self\_att   & (0.2829) & 0.1395 & 0.3706 & 0.5785 & 0.0295 & 1.2837 & 2.0039 \\
sparse\_att & (0.2483) & 0.1498 & 0.3677 & 0.5204 & 0.0320 & 1.2737 & 1.8028 \\
BHV    & (0.4513) & 0.1337 & 0.2673 & 0.3263 & 0.0408 & 0.9260 & 1.1303 \\
\midrule
\multicolumn{8}{c}{\textbf{Period Including COVID-19 and One-Year After (2212)}} \\
\midrule
B-Additive        & (0.2996) & 0.1297 & 0.2757 & 0.3970 & 0.0380 & 0.9551 & 1.3754 \\
GRU         & (0.2915) & 0.1241 & 0.2745 & 0.3952 & 0.0364 & 0.9507 & 1.3690 \\
L-Concat          & (0.3354) & 0.1271 & 0.2599 & 0.3615 & 0.0398 & 0.9003 & 1.2524 \\
L-DotProd          & (0.2691) & 0.1308 & 0.2883 & 0.4195 & 0.0364 & 0.9985 & 1.4533 \\
L-General          & (0.2989) & 0.1268 & 0.2742 & 0.3858 & 0.0373 & 0.9500 & 1.3366 \\
LSTM        & (0.3410) & 0.1301 & 0.2623 & 0.3585 & 0.0404 & 0.9087 & 1.2420 \\
RNN         & (0.3113) & 0.0773 & 0.2295 & 0.3120 & 0.0266 & 0.7951 & 1.0807 \\
self\_att   & (0.2581) & 0.1240 & 0.3120 & 0.4985 & 0.0315 & 1.0808 & 1.7269 \\
sparse\_att  & (0.2462) & 0.1354 & 0.3070 & 0.4523 & 0.0351 & 1.0636 & 1.5668 \\
BHV    & (0.4513) & 0.1156 & 0.2311 & 0.3031 & 0.0438 & 0.8006 & 	1.0500 \\
\bottomrule
\end{tabularx}
\caption[Value-weighted portfolio performance considering static transaction cost.]{Value-weighted portfolio performance considering static transaction cost.`Ann' is the notation for annulization, while `SR' and `SO' are short for Sharpe ratio and Sortino ratio respectively. The notation of `Std' means standard deviation, and `BHE' means buy-and-hold strategy in equal-weighted portfolios. The values in the parentheses indicate negative values. `Ann Return' and `Max Drawdown' can be transformed to a percentage presentation by multiplying by 100.}
\label{tab:value_weight_perf}
\end{table}

\begin{figure}[htbp!]
    \centering
    
    \begin{subfigure}{\textwidth}
        \centering
        \includegraphics[width=0.9\linewidth]{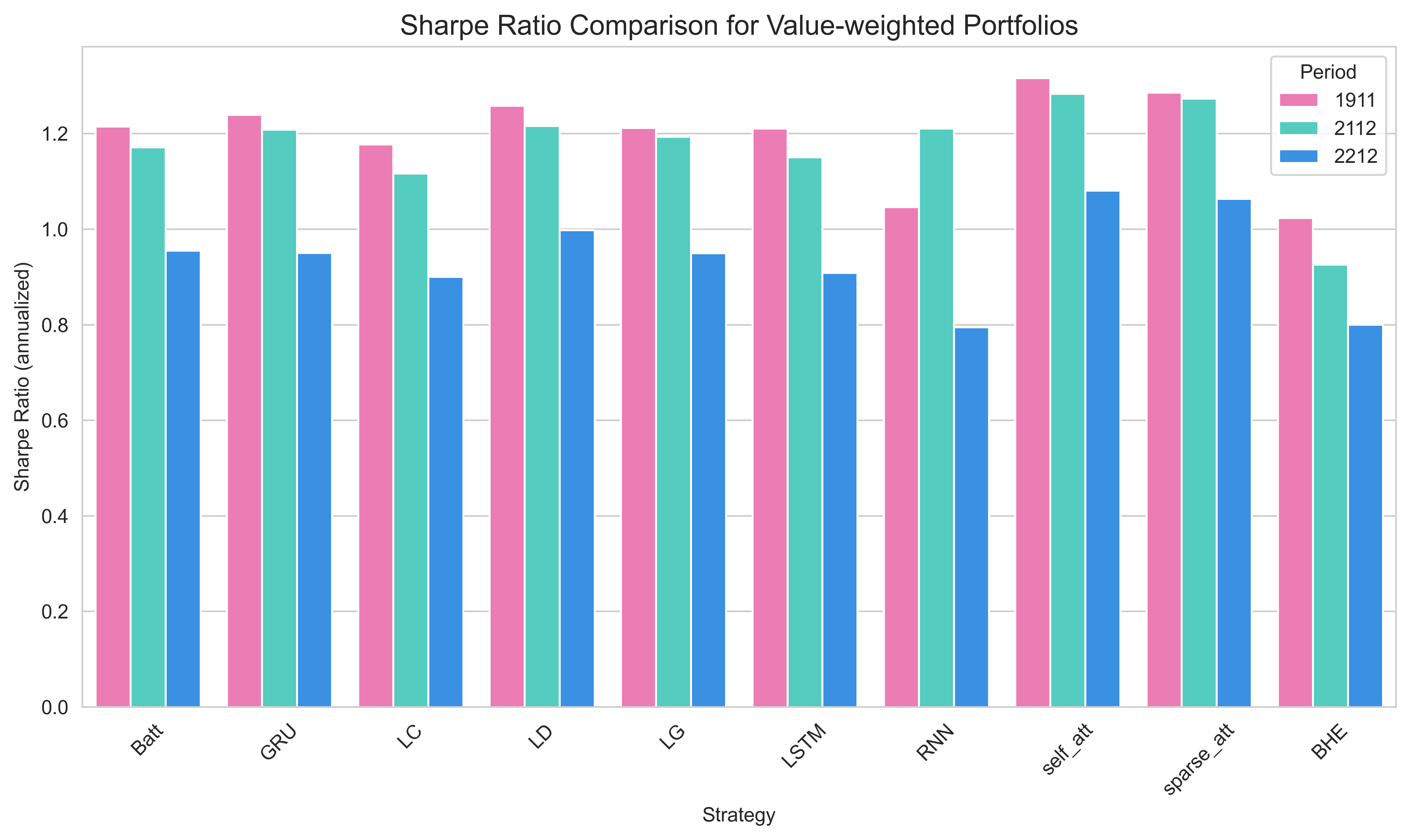}
        \caption{Annualized Sharpe ratio comparison}
        \label{fig:sr_bar_ew_ch2}
    \end{subfigure}
    
    \vspace{0.1cm} 
    
    \begin{subfigure}{\textwidth}
        \centering
        \includegraphics[width=0.9\linewidth]{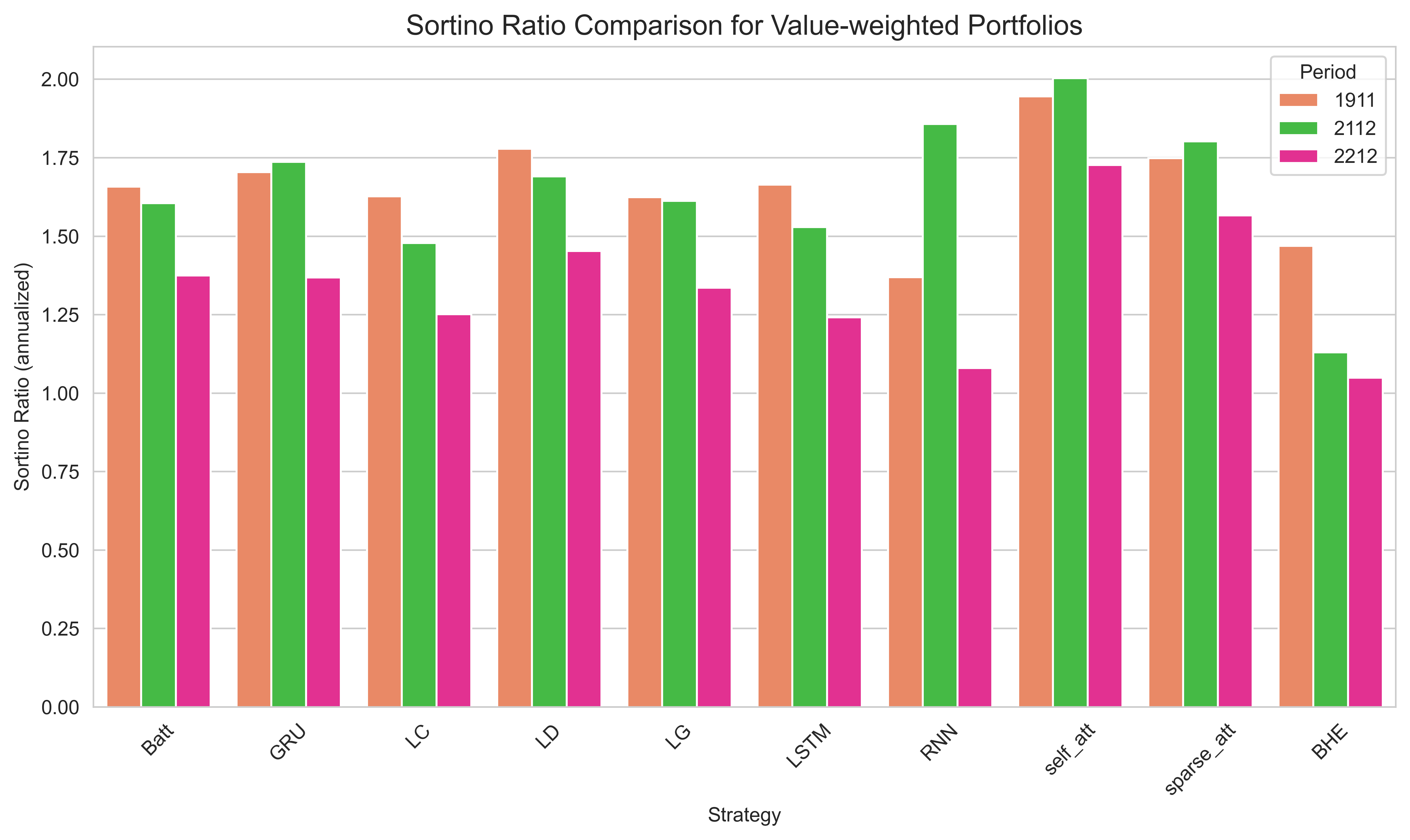}
        \caption{Annualized Sortino ratio comparison}
        \label{fig:so_bar_vw_ch2}
    \end{subfigure}

    \caption[Annualized Sharpe ratio and Sortino ratio for value-weighted considering the static transaction cost.]{Annualized Sharpe ratio and Sortino ratio for value-weighted considering the static transaction cost. `Batt' indicates the additive attention model (B-Additive), `LC', `LD' and `LG' mean L-Concat, L-DotProd and L-General attention model, respectively.}
    \label{fig:sr_so_vw_ch3}
\end{figure}

The comparison of the Sharpe ratios and the Sortino ratios for value-weighted portfolios is exhibited in Figure~\ref{fig:sr_so_vw_ch3}. The value-weighted method narrows the difference in periodic performance between models. In the Sharpe ratio diagram, the findings are consistent with those in the equal-weighted method diagram. The self\_att model outperforms all alternatives, followed by the sparse\_att model. And the difference between the proposed models is negligible. All models perform similarly in Period 1911 and Period 2112, which indicates less advantage shows up for these models on larger-cap stocks when the market has a mild or sharp uptrend. However, the Sharpe ratios are significantly reduced when the market slumps. This phenomenon is less significant in the equal-weighted method. This provides indirect evidence that these models exhibit stronger effectiveness in moderating downside risks for volatile (smaller-cap) stocks. The information from the diagrams of the Sortino ratios in both portfolio weighting methods agrees with this stance. However, the noticeably high sortino ratio of the vanilla RNN model in Period 2112 reflects the sensitivity of the vanilla RNN model towards the downside risks caused by the extreme market turbulence when larger market capitalization presents. Model GRU shows similar characteristics.\\

From the cumulative return plot Figure~\ref{fig:cum_ret_vw_ch3} in the value-weighted method considering static transaction cost, we can see value-weighted portfolios narrow the difference between models to a degree during the pandemic and post-pandemic period, except for the vanilla RNN model, self\_att model and buy-and-hold (BHV) benchmark. The performance of the proposed sparse\_att model is stable in both the equal-weighted portfolios and the value-weighted portfolios, and it ranks as the best model after 2016 due to the best model in the equal-weighted portfolios, self\_att, dropping behind all RNN attention benchmark models and RNN variation models and losing its advantage when heavy weights are added to the larger-cap stocks. This phenomenon implies that the self\_att model latently performs better on smaller-cap stocks, and it is sensitive to variations in time series. The variation is larger, the performance of the self\_att is better. A similar phenomenon happens on the GRU model, while it drops right before the rank of the self\_att model during the market turbulence in the value-weighted case. However, the other RNN variation model, LSTM, presents a reasonable capability in profitability among the larger-cap stocks, which ranks in first place before 2016 and second place after 2016. Likewise, the L-DotProd model remains in third place in the value-weighted case, which means Model sparse\_att and Model L-DotProd are less likely to be affected by the size of stocks' market capitalization. They consistently possess great back-testing performance, which implies they are suitable for stocks with different levels of volatility. It is also interesting to notice that the vanilla RNN model performs worse in cumulative return plots. It shows the capacity for hedging downside risks instead of deriving the absolute returns, but it does not work well during market turbulence when heavy weights are given to larger-cap stocks. It indicates a preference for volatile stocks. \\

\begin{figure}[htbp!]
    \centering
    
    \begin{subfigure}{\textwidth}
        \centering
        \includegraphics[width=0.9\linewidth]{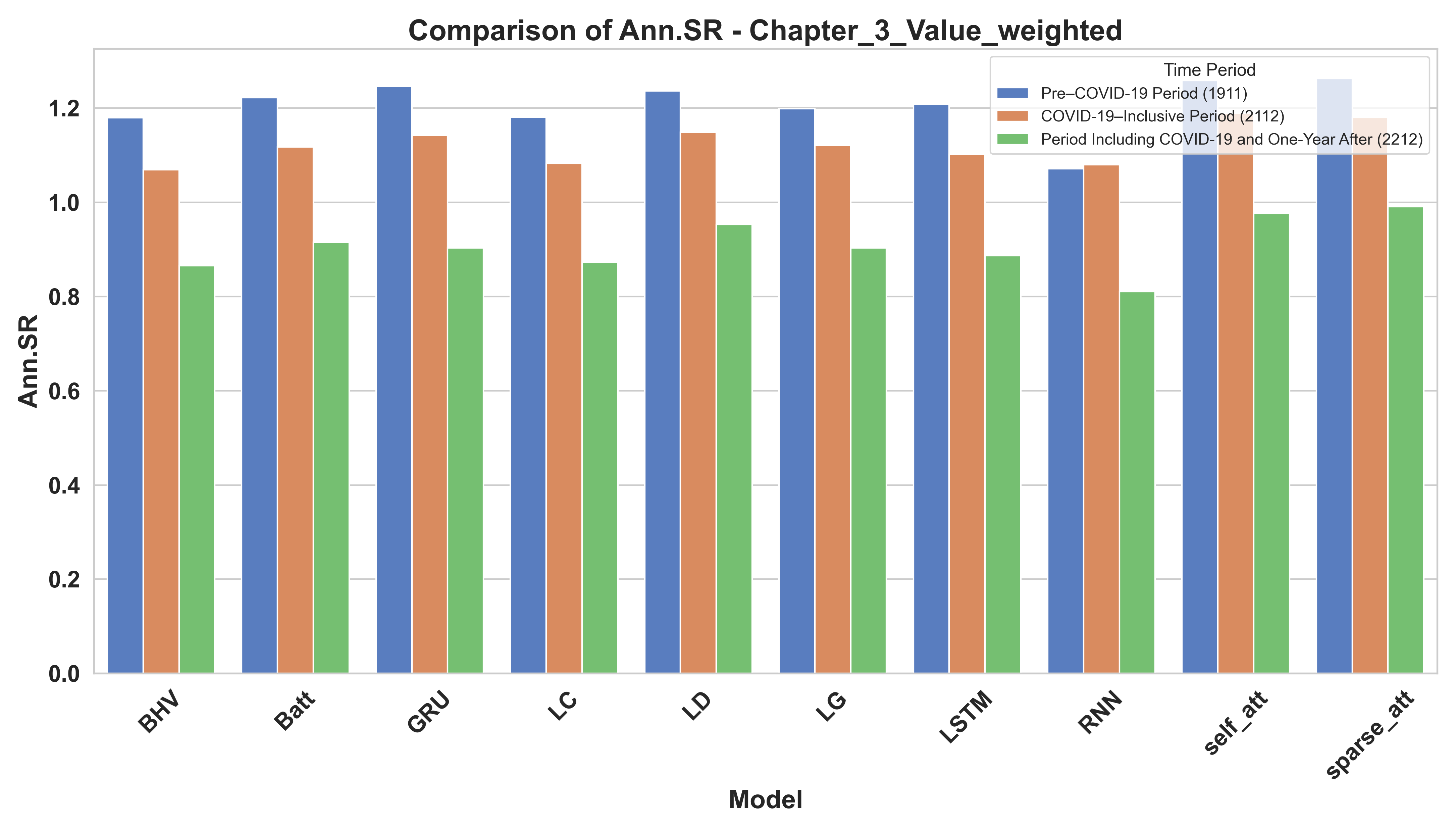}
        \caption{Annualized Sharpe ratio comparison}
        \label{fig:sr_bar_vw_ch2}
    \end{subfigure}
    
    \vspace{0.3cm} 
    
    \begin{subfigure}{\textwidth}
        \centering
        \includegraphics[width=0.9\linewidth]{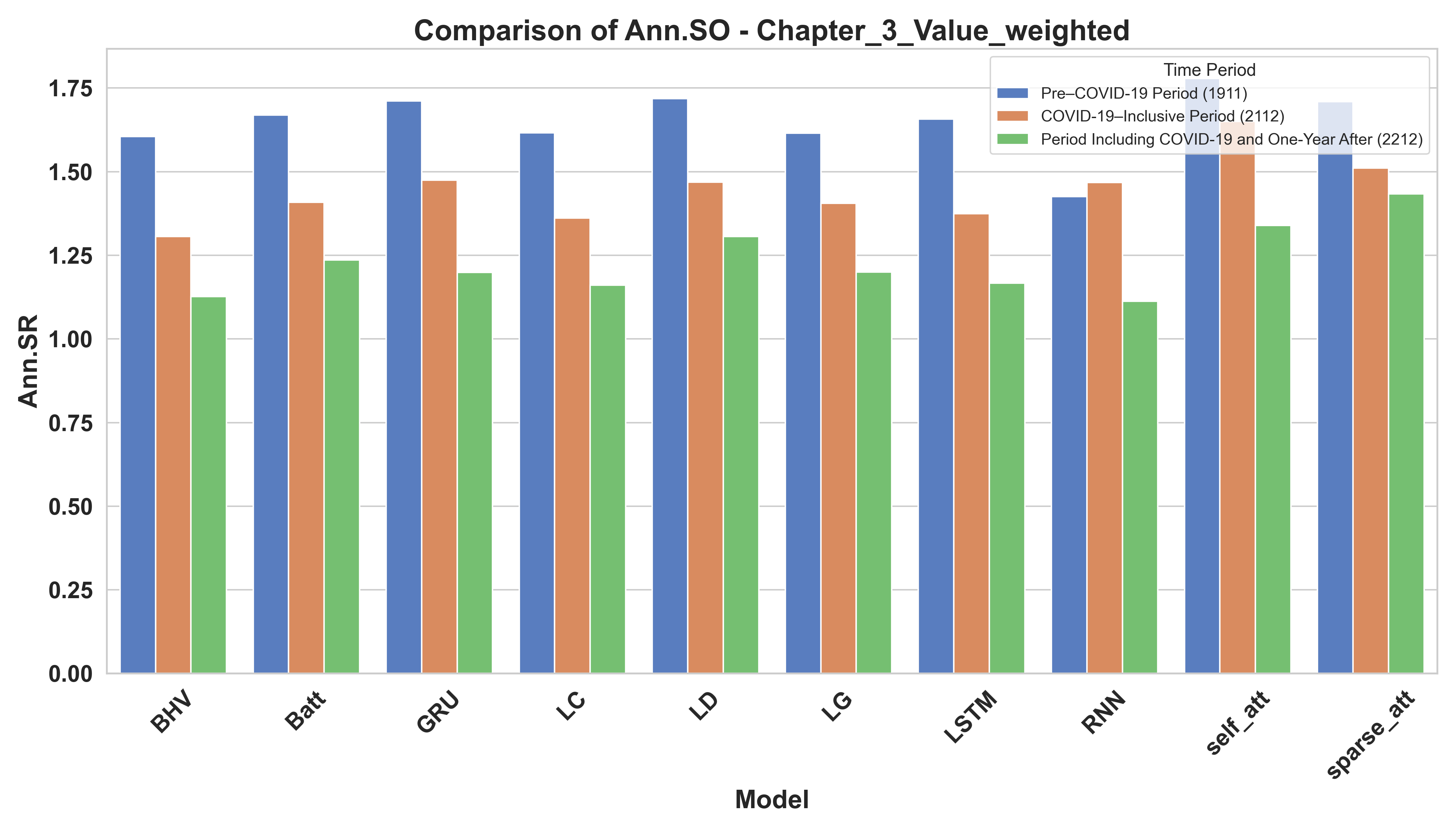}
        \caption{Annualized Sortino ratio comparison}
        \label{fig:so_bar_vw_ch2}
    \end{subfigure}

    \caption[Annualized Sharpe ratio and Sortino ratio for Value-weighted portfolios considering the dynamic transaction cost.]{Annualized Sharpe ratio and Sortino ratio for Value-weighted portfolios considering dynamic transaction cost (turnover and 20bps). `Batt' indicates the additive attention model (B-Additive), `LC', `LD' and `LG' mean L-Concat, L-DotProd and L-General attention model, respectively.}
    \label{fig:sr_so_vw_turnover_ch3}
    
\end{figure}

\begin{sidewaysfigure}[htbp!]
\centering
\includegraphics[width=1\columnwidth]{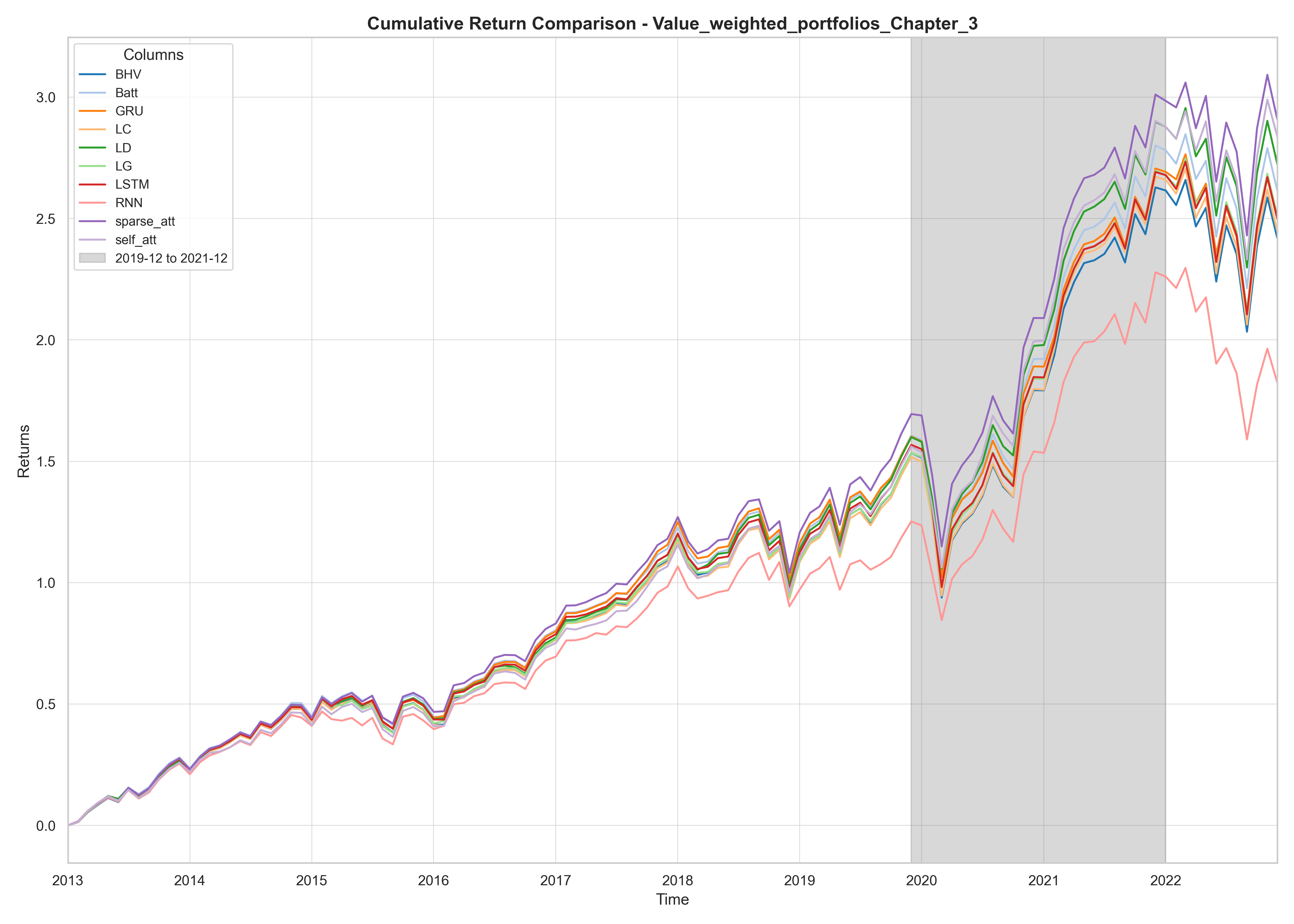}
\caption[Cumulative return plot for Value-weighted portfolios considering the dynamic transaction cost.]{Cumulative return plot for Value-weighted portfolios considering the dynamic transaction cost (turnover and 20bps). `Batt' indicates the additive attention model (B-Additive), `LC', `LD' and `LG' mean L-Concat, L-DotProd and L-General attention model, respectively.}
\label{fig:cum_ret_vw_turnover_ch2}
\end{sidewaysfigure}

The value-weighted method is also examined for its robustness by considering its dynamic transaction costs (turnover and 20bps for large-cap). Table~\ref{tab:value_weight_perf_turnover_ch3} presents the backtesting results of value-weighted portfolios considering the dynamic transaction cost. Figure~\ref{fig:cum_ret_vw_turnover_ch2} shows the variation of cumulative returns over time with the consideration of turnover rate. Figure ~\ref{fig:sr_so_vw_turnover_ch3} visualizes the annualized Sharpe and Sortino ratios of three periods as a comparison of the profitability in terms of risks for each model-based strategy in three market regimes. The robustness test results support the conclusion drawn from the equal-weighted methods' robustness test.\\

\begin{table}[htbp]
  \centering
  \small 
  \begin{tabular}{lccccccc}
    \toprule
    \textbf{Model} & \textbf{AR} & \textbf{Ann.SR} & \textbf{SR} & \textbf{Ann.SO} & \textbf{SO} & \textbf{MDD} & \textbf{Turnover} \\
    \midrule
    \multicolumn{8}{c}{\textbf{Pre-COVID-19 Period (1911)}} \\
    \midrule
    BHV        & 0.1388 & 1.1795 & 0.3405 & 1.6054 & 0.4634 & -0.1297 & 0.0120 \\
    B-Additive       & 0.1445 & 1.2223 & 0.3529 & 1.6689 & 0.4818 & -0.1311 & 0.0589 \\
    GRU        & 0.1446 & 1.2462 & 0.3598 & 1.7117 & 0.4941 & -0.1265 & 0.0693 \\
    L-Concat         & 0.1394 & 1.1812 & 0.3410 & 1.6162 & 0.4666 & -0.1317 & 0.0434 \\
    L-DotProd         & 0.1458 & 1.2361 & 0.3568 & 1.7188 & 0.4962 & -0.1218 & 0.0747 \\
    L-General         & 0.1399 & 1.1986 & 0.3460 & 1.6147 & 0.4661 & -0.1322 & 0.0656 \\
    LSTM       & 0.1425 & 1.2075 & 0.3486 & 1.6573 & 0.4784 & -0.1286 & 0.0446 \\
    RNN        & 0.1152 & 1.0714 & 0.3093 & 1.4257 & 0.4116 & -0.1046 & 0.4718 \\
    self\_att  & 0.1459 & 1.2586 & 0.3633 & 1.7786 & 0.5134 & -0.1150 & 0.1142 \\
    sparse\_att& 0.1475 & 1.2627 & 0.3645 & 1.7094 & 0.4935 & -0.1256 & 0.0847 \\
    \midrule
    \multicolumn{8}{c}{\textbf{COVID-19-Inclusive Period (2112)}} \\
    \midrule
    BHV        & 0.1540 & 1.0688 & 0.3085 & 1.3057 & 0.3769 & -0.2352 & 0.0093 \\
    B-Additive       & 0.1601 & 1.1172 & 0.3225 & 1.4088 & 0.4067 & -0.2244 & 0.0516 \\
    GRU        & 0.1597 & 1.1421 & 0.3297 & 1.4750 & 0.4258 & -0.2133 & 0.0922 \\
    L-Concat         & 0.1555 & 1.0829 & 0.3126 & 1.3614 & 0.3930 & -0.2273 & 0.0425 \\
    L-DotProd         & 0.1630 & 1.1487 & 0.3316 & 1.4691 & 0.4241 & -0.2182 & 0.0674 \\
    L-General         & 0.1566 & 1.1208 & 0.3235 & 1.4051 & 0.4056 & -0.2151 & 0.0599 \\
    LSTM       & 0.1578 & 1.1020 & 0.3181 & 1.3744 & 0.3968 & -0.2262 & 0.0450 \\
    RNN        & 0.1386 & 1.0797 & 0.3117 & 1.4675 & 0.4236 & -0.1842 & 0.4679 \\
    self\_att  & 0.1624 & 1.1891 & 0.3433 & 1.6505 & 0.4765 & -0.1971 & 0.1168 \\
    sparse\_att& 0.1652 & 1.1802 & 0.3407 & 1.5108 & 0.4361 & -0.2118 & 0.0901 \\
    \midrule
    \multicolumn{8}{c}{\textbf{Period Including COVID-19 and One-Year After (2212)}} \\
    \midrule
    BHV        & 0.1308 & 0.8654 & 0.2498 & 1.1270 & 0.3253 & -0.2352 & 0.0083 \\
    B-Additive       & 0.1371 & 0.9155 & 0.2643 & 1.2361 & 0.3568 & -0.2216 & 0.0540 \\
    GRU        & 0.1333 & 0.9033 & 0.2608 & 1.1988 & 0.3461 & -0.2224 & 0.0990 \\
    L-Concat         & 0.1317 & 0.8727 & 0.2519 & 1.1607 & 0.3351 & -0.2270 & 0.0392 \\
    L-DotProd         & 0.1405 & 0.9526 & 0.2750 & 1.3059 & 0.3770 & -0.2086 & 0.0652 \\
    L-General         & 0.1338 & 0.9033 & 0.2608 & 1.1996 & 0.3463 & -0.2182 & 0.0548 \\
    LSTM       & 0.1334 & 0.8867 & 0.2560 & 1.1663 & 0.3367 & -0.2282 & 0.0397 \\
    RNN        & 0.1094 & 0.8103 & 0.2339 & 1.1122 & 0.3211 & -0.2144 & 0.4904 \\
    self\_att  & 0.1439 & 0.9905 & 0.2859 & 1.4336 & 0.4138 & -0.1950 & 0.1101 \\
    sparse\_att & 0.1460 & 0.9765 & 0.2819 & 1.3391 & 0.3866 & -0.2023 & 0.0810 \\
    \bottomrule
  \end{tabular}
\caption[Value-weighted portfolio performance considering the dynamic transaction cost.]{Value-weighted portfolio performance considering the dynamic transaction cost (turnover and 20bps). `Ann' is the notation for annulization, while `SR' and `SO' are short for Sharpe ratio and Sortino ratio respectively. The notation of `MDD' means Maximum drawdown, and `BHV' means buy-and-hold strategy in value-weighted portfolios. The Annulized return (AR) and `MDD' can be transformed to a percentage presentation by multiplying by 100.}
\label{tab:value_weight_perf_turnover_ch3}
\end{table}

\clearpage
\section{Conclusion}\label{sec:Conclusion_ch2}
This study investigates the pre-trained RNN attention models with the mainstream attention mechanisms such as additive attention \citep{Bahdanau2015NeuralTranslate}, Luong's three attentions \citep{Luong2015EffectiveTranslation}, global self-attention \citep{Vaswani2017AttentionNeed} and sliding window sparse attention \citep{beltagy2020longformer} for the empirical asset pricing assignment. The research emphasises the pre-trained RNN global self-attention model (self\_att) and the pre-trained RNN sliding window sparse attention model (sparse\_att) as they are the innovative structures in different domains, especially for NLP. Less literature systematically investigates the role of attention mechanisms in the context of asset pricing. This is the first empirical asset pricing research which employs the RNN frames with the global self-attention mechanism, sliding window sparse attention mechanism and Luong's three attention mechanisms \citep{Luong2015EffectiveTranslation}. They solve or moderate the issues in classic ML model-based empirical asset pricing research, such as sequence or time series data recognition, models' financial or economic interpretability and overfitting caused by over-parameterization of the ML models. It also investigates the necessity of the attention mechanisms in the context of ML-based asset pricing by constructing the benchmark models of the pre-trained RNN additive attention model (B-Additive), Luong's three attention models (L-General, L-DotProd, L-Concat) and no-attention RNN models (RNN, LSTM, GRU). Moreover, all models are evaluated and portfolio-wise backtested in three periods: Pre–COVID-19 Period (1911), COVID-19–Inclusive Period (2112), and Period Including COVID-19 and One-Year After (2212) to prove whether the model has the capability to handle extreme market conditions. These capabilities include absolute and risk-considered profitability during extreme market turbulence. In addition, research in this chapter also investigated how the turnover rate could affect the profitability of these models via the dynamic transaction cost robustness examination. In summary, this study pursues two objectives across two dimensions: 1) to examine whether the factors in the proposed and alternative structures can better explain the stock excess returns and satisfy the no-arbitrage assumption in the background of `factor zoo', which is one direction of empirical asset pricing research. 2) to examine the predictability and trading strategy-wise profitability of these models, which is the goal of the factor investing studies for practitioners. \\

The results from model performance evaluation find that the vanilla RNN model has the highest model out-of-sample fitness, followed by Model self\_att and Model sparse\_att, during all periods. The proposed models present stability during the extreme market fluctuations, especially for the sparse\_att model. This confirms the predictability of the vanilla RNN model and the proposed models to some extent. The average positive values of the residual $\alpha$ and the distribution plots of $\alpha$ provide strong evidence of market inefficiencies and excess predictability across the three periods. These anomalies suggest uncaptured non-linear risk premiums, which is where factor investors can potentially benefit from. These are anomalies caused by mispricing in empirical asset pricing literature, and it is where factor investors can benefit from. The significant increase in the average $\alpha$ of all models in Period 2112, which follows the trend of the price index, along with the modest differences between models within each period, further reduces the likelihood of model misspecification. This study successfully explores the meaning of the residual $\alpha$ in ML-based asset pricing.\\

From portfolio-wise backtesting, the self\_att model outperforms all alternative models when returns are considered in conjunction with risks in both equal-weighted portfolios and value-weighted portfolios, followed by the sparse\_att model. However, when returns are not considered with risks, Model self\_att is the best in equal-weighted portfolios, but deteriorates in value-weighted portfolios. Instead, Model sparse\_att shows some stability, which is in the second place of equal-weighted portfolios but is the best in the value-weighted portfolios. This implies that the self\_att model performs better on smaller market-cap stocks. This verifies the conclusions from the self-attention literature, which includes the classic Transformer literature, that the self-attention mechanism has higher sensitivity to the large variation between time steps, which makes it capable of modelling the highly volatile data. On the other hand, the sparse\_att model is not affected by the stock market capitalisation. It stably provides excellent performance without considering the volatility of the stocks. Namely, it is more generalized than the self\_att model, which is a notable advantage for both practitioners and academic scholars. Additionally, the benchmark RNN attention models in equal-weighted portfolios generally equal or outperform the models without an attention mechanism, apart from the RNN Luong's concatenate attention model (L-Concat). Nevertheless, it varies in value-weighted portfolios. The market capitalization weights seem to significantly reduce the benefit of attention mechanisms if only considering the absolute capital gain. Moreover, value-weighted portfolios narrow the difference between models. Thus, the advantage of the attention mechanism is not as significant as in the equal-weighted portfolios. The overall performance of value-weighted portfolios is higher than that of equal-weighted portfolios. Approximately, all models somehow show the capability of handling the downside risks during the period with pandemic effect, while all models outperform the buy-and-hold strategy in both portfolio groups, excluding the vanilla RNN model. And it is also interesting to notice that this model fails in the backtesting model competition, especially on the absolute returns of value-weighted portfolios. This indicates that it is easily affected by the stock volatility level. From the perspective of the Sharpe ratios, this work supports the findings from \citet{Chatigny2021AssetLearning} that in value-weighted portfolios, the large-cap stocks achieve the annualized Sharpe ratio above 1.0, but it does not persist during the period that the market was experiencing the sideways movement with extreme fluctuations. Furthermore, the transaction cost robustness examination illustrates that controlling turnover is critical for maintaining the profitability of these models. The dot-product-based attention mechanisms (L-DotProd, self\_att, sparse\_att) have significant advantages in generating more stable and selective trading signals. However, although the RNN model shows persistence in risk control, its profitability significantly decays due to the high turnover. Thus, high forecasting accuracy does not guarantee absolute profitability.\\

This study also finds some interesting research questions for further discussion. Although this work covers large-scale factors, these factors are pre-selected, which indicates that the issue of omitted factors may exist. The positive $\alpha$ values also imply this issue. Thus, alternative methods can be proposed as a substitute for the pre-selecting factors. Moreover, literature on ML-based asset pricing neglects the discussion on how to decide the number of abstracted or transformed factors. It is worth exploring the method for determining factor numbers via the ML approaches. Furthermore, it could be interesting to explore the structure of $\alpha$s and how to interpret these $\alpha$s in the ML-based asset pricing research. Finally, this study only adopts the simple asset allocation strategies, equal-weighted and value-weighted, but intuitively, ML methods can be deployed for asset allocation to further improve the strategic returns of portfolios. In addition, to further improve the financial and economic interpretability of these neural network models, the DM test can be applied to individual stocks to investigate the model's preference for an industry or sector, while investigating further on MSE and OOS $R^2$ for the contribution of characters, for example, which industry sectors have higher predictive power in a certain framework, and what is the impact of market capitalization size. Also, for attention models, attention weights variation over time can be examined to investigate which factors are emphasized during the pandemic period, and conduct subset factors from the factor pool as factor sensitivity analysis. As future work, these advanced neural network models can also be upgraded by altering the specific modules, such as adding an additional MLP structure within the Transformer core structures. Moreover, these structures can be applied to the volatility modelling for asset allocation and risk controlling, exemplified by \citet{Ma2023AttentionApproach}.  \\ 

To conclude, `attention' is something worth exploring further, but far from 'all we need' in the context of the empirical asset pricing.\\


\bmhead{Acknowledgements}

I would like to express my sincere gratitude to my supervisor, Professor Peter N. Smith (University of York), for his patient guidance, constant encouragement, and invaluable support throughout this project. I am also grateful to Dr Jiaxing Peng (University of York) for the advice on the data source. This research was supported by the Viking Computing Centre (University of York) through access to its GPU-based high-performance computing (HPC) cluster.\\

Constructive feedback is greatly appreciated.





\clearpage
\bibliography{references}

\clearpage
\begin{appendices}

\begin{sidewaysfigure}[H]
\section{Out-of-sample $\alpha$ distributions for alternative models}\label{sec:AF_alpha_ch2}
\centering
\begin{subfigure}{0.32\textwidth}
  \includegraphics[width=\linewidth]{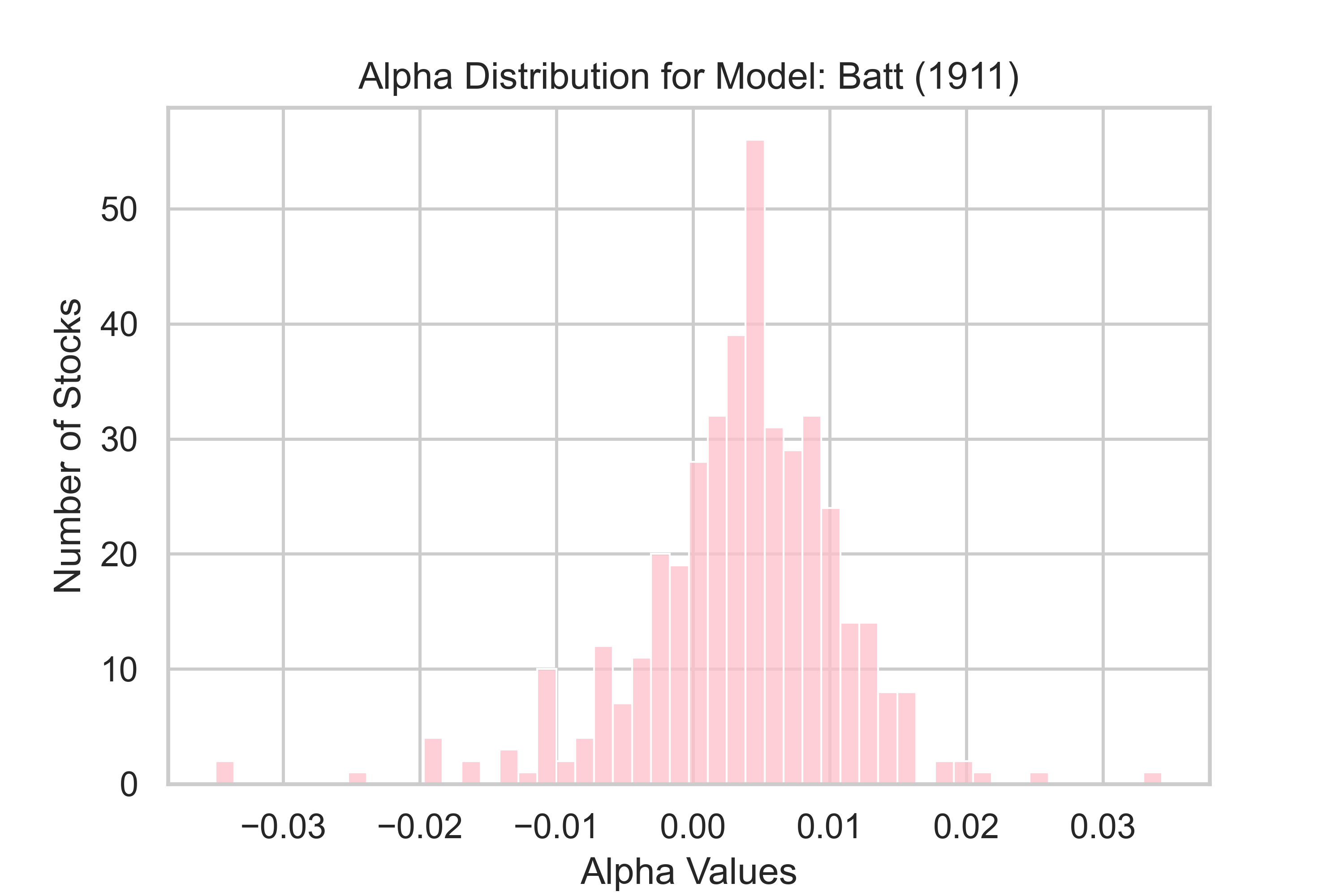}
  \caption{B-Additive(1911)}
\end{subfigure}
\hfill
\begin{subfigure}{0.32\textwidth}
  \includegraphics[width=\linewidth]
  {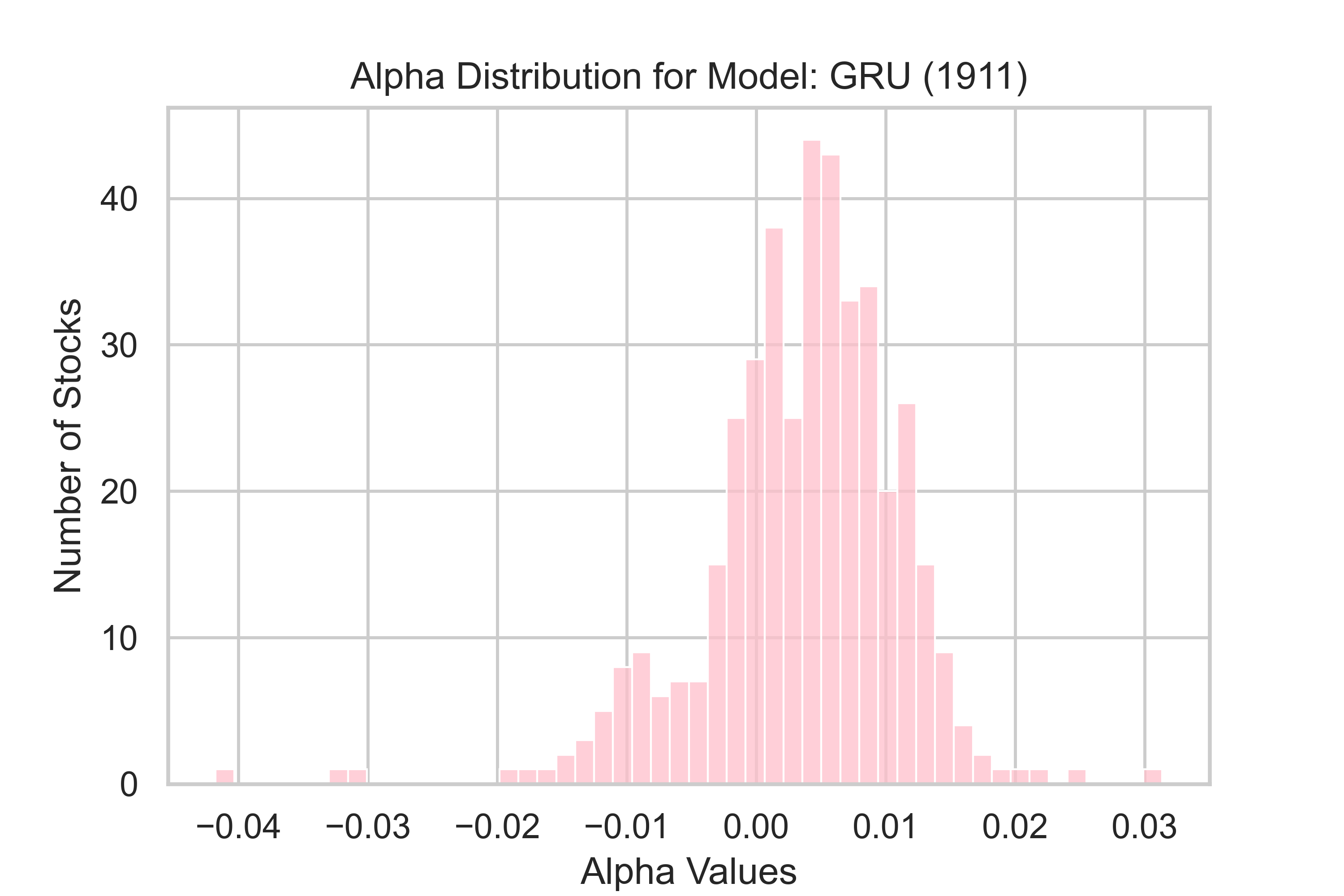}
  \caption{GRU(1911)}
\end{subfigure}
\hfill
\begin{subfigure}{0.32\textwidth}
  \includegraphics[width=\linewidth]{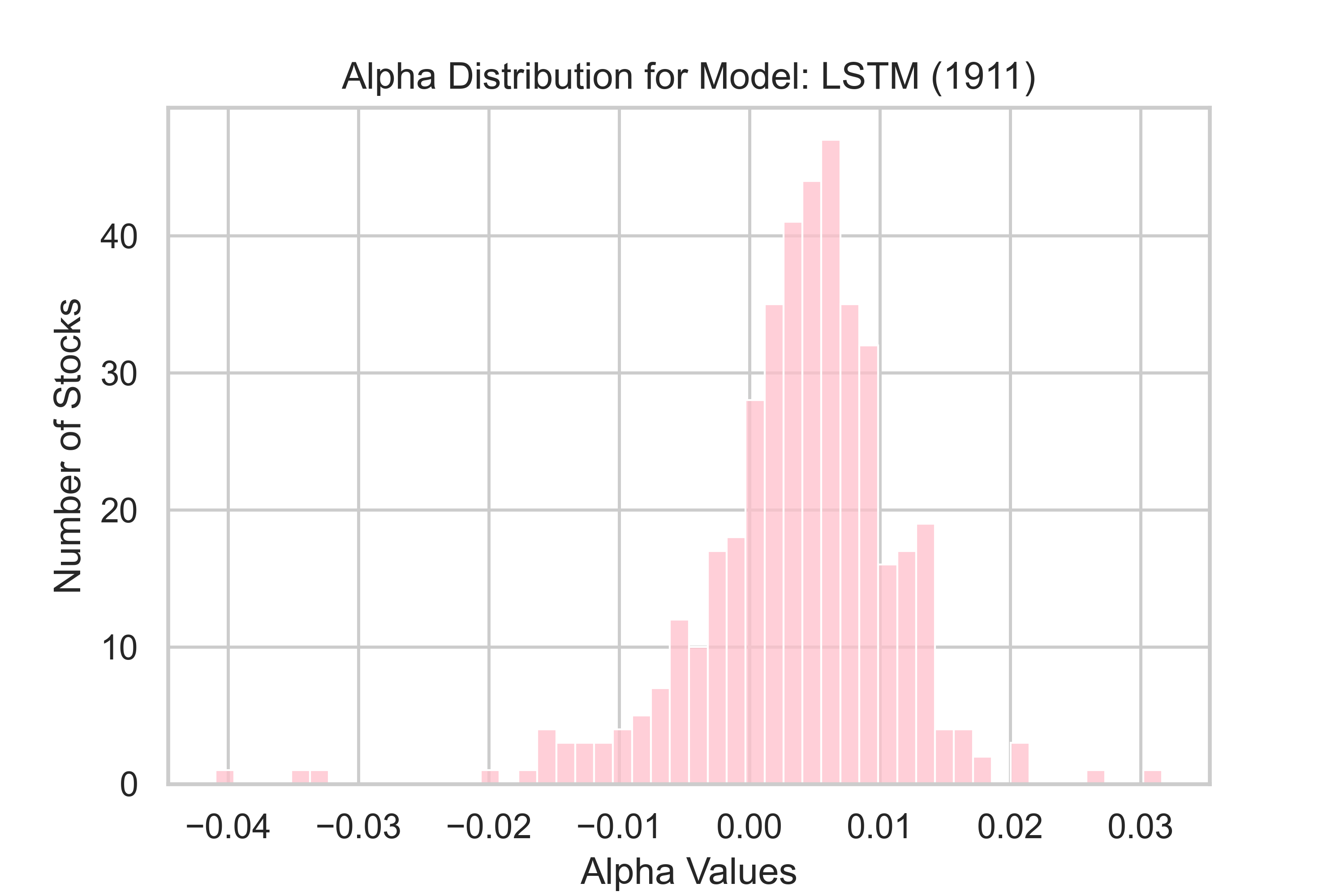}
  \caption{LSTM(1911)}
\end{subfigure}

\vspace{0.3em}

\begin{subfigure}{0.32\textwidth}
  \includegraphics[width=\linewidth]
  {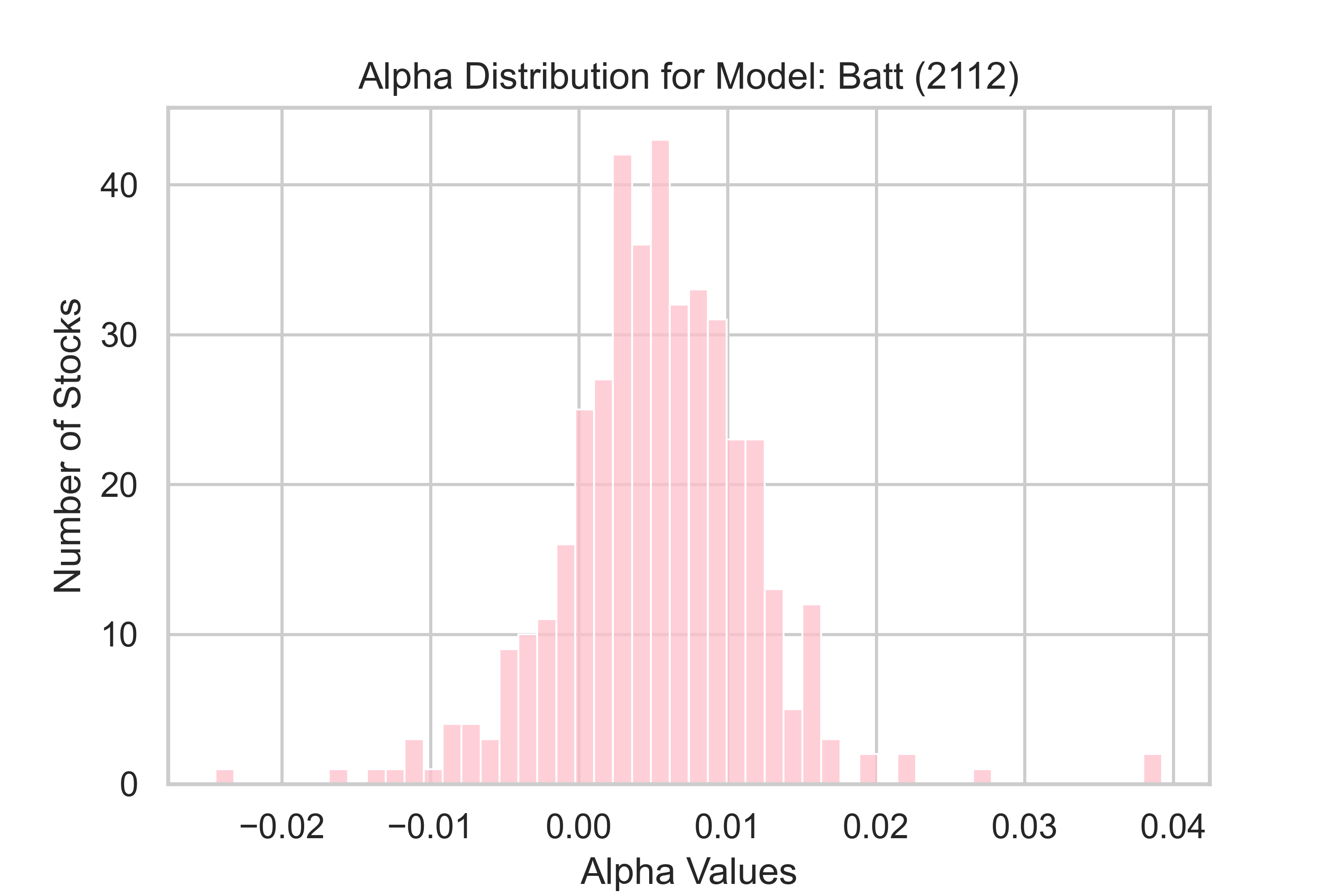}
  \caption{B-Additive(2112)}
\end{subfigure}
\hfill
\begin{subfigure}{0.32\textwidth}
  \includegraphics[width=\linewidth]
  {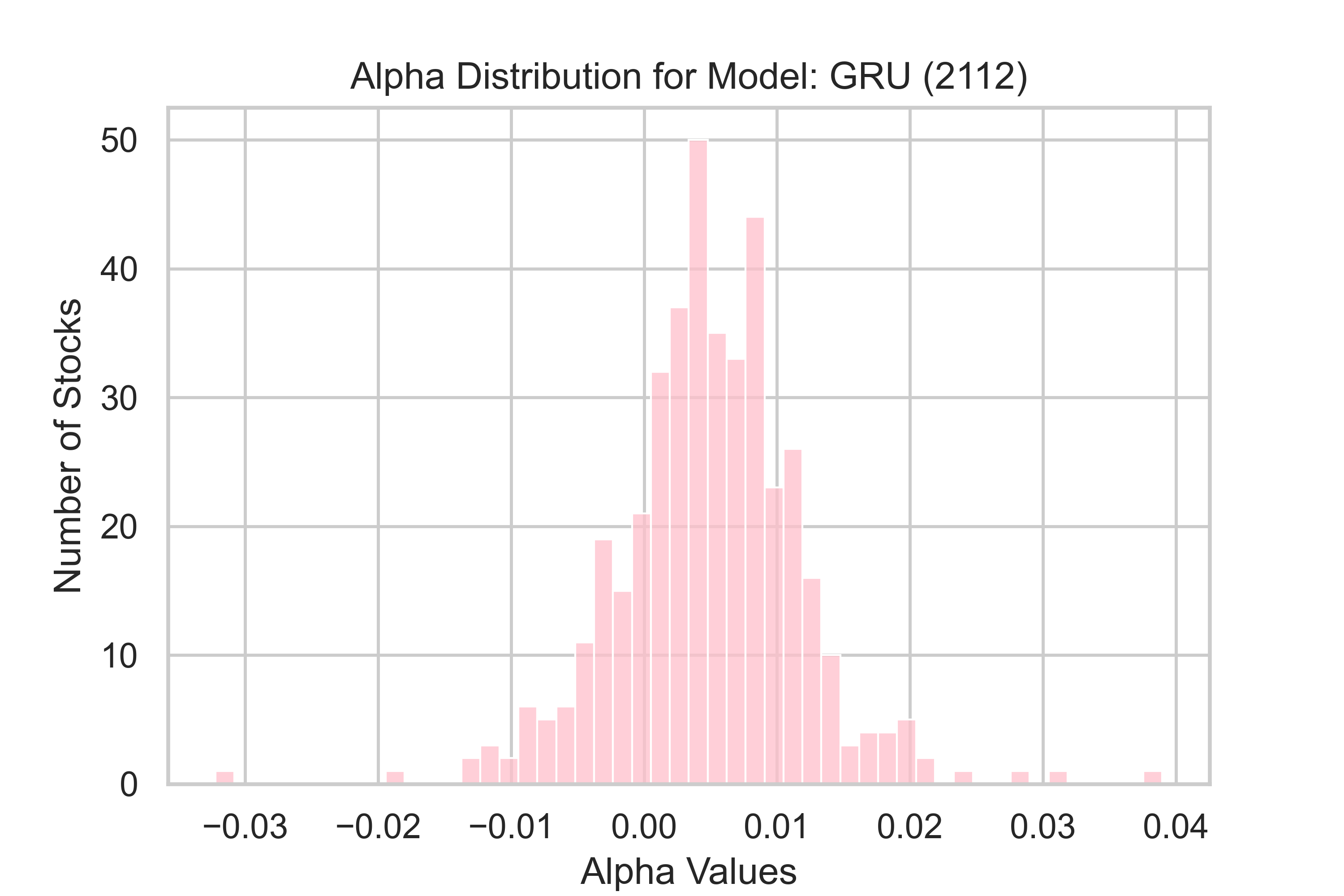}
  \caption{GRU(2112)}
\end{subfigure}
\hfill
\begin{subfigure}{0.32\textwidth}
  \includegraphics[width=\linewidth]{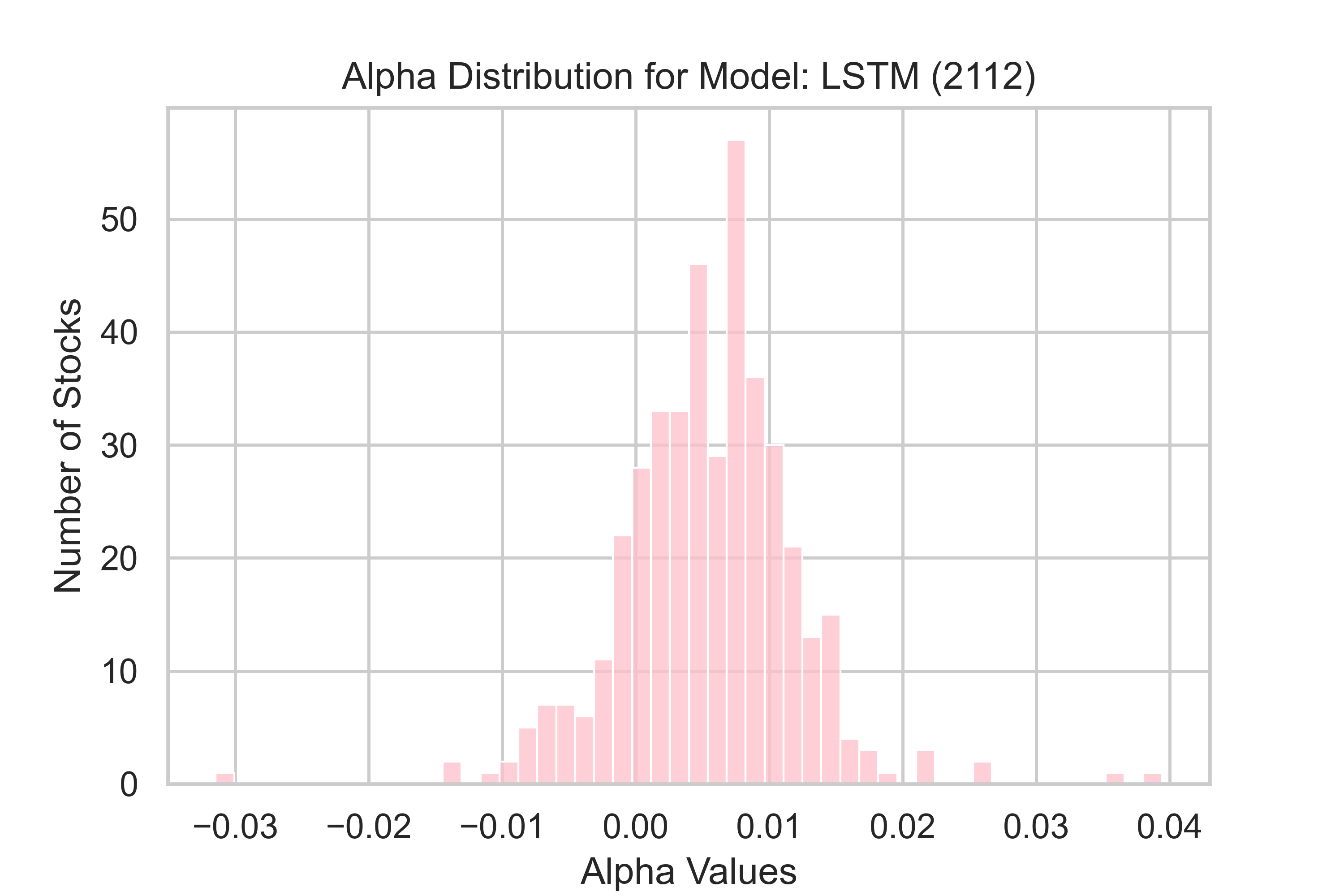}
  \caption{LSTM(2112)}
\end{subfigure}

\vspace{0.3em}

\begin{subfigure}{0.32\textwidth}
  \includegraphics[width=\linewidth]{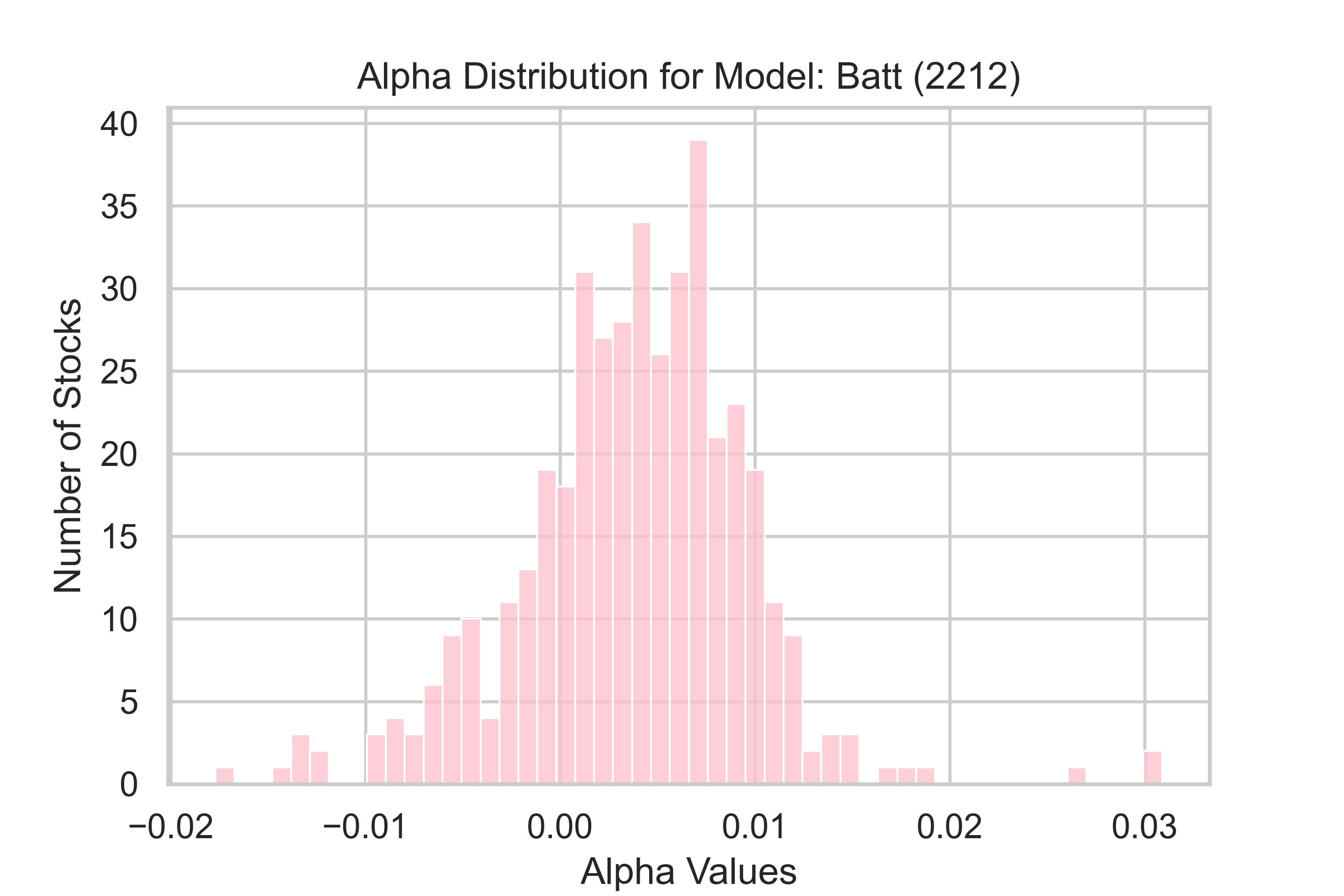}
  \caption{B-Additive(2212)}
\end{subfigure}
\hfill
\begin{subfigure}{0.32\textwidth}
  \includegraphics[width=\linewidth]
  {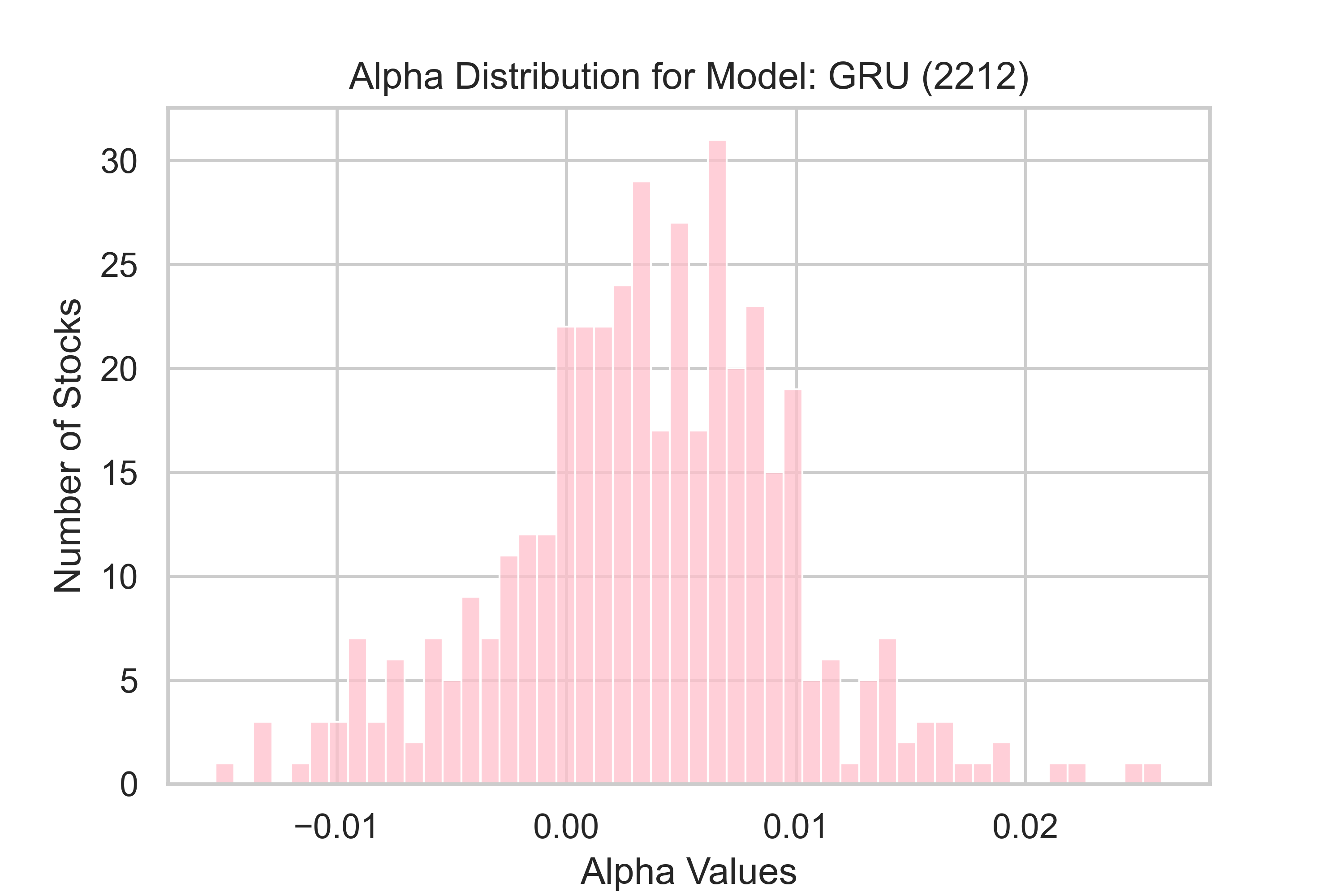}
  \caption{GRU(2212)}
\end{subfigure}
\hfill
\begin{subfigure}{0.32\textwidth}
  \includegraphics[width=\linewidth]{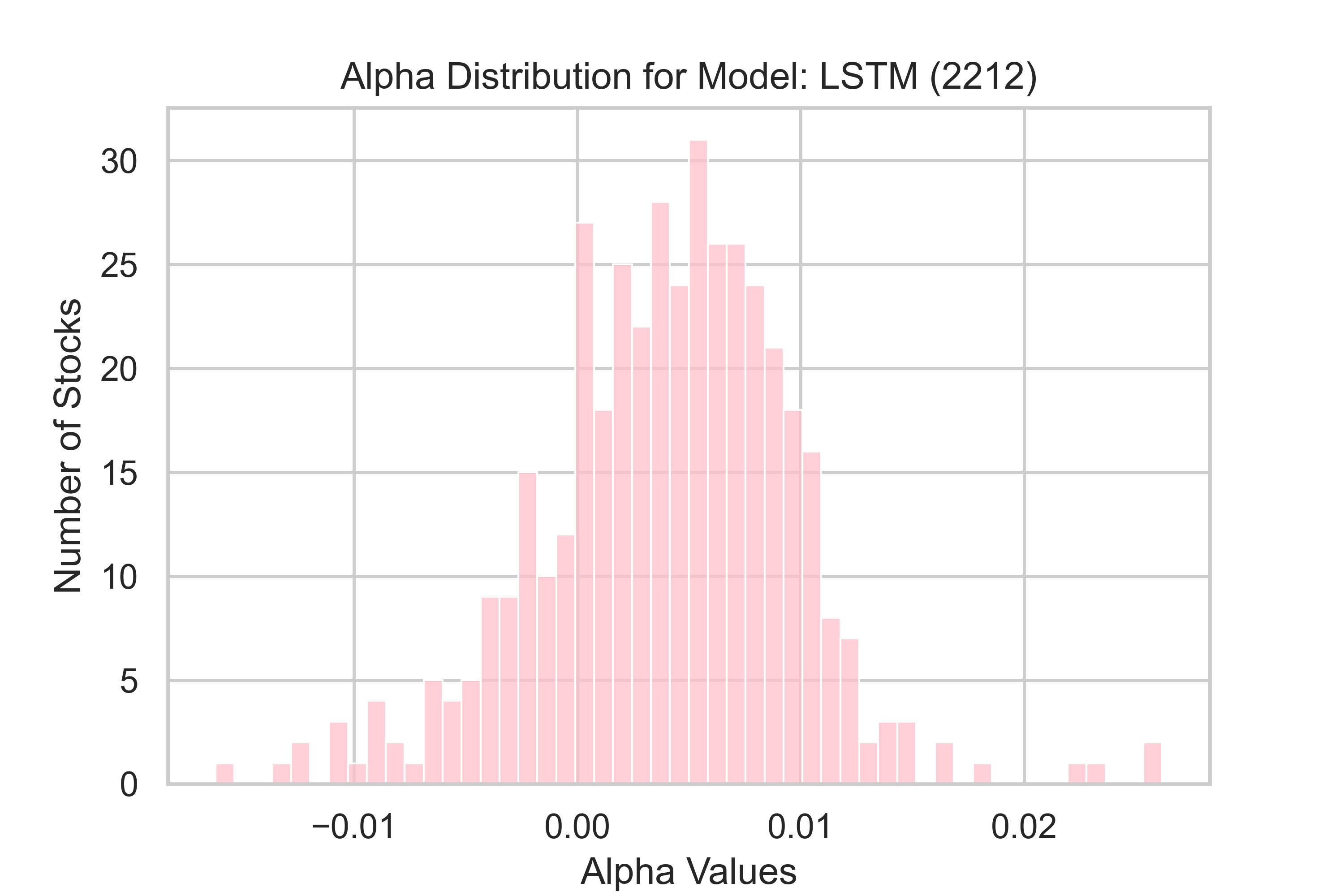}
  \caption{LSTM(2212)}
\end{subfigure}
\caption[OOS $\alpha$ distribution of the alternative models Part 1.]{OOS $\alpha$ distribution of the alternative models Part 1}
\label{fig:alpha_distribution_alter1_ch2}
\end{sidewaysfigure}

\begin{sidewaysfigure}[H]
\centering
\begin{subfigure}{0.32\textwidth}
  \includegraphics[width=\linewidth]{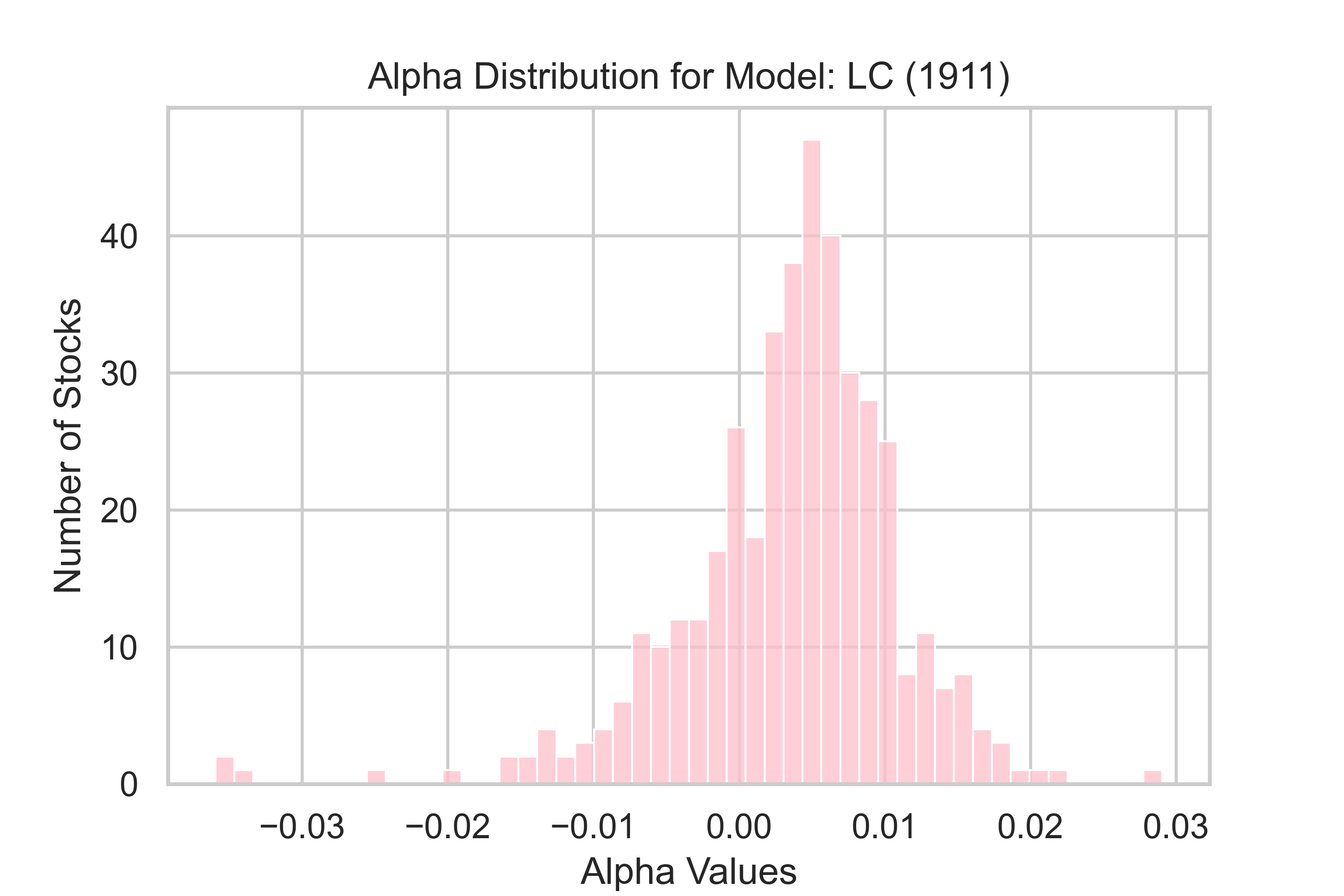}
  \caption{L-Concat(1911)}
\end{subfigure}
\hfill
\begin{subfigure}{0.32\textwidth}
  \includegraphics[width=\linewidth]
  {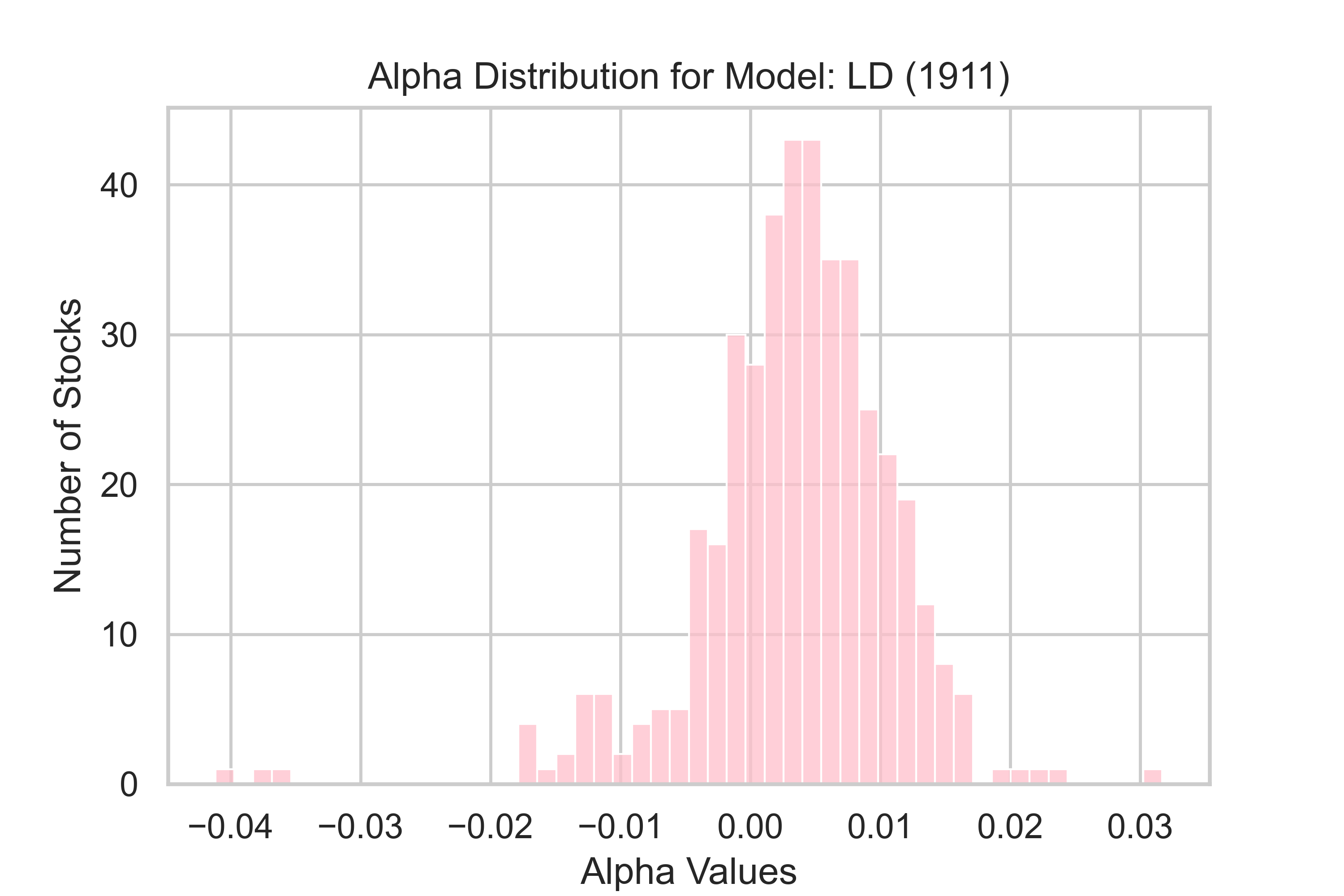}
  \caption{L-DotProd(1911)}
\end{subfigure}
\hfill
\begin{subfigure}{0.32\textwidth}
  \includegraphics[width=\linewidth]{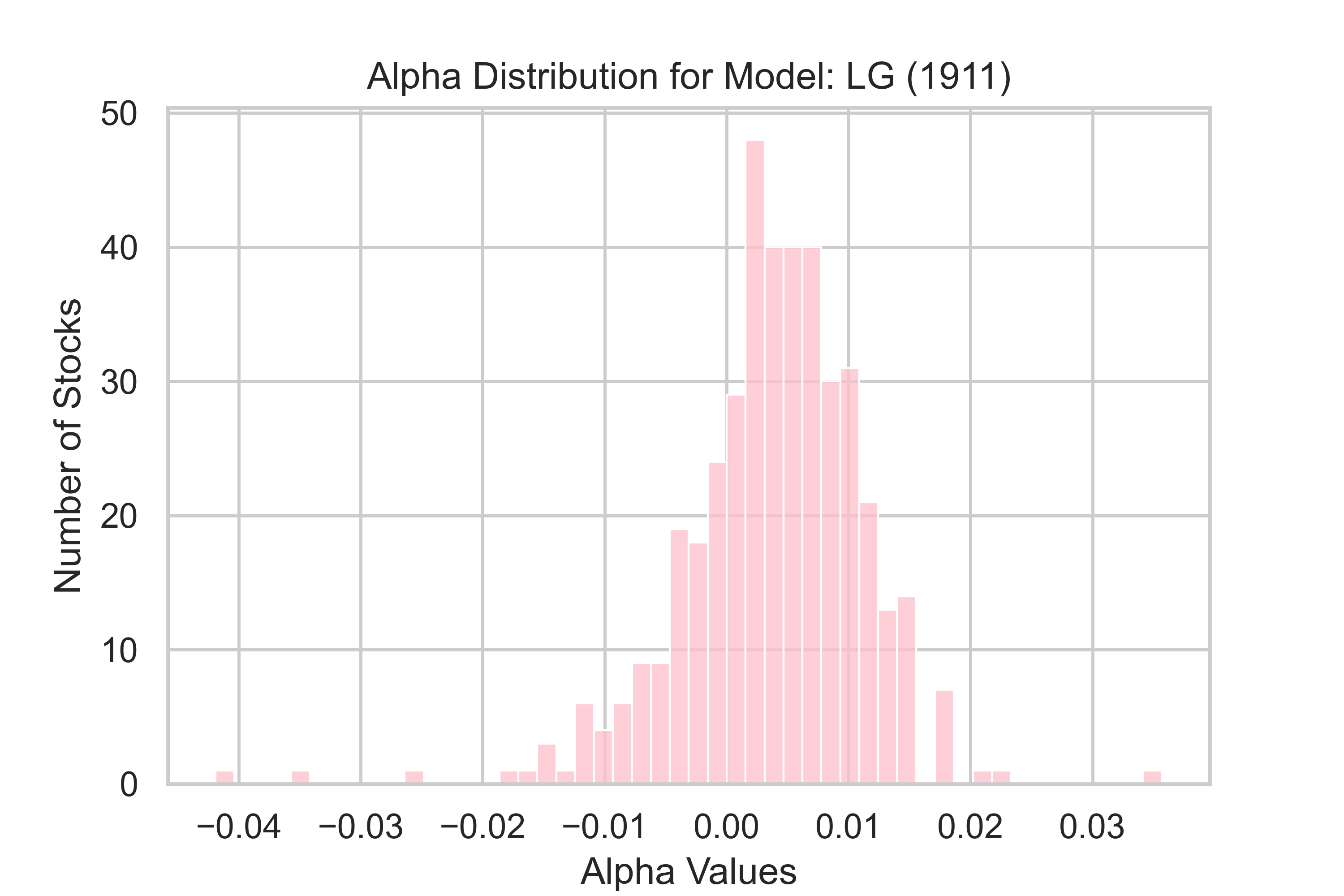}
  \caption{L-General(1911)}
\end{subfigure}

\vspace{0.3em}

\begin{subfigure}{0.32\textwidth}
  \includegraphics[width=\linewidth]
  {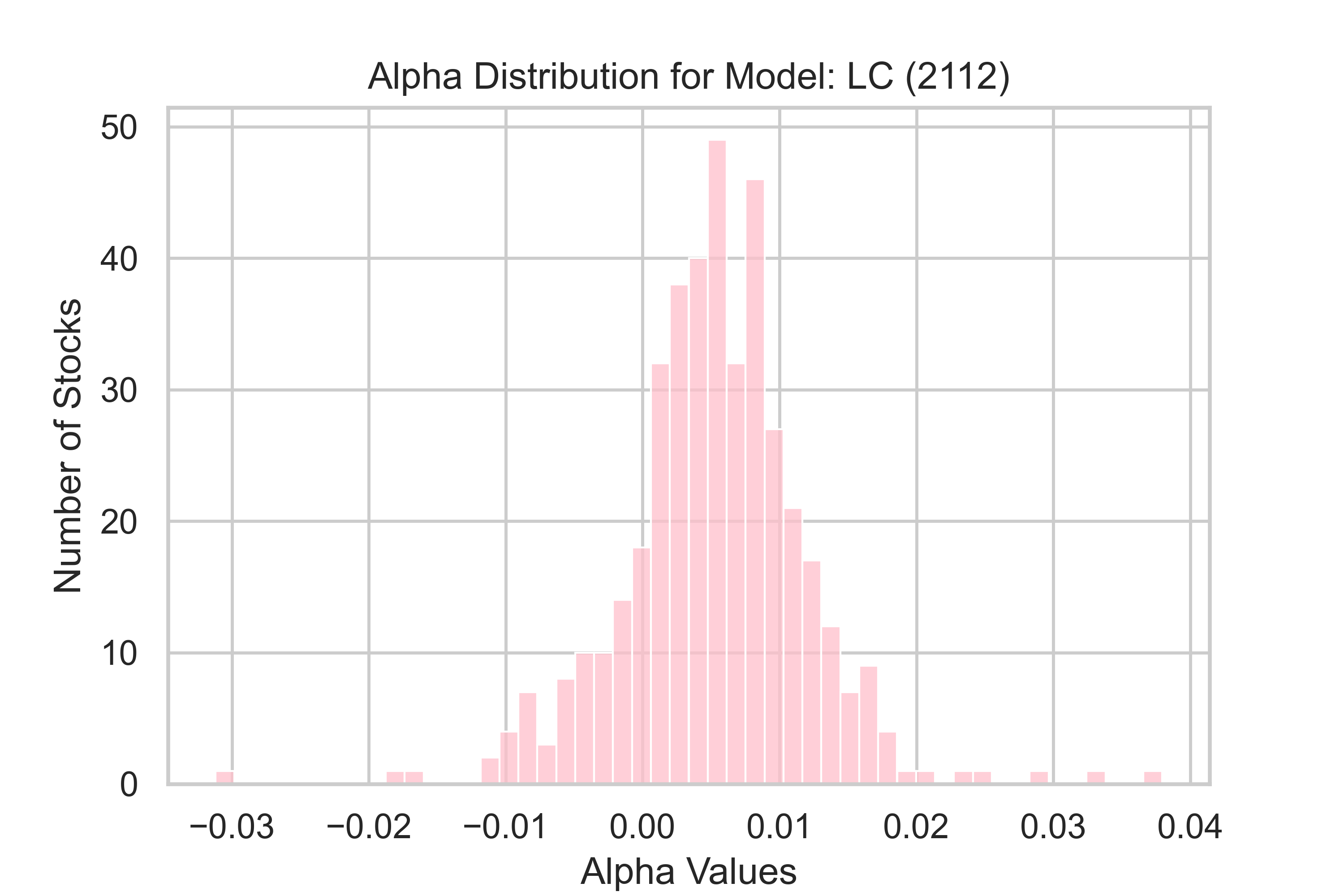}
  \caption{L-Concat(2112)}
\end{subfigure}
\hfill
\begin{subfigure}{0.32\textwidth}
  \includegraphics[width=\linewidth]
  {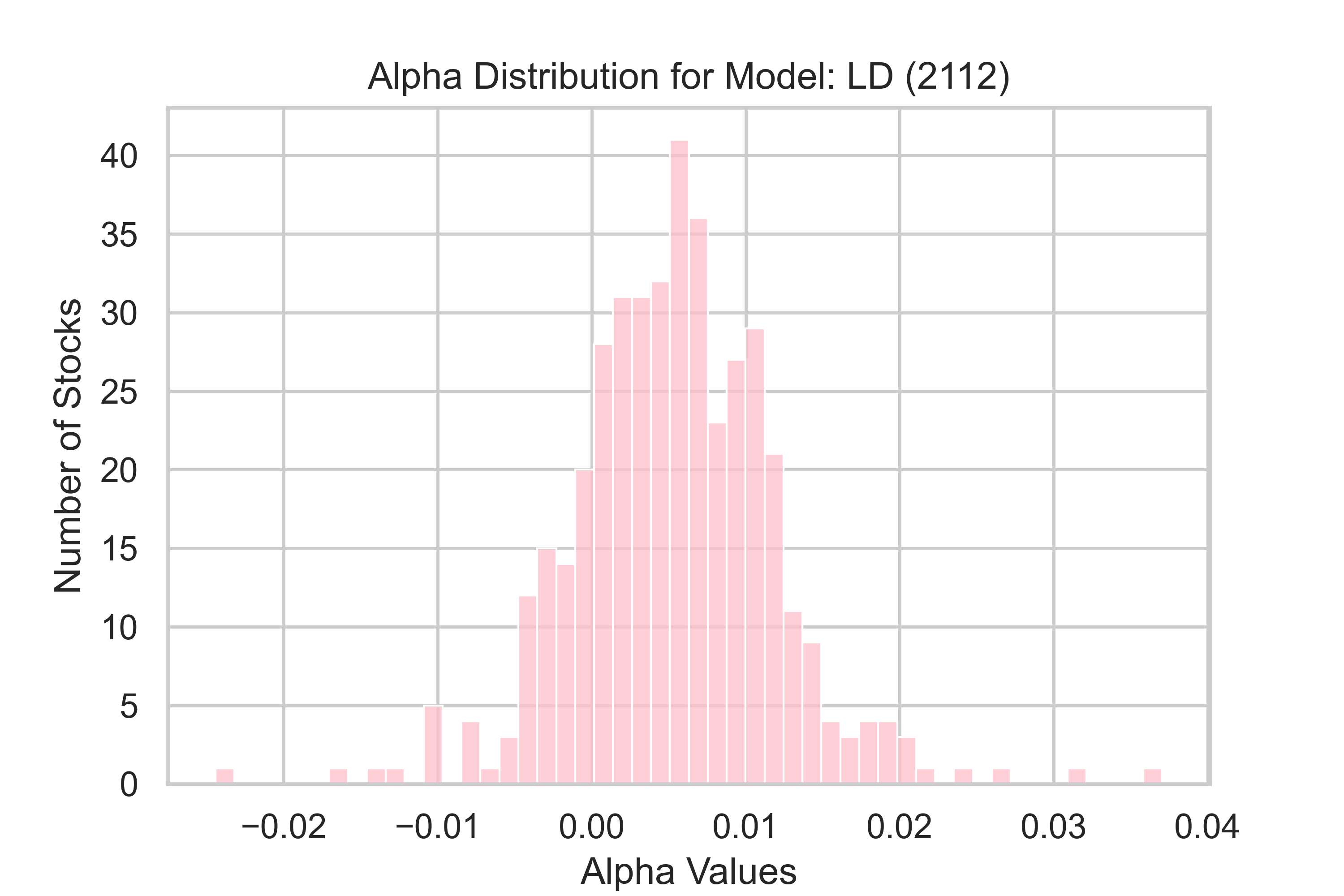}
  \caption{L-DotProd(2112)}
\end{subfigure}
\hfill
\begin{subfigure}{0.32\textwidth}
  \includegraphics[width=\linewidth]{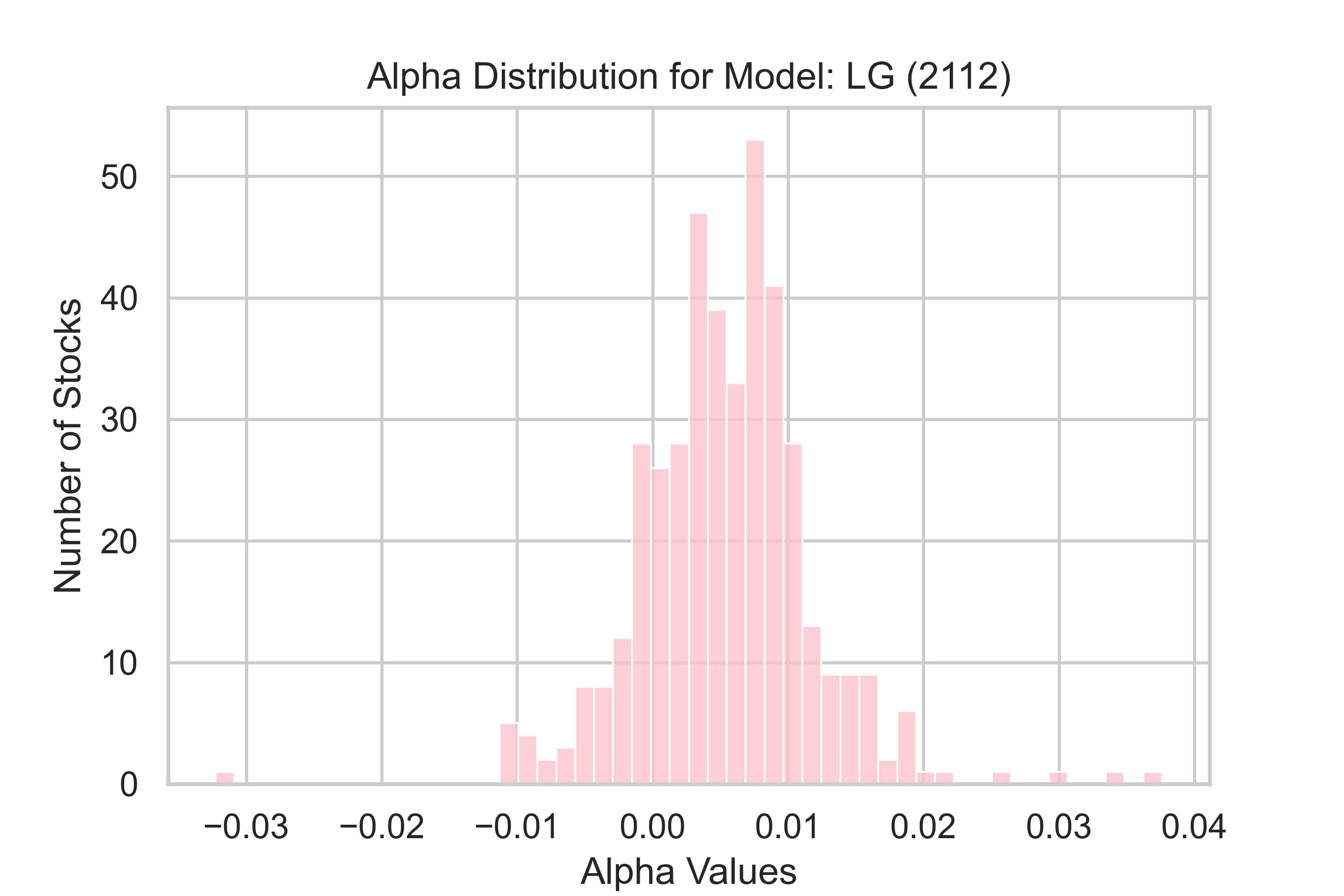}
  \caption{L-General(2112)}
\end{subfigure}

\vspace{0.3em}

\begin{subfigure}{0.32\textwidth}
  \includegraphics[width=\linewidth]{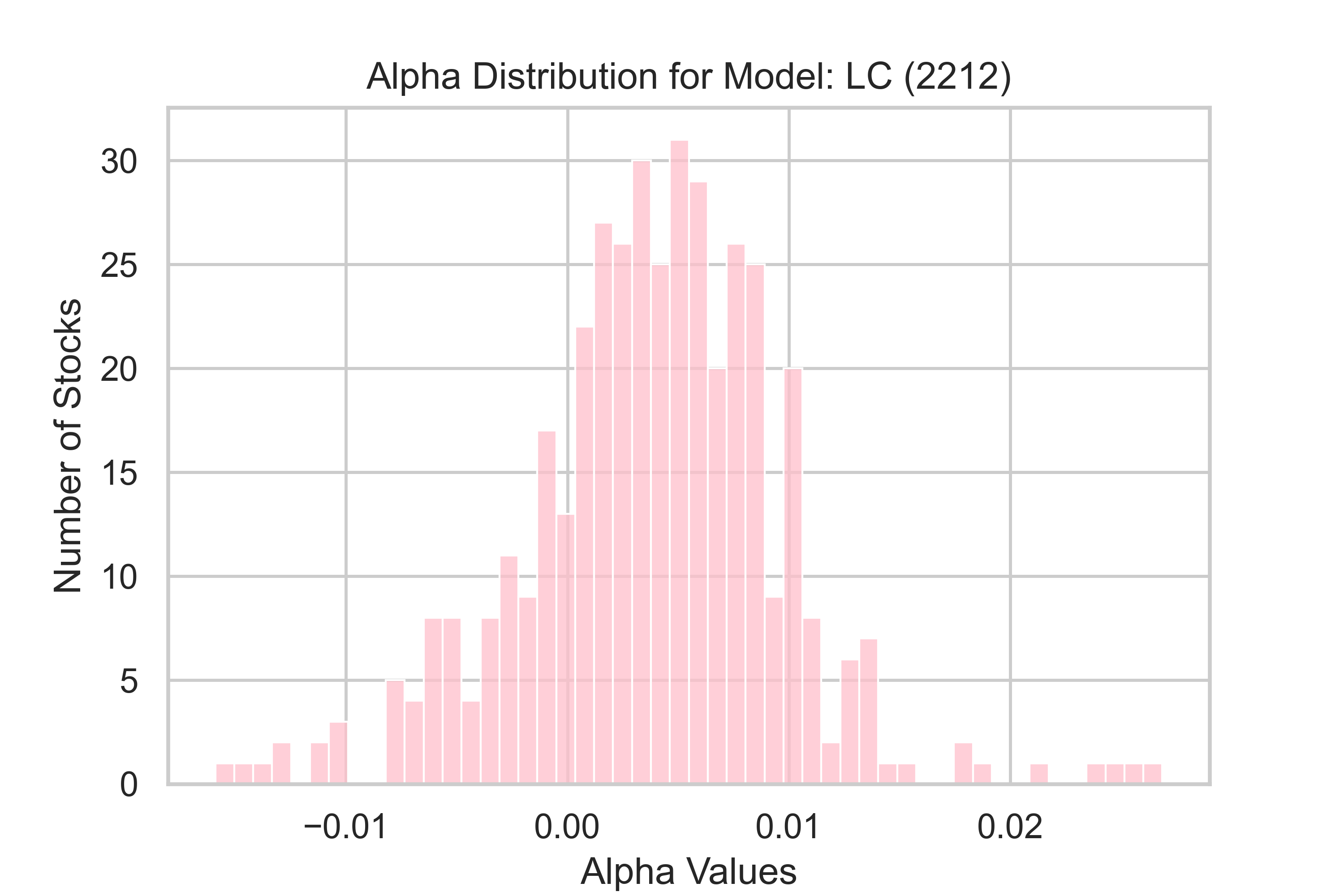}
  \caption{L-Concat(2212)}
\end{subfigure}
\hfill
\begin{subfigure}{0.32\textwidth}
  \includegraphics[width=\linewidth]
  {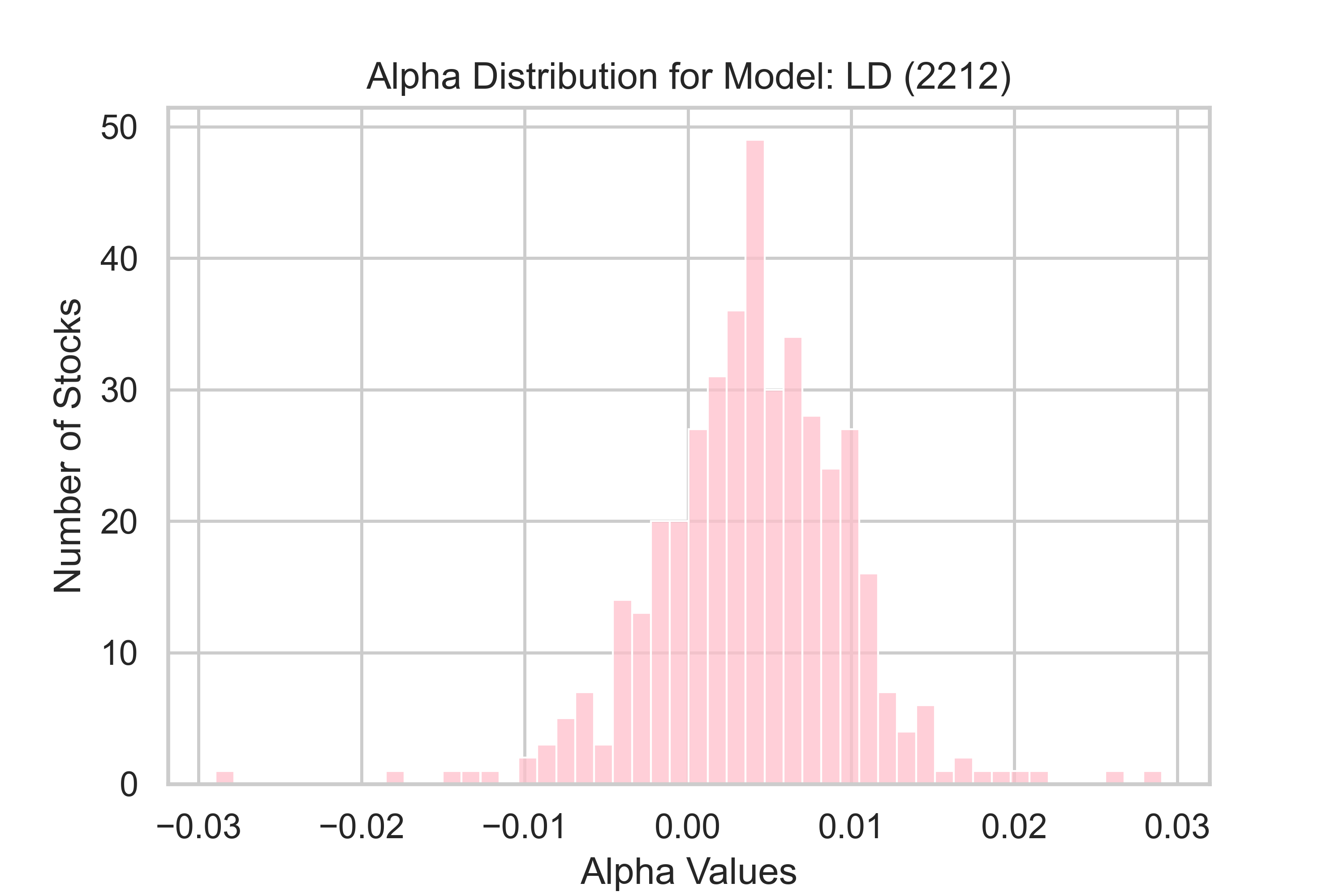}
  \caption{L-DotProd(2212)}
\end{subfigure}
\hfill
\begin{subfigure}{0.32\textwidth}
  \includegraphics[width=\linewidth]{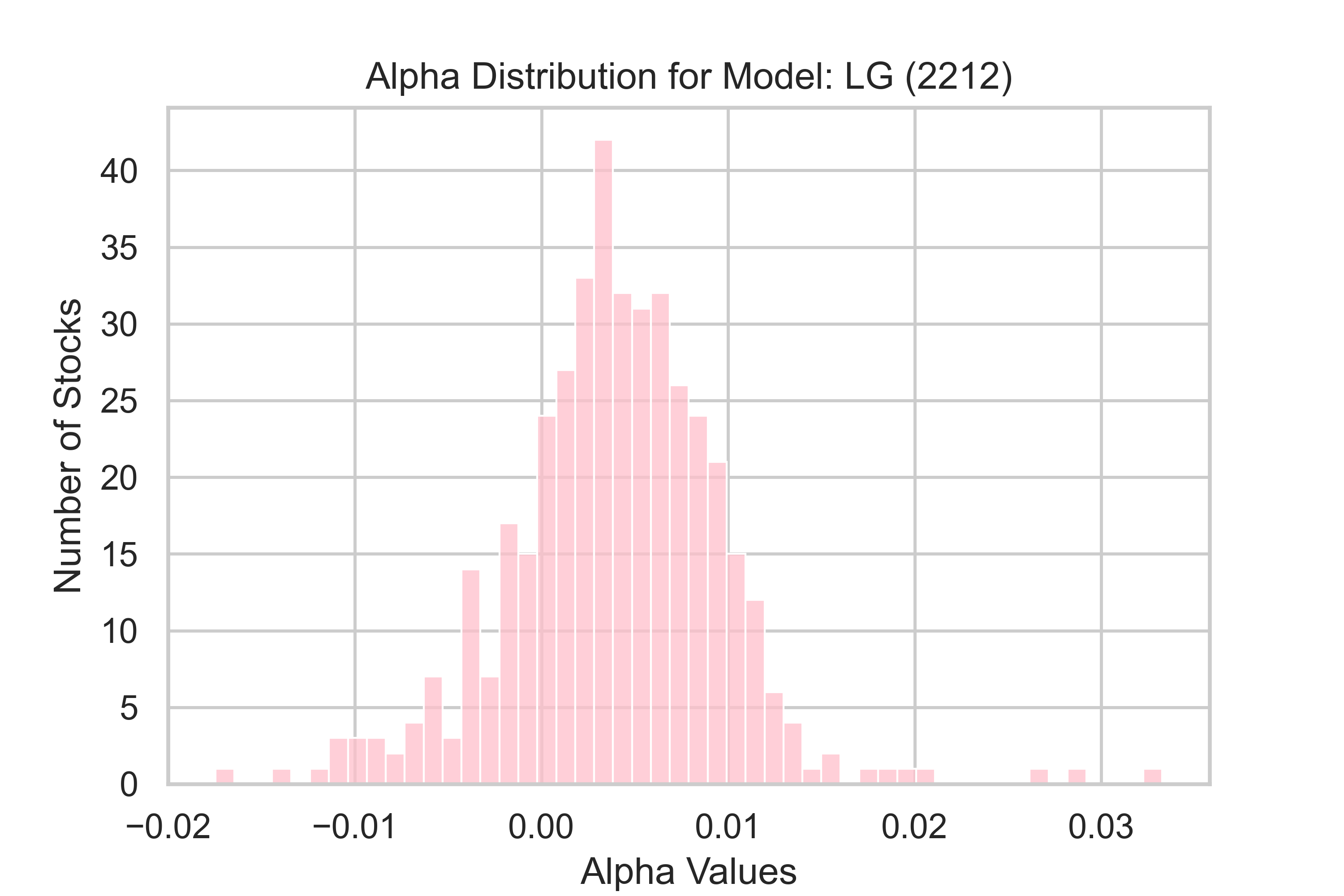}
  \caption{L-General(2212)}
\end{subfigure}
\caption[OOS $\alpha$ distribution of the alternative models Part 2.]{OOS $\alpha$ distribution of the alternative models Part 2}
\label{fig:alpha_distribution_alter1_ch2}
\end{sidewaysfigure}

\begin{algorithm}[H]
\section{Adam optimizer}\label{sec:AB_adam_ch2}
Adam updates parameters by maintaining the first and second moments of the gradients.
\caption{Adam Optimization}
\begin{algorithmic}[1]
\State \textbf{Initialize:} $m_0 = 0$, $v_0 = 0$, $l = 0$
\State \textbf{Set hyperparameters:} learning rate $\eta = 0.001$, $\beta_1 = 0.9$, $\beta_2 = 0.999$, $\epsilon = 10^{-8}$
\While{$\theta_l$ not converged}
    \State $l \gets l + 1$
    \State Compute gradient: $g_l = \nabla_\theta L(\theta_{l-1})$
    \State $m_l \gets \beta_1 m_{l-1} + (1 - \beta_1) g_l$
    \State $v_l \gets \beta_2 v_{l-1} + (1 - \beta_2) g_l \odot g_l$ \Comment{$\odot$: element-wise multiplication}
    \State $\hat{m}_l \gets \frac{m_l}{1 - \beta_1^l}$ \Comment{Bias correction}
    \State $\hat{v}_l \gets \frac{v_l}{1 - \beta_2^l}$
    \State $\theta_l \gets \theta_{l-1} - \eta \cdot \frac{\hat{m}_l}{\sqrt{\hat{v}_l} + \epsilon}$
\EndWhile
\State \textbf{Return:} $\theta_l$
\end{algorithmic}
\begin{flushleft}
\textit{Source: Adapted from \citet{Gu2020EmpiricalLearning}}
\end{flushleft}
\end{algorithm}
where $g_t$ and $\theta_t$ represent the gradients and parameters computed from previous sections respectively.

\section{Early Stopping}\label{sec:AC_Early_Stop_ch2}
\begin{algorithm}[H]
\caption{Early Stopping}
\begin{algorithmic}[1]
\State \textbf{Initialize:} $j = 0$, $\epsilon = \infty$, select patience parameter $p$.
\While{$j < p$}
    \State Update $\theta$ using the training algorithm (e.g., for $h$ steps).
    \State Calculate the prediction error from the validation sample, denoted as $\epsilon'$.
    \If{$\epsilon' < \epsilon$}
        \State $j \gets 0$
        \State $\epsilon \gets \epsilon'$
        \State $\theta' \gets \theta$
    \Else
        \State $j \gets j + 1$
    \EndIf
\EndWhile
\State \textbf{Return:} $\theta'$
\end{algorithmic}
\begin{flushleft}
\textit{Source: Adapted from \citet{Gu2020EmpiricalLearning}}
\end{flushleft}
\end{algorithm}

\section{List of Abbreviations}\label{sec:abbr_ch3}
\begin{tabular}{ll}
RNN   & pre-trained vanilla recurrent neural network model\\
GRU   & pre-trained RNN with Gated Recurrent Unit \\
LSTM  & pre-trained RNN with Long Short-Term Memory \\
B-Additive  & pre-trained RNN additive attention model\\
L-Concat    & pre-trained RNN Luong's cancatenate attention model\\
L-DotProd    & pre-trained RNN Luong's dot product attention model\\
L-General    & pre-trained RNN Luong's general attention model\\
self\_att    & pre-trained RNN global self-attention model\\
sparse\_att    & pre-trained RNN sliding window sparse attention model\\
\end{tabular}

\begin{sidewaysfigure}[htbp]
\section{Variable Importance}\label{sec:AA_ch2}
\centering
\begin{subfigure}{0.32\textwidth}
  \includegraphics[width=\linewidth]{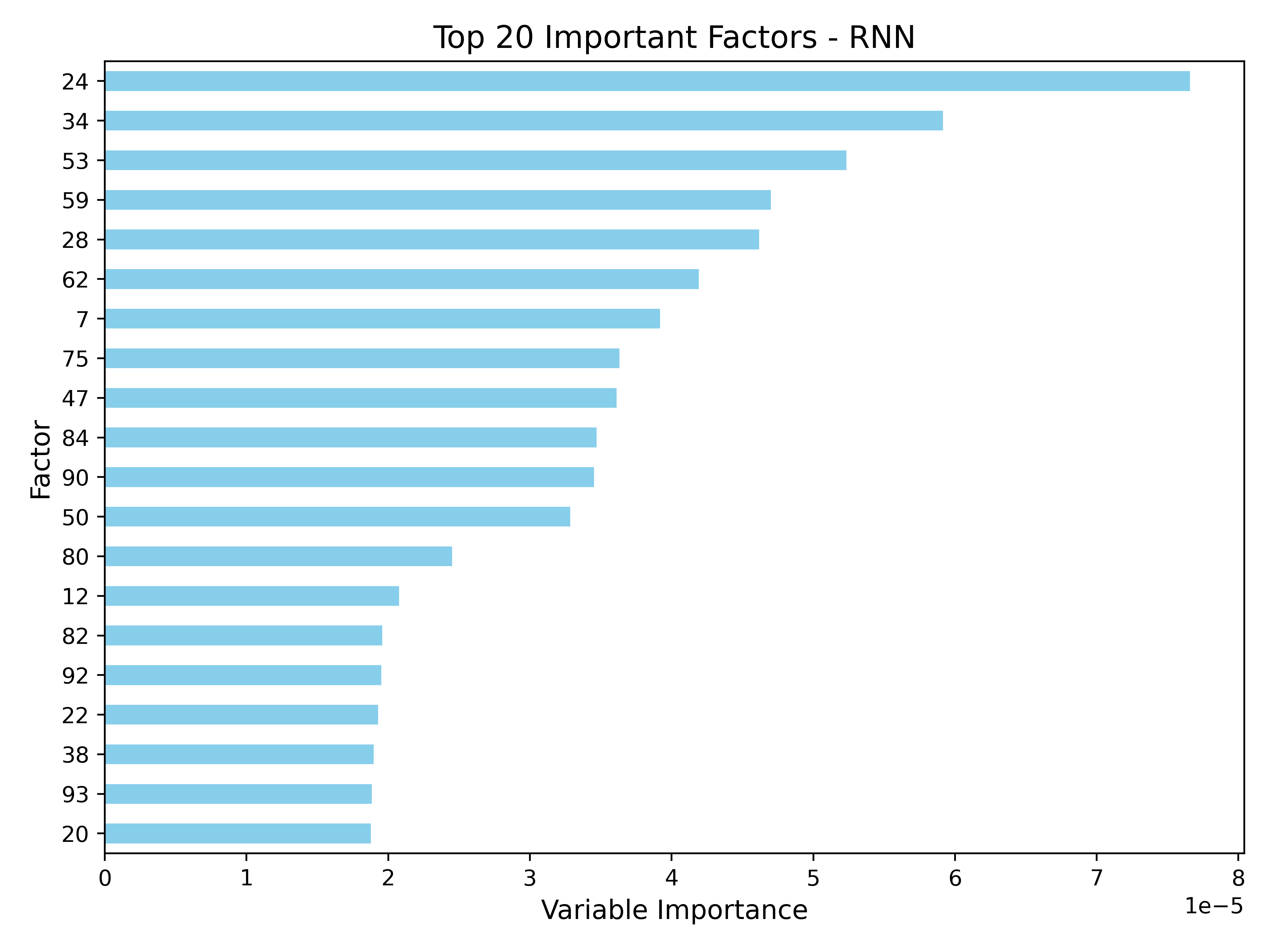}
  \caption{RNN}
\end{subfigure}
\hfill
\begin{subfigure}{0.32\textwidth}
  \includegraphics[width=\linewidth]{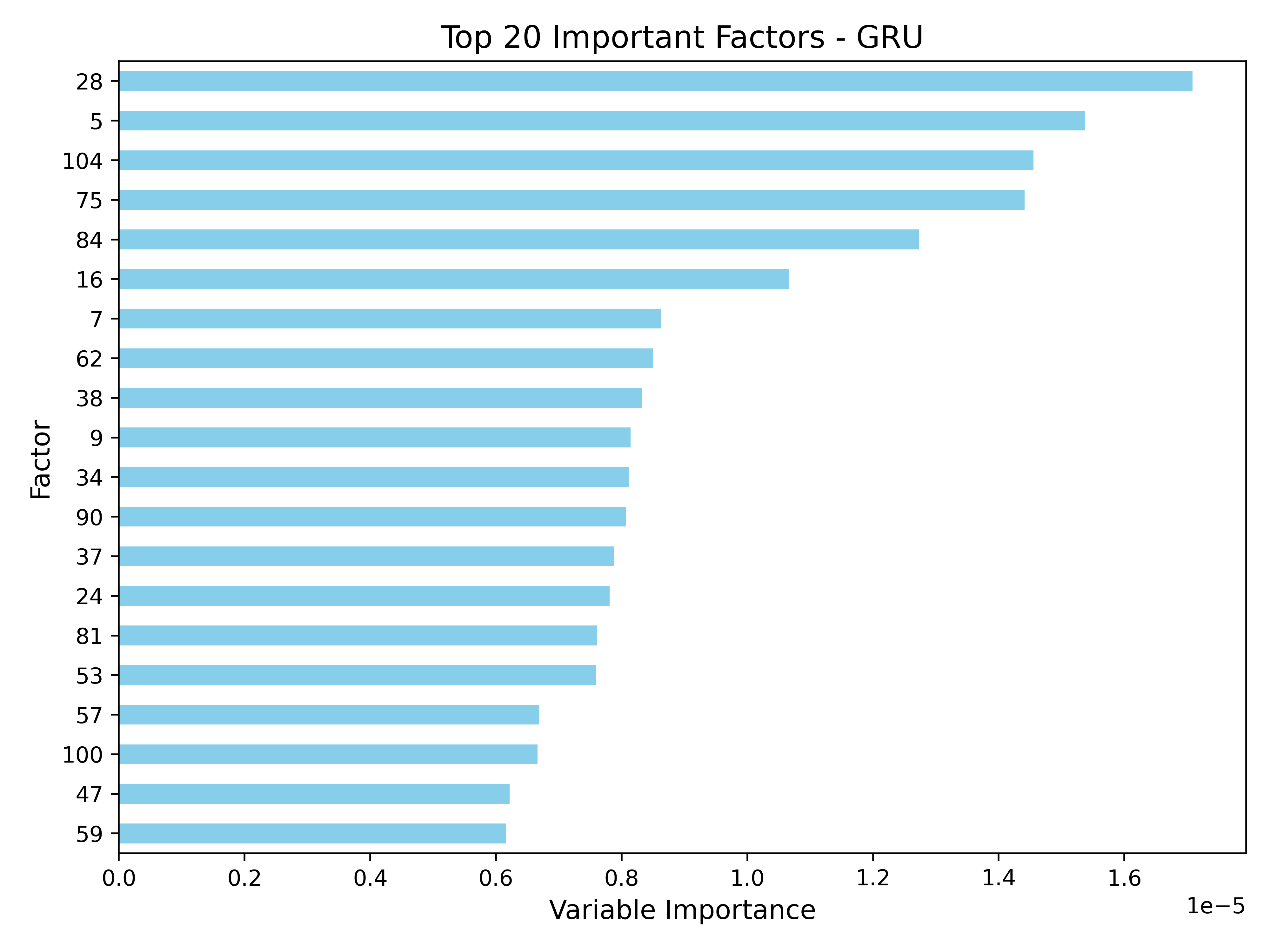}
  \caption{GRU}
\end{subfigure}
\hfill
\begin{subfigure}{0.32\textwidth}
  \includegraphics[width=\linewidth]{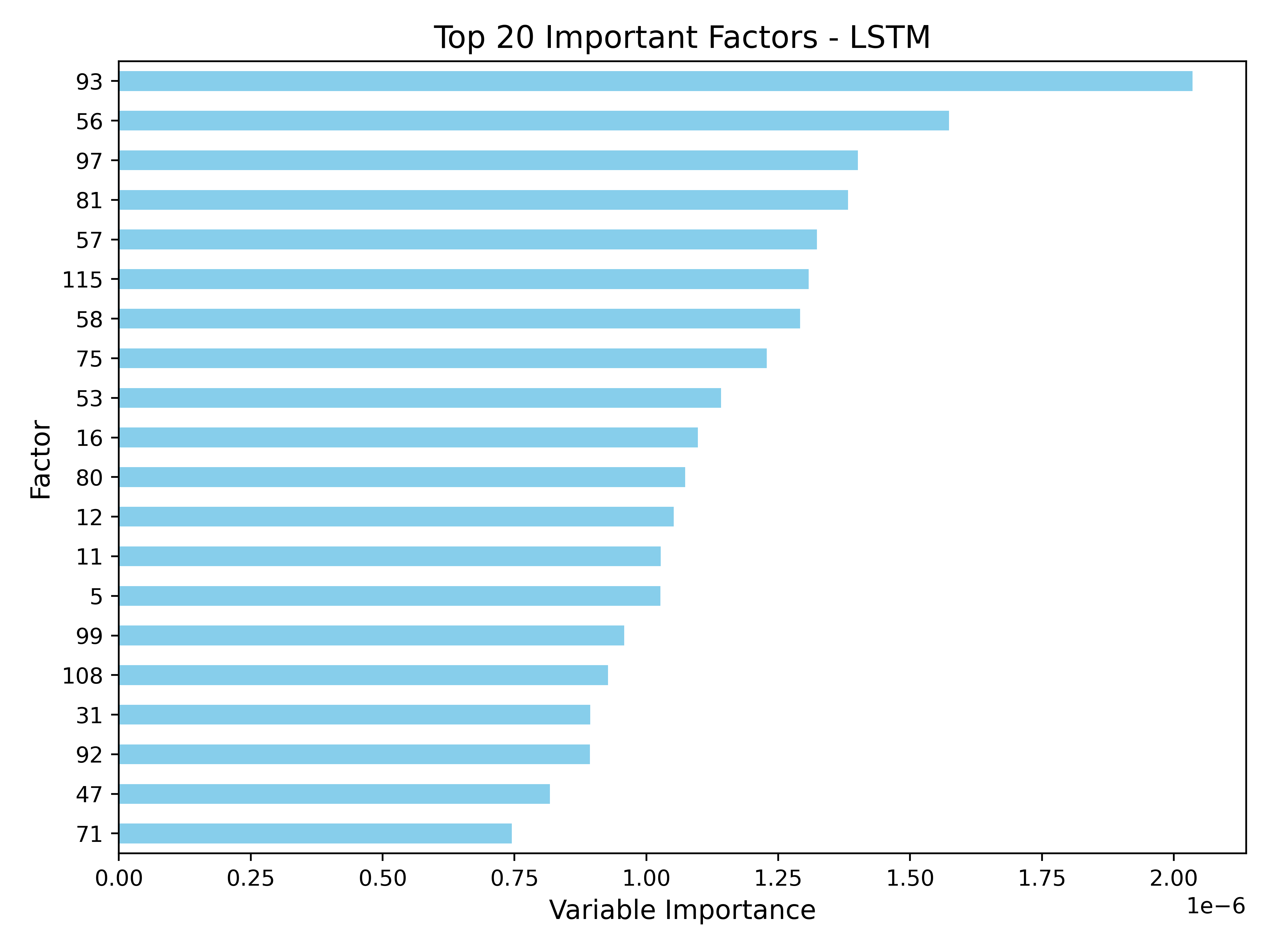}
  \caption{LSTM}
\end{subfigure}

\vspace{0.3em}

\begin{subfigure}{0.32\textwidth}
  \includegraphics[width=\linewidth]{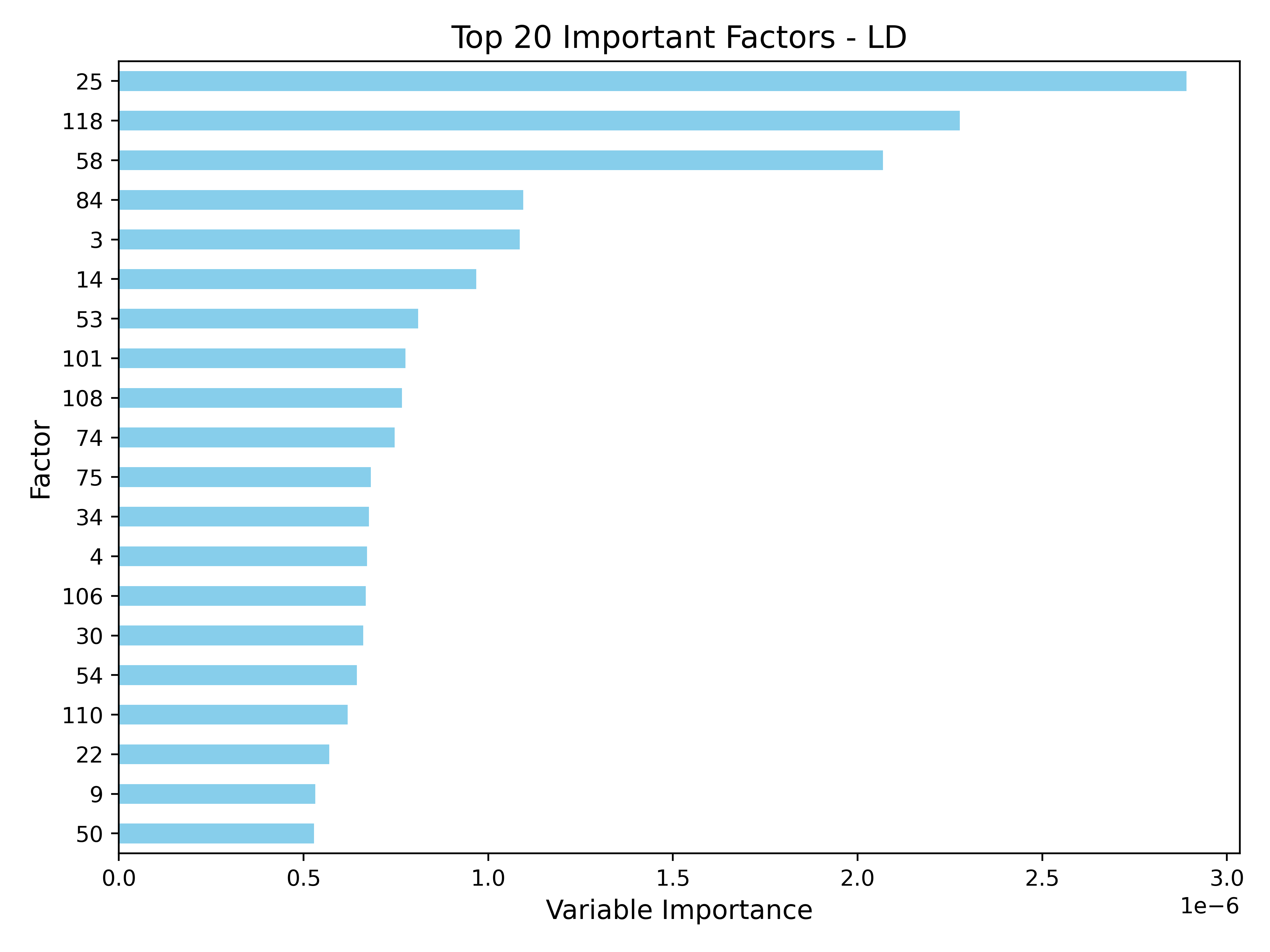}
  \caption{L-DotProd}
\end{subfigure}
\hfill
\begin{subfigure}{0.32\textwidth}
  \includegraphics[width=\linewidth]{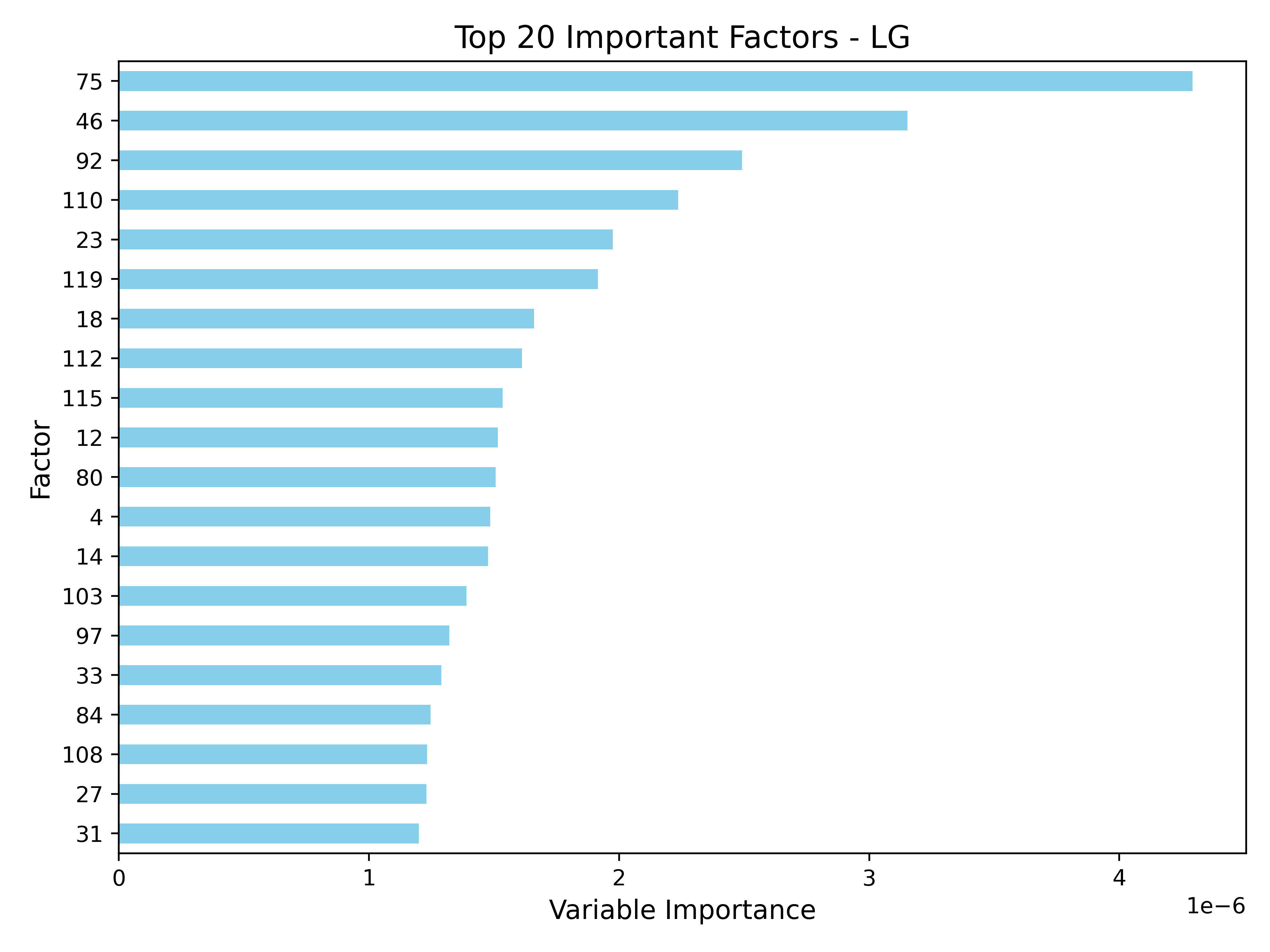}
  \caption{L-General}
\end{subfigure}
\hfill
\begin{subfigure}{0.32\textwidth}
  \includegraphics[width=\linewidth]{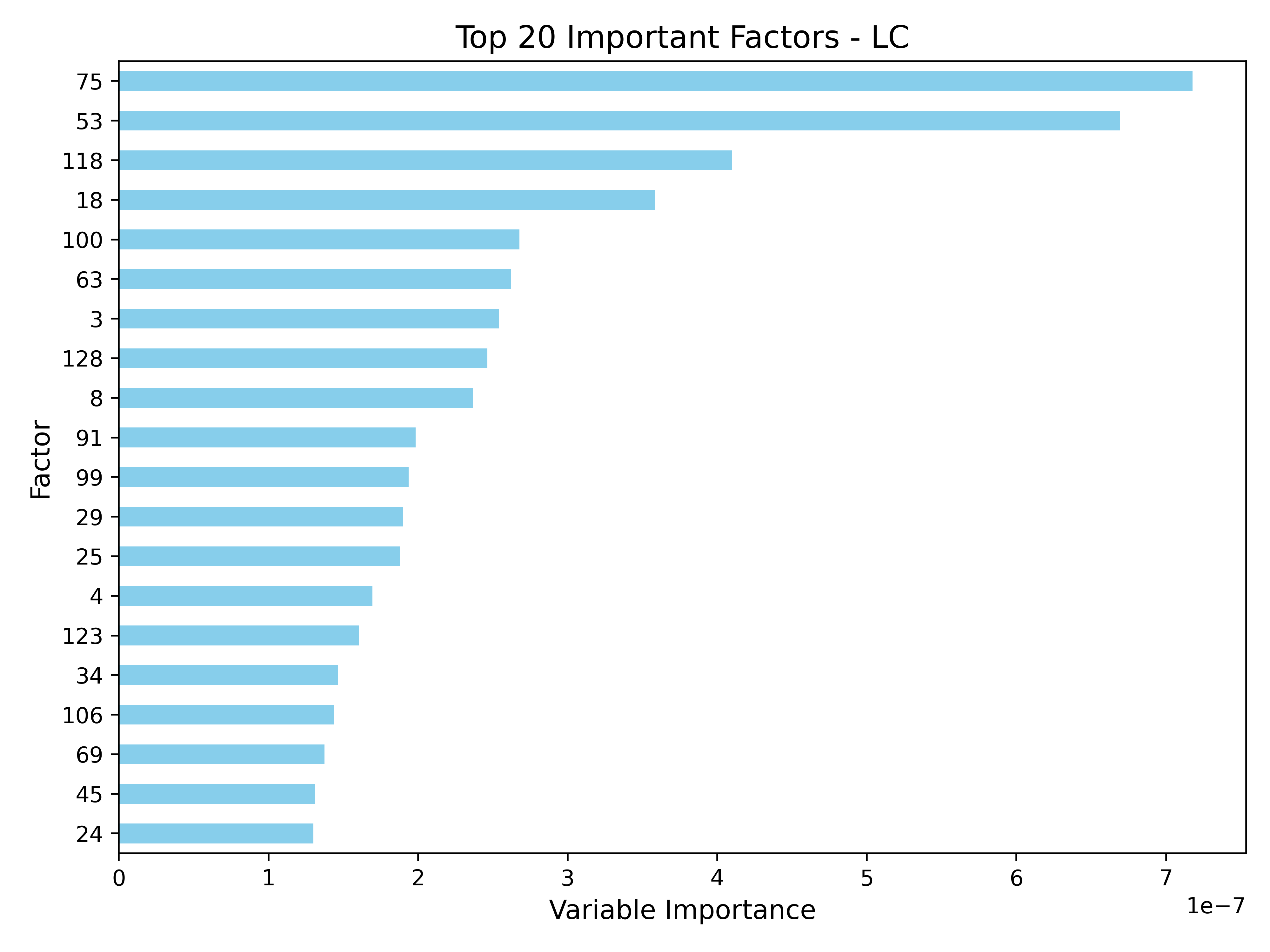}
  \caption{L-Concat}
\end{subfigure}

\vspace{0.3em}

\begin{subfigure}{0.32\textwidth}
  \includegraphics[width=\linewidth]{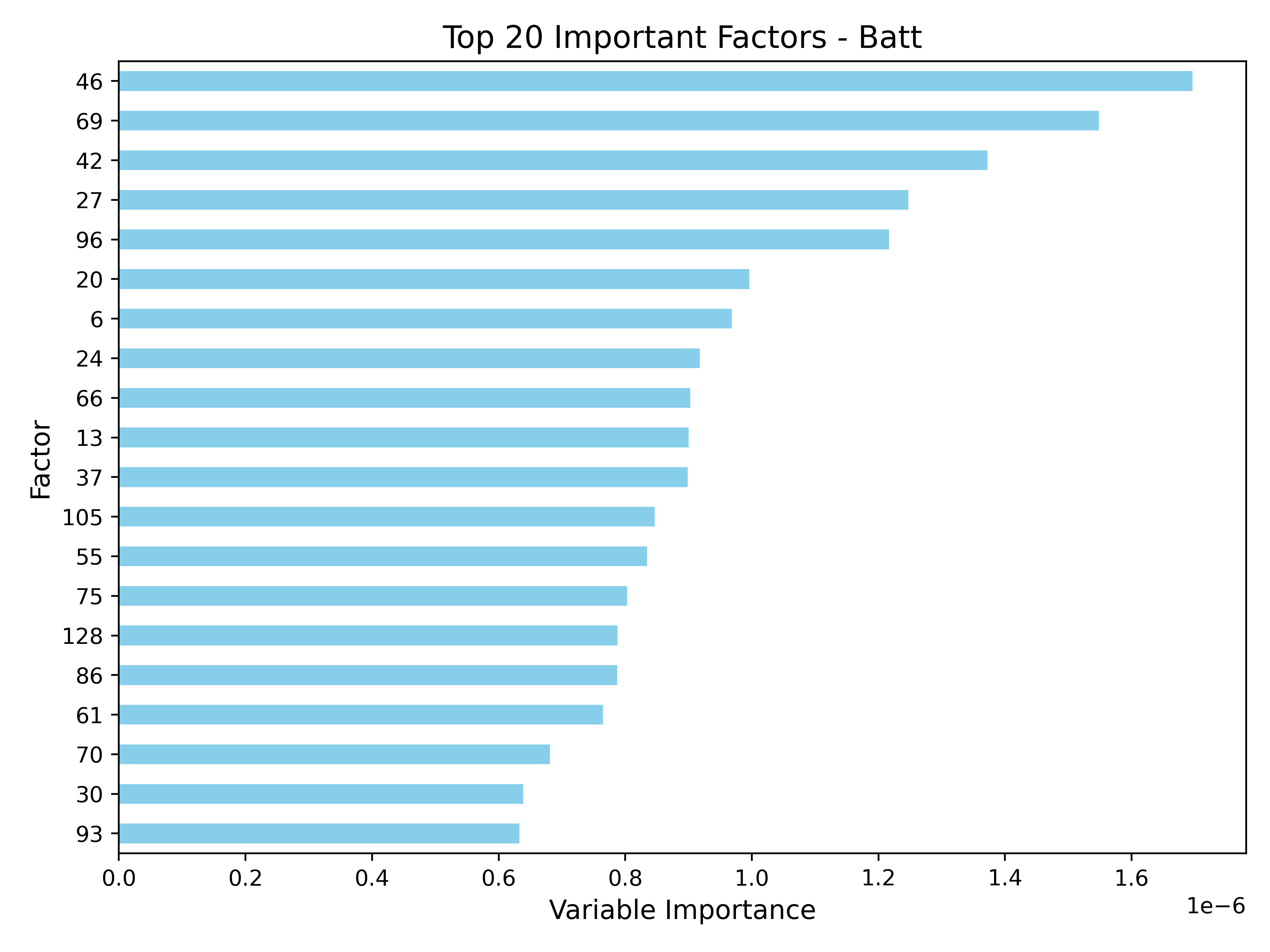}
  \caption{B-Additive}
\end{subfigure}
\hfill
\begin{subfigure}{0.32\textwidth}
  \includegraphics[width=\linewidth]{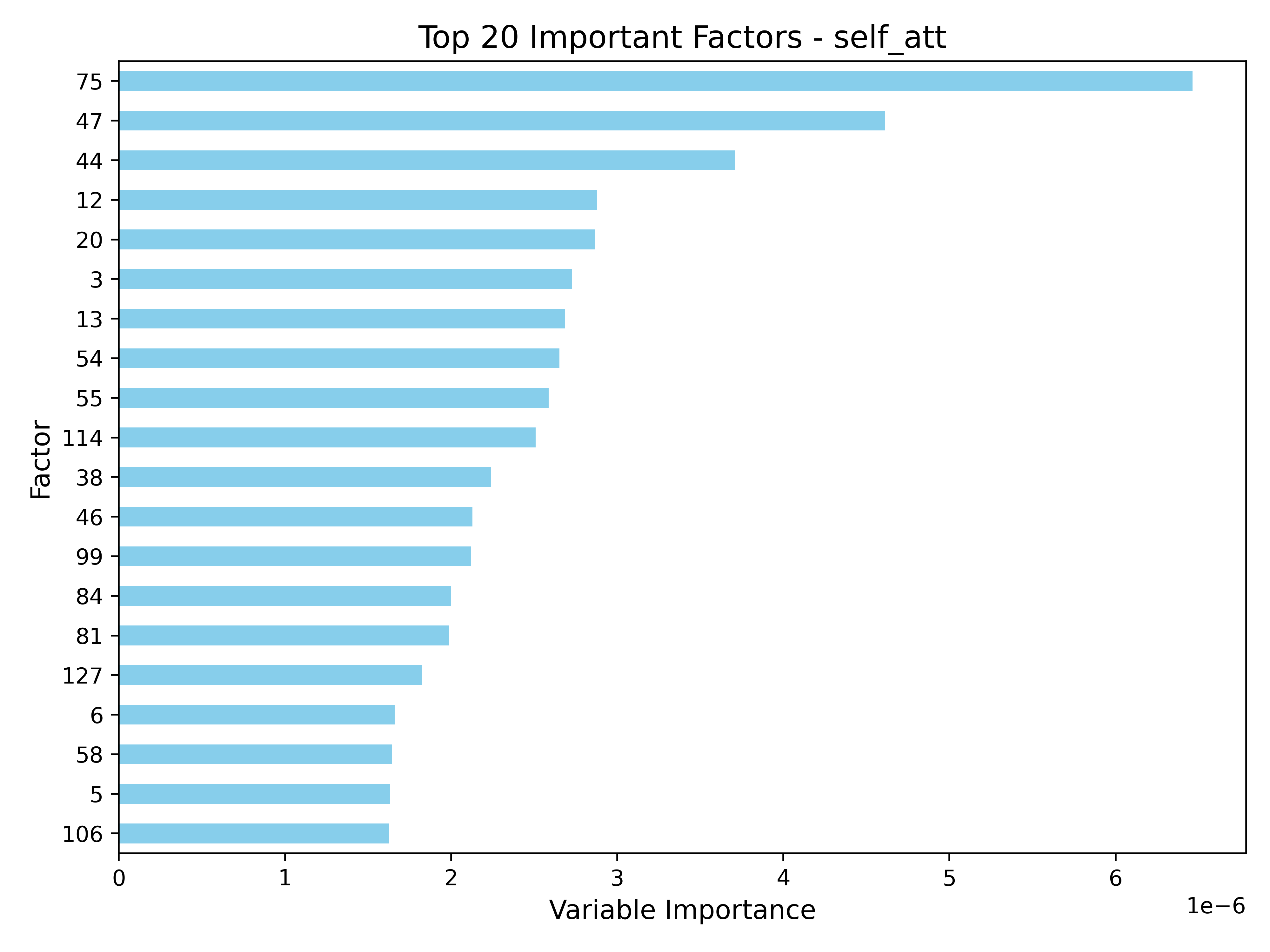}
  \caption{Self\_att}
\end{subfigure}
\hfill
\begin{subfigure}{0.32\textwidth}
  \includegraphics[width=\linewidth]{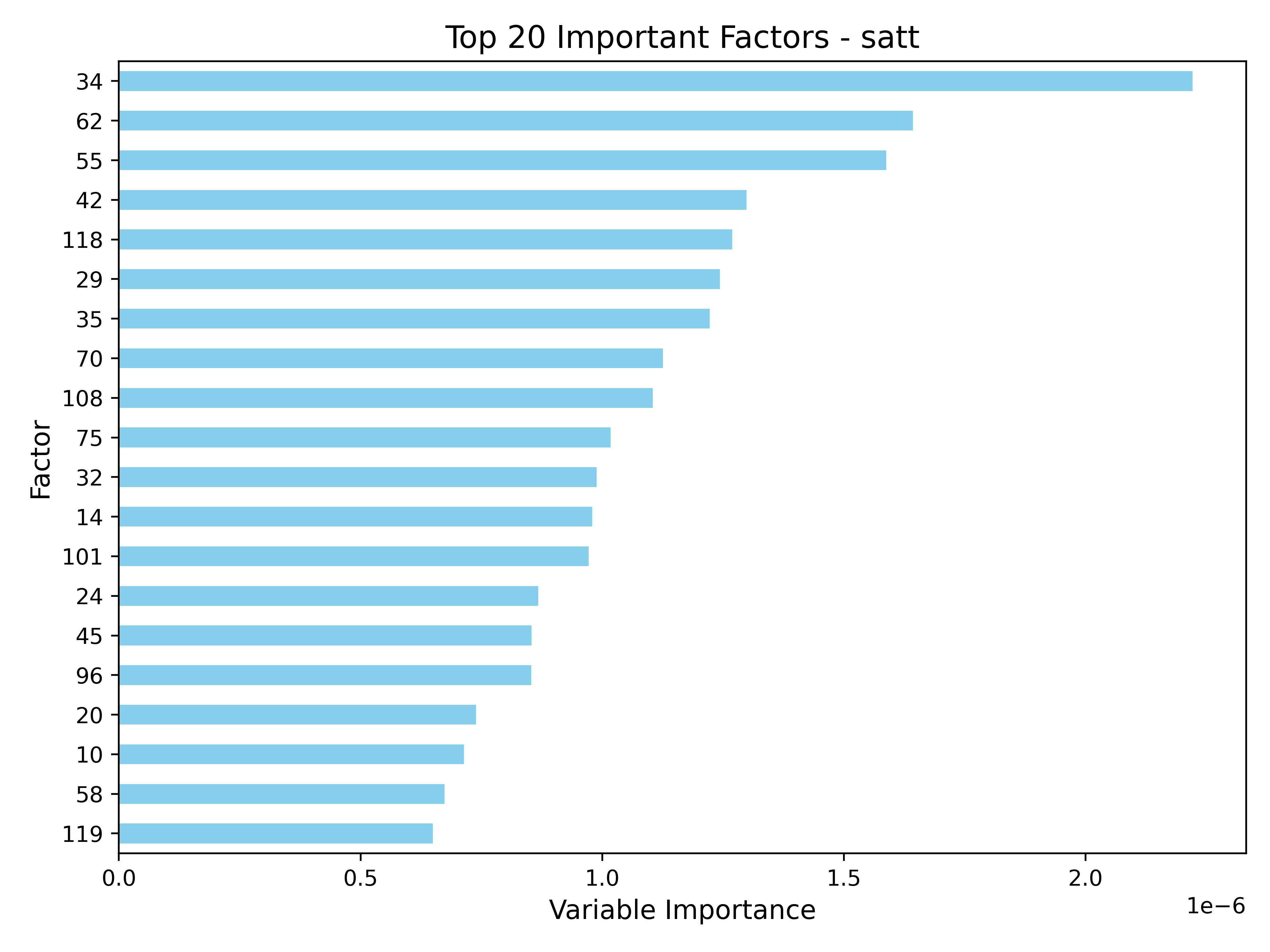}
  \caption{Sparse\_att}
\end{subfigure}

\caption[Factor importance of different models.]{Factor importance of different models}
\label{fig:factor_importance}
\end{sidewaysfigure}

\clearpage
\section{The relationship between $\alpha$ in traditional linear factor model and in ML models}\label{sec:app_alpha_ch2}
Concretely, according to \citet{Gu2020EmpiricalLearning}, the machine learning-based empirical asset pricing factor model can be presented as:
\begin{equation}
r_{j}-r_f = g\left(f_{k}; \theta\right) + \varepsilon_{j}
\label{eq:ML_general_intro}
\end{equation}
where $g(\cdot)$ is the function form of ML algorithms, $\theta$ is the parameter matrix, which is the weight matrix as the terminology of machine learning. $\varepsilon_{j}$ is the residual of the equation, which contains bias and zero-mean noise in the context of machine learning. Specifically, it can be decomposed as:
\begin{equation}
r_{j,t} - r_{f,t} = \underbrace{g(f_{k,t}; \theta)}_{\text{ML structure}} + \underbrace{b_j}_{\text{Bias (ML residual alpha)}} + \underbrace{\eta_{j,t}}_{\text{Zero-mean Noise}}
\end{equation}
Thus, the bias $b_j$ could be expressed as:
\begin{equation}
    b_j = E[R_{i,t}] - E[F_t^{ML}]
\end{equation}
If $\alpha$ in the traditional linear factor model is:
\begin{equation}
\alpha_{trad} = E[R_{j,t}] - \beta_j' E[F_t]
\end{equation}
The bias $b_j$ (ML residual $\alpha$), can be transformed to:
\begin{equation}
b_j = \alpha_i^{trad} + \left( \beta_j' E[F_t] - E[F_t^{ML}] \right)
\end{equation}
Therefore, the residual $\alpha$ includes traditional linear $\alpha$ and the risk premium of the linear and non-linear model structures.\\

\clearpage
\section{Revised Python code of OOS $R^2$ and MSE calculation}\label{sec:app_python_r2_mse_ch2}
Please note that the input actual return form and predicted return form, the row labels are the stock permno, the column labels are the time steps.\\

\begin{lstlisting}
import pandas as pd
import numpy as np
import os
import glob

path_history = r"x:\xxx\xxxx\xx.csv"
r_history_df = pd.read_csv(path_history, index_col=0)

stock_benchmarks = r_history_df.iloc[:, :-120].mean(axis=1)

path_actual_oos = r"x:\xxx\xxxx\xx.csv"
actual_all = pd.read_csv(path_actual_oos, index_col=0)

actual_sub_df = actual_all.iloc[:, -120:]

actual_values = actual_sub_df.values 

stock_benchmarks = stock_benchmarks.reindex(actual_sub_df.index)

benchmark_vector = stock_benchmarks.values.reshape(-1, 1)

input_folder = r"x:\xxx\xxxx\xx"

if not os.path.exists(details_folder):
    os.makedirs(details_folder)

summary_list = []

for file_name in os.listdir(input_folder):
    if file_name.endswith('.csv'):
        file_path = os.path.join(input_folder, file_name)
        
        pred_df = pd.read_csv(file_path, index_col=0)
        pred_df = pred_df.reindex(actual_sub_df.index)
        
        pred_df_sub = pred_df
        pred_values = pred_df_sub.values
        
        errors = actual_values - pred_values
        numerator = np.sum(errors**2, axis=1)
        
        benchmark_errors = actual_values - benchmark_vector
        
        denominator = np.sum(benchmark_errors**2, axis=1)

        stock_mse = np.mean(errors**2, axis=1)
        
        with np.errstate(divide='ignore', invalid='ignore'):
            stock_r2 = 1 - (numerator / denominator)
            stock_r2[~np.isfinite(stock_r2)] = np.nan
        
        stock_details = pd.DataFrame({
            'permno': actual_sub_df.index,
            'Benchmark_Used': stock_benchmarks.values,
            'OOS_R2': stock_r2,
            'OOS_MSE': stock_mse
        })
        
        detail_file_name = f"Detail_{file_name}"
        stock_details.to_csv(os.path.join(details_folder, detail_file_name), index=False)
        
        avg_r2 = np.nanmean(stock_r2)
        avg_mse = np.nanmean(stock_mse)
        
        summary_list.append({
            'Model': file_name,
            'Avg_OOS_R2': avg_r2,
            'Avg_OOS_MSE': avg_mse
        })
\end{lstlisting}

\end{appendices}

\end{document}